
%
\def\unredoffs{}
\tolerance=1000\hfuzz=2pt
\catcode`\@=11 
\ifx\hyperdef\UNd@FiNeD\def\hyperdef#1#2#3#4{#4}\def\hyperref#1#2#3#4{#4}\def\href#1#2{#2}\fi
\magnification=1200\unredoffs\baselineskip=16pt plus 2pt minus 1pt
\def\Date#1{\vfill\leftline{#1}\tenpoint\supereject%
\footline={\hss\tenrm\hyperdef\hypernoname{page}\folio\folio\hss}}%

{\count255=\time\divide\count255 by 60 \xdef\hourmin{\number\count255}
 \multiply\count255 by-60\advance\count255 by\time
 \xdef\hourmin{\hourmin:\ifnum\count255<10 0\fi\the\count255}
}
\def\date{\number\day.\number\month.\number\year\ at \hourmin}


\def\nolabels{\def\wrlabeL##1{}\def\eqlabeL##1{}\def\reflabeL##1{}}
\def\writelabels{\def\wrlabeL##1{\leavevmode\vadjust{\rlap{\smash%
{\line{{\escapechar=` \hfill\rlap{\sevenrm\hskip.03in\string##1}}}}}}}%
\def\eqlabeL##1{{\escapechar-1\rlap{\sevenrm\hskip.05in\string##1}}}%
\def\reflabeL##1{\noexpand\llap{\noexpand\sevenrm\string\string\string##1}}}
\nolabels

\global\newcount\secno \global\secno=0
\global\newcount\meqno \global\meqno=1
\def\s@csym{}

\def\newsec#1\par{\global\advance\secno by1%
{\toks0{#1}\message{(\the\secno. \the\toks0)}}%
\global\subsecno=0\eqnres@t\let\s@csym\secsym\xdef\secn@m{\the\secno}\noindent
{\bf\hyperdef\hypernoname{section}{\the\secno}{\the\secno.} #1}%
\writetoca{{\string\hyperref{}{section}{\the\secno}{\bf \the\secno\quad}} {\bf #1}}%
\par\nobreak\medskip\nobreak\noindent\ignorespaces}
\def\eqnres@t{\xdef\secsym{\the\secno.}\global\meqno=1\bigbreak\bigskip}
\def\sequentialequations{\def\eqnres@t{\bigbreak}}\xdef\secsym{}

\global\newcount\subsecno \global\subsecno=0
\def\subsec#1\par{\global\advance\subsecno by1%
{\toks0{#1}\message{(\s@csym\the\subsecno. \the\toks0)}}%
\global\subsubsecno=0%
\ifnum\lastpenalty>9000\else\bigbreak\fi
\noindent{\it\hyperdef\hypernoname{subsection}{\secn@m.\the\subsecno}%
{\secn@m.\the\subsecno.} #1}\writetoca{\string\hskip1.45cm
{\string\hyperref{}{subsection}{\secn@m.\the\subsecno}{\secn@m.\the\subsecno.}}
{#1}}\par\nobreak\medskip\nobreak\noindent\ignorespaces}

\def\appendix#1#2{\global\meqno=1\global\subsecno=0\xdef\secsym{\hbox{#1.}}%
\bigbreak\bigskip\noindent{\bf Appendix \hyperdef\hypernoname{appendix}{#1}%
{#1.} #2}{\toks0{(#1. #2)}\message{\the\toks0}}%
\xdef\s@csym{#1.}\xdef\secn@m{#1}%
\writetoca{{\string\hyperref{}{appendix}{#1}{\bf {#1}\quad}} {\bf #2}}%
\par\nobreak\medskip\nobreak}

%
\def\checkm@de#1#2{\ifmmode{\def\f@rst##1{##1}\hyperdef\hypernoname{equation}%
{#1}{#2}}\else\hyperref{}{equation}{#1}{#2}\fi}
\def\eqnn#1{\DefWarn#1\xdef #1{(\noexpand\relax\noexpand\checkm@de%
{\s@csym\the\meqno}{\secsym\the\meqno})}%
\wrlabeL#1\writedef{#1\leftbracket#1}\global\advance\meqno by1}
\def\f@rst#1{\c@t#1a\em@ark}\def\c@t#1#2\em@ark{#1}
\def\eqna#1{\DefWarn#1\wrlabeL{#1$\{\}$}%
\xdef #1##1{(\noexpand\relax\noexpand\checkm@de%
{\s@csym\the\meqno\noexpand\f@rst{##1}1}{\hbox{$\secsym\the\meqno##1$}})}
\writedef{#1\numbersign1\leftbracket#1{\numbersign1}}\global\advance\meqno by1}
\def\eqn#1#2{\DefWarn#1%
\xdef #1{(\noexpand\hyperref{}{equation}{\s@csym\the\meqno}%
{\secsym\the\meqno})}$$#2\eqno(\hyperdef\hypernoname{equation}%
{\s@csym\the\meqno}{\secsym\the\meqno})\eqlabeL#1$$%
\writedef{#1\leftbracket#1}\global\advance\meqno by1}
\def\xeqn{\expandafter\xe@n}\def\xe@n(#1){#1}
\def\xeqna#1{\expandafter\xe@n#1}
\def\eqns#1{(\e@ns #1{\hbox{}})}
\def\e@ns#1{\ifx\UNd@FiNeD#1\message{eqnlabel \string#1 is undefined.}%
\xdef#1{(?.?)}\fi{\let\hyperref=\relax\xdef\next{#1}}%
\ifx\next\em@rk\def\next{}\else%
\ifx\next#1\xeqn#1\else\def\n@xt{#1}\ifx\n@xt\next#1\else\xeqna#1\fi
\fi\let\next=\e@ns\fi\next}

\def\DefWarn#1{\ifx\UNd@FiNeD#1\else
\immediate\write16{*** WARNING: the label \string#1 is already defined ***}\fi}
%
\newskip\footskip\footskip14pt plus 1pt minus 1pt 
\def\footnotefont{\ninepoint}\def\f@t#1{\footnotefont #1\@foot}
\def\f@@t{\baselineskip\footskip\bgroup\footnotefont\aftergroup\@foot\let\next}
\setbox\strutbox=\hbox{\vrule height9.5pt depth4.5pt width0pt}
\global\newcount\ftno \global\ftno=0
\def\foot{\global\advance\ftno by1\def\foot@rg{\hyperref{}{footnote}%
{\the\ftno}{\the\ftno}\xdef\foot@rg{\noexpand\hyperdef\noexpand\hypernoname%
{footnote}{\the\ftno}{\the\ftno}}}\footnote{$^{\foot@rg}$}}
%
%
%
\global\newcount\refno \global\refno=1
\newwrite\rfile
\def\ref{[\hyperref{}{reference}{\the\refno}{\the\refno}]\nref}
\def\nref#1{\DefWarn#1%
\xdef#1{[\noexpand\hyperref{}{reference}{\the\refno}{\the\refno}]}%
\writedef{#1\leftbracket#1}%
\ifnum\refno=1\immediate\openout\rfile=\jobname.refs\fi
\chardef\wfile=\rfile\immediate\write\rfile{\noexpand\item{[\noexpand\hyperdef%
\noexpand\hypernoname{reference}{\the\refno}{\the\refno}]\ }%
\reflabeL{#1\hskip.31in}\pctsign}\global\advance\refno by1\findarg}
\def\findarg#1#{\begingroup\obeylines\newlinechar=`\^^M\pass@rg}
{\obeylines\gdef\pass@rg#1{\writ@line\relax #1^^M\hbox{}^^M}%
\gdef\writ@line#1^^M{\expandafter\toks0\expandafter{\striprel@x #1}%
\edef\next{\the\toks0}\ifx\next\em@rk\let\next=\endgroup\else\ifx\next\empty%
\else\immediate\write\wfile{\the\toks0}\fi\let\next=\writ@line\fi\next\relax}}
\def\striprel@x#1{} \def\em@rk{\hbox{}}
\def\lref{\begingroup\obeylines\lr@f}
\def\lr@f#1#2{\DefWarn#1\gdef#1{\let#1=\UNd@FiNeD\ref#1{#2}}\endgroup\unskip}
\def\semi{;\hfil\break}
\def\addref#1{\immediate\write\rfile{\noexpand\item{}#1}} 
\def\listrefs{\vfill\supereject\immediate\closeout\rfile\writestoppt
\baselineskip=\footskip\centerline{{\bf References}}\bigskip{\parindent=20pt%
\frenchspacing\escapechar=` \input \jobname.refs\vfill\eject}\nonfrenchspacing}
\def\startrefs#1{\immediate\openout\rfile=\jobname.refs\refno=#1}
\def\xref{\expandafter\xr@f}\def\xr@f[#1]{#1}
\def\refs#1{\count255=1[\r@fs #1{\hbox{}}]}
\def\r@fs#1{\ifx\UNd@FiNeD#1\message{reflabel \string#1 is undefined.}%
\nref#1{need to supply reference \string#1.}\fi%
\vphantom{\hphantom{#1}}{\let\hyperref=\relax\xdef\next{#1}}%
\ifx\next\em@rk\def\next{}%
\else\ifx\next#1\ifodd\count255\relax\xref#1\count255=0\fi%
\else#1\count255=1\fi\let\next=\r@fs\fi\next}
%

%
\newwrite\ffile\global\newcount\figno \global\figno=1
\def\fig{fig.~\hyperref{}{figure}{\the\figno}{\the\figno}\nfig}
\def\nfig#1{\DefWarn#1%
\xdef#1{fig.~\noexpand\hyperref{}{figure}{\the\figno}{\the\figno}}%
\writedef{#1\leftbracket fig.\noexpand~\xfig#1}%
\ifnum\figno=1\immediate\openout\ffile=\jobname.figs\fi\chardef\wfile=\ffile%
{\let\hyperref=\relax
\immediate\write\ffile{\noexpand\medskip\noexpand\item{Fig.\ %
\noexpand\hyperdef\noexpand\hypernoname{figure}{\the\figno}{\the\figno}. }
\reflabeL{#1\hskip.55in}\pctsign}}\global\advance\figno by1\findarg}
\def\xfig{\expandafter\xf@g}\def\xf@g fig.\penalty\@M\ {}
\def\figs#1{figs.~\f@gs #1{\hbox{}}}
\def\f@gs#1{{\let\hyperref=\relax\xdef\next{#1}}\ifx\next\em@rk\def\next{}\else
\ifx\next#1\xfig #1\else#1\fi\let\next=\f@gs\fi\next}
%
\def\figin{\epsfcheck\figin}\def\figins{\epsfcheck\figins}
\def\epsfcheck{\ifx\epsfbox\UnDeFiNeD
\message{(NO epsf.tex, FIGURES WILL BE IGNORED)}
\gdef\figin##1{\vskip2in}\gdef\figins##1{\hskip.5in}
\else\message{(FIGURES WILL BE INCLUDED)}%
\gdef\figin##1{##1}\gdef\figins##1{##1}\fi}
\def\DefWarn#1{}
\def\figinsert{\goodbreak\topinsert}
\def\ifig#1#2#3{\DefWarn#1\xdef#1{fig.~\the\figno}
\writedef{#1\leftbracket fig.\noexpand~\the\figno}%
\figinsert\figin{\centerline{#3}}
\smallskip
\leftskip=20pt \rightskip=20pt
\baselineskip12pt\noindent
{{\bf Fig.~\the\figno}\ \ninepoint #2}
\medskip
\global\advance\figno by1\par\endinsert}
\newwrite\lfile
{\escapechar-1\xdef\pctsign{\string\%}\xdef\leftbracket{\string\{}
\xdef\rightbracket{\string\}}\xdef\numbersign{\string\#}}
\def\writedefs{\immediate\openout\lfile=label.defs \def\writedef##1{%
{\let\hyperref=\relax\let\hyperdef=\relax\let\hypernoname=\relax
 \immediate\write\lfile{\string\def\string##1\rightbracket}}}}%
\def\writestop{\def\writestoppt{\immediate\write\lfile{\string\pageno
 \the\pageno\string\startrefs\leftbracket\the\refno\rightbracket
 \string\def\string\secsym\leftbracket\secsym\rightbracket
 \string\secno\the\secno\string\meqno\the\meqno}\immediate\closeout\lfile}}
\def\writestoppt{}\def\writedef#1{}

\def\seclab#1{\DefWarn#1%
\xdef #1{\noexpand\hyperref{}{section}{\the\secno}{\the\secno}}%
\writedef{#1\leftbracket#1}\wrlabeL{#1=#1}}
\def\subseclab#1{\DefWarn#1%
\xdef #1{\noexpand\hyperref{}{subsection}{\the\secno.\the\subsecno}%
{\the\secno.\the\subsecno}}\writedef{#1\leftbracket#1}\wrlabeL{#1=#1}}
\def\applab#1{\DefWarn#1%
\xdef #1{\noexpand\hyperref{}{appendix}{\secn@m}{\secn@m}}%
\writedef{#1\leftbracket#1}\wrlabeL{#1=#1}}
\newwrite\tfile \def\writetoca#1{}
\def\leaderfill{\leaders\hbox to 1em{\hss.\hss}\hfill}
\def\writetoc{\immediate\openout\tfile=\jobname.toc
   \def\writetoca##1{{\edef\next{\write\tfile{\noindent ##1
   \string\leaderfill{
   \string\hyperref{}{page}{\noexpand\number\pageno}%
   {\noexpand\number\pageno}} \par}}\next}}
}
\newread\ch@ckfile
\def\listtoc{\immediate\closeout\tfile\immediate\openin\ch@ckfile=\jobname.toc
\ifeof\ch@ckfile\message{no file \jobname.toc, no table of contents this pass}%
\else\closein\ch@ckfile\centerline{\bf Contents}\nobreak\medskip%
{\baselineskip=16pt\footnotefont\parskip=0pt\catcode`\@=11\input\jobname.toc
\catcode`\@=12\bigbreak\bigskip}\fi}
\catcode`\@=12 
\def\tenpoint{\def\rm{\fam0\tenrm}
\textfont0=\tenrm \scriptfont0=\sevenrm \scriptscriptfont0=\fiverm
\textfont1=\teni  \scriptfont1=\seveni  \scriptscriptfont1=\fivei
\textfont2=\tensy \scriptfont2=\sevensy \scriptscriptfont2=\fivesy
\textfont\itfam=\tenit \def\it{\fam\itfam\tenit}\def\footnotefont{\ninepoint}%
\textfont\bffam=\tenbf \def\bf{\fam\bffam\tenbf}\def\sl{\fam\slfam\tensl}\rm}
\font\ninerm=cmr9 \font\sixrm=cmr6 \font\ninei=cmmi9 \font\sixi=cmmi6
\font\ninesy=cmsy9 \font\sixsy=cmsy6 \font\ninebf=cmbx9
\font\nineit=cmti9 \font\ninesl=cmsl9 \skewchar\ninei='177
\skewchar\sixi='177 \skewchar\ninesy='60 \skewchar\sixsy='60
\def\ninepoint{\def\rm{\fam0\ninerm}
\textfont0=\ninerm \scriptfont0=\sixrm \scriptscriptfont0=\fiverm
\textfont1=\ninei \scriptfont1=\sixi \scriptscriptfont1=\fivei
\textfont2=\ninesy \scriptfont2=\sixsy \scriptscriptfont2=\fivesy
\textfont\itfam=\ninei \def\it{\fam\itfam\nineit}\def\sl{\fam\slfam\ninesl}%
\textfont\bffam=\ninebf \def\bf{\fam\bffam\ninebf}\rm}
%
\hyphenation{anom-aly anom-alies coun-ter-term coun-ter-terms}

\global\newcount\subsubsecno \global\subsubsecno=0
\def\subsubsec#1\par{\global\advance\subsubsecno by1%
{\toks0{#1}\message{(\the\secno\the\subsecno\the\subsubsecno. \the\toks0)}}%
\ifnum\lastpenalty>9000\else\bigbreak\fi
\noindent{\it\hyperdef\hypernoname{subsubsection}{\the\secno.\the\subsecno\the\subsubsecno}%
{\the\secno.\the\subsecno.\the\subsubsecno.} #1}
\par\nobreak\medskip\nobreak\noindent\ignorespaces}

\def\DefWarn#1{}
\def\tikzcaption#1#2{\DefWarn#1\xdef#1{Fig.~\the\figno}
\writedef{#1\leftbracket Fig.\noexpand~\the\figno}%
{
\smallskip
\leftskip=20pt \rightskip=20pt \baselineskip12pt\noindent
{{\bf Fig.~\the\figno}\ \ninepoint #2}
\bigskip
\global\advance\figno by1 \par}}

\def\ntoalpha#1{%
\ifcase#1%
@%
\or A\or B\or C\or D\or E\or F\or G\or H\or I
\fi
}

\global\newcount\appno \global\appno=1
\def\applab#1{\xdef #1{\ntoalpha\appno}\writedef{#1\leftbracket#1}\wrlabeL{#1=#1}
\global\advance\appno by1}

\def\preprint#1 #2\par{\rightline{\vbox{\baselineskip12pt\hbox{#1}\hbox{#2}}}\vskip2cm}
%
\def\title#1\par{\centerline{\bf #1}\nopagenumbers\pageno=0}
\def\author#1\par{\bigskip\bigskip\centerline{#1}}

\newcount\addressno

\def\email#1#2{\unskip$^#1$\footnote{\null}{\kern-\parindent \llap{$^#1$\hskip1pt}email: #2}}

\def\startcenter{%
  \par
  \begingroup
  \leftskip=0pt plus 1fil
  \rightskip=\leftskip
  \parindent=0pt
  \parfillskip=0pt
}
\def\stopcenter{\endgroup}

\def\address{\bigskip%
  \ifnum\the\addressno=0\else\stopcenter\endgroup\fi
  \advance\addressno by 1%
  \begingroup
  \startcenter
  \it
  \obeylines
  \addressAux
}
\def\addressAux#1{#1}

\def\abstract{\stopcenter\endgroup\bigskip\bigskip\noindent}

\def\Dsl{\,\raise.15ex\hbox{/}\mkern-13.5mu D} 
\def\dsl{\raise.15ex\hbox{/}\kern-.57em\partial}
 
\def\boxeqn#1{\vcenter{\vbox{\hrule\hbox{\vrule\kern3pt\vbox{\kern3pt
	\hbox{${\displaystyle #1}$}\kern3pt}\kern3pt\vrule}\hrule}}}


\def\a{\alpha}
\def\b{{\beta}}
\def\g{{\gamma}}
\def\d{{\delta}}
\def\e{{\epsilon}}
\def\l{\lambda}
\def\k{{\kappa}}
\def\s{{\sigma}}
\def\t{{\theta}}

\def\lb{{\overline\lambda}}

\def\half{{1\over 2}}
\def\p{{\partial}}

\def\bar{\overline}
\def\({\left(}
\def\){\right)}
\def\cF{{\cal F}}
\def\cW{{\cal W}}
\def\cY{{\cal Y}}
\def\cA{{\cal A}}
\def\cV{{\cal V}}
\def\cJ{{\cal J}}

\font\tenshuffle=shuffle10 \font\sevenshuffle=shuffle7 \font\fiveshuffle=shuffle7 at 5pt
\def\shuffle{{%
\def\Dshuffle{\mathbin{\hbox{\tenshuffle\char'001}}}%
\def\Sshuffle{\mathbin{\hbox{\sevenshuffle\char'001}}}%
\def\SSshuffle{\mathbin{\hbox{\fiveshuffle\char'001}}}%
\mathchoice{\Dshuffle}{\Dshuffle}{\Sshuffle}{\SSshuffle}}}


\def\qed{\hbox{\hskip 3pt
\vbox{\hrule\hbox to 7pt{\vrule height 7pt\hfill\vrule}
\hrule}}\hskip3pt}

\overfullrule=0pt\relax

\frenchspacing

\newread\instream \openin\instream= label.defs
\ifeof\instream \message{No labels in advance yet. Wait till next pass.}
\else \closein\instream \input label.defs
\fi
\writedefs

\def\arXiv:#1].{\hepthStrip#1 \nil}
\def\hepthStrip#1 #2\nil{\href{http://arxiv.org/abs/#1}{arXiv:#1 #2\unskip}].}

\input epsf

\newdimen\pageremains\newdimen\pdepth
\newdimen\figwidth
\newdimen\figheight
\newcount\figlines
\newcount\flevel

\def\figflow#1#2#3{
\ifnum\flevel>0
\message{******Figure collision. Ignoring second figure.******}
\else
\figwidth=#1
\figheight=#2
\def\contents{#3}
\def\figure{\let\temp=\par \let\par=\plainpar
  \line{\overfullrule=0pt
   \ifdim \figwidth<0pt \hsize=-\figwidth \hss\else \hsize=\figwidth\fi
   \advance \hsize by -15pt
   \vbox to \figheight{\vfil\noindent\contents\vfill}
   \ifdim \figwidth>0pt \hss\fi
  } \vskip-\figheight \vskip-5pt
  \let\par=\temp%
}
\advance\figheight by \baselineskip
\divide\figheight by \baselineskip
\figlines=\figheight \multiply\figheight by \baselineskip
\begingroup\overfullrule=0pt
\tolerance=1000
\flevel=1
\let\plainpar=\par
\def\par{
  \ifnum\flevel=1
   \plainpar
   \pageremains=\pagegoal \advance\pageremains by -\pagetotal
   \ifdim\pageremains<\figheight \message{Moving figure...}
   \else
      \pdepth=\prevdepth
      \nointerlineskip
      \figure
      \hangindent \figwidth \hangafter -\figlines \hfuzz 5 pt
      \flevel=2
      \prevgraf=0
      \figheight=\baselineskip
   \fi
  \else
   \ifnum\flevel=2
    \ifdim\figheight<\parskip
       \advance\figlines -1 \advance\hangafter 1
       \advance\figheight\baselineskip
    \else
       \advance\figheight -\parskip
    \fi
    \hangcarrypar\relax
   \fi
  \fi
}
\par
\vskip-\pdepth
\fi
}
\def\endflow{\global\let\par=\plainpar\endgroup}
\def\hangcarrypar{
\edef\next{\hangafter=\the\hangafter\hangindent=\the\hangindent}
\plainpar\next
\edef\next{\prevgraf=\the\prevgraf}
\ifnum\prevgraf>0
   \ifnum\prevgraf>\figlines \endflow \flevel=0
   \else
     \message{FIGFLOW: line \the\prevgraf, of \the\figlines.}
     \leavevmode
     \next
   \fi
\fi
}

\def\Tijk,#1,#2,#3{T_{#1,#2,#3}}

\def\mod#1{|\mkern-1.7mu #1\mkern-3.5mu|}

\preprint AEI--2014--037 DAMTP--2014--46

\title Cohomology foundations of one-loop amplitudes in pure spinor superspace

\author Carlos R. Mafra\email{\dagger}{c.r.mafra@damtp.cam.ac.uk} and
	Oliver Schlotterer\email{\ddagger}{olivers@aei.mpg.de}

\address
$^\dagger$DAMTP, University of Cambridge
Wilberforce Road, Cambridge, CB3 0WA, UK

\address
$^\ddagger$Max--Planck--Institut f\"ur Gravitationsphysik, Albert--Einstein--Institut

Am M\"uhlenberg 1, 14476 Potsdam, Germany

\abstract

We describe a pure spinor BRST cohomology framework to compactly represent ten-dimensional one-loop amplitudes involving any number of massless
open- and closed-string states. The method of previous work
to construct scalar and vectorial BRST invariants in pure spinor superspace signals the appearance of the hexagon
gauge anomaly when applied to tensors.
We study the systematics of the underlying BRST anomaly by defining the notion of pseudo-cohomology.
This leads to a rich network of pseudo-invariant superfields of arbitrary tensor
rank whose behavior under traces and contractions with external momenta is determined from cohomology manipulations.
Separate papers will illustrate the virtue of the superfields in this work to represent
one-loop amplitudes of the superstring and of ten-dimensional super-Yang--Mills theory.

\Date {August 2014}

\newif\iffig
\figfalse


\lref\eombbs{
	C.R.~Mafra and O.~Schlotterer,
  	``Multiparticle SYM equations of motion and pure spinor BRST blocks,''
	JHEP {\bf 1407}, 153 (2014).
	[arXiv:1404.4986 [hep-th]].

}

\lref\theorems{
	N.~Berkovits,
	``New higher-derivative $R^4$ theorems,''
	Phys.\ Rev.\ Lett.\  {\bf 98}, 211601 (2007).
	[hep-th/0609006].
}

\lref\oneloopbb{
	C.R.~Mafra and O.~Schlotterer,
	``The Structure of n-Point One-Loop Open Superstring Amplitudes,''
	[arXiv:1203.6215 [hep-th]].
}
\lref\nptMethod{
	C.R.~Mafra, O.~Schlotterer, S.~Stieberger and D.~Tsimpis,
	``A recursive method for SYM n-point tree amplitudes,''
	Phys.\ Rev.\ D {\bf 83}, 126012 (2011).
	[arXiv:1012.3981 [hep-th]].
}

\lref\nptTree{
	C.~R.~Mafra, O.~Schlotterer and S.~Stieberger,
	``Complete N-Point Superstring Disk Amplitude I. Pure Spinor Computation,''
	Nucl.\ Phys.\ B {\bf 873}, 419 (2013).
	[arXiv:1106.2645 [hep-th]].
}

\lref\nptTreeII{
	C.R.~Mafra, O.~Schlotterer and S.~Stieberger,
  	``Complete N-Point Superstring Disk Amplitude II. Amplitude and Hypergeometric Function Structure,''
	Nucl.\ Phys.\ B {\bf 873}, 461 (2013).
	[arXiv:1106.2646 [hep-th]].
}
\lref\wittentwistor{
	E.Witten,
        ``Twistor-Like Transform In Ten-Dimensions''
        Nucl.Phys. B {\bf 266}, 245~(1986)
}
\lref\SiegelYI{
	W.~Siegel,
	``Superfields in Higher Dimensional Space-time,''
	Phys.\ Lett.\ B {\bf 80}, 220 (1979).
}
\lref\psf{
 	N.~Berkovits,
	``Super-Poincar\'e covariant quantization of the superstring,''
	JHEP {\bf 0004}, 018 (2000)
	[arXiv:hep-th/0001035].
}
\lref\multiloop{
        N.~Berkovits,
	``Multiloop amplitudes and vanishing theorems using the pure spinor formalism for the superstring,''
	JHEP {\bf 0409}, 047 (2004).
	[hep-th/0406055].
}

\lref\Figueroa{
	J.M.~Figueroa-O'Farrill,
	``N=2 structures in all string theories,''
	J.\ Math.\ Phys.\  {\bf 38}, 5559 (1997).
	[hep-th/9507145].
}

\lref\oneloopMichael{
	M.B.~Green, C.R.~Mafra and O.~Schlotterer,
  	``Multiparticle one-loop amplitudes and S-duality in closed superstring theory,''
	JHEP {\bf 1310}, 188 (2013).
	[arXiv:1307.3534].
}

\lref\BerendsME{
	F.~A.~Berends and W.T.~Giele,
  	``Recursive Calculations for Processes with n Gluons,''
	Nucl.\ Phys.\ B {\bf 306}, 759 (1988).
}

\lref\anomaly{
	N.~Berkovits and C.R.~Mafra,
	``Some Superstring Amplitude Computations with the Non-Minimal Pure Spinor Formalism,''
	JHEP {\bf 0611}, 079 (2006).
	[hep-th/0607187].
}
\lref\BCJ{
	Z.~Bern, J.J.M.~Carrasco and H.~Johansson,
	``New Relations for Gauge-Theory Amplitudes,''
	Phys.\ Rev.\ D {\bf 78}, 085011 (2008).
	[arXiv:0805.3993 [hep-ph]].
}
\lref\oldMomKer{
	Z.~Bern, L.~J.~Dixon, M.~Perelstein and J.~S.~Rozowsky,
	``Multileg one loop gravity amplitudes from gauge theory,''
	Nucl.\ Phys.\ B {\bf 546}, 423 (1999).
	[hep-th/9811140].
}
\lref\MomKer{
	N.~E.~J.~Bjerrum-Bohr, P.~H.~Damgaard, T.~Sondergaard and P.~Vanhove,
	``The Momentum Kernel of Gauge and Gravity Theories,''
	JHEP {\bf 1101}, 001 (2011).
	[arXiv:1010.3933 [hep-th]].
}
\lref\Polylogs{
	J.~Broedel, O.~Schlotterer and S.~Stieberger,
	``Polylogarithms, Multiple Zeta Values and Superstring Amplitudes,''
	Fortsch.\ Phys.\  {\bf 61}, 812 (2013).
	[arXiv:1304.7267 [hep-th]].
}
\lref\BroedelAZA{
	J.~Broedel, O.~Schlotterer, S.~Stieberger and T.~Terasoma,
	``All order alpha'-expansion of superstring trees from the Drinfeld associator,''
	Phys.\ Rev.\ D {\bf 89}, 066014 (2014).
	[arXiv:1304.7304 [hep-th]].
}
\lref\KKref{
	R.~Kleiss and H.~Kuijf,
	``Multi - Gluon Cross-sections and Five Jet Production at Hadron Colliders,''
	Nucl.\ Phys.\ B {\bf 312}, 616 (1989).
\semi
	V.~Del Duca, L.J.~Dixon and F.~Maltoni,
	``New color decompositions for gauge amplitudes at tree and loop level,''
	Nucl.\ Phys.\ B {\bf 571}, 51 (2000).
	[hep-ph/9910563].
}
\lref\PSS{
	C.R.~Mafra,
	``PSS: A FORM Program to Evaluate Pure Spinor Superspace Expressions,''
	[arXiv:1007.4999 [hep-th]].
}
\lref\FORM{
	J.A.M.~Vermaseren,
	``New features of FORM,''
	arXiv:math-ph/0010025.
\semi
	M.~Tentyukov and J.A.M.~Vermaseren,
	``The multithreaded version of FORM,''
	arXiv:hep-ph/0702279.
}
\lref\GSanomaly{
	M.~B.~Green and J.~H.~Schwarz,
	``Anomaly Cancellation in Supersymmetric D=10 Gauge Theory and Superstring Theory,''
	Phys.\ Lett.\ B {\bf 149}, 117 (1984).
\semi
	M.B.~Green and J.H.~Schwarz,
	``The Hexagon Gauge Anomaly in Type I Superstring Theory,''
	Nucl.\ Phys.\ B {\bf 255}, 93 (1985).
}
\lref\LiEprogram{
M.A.A. van Leeuwen, A.M. Cohen and B. Lisser,
``LiE, A Package for Lie Group Computations'', Computer Algebra Nederland, Amsterdam, ISBN 90-74116-02-7, 1992
}
\lref\wipH{
C.R.~Mafra and O.~Schlotterer, work in progress.
}
\lref\wipG{
M.B.~Green, C.R.~Mafra, O.~Schlotterer, work in progress.
}

\lref\twolooptwo{
	N.~Berkovits and C.R.~Mafra,
	  ``Equivalence of two-loop superstring amplitudes in the pure spinor and RNS formalisms,''
	Phys.\ Rev.\ Lett.\  {\bf 96}, 011602 (2006).
	[hep-th/0509234].
}

\lref\anguelova{
	L.~Anguelova, P.A.~Grassi and P.~Vanhove,
	``Covariant one-loop amplitudes in D=11,''
	Nucl.\ Phys.\ B {\bf 702}, 269 (2004).
	[hep-th/0408171].
}

\lref\RichardsJG{
	D.~M.~Richards,
	``The One-Loop Five-Graviton Amplitude and the Effective Action,''
	JHEP {\bf 0810}, 042 (2008).
	[arXiv:0807.2421 [hep-th]].
}
\lref\NMPS{
	N.~Berkovits,
	  ``Pure spinor formalism as an N=2 topological string,''
	JHEP {\bf 0510}, 089 (2005).
	[hep-th/0509120].
}
\lref\twoloop{
	N.~Berkovits,
	  ``Super-Poincar\'e covariant two-loop superstring amplitudes,''
	JHEP {\bf 0601}, 005 (2006).
	[hep-th/0503197].
}
\lref\explaining{
	N.~Berkovits,
	``Explaining Pure Spinor Superspace,''
	[hep-th/0612021].
}
\lref\mafraids{
	C.R.~Mafra,
  	``Pure Spinor Superspace Identities for Massless Four-point Kinematic Factors,''
	JHEP {\bf 0804}, 093 (2008).
	[arXiv:0801.0580 [hep-th]].
}
\lref\towards{
	C.R.~Mafra,
  	``Towards Field Theory Amplitudes From the Cohomology of Pure Spinor Superspace,''
	JHEP {\bf 1011}, 096 (2010).
	[arXiv:1007.3639 [hep-th]].
}
\lref\RNS{
    P.~Ramond,
    ``Dual Theory for Free Fermions,''
    Phys.\ Rev.\  D {\bf 3}, 2415 (1971)
    \semi
    A.~Neveu and J.~H.~Schwarz,
    ``Factorizable dual model of pions,''
    Nucl.\ Phys.\  B {\bf 31} (1971) 86
    \semi
    A.~Neveu and J.~H.~Schwarz,
    ``Quark Model of Dual Pions,''
    Phys.\ Rev.\  D {\bf 4}, 1109 (1971).
}
\lref\GS{
    M.~B.~Green and J.~H.~Schwarz,
    ``Covariant Description Of Superstrings,''
    Phys.\ Lett.\  B {\bf 136}, 367 (1984)
    \semi
     M.~B.~Green and J.~H.~Schwarz,
    ``Supersymmetrical String Theories,''
    Phys.\ Lett.\  B {\bf 109}, 444 (1982).
}

\lref\bigHowe{
  P.S.~Howe,
  ``Pure Spinors Lines In Superspace And Ten-Dimensional Supersymmetric
  Theories,''
  Phys.\ Lett.\  B {\bf 258}, 141 (1991)
  [Addendum-ibid.\  B {\bf 259}, 511 (1991)].
\semi
  P.S.~Howe,
  ``Pure Spinors, Function Superspaces And Supergravity Theories In
  Ten-Dimensions And Eleven-Dimensions,''
  Phys.\ Lett.\  B {\bf 273}, 90 (1991).
}
\lref\NilssonCM{
	B.E.W.~Nilsson,
  	``Pure Spinors as Auxiliary Fields in the Ten-dimensional Supersymmetric {Yang-Mills} Theory,''
	Class.\ Quant.\ Grav.\  {\bf 3}, L41 (1986).
}

\lref\WWW{
	C.R.~Mafra, O.~Schlotterer,
http://www.damtp.cam.ac.uk/user/crm66/SYM/pss.html
}

\lref\WittenBghost{
	E.~Witten,
  	``More On Superstring Perturbation Theory,''
	[arXiv:1304.2832 [hep-th]].
}

\lref\thetaSYM{
  	J.P.~Harnad and S.~Shnider,
	``Constraints And Field Equations For Ten-Dimensional Superyang-Mills
  	Theory,''
  	Commun.\ Math.\ Phys.\  {\bf 106}, 183 (1986)
\semi
	P.A.~Grassi and L.~Tamassia,
        ``Vertex operators for closed superstrings,''
        JHEP {\bf 0407}, 071 (2004)
        [arXiv:hep-th/0405072].
\semi
	G.~Policastro and D.~Tsimpis,
	``$R^4$, purified,''
	Class.\ Quant.\ Grav.\  {\bf 23}, 4753 (2006)
	[arXiv:hep-th/0603165].
}
\lref\motivic{
	O.~Schlotterer and S.~Stieberger,
  	``Motivic Multiple Zeta Values and Superstring Amplitudes,''
	J.\ Phys.\ A {\bf 46}, 475401 (2013).
	[arXiv:1205.1516 [hep-th]].
}
\lref\associator{
	J.M.~Drummond and E.~Ragoucy,
	``Superstring amplitudes and the associator,''
	JHEP {\bf 1308}, 135 (2013).
	[arXiv:1301.0794 [hep-th]].
}
\lref\deligne{
	S.~Stieberger,
  	``Closed superstring amplitudes, single-valued multiple zeta values and the Deligne associator,''
	J.\ Phys.\ A {\bf 47}, 155401 (2014).
	[arXiv:1310.3259 [hep-th]].
}

\lref\GreenFT{
	M.B.~Green, J.H.~Schwarz and L.~Brink,
	``N=4 Yang-Mills and N=8 Supergravity as Limits of String Theories,''
	Nucl.\ Phys.\ B {\bf 198}, 474 (1982).
}

\listtoc
\writetoc
\filbreak

\newsec Introduction

Pure spinors are known to facilitate the superspace description of ten-dimensional super-Yang--Mills (SYM) \refs{\NilssonCM,
\bigHowe} which descends from the pure spinor superstring \psf.
As will be explained below, ten-dimensional pure
spinor superspace allows to take advantage of BRST symmetry to provide valuable guidance for the construction of
scattering amplitudes in both string- and field-theory.

\subsec Amplitudes as expressions in pure spinor superspace

The prescription to compute multiloop superstring amplitudes in the pure spinor formalism \refs{\psf,\multiloop,\NMPS}
is considerably simpler than in the Ramond--Neveu--Schwarz (RNS) \RNS\ and Green--Schwarz (GS) \GS\ formulations of superstring theory.
Unlike in the RNS, spacetime supersymmetry is manifest and there is no need to sum over spin structures since there
are no worldsheet spinors. And in contrast to the GS, the amplitudes are computed in a manifestly super-Poincar\'e
covariant manner. These two features combined allow to bypass the technical challenges associated with amplitude
computations in the RNS and GS. However, there is another feature in the pure spinor setup which is not as prominently
stressed but is of equal importance: The result of amplitude computations belongs to the BRST cohomology of pure
spinor superspace expressions at ghost number three.

Pure spinor superspace\foot{The superspace defined here is the {\it minimal} pure spinor superspace associated with
the original formulation in \psf. The {\it non-minimal} superspace appropriate in the context of the non-minimal
formalism of \NMPS\ also contains $\lb_\a$ variables and is not the subject of the present paper.}  is defined in terms of the standard ten-dimensional superspace variables $(x^m, \t^\alpha)$ and
the pure spinor $\l^\a$ (of ghost number one) satisfying $\l^\a \g^m_{\a\b}\l^\b = 0$ \explaining.
As will be
explained below, it turns out that the kinematic factors\foot{Kinematic factors are understood as the
polarization-dependent parts of amplitudes accompanying a basis of worldsheet integrals.} of {\it multiloop} amplitudes
can be written as pure spinor superspace expressions of the form
\eqn\PSSex{
K = \l^\a\l^\b\l^\g f_{\a\b\g}(x,\t)\,,
}
where $f_{\a\b\g}(x,\t)$ represents a function of ten-dimensional superspace and includes the dependence on polarizations and momenta.
This novel type of superspace was shown in \psf\ to encode the results of {\it tree-level} string amplitudes and proven
to be supersymmetric and gauge invariant when $K$ is in the
cohomology of the pure spinor BRST charge.
Furthermore, in order to extract the precise contractions of polarizations and momenta from a superspace expression $K$,
one computes its pure spinor bracket $\langle K \rangle$ defined by $\langle
(\l\g^m\t)(\l\g^n\t)(\l\g^p\t)(\t\g_{mnp}\t)\rangle = 1$ \psf. Since component expansions are straightforward to
evaluate and can be automated \refs{\PSS,\FORM}, the real challenge in computing string scattering amplitudes consists of obtaining their
corresponding superspace expressions in the BRST cohomology, which motivates the studies presented in this paper.

\subsec Multiloop amplitude prescription and pure spinor superspace

The  prescription to compute multiloop amplitudes in the pure spinor formalism was presented in
\refs{\multiloop,\NMPS} and integrates out {\it all} eleven pure spinor components $\l^\a$ and {\it all} sixteen $\t^\a$ variables.
Performing those integrations leads to awkward expressions which are
hard to manipulate. As observed at one-loop \anguelova\ and emphasized in the two-loop computations of \twolooptwo,
one can equivalently rewrite
those complicated-looking expressions by reinstating three pure spinors
$\l^\a$ in such a way as to obtain the same type of {\it tree-level} pure spinor superspace expressions \PSSex\ discussed
above. To illustrate the above point, the two-loop kinematic factor of \twoloop\ after performing the path integral over
all variables is written as
\eqn\complicated{
(T^{-1})^{\alpha\beta\gamma}_{\rho_1{\ldots}\rho_{11}}
\epsilon^{\rho_1{\ldots} \rho_{16}}
{\p\over{\p\theta^{\rho_{12}}}}{\ldots}{\p\over{\partial
\theta^{\rho_{16}}}}
(\gamma^{mnpqr})_{\alpha\beta}\gamma^{s}_{\gamma\delta}
F^1_{mn}(\theta)F^2_{pq}(\theta)F^3_{rs}(\theta)
W^{4\delta}(\theta) \ .
}
The superfields $F_{mn}(\theta)$ and $W^{\delta}(\theta)$ represent the gauge multiplet of ten-dimensional SYM,
and the tensor $(T^{-1})^{\a\b\g}_{\rho_1 ...\rho_{11}}$ is proportional to a complicated combination of gamma
matrices, $\e_{\rho_1 ...\rho_{16}}(\g^m)^{\k\rho_{12}}
(\g^n)^{\s\rho_{13}}
(\g^p)^{\tau\rho_{14}}
(\g_{mnp})^{\rho_{15}\rho_{16}} (\d^{(\a}_{\k}\d^\b_\s\d^{\g)}_{\tau}
-{1\over{40}}\g_q^{(\a\b}\d^{\g)}_\k \g^q_{\s\tau})$.
However, the kinematic factor \complicated\ can be equivalently written as the tree-level pure spinor superspace expression
$\langle K\rangle$ after reinstating three pure spinors, where
\eqn\simpler{
K = (\l\g^{mnpqr}\l)(\l\g^s W^4)F^1_{mn} F^2_{pq} F^3_{rs}\,.
}
In writing the kinematic factor \complicated\ as \simpler, its BRST invariance becomes easier to prove;
using standard manipulations of gamma matrices, the pure spinor
constraint and SYM equations of motion for $D_{\alpha} W^{\delta}$ and $D_{\alpha} F^{mn}$, it follows that
\eqn\twocoho{
Q\big[(\l\g^{mnpqr}\l)(\l\g^s W^4)F^1_{mn} F^2_{pq} F^3_{rs}\bigr] = 0\,.
}
Furthermore, one can also show that the kinematic factor \simpler\ is in the cohomology of the BRST charge\foot{A
superspace proof that \simpler\ is not BRST exact requires a combination of the identities in \mafraids\ and section
\secnine\ of this paper.}. The compact nature of pure spinor superspace expressions as exemplified by \simpler\
compared to \complicated\ and the observation \twocoho\ constitute the central pillars in the study of multiloop
string scattering amplitudes as objects in the BRST cohomology of {\it tree-level} pure spinor superspace.

\subsec BRST cohomology considerations as a method to simplify computations

Following observations based on the BRST structure of explicit lower-point results
it was suggested in \towards\ that the field-theory amplitudes at
tree-level could be uniquely obtained as pure spinor superspace expressions in the BRST cohomology. Indeed, the
color-ordered $N$-point tree amplitude of SYM can be compactly written as \nptMethod
\eqn\nptsym{
A(1,2, \ldots,N) = \langle V_1 E_{23 \ldots N}\rangle,
}
where $E_{23 \ldots N}$ is a superfield in the BRST cohomology. The pursuit of the general expression of the SYM tree amplitude as the solution of a cohomology
problem in pure spinor superspace led to the discovery of interesting mathematical objects such as the BRST blocks and supersymmetric Berends--Giele
currents reviewed in section \sectwo. These BRST-covariant objects also played an essential role in the derivation of the
general $N$-point open superstring tree-level amplitude in \nptTree. And as a byproduct of the BRST-covariant
organization of the string tree-level amplitudes, the worldsheet integrals conspire to a particularly symmetric form which was later
exploited to find interesting patterns in their $\a'$ expansion \refs{\nptTreeII\motivic\associator\Polylogs
\BroedelAZA{--}\deligne}.

\subsubsec Challenges at one-loop

Studying the one-loop open superstring amplitudes as a BRST cohomology problem was firstly put forward in \oneloopbb.
Using the multiloop prescription of \multiloop\ as a guide to obtain the patterns of zero-mode saturation, the
kinematic factors could be expressed in pure spinor superspace. Furthermore, integration by parts identities among the
worldsheet integrals built up BRST-closed linear combinations of those kinematic factors, denoted $C_{i|A,B,C}$ and
reviewed in section \twofour.

This cohomology setup led to a general and manifestly BRST-invariant expression for the $N$-point amplitude. For example, the six-point amplitude of open superstrings was
found to be a worldsheet integral over\foot{The objects $X_{ij}$ are related to the one-loop worldsheet
Green function $G(z)$, and the Koba--Nielsen factor $\prod_{i<j} e^{\alpha' (k_i \cdot k_j) G(z_i-z_j)}$ is suppressed \oneloopbb.}
\eqn\sixpoint{
X_{23}  X_{34}\langle C_{1|234,5,6}\rangle + X_{23} X_{45} \langle C_{1|23,45,6}\rangle + {\rm permutations} \,.
}
However, there is one subtlety in the one-loop BRST cohomology program outlined above, the bare one-loop amplitudes
are in general anomalous and therefore not BRST invariant. The cancelation of the anomaly as described in \GSanomaly\
involves a sum over amplitudes with different worldsheet topologies, but the composing amplitudes are still anomalous
when the number of external particles is six or higher. The BRST-invariant expression \sixpoint\ could not be the
whole story since it is non-anomalous\foot{We thank Michael Green for insisting on a clarification of this point.}.

It is clear that in order to study the missing pieces of the one-loop amplitudes associated to the anomaly in a
BRST cohomology setup one needs to relax the condition of BRST invariance. So in this work, among other things, we introduce the
notion of a {\it pseudo} BRST cohomology which meets this criterium. The essential idea behind the pseudo BRST
cohomology goes back to the pure spinor analysis of the gauge anomaly in \anomaly. It was shown that the gauge
variation of the six-point amplitude w.r.t particle one is proportional to the pure spinor superspace expression,
\eqn\anom{
\langle (\l\g^m W^2)(\l\g^n W^3)(\l\g^p W^4)(W^5\g_{mnp}W^6)\rangle,
}
whose component expansion correctly reproduces the known form of the anomaly, $\epsilon_{10}F^5$. As discussed
in section \secthree, one can recursively construct objects whose BRST variation is proportional to the anomalous
superfield \anom. It will be shown in a subsequent work that these pseudo BRST invariants correctly capture
the anomalous parts of the one-loop amplitudes which were not considered in \oneloopbb.

So as the main focus of this work, we will study the (pseudo-)cohomology properties of various superfields expected to
appear in one-loop amplitudes of open- and closed-strings. We introduce a grid of superspace kinematic factors which
naturally describe the BRST cohomology properties of one-loop amplitudes. The axes of this grid are set by the number
of free vector indices $m,n,p,\ldots$ and the number of multiparticle slots $A,B,\ldots,G$ which are interpreted as
representing external tree-level subdiagrams. This leads to the arrangement in \figoverview, and we will derive
recursion relations whose flow is indicated by the diagonal arrows. The tensorial superfields therein play an
important role in two different contexts:
\bigskip

\ifig\figoverview{Overview of (pseudo-)invariants. The arrows indicate whenever superfields of different type enter
the recursion for the pseudoinvariants on their right.}
{\epsfxsize=0.80\hsize\epsfbox{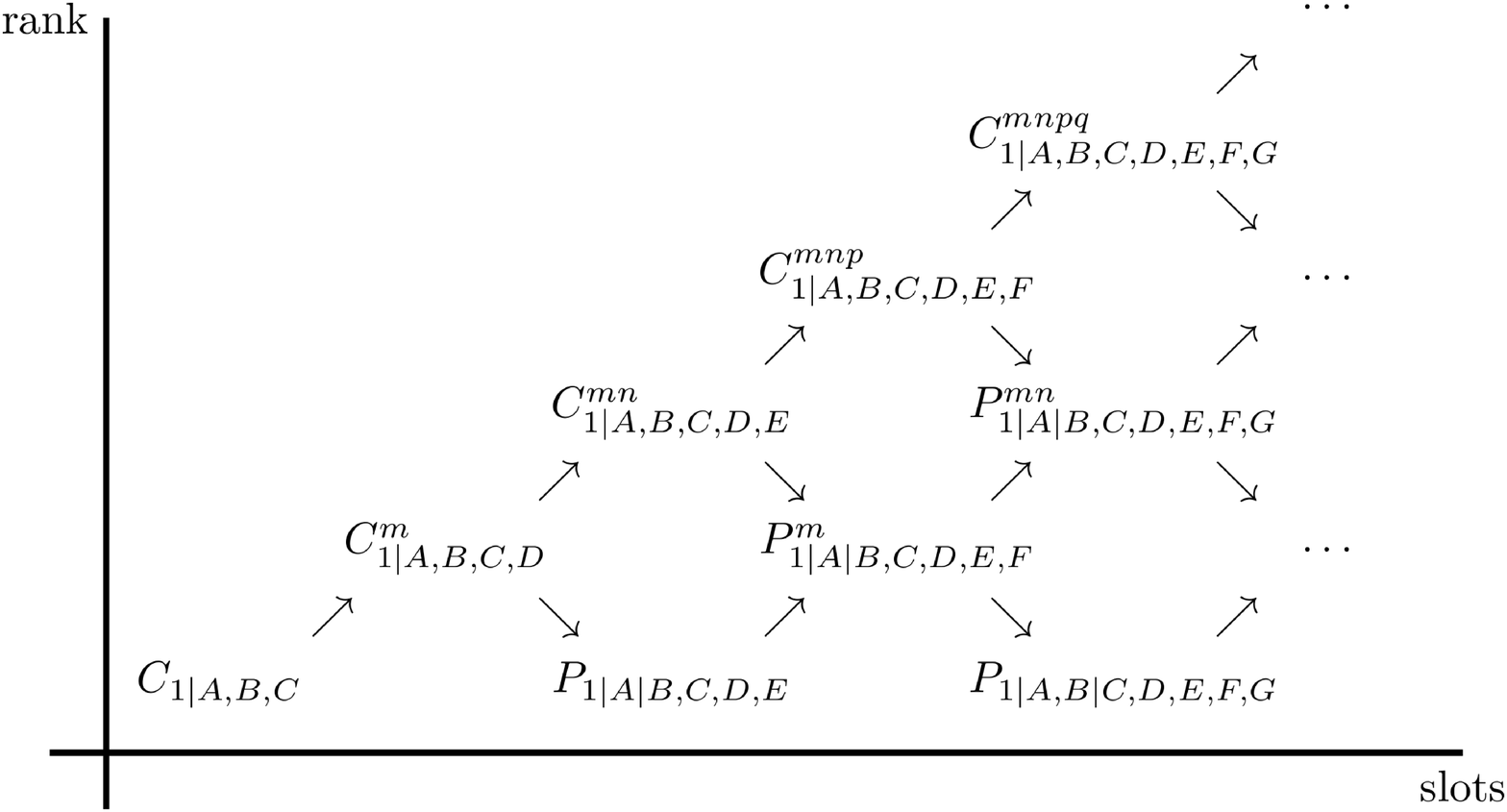}}


\item{(i)} Closed string amplitudes involving five and more external legs allow for vector contractions between left-
and right-moving degrees of freedom, see e.g. \refs{\RichardsJG, \oneloopMichael}. They originate from the zero modes of the worldsheet fields $\partial x^m$ and $\bar
\partial x^m$. In a manifestly BRST-invariant representation of the five- and six-point torus amplitude, the left-right
contractions enter in the form $C^m_{1|2,3,4,5}\tilde C^m_{1|2,3,4,5}$ and $C^{mn}_{1|2,3,4,5,6}\tilde
C^{mn}_{1|2,3,4,5,6}$, details will be elaborated in \wipG. Accordingly, scattering of
$N$ closed strings requires tensors of rank $r \leq N-4$.

\item{(ii)} The field theory limit of open and closed string amplitudes reproduces $n$-gon Feynman integrals \GreenFT\ where the loop momentum
$\ell^m$ may contract kinematic factors. In a manifestly BRST invariant form of the five-point amplitude, this loop momentum dependence enters
in the form $\ell_m C^m_{1|2,3,4,5}$. At six-points, the significance of the tensor hexagon $\ell_m \ell_n C^{mn}_{1|2,3,4,5,6}$
for the gauge anomaly of SYM will be clarified in \wipG. More generally, the systematic association of tensorial
Feynman integrals with the superfields in \figoverview\ is discussed in \wipH.

\medskip
\noindent The present work is devoted to the cohomology foundations of one-loop amplitudes in string- and
field-theory. The key definitions and results are formulated in generality to describe any number of external legs.
Applications to six-point string amplitudes and to field-theory amplitudes at multiplicity $\leq 7$ are
given in upcoming work \refs{\wipG,\wipH}, and the generalization to arbitrary multiplicity is left for the future.

\subsec The anatomy of one-loop amplitudes

One can embed the pseudoinvariants listed in \figoverview\ into a broader context. BRST cohomology methods are of crucial importance
 to decompose the computation of scattering amplitudes into smaller and more manageable problems. As we will see in various
 places of this work, pseudo-invariance of the superfields $C^{m\ldots}_{1|\ldots}$, $P^{m\ldots}_{1|\ldots}$ in \figoverview\
 requires BRST-covariant substructures, which in turn
furnish systematic arrangements of smaller constituents.
The different hierarchy levels of this decomposition are made more specific in fig.2.
The figure applies universally to one-loop scattering amplitudes involving SYM or supergravity states in maximally
supersymmetric string- and field-theory.
\bigskip

\advance\figno by 1

\figflow{-2.8 truein}{4.6 truein}{{\epsfxsize=1.00\hsize\epsfbox{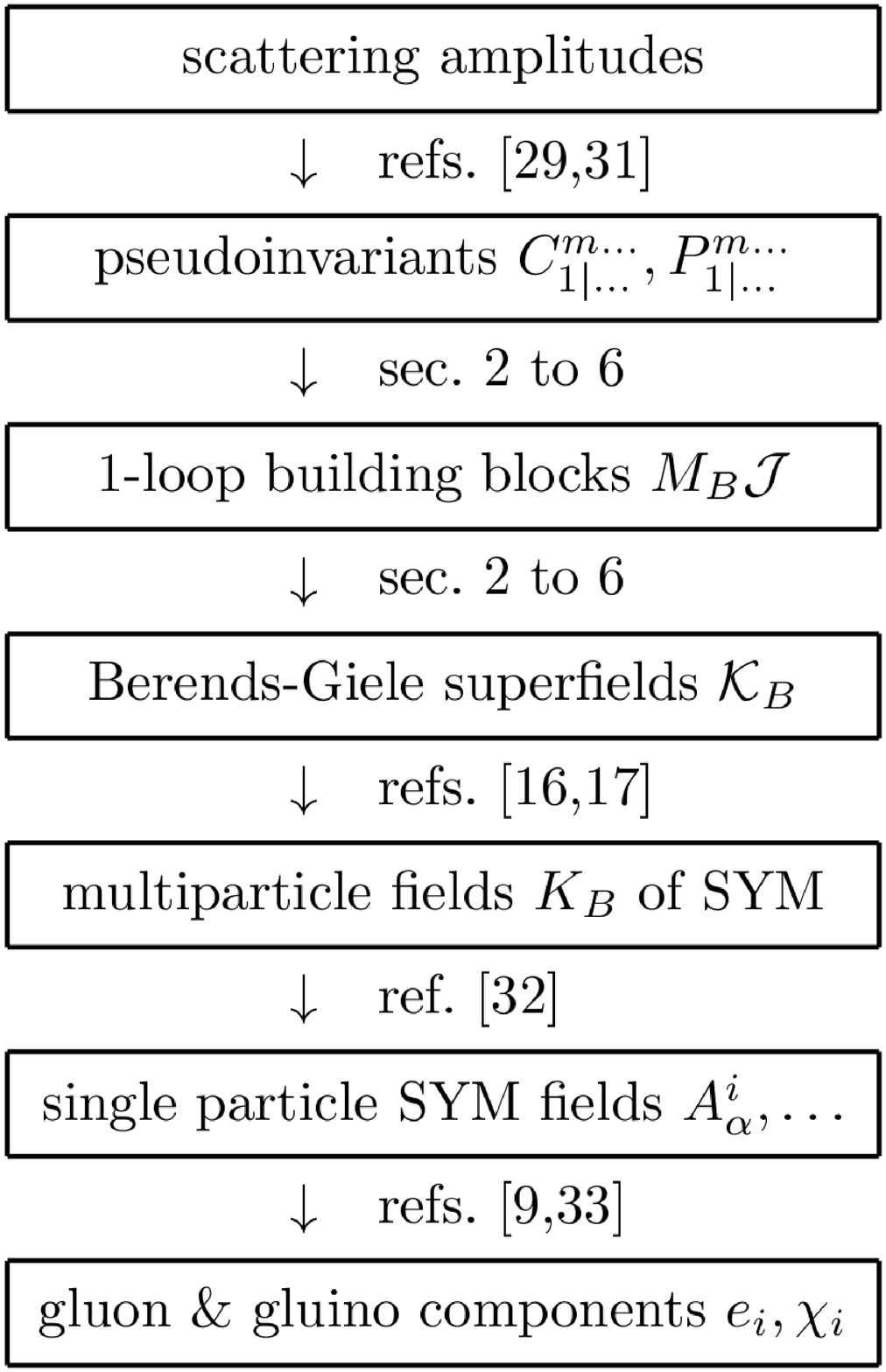}}\vfill\noindent {\bf Fig.~2}.\baselineskip=10pt \ninerm 
The seven hierarchy levels describing the anatomy of one-loop amplitudes in a BRST cohomology setup.
For each step, details can be found in the references alongside the arrows.}
These classes of one-loop amplitudes are claimed to have a beautiful representation in terms of pseudoinvariants
$C^{m\ldots}_{1|\ldots}$ and $ P^{m\ldots}_{1|\ldots}$ (or their holomorphic squares in case of supergravity and
closed-string amplitudes). Their composition rules in terms of integrals
over a loop momentum or over worldsheet moduli are the subject of upcoming work \refs{\wipG,\wipH}. 
As detailed in the first six sections, pseudoinvariants are built from
Berends--Giele currents $M_B$ (whose trilinears exhaust tree amplitudes of SYM \nptMethod\ and the open superstring \nptTree)
and ghost-number-two superfields ${\cal J}$ which are specific to the one-loop order. These ${\cal J}$ superfields in turn encompass
various numbers of further Berends-Giele superfields ${\cal K}_B$ with kinematic poles
in external momenta \eombbs. Both $M_B$ and ${\cal K}_B$ represent external tree subamplitudes which
can be expanded in terms of (products of) external propagators. Their numerators are multiparticle superfields
of SYM $K_B \in \{ A_\alpha^B,A_B^m,W_B^{\alpha}, F^{mn}_{B} \}$ which have been recursively constructed in \eombbs.
They encompass the degrees of freedom of several standard superfields $A_\alpha^i$, $A_i^m$, $W_i^{\alpha}$, $F^{mn}_{i}$
describing a single particle~$i$. Finally, the components of the supersymmetry multiplet -- a gluon with
polarization vector $e_i$ and a gaugino with spinor wavefunction $\chi_i$ -- furnish the lowest hierarchy level.
They are incorporated into the expansion of the superfields in terms of the Grassmann coordinate $\theta$
of pure spinor superspace \thetaSYM, e.g.
\eqn\extheta{
A_{\a}(x,\t) = \Big({1\over 2}e_m(\g^m\t)_\a -{1\over 3}(\chi \g_m\t)(\g^m\t)_\a
-{1\over 16}k_{m} e_{n}(\g_p\t)_\a (\t\g^{mnp}\t) + \cdots \Big) e^{ik\cdot x}\,.
}

As the number of external legs increases, every intermediate structure in fig.~2 reduces the complexity of
amplitudes by more and more orders of magnitude. And it should be stressed that the four lower hierarchy levels in
fig.~2 -- from Berends--Giele superfields $M_B,{\cal K}_B$ to the components $e_i,\chi_i$ -- are expected to play a
universal role at any loop order.

\subsec Outline

The main body of this work begins with a review of multiparticle SYM superfields \eombbs\ in section \sectwo. This
sets the stage to define the notion of {\it anomalous} superfields and BRST {\it pseudo-cohomology} in section \secthree. The
introductory examples are then generalized to arbitrary tensor rank in section \secfour. The resulting tensor traces
are shown to involve several constituents (indicated by $P$ in \figoverview) which are separately BRST
pseudoinvariant, see sections \secfive\ and \secsix. This completes the construction of the pseudo cohomology at ghost
number three which is visualized in \figoverview.

In section \secseven, we point out a close parallel between the superfields in anomalous BRST variations and the
previously-constructed pseudoinvariants. The more abstract viewpoint on this connection is opened up in section
\seceight\ and rewarded by manifold relations between superfields at different rank, see sections \secnine\ and
\secten. The same approach leads to the proof in section \seceleven\ that -- up to anomaly subtleties -- the span of the pseudoinvariants in
\figoverview\ is independent on the choice of reference leg $1$ which descends from the choice of unintegrated vertex
operator $V_1$ in the one-loop string amplitude prescription \multiloop.

Some appendices supplement the discussion by examples or serve to outsource technical aspects from the main
body. For example, appendix~\appA\ displays the expansions of superfields in \figoverview\ at higher multiplicity, and
appendix~\appB\ provides the prerequisites to extract anomalous gauge variations of pseudoinvariants from their BRST
transformations given in section~\secseven.

\newsec Review and conventions

\seclab\sectwo
This section provides a brief review of multiparticle SYM superfields introduced in \eombbs\ as well as their simplest
applications to one-loop kinematic factors. It also introduces notation and conventions used in the
rest of this work.

\subsec Diagrammatic introduction of BRST blocks

\subseclab\twoone
Linearized super-Yang--Mills (SYM) theory in ten dimensions can be described using the
superfields\foot{It is customary to use a calligraphic letter for the superfield field-strength. However in this
paper calligraphic letters will denote the Berends--Giele currents associated to the superfields, see section 4.}
$A^i_\a(x,\t)$, $A^i_m(x,\t)$, $W_i^\a(x,\t)$ and $F^i_{mn}(x,\t)$ encoding the on-shell degrees of
freedom of one external particle $i$.

They satisfy equations of motion \refs{\wittentwistor,\SiegelYI}
\eqn\RankOneEOM{
\eqalign{
2 D_{(\a} A^i_{\b)} & = \g^m_{\a\b} A^i_m\cr
D_\a F^i_{mn} & = 2k^i_{[m} (\g_{n]} W_i)_\a
}\qquad\eqalign{
D_\a A^i_m &= (\g_m W_i)_\a + k_m A^i_\a  \cr
D_\a W_i^{\b} &= {1\over 4}(\g^{mn})^{\phantom{m}\b}_\a F^i_{mn}
}}
with light-like momentum $k_i$ and gauge transformations $\d_i A^i_\a = D_\a \omega_i$ as well as $\d_i A^i_m = k^i_m
\omega_i$ for some scalar superfield $\omega_i$. The fermionic operator
\eqn\covder{
D_\a \equiv {\p \over \p \theta^\a} + \half k^m(\g_m\t)_\a
}
denotes the standard superspace covariant derivative.
Note that if a superfield $K(x,\t)$ depends only on the zero-modes of $\t$,
the action of the pure spinor BRST charge $Q$ is given by a covariant derivative:
\eqn\BRSTcharge{
Q \equiv \oint \l^\a d_\a  \ \ \Longrightarrow \ \ QK = \l^\a D_\a K \ .
}
This fact allows one to use
the equations of motion \RankOneEOM\ in combination with BRST cohomology manipulations to
simplify expressions considerably.

In previous work \eombbs, these superfields were promoted to
multiparticle versions
\eqn\prom{
K_i \in \{ A^i_\a, A^i_m , W_i^\a ,F^i_{mn} \} \ \rightarrow \ K_B \in \{ A^B_\a, A^B_m , W_B^\a ,F^B_{mn} \} \ ,
}
where the multiparticle label $B=b_1b_2{\ldots} b_p$ describes $|B| \equiv p$ external particles attached to a tree
subdiagram as shown in \figtwo. The off-shell leg indicated by the $\ldots$ in the figure reflects that the overall
momentum $k_B^m \equiv \sum_{i=1}^{p} k_{b_i}^m$ is no longer lightlike in general, $k_B^2 \neq 0$.

\ifig\figtwo{Four superfield realizations $K_B\in \{A^B_\a, A^B_m , W_B^\a ,F^B_{mn}\}$ of cubic
tree graphs $B = b_1b_2\ldots b_p$.}
{\epsfxsize=0.60\hsize\epsfbox{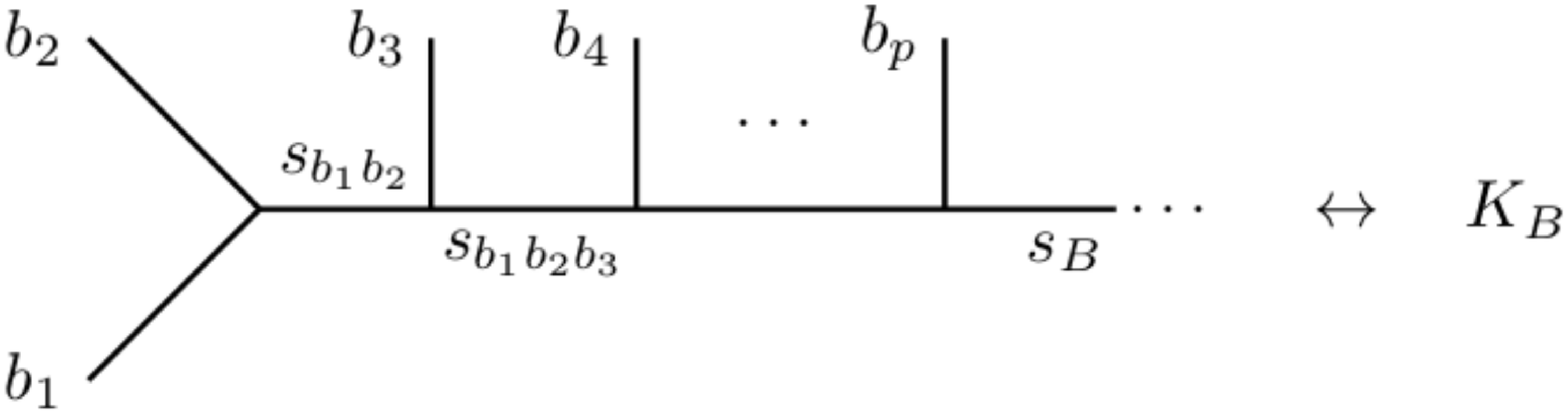}}

 
As detailed in \eombbs, the diagrammatic interpretation is supported by the Lie symmetries of $K_B$ matching with the color representative of the diagram in \figtwo,
\eqn\newstruCte{
K_{1234 \ldots p} \leftrightarrow\; f^{12 a_2} \, f^{a_2 3 a_3} \, f^{a_3 4 a_4} \ldots f^{a_{p-1} p a_p}\ ,
}
subject to antisymmetry $f^{abc}=f^{[abc]}$ and Jacobi identities $f^{e[ab} f^{c]de}=0$. For example,
\eqnn\Lietwothree
\eqnn\Liefour
$$\eqalignno{
0 &= K_{12} + K_{21} \ , \ \ \ 0 = K_{123} + K_{231} + K_{312} &\Lietwothree\cr
0 &= K_{1234} - K_{1243} + K_{3412} - K_{3421} &\Liefour
}$$
furnish the kinematic analogue of $f^{12a} = -f^{21a}$ and Jacobi identities among permutations of $f^{12a} f^{a3b}$ and $f^{12a} f^{a3b} f^{b4c}$.
 
\subsec Recursive construction of BRST blocks

\subseclab\twotwo
A recursive procedure was described in \eombbs\ to construct BRST blocks at arbitrary multiplicity from the elementary SYM superfields.
The definition of the multiparticle fields \prom\ is inspired by the OPE among integrated massless vertex operators
in the pure spinor formalism \psf. For two particles, this directly leads to
\eqnn\Atwo
$$\eqalignno{
A^{12}_\a &= - \half\bigl[ A^1_\a (k^1\cdot A^2) + A^1_m (\g^m W^2)_\a - (1\leftrightarrow 2)\bigr]
&\Atwo \cr
A^{12}_m &=  \half\Bigl[ A^1_p F^2_{pm} - A^1_m(k^1\cdot A^2) + (W^1\g_m W^2) - (1\leftrightarrow 2)\Bigr] \cr
W_{12}^\a &= {1\over 4}(\g^{mn}W^2)^\a F^1_{mn} + W_2^\a (k^2\cdot A^1) - (1\leftrightarrow 2) \cr
F^{12}_{mn} &= k^{12}_m A^{12}_n - k^{12}_n A^{12}_m - (k^1\cdot k^2)(A^1_m A^2_n -A^1_n A^2_m)\,,
}$$
compatible with antisymmetry $K_{12}=-K_{21}$. Remarkably, the equations of motion for these two-particle superfields
take the same form as their single-particle equations \RankOneEOM\ with the addition of contact terms,
\eqnn\EOMAtwo
$$\eqalignno{
2 D_{(\a} A^{12}_{\b)} &= \g^m_{\a\b}A^{12}_m + (k^1\cdot k^2)(A^1_\a A^2_\b + A^1_\b A^2_\a) &\EOMAtwo\cr
D_\a A^{12}_m &= (\g_m W^{12})_\a + k^{12}_m A^{12}_\a + (k^1\cdot k^2)(A^1_\a A^2_m - A^2_\a A^1_m)  \cr
D_\a W^\b_{12} &= {1\over 4}(\g^{mn})_\a{}^\b F^{12}_{mn} + (k^1\cdot k^2)(A^1_\a W_2^\b - A^2_\a W^\b_1) \cr
D_\a F^{12}_{mn} &= k^{12}_m (\g_n W^{12})_\a - k^{12}_n (\g_m W^{12})_\a + (k^1\cdot k^2)(A^1_\a F^2_{mn} - A^2_\a
F^1_{mn})   \cr
& \ \  + (k^1\cdot k^2)( A^{1}_{n} (\g_{m} W^2)_\a - A^{2}_{n} (\g_{m} W^1)_\a - A^{1}_{m} (\g_{n} W^2)_\a + A^{2}_{m}
(\g_{n} W^1)_\a)\,. \cr
}$$
Starting from multiplicity three, application of the recursion \Atwo\ yields superfields
\eqnn\Athree
$$\eqalignno{
\widehat A^{123}_\a &= - \half\bigl[ A^{12}_\a (k^{12}\cdot A^3) + A^{12}_m (\g^m W^3)_\a - (12\leftrightarrow 3)\bigr] &\Athree
\cr
\widehat A^{123}_m &= \half\Bigl[ A_{12}^p F^3_{pm} - A^{12}_m (k^{12}\cdot A^3) + (W^{12}\g_m W^3) - (12\leftrightarrow 3)\Bigr]
\cr
}$$
which require redefinitions
\eqnn\redefsthree
\eqnn\Rthree
$$\eqalignno{
A^{123}_m &= \widehat A^{123}_m - k^{123}_m H^{123} \,, \ \ \ A_\a^{123} = \widehat A_\a^{123} - D_\a H^{123} &\redefsthree\cr
H^{123} &= {1\over 6}\bigl[(A^1\cdot A^{23}) - (k^2_p - k^3_p)A^p_1 (A^2\cdot A^3) +  {\rm cyclic}(123)\bigr] \, &\Rthree
}$$
by some scalar superfield $H_{ijk}$ \eombbs\ before they satisfy the Lie symmetries in \Lietwothree\ and qualify as BRST blocks. The three-particle set of BRST
blocks $K_{123} \in \{ A_\a^{123}, A^{123}_m ,W_{123}^\a ,F^{123}_{mn}\}$ is completed by field strengths
\eqnn\moreAthree
$$\eqalignno{
W_{123}^\a &= \Big[ {1\over 4}(\g^{rs}W^3)^\a F^{12}_{rs} + W_3^\a (k^{3}\cdot A^{12})  - (12 \leftrightarrow 3)  \Big]+ \half (k^1\cdot k^2) \bigl[ W_2^\a (A^1\cdot A^3) - (1 \leftrightarrow 2)\bigr] 
\cr
F^{123}_{mn} &= k^{123}_m A^{123}_n - k^{123}_n A^{123}_m  - (k^1\cdot k^2)\bigl[ 2 A^1_{[m} A^{23}_{n]} - (1\leftrightarrow 2)\bigr] - (k^{12}\cdot k^3) 2 A^{12}_{[m} A^3_{n]} \ . &\moreAthree 
}$$
As shown in \eombbs, the equations of motion for the $K_{123}$ reproduce the universal structure of \RankOneEOM\ and
\EOMAtwo\ and incorporate a richer set of contact terms $\sim (k^1 \cdot k^2) $ and $(k^{12} \cdot k^3)$:
\eqnn\EOMAthree
$$\eqalignno{
2D_{(\a}  A^{123}_{\b)}   &= \g^m_{\a\b} A^{123}_m + (k^{12}\cdot k^3)\bigl[A^{12}_\a A^3_\b - (12\leftrightarrow 3)\bigr]  &\EOMAthree  \cr
&\quad{} + (k^1\cdot k^2)\bigl[A^1_\a A^{23}_\b + A^{13}_\a A^2_\b - (1\leftrightarrow 2)\bigr]  \cr
D_\a   A^{123}_m &= (\g_m W^{123})_\a + k^{123}_m  A^{123}_\a + (k^{12}\cdot k^3)(A^{12}_\a A^3_m - A^3_\a A^{12}_m) \cr
&\quad{} + (k^1\cdot k^2)\bigl[ A^1_\a A^{23}_m + A^{13}_\a A^2_m - A^{23}_\a A^1_m - A^2_\a A^{13}_m\bigr]  \cr
D_\a W^\b_{123} &= {1\over 4}(\g^{mn})_\a{}^\b F^{123}_{mn} + (k^{12}\cdot k^3) \bigl[ A^{12}_\a W_3^\b - (12 \leftrightarrow 3)\bigr]  \cr
&\quad{} + (k^1\cdot k^2)\big[A^1_\a W_{23}^\b + A^{13}_\a W_2^\b - (1 \leftrightarrow 2)\bigr]\cr
D_\a F^{123}_{mn} &= 2 k^{123}_{[m} (\g_{n]} W^{123})_\a + (k^{12}\cdot k^3)\bigl[ A^{12}_\a F^3_{mn} - (12\leftrightarrow 3)\bigr]  \cr
&\quad{} + (k^{12}\cdot k^3)\bigl[  2 A^{12}_{[n} (\g_{m]} W^3)_\a - (12\leftrightarrow 3)\bigr]  \cr
&\quad{} + (k^{1}\cdot k^2)\bigl[ A^1_\a F^{23}_{mn} + A^{13}_\a F^2_{mn} - (1\leftrightarrow 2)\big]\cr
&\quad{} + (k^{1}\cdot k^2)\bigl[ 2 A^1_{[n} (\g_{m]} W^{23})_\a + 2 A^{13}_{[n} (\g_{m]} W^{2})_\a -(1\leftrightarrow 2)\bigr] \ .
}$$
The starting point towards BRST blocks at higher multiplicity $p$ is the recursion
\eqn\Arankp{
\widehat A^{12\ldots p}_\a = - \half\bigl[ A^{12\ldots p-1}_\a (k^{12\ldots p-1}\cdot A^p) + A^{12\ldots p-1}_m (\g^m W^p)_\a - (12\ldots p-1\leftrightarrow p)\bigr]
}
with manifest Lie symmetries \newstruCte\ in the first $p-1$ labels. Then, an algorithmic redefinition along the lines
of \redefsthree\ enforces the remaining symmetry \newstruCte\ involving the last label $p$, see \eombbs\ for details. 

The equations of motion at any multiplicity combine the single-particle structure from \RankOneEOM\ with a growing
tail of contact terms $\sim (k^{12\ldots j-1} \cdot k^j)$ generalizing the three-particle example \EOMAthree. Their
explicit forms can be found in \eombbs.

\subsec Berends--Giele currents

\subseclab\twothree
BRST blocks $K_B$ of multiplicity $|B|$ are diagrammatically interpreted as off-shell cubic graphs shown in \figtwo.
This suggests to assemble
diagrams to a color-ordered SYM $(|B|+1)$-point tree amplitude where one of the legs is off-shell, as schematically
depicted in \figthree. The precise form of this diagrammatic construction was explained in the appendix A of \eombbs,
and the result is
the promotion of BRST blocks $K_B$ to Berends--Giele currents ${\cal K}_B$,
\eqn\promBG{
K_B \in \{ A^{B}_\alpha, A^m_{B} ,
W_{B} ^\alpha, F_{B} ^{mn} \} \ \rightarrow \ {\cal K}_B \in \{  \cA^{B}_\alpha, \cA^m_{B} , \cW_{B} ^\alpha, \cF_{B} ^{mn} \} \,,
}
The name goes back to Berends and Giele who recursively constructed gluonic currents which were then used to compute
tree-level amplitudes \BerendsME. From now on the ordered subsets $B=b_1 b_2 \ldots b_{|B|}$ of
external particle labels which appear along with Berends--Giele currents ${\cal K}_B$ will be denoted ``words''.

\ifig\figthree{From cubic diagrams $K_A$ to Berends--Giele currents ${\cal K}_A$.}
{\epsfxsize=0.75\hsize\epsfbox{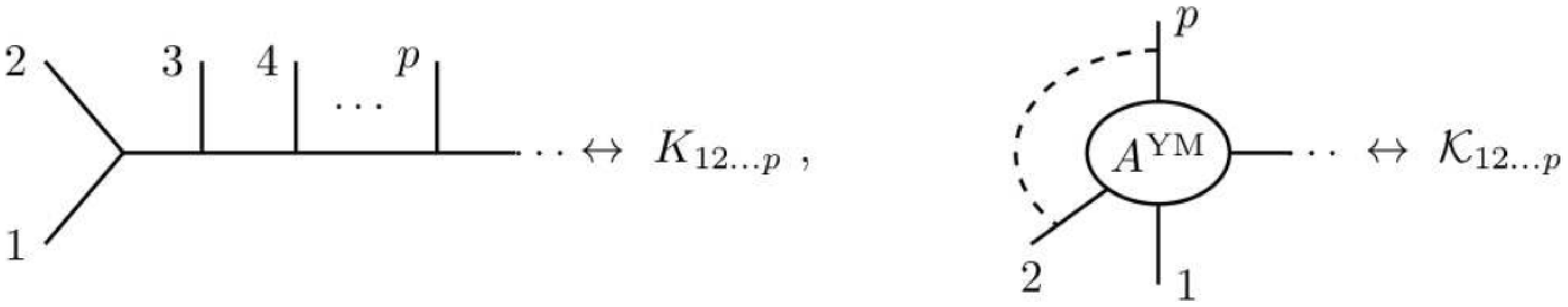}}


The first examples of Berends--Giele currents ${\cal K}_B$ are given by,
\eqnn\BGtwothree
\eqnn\BGfour
$$\eqalignno{
{\cal K}_{12} &= { K_{12} \over s_{12}}, \quad \quad {\cal K}_{123} = {K_{123} \over s_{12} s_{123}} + { K_{321} \over s_{23} s_{123}} &\BGtwothree \cr
{\cal K}_{1234} &= {1 \over s_{1234}} \Big( {K_{1234}\over s_{12}s_{123} } 
 + {K_{3214}\over s_{23}s_{123} }+ {K_{3421} \over s_{34}s_{234} } + {K_{3241} \over s_{23}s_{234} } 
 + {2K_{12[34]}\over s_{12}s_{34} }\Big)  \ ,
&\BGfour
}$$
where the conventions for the generalized Mandelstam invariants is
\eqn\mandvar{
s_{12\ldots p} =  \sum_{1\leq i<j}^{p} (k_i\cdot k_j) 
= {1\over 2} k_{12\ldots p}^2
\,.
}
It turns out that they enjoy simplified BRST variations compared to their corresponding BRST blocks $K_B$. In particular, the Berends--Giele version
of the unintegrated\foot{According to the calligraphic-letter convention of \promBG\ the Berends--Giele current
associated to $V_B$ would be denoted $\cV_B$. However, the definition
\QBGV\ is used for historic reasons.} multiparticle vertex $V_B\equiv \lambda^{\alpha} A^B_{\alpha}$,
\eqn\QBGV{
M_B \equiv \lambda^{\alpha} {\cal A}^B_{\alpha} \ ,
}
satisfies
\eqn\QBG{
QM_{B} = \sum_{XY=B} M_X M_Y = \sum^{|B|-1}_{j=1} M_{b_1b_2\ldots b_j}M_{b_{j+1}\ldots b_p}\, .
}
This has been exploited in \refs{\nptMethod,\nptTree} to construct tree amplitudes of ten-dimensional SYM and of the
open superstring. Throughout this paper the notation $XY=B$ (as in \QBG) denotes a sum over all
deconcatenations of the word $B$ into smaller (non-empty) words $X=b_1b_2\ldots b_j$ and $Y = b_{j+1}\ldots b_p$ with
$j=1,2,\ldots,|B|-1$.

As partially used in \oneloopbb\ and generalized in \eombbs, the equations of motion for the remaining ${\cal K}_B$ representatives
are given by,
\eqnn\QBGs
$$\eqalignno{
Q\cA^m_{B} &= (\lambda \gamma^m \cW_{B})
+ k_{B}^m M_{B} + \sum_{XY=B} (M_{X} \cA^m_{Y}
-  M_{Y} \cA^m_{X})  \cr
Q\cW^\alpha_{B} &={1\over 4}(\lambda \gamma_{mn})^\alpha \cF^{mn}_{B}
+\sum_{XY=B} (M_{X}\cW^\alpha_{Y}-M_{Y}\cW^\alpha_{X}) &\QBGs \cr
Q\cF^{mn}_{B} &=2 k^{[m}_{B} (\lambda \gamma^{n]} \cW_{B})
+ \sum_{XY=B} (M_{X}\cF^{mn}_{Y}-M_{Y}\cF^{mn}_{X})\cr
&\quad{}+ \sum_{XY=B} 2\big[ \cA^{[n}_{X} (\lambda \gamma^{m]} \cW_{Y})
-\cA_{Y}^{[n} (\lambda \gamma^{m]} \cW_{X}) \big] \,,\cr
}$$
where the contact terms present in $QK_B$ are traded by deconcatenations as the result of the Berends--Giele map \promBG.
At general multiplicity, the transformation matrix between BRST blocks and their Berends--Giele currents was identified in \Polylogs\ to be the momentum
kernel \refs{\oldMomKer,\MomKer}, see \eombbs\ for further details.

According to their definition as off-shell SYM amplitudes, the Lie symmetries of the BRST-blocks $K_B$ translate into
Kleiss--Kuijf relations \KKref\ among their Berends--Giele counterparts ${\cal K}_B$ \refs{\nptMethod,\nptTree}. Up to multiplicity four, these are
\eqnn\KKs
$$\eqalignno{
0&={\cal K}_{12} + {\cal K}_{21}  \,, \ \ \ \  0= {\cal K}_{123}  -  {\cal K}_{321}  = {\cal K}_{123} + {\cal K}_{231} + {\cal K}_{312} &\KKs \cr
0&= {\cal K}_{1234} + {\cal K}_{4321} = {\cal K}_{1234} +{\cal K}_{2134} +{\cal K}_{2314} +{\cal K}_{2341} \,, 
}$$
and higher multiplicity generalizations are most compactly written as\foot{We follow the convention ${\cal K}_{\ldots A \shuffle B \ldots} \equiv \sum_{C\in A\shuffle B}
{\cal K}_{\ldots C \ldots}$.}
\eqn\BGsym{
{\cal K}_{B1A} = (-1)^{|B|} {\cal K}_{1(A \shuffle B^T)}\,.
}
The superscript along with $B^T$ denotes the reversal of the word $B$ in external particles $b_j$ such as
$(b_1b_2\ldots b_{|B|})^T = (b_{|B|} \ldots b_2 b_1)$, and $\shuffle$ denotes the shuffle product.

\subsec One--loop building blocks

\subseclab\twofour
The saturation of fermionic
zero modes in the pure spinor formalism \psf\ imposes tight constraints on contributions to loop amplitudes, see e.g.
\refs{\multiloop,\theorems}. As argued in \oneloopbb, the one-loop prescription in the minimal version of the formalism
requires zero modes $d_\alpha d_\beta N^{mn}$ from the external vertices, leaving behind $(\lambda \gamma^{[m})_\alpha
(\lambda \gamma^{n]})_\beta$. This effective rule leads to the BRST-closed expression $(\l\g^m W^i)(\l\g^n W^j) F^k_{mn}$
in the four-point amplitude \multiloop\ and motivates the following higher-point generalization \refs{\oneloopbb,\eombbs}
\eqn\TABC{
T_{A,B,C} \equiv  {1 \over 3}(\l\g_m W_A)(\l\g_n W_B) F^{mn}_C + (C\leftrightarrow A,B)\,,
}
as well as its associated Berends--Giele current
\eqn\BRSTMi{
M_{A,B,C} \equiv  {1 \over 3}(\l\g_m \cW_A)(\l\g_n \cW_B) \cF^{mn}_C + (C\leftrightarrow A,B)\,.
}
From now on, the Berends--Giele versions of various superfield combinations will be emphasized since
explicit results for BRST (pseudo-)invariants and amplitudes simplify in this basis. Furthermore,
their BRST-block counterparts such as $T_{A,B,C}$ can always be trivially recovered by using the superfields $A_B, W_B, F_B$ instead of  ${\cal A}_B,{\cal W}_B ,{\cal F}_B$.

The universal form \QBGs\ of $Q {\cal W}^\a_{B}$ and $Q {\cal F}^{mn}_B$ gives rise to the BRST-covariant\foot{Due to the tensor $(\lambda \gamma^{[m})_\alpha
(\lambda \gamma^{n]})_\beta$ in \BRSTMi, the pure spinor
constraint projects out all terms in \QBGs\ with an explicit appearance of $\l^\a$, regardless of the words $A,B$ and $C$.} transformation \refs{\oneloopbb,\eombbs}
\eqn\QMABC{
Q  M_{A,B,C}  =
\sum_{XY=A}   \big( M_{X}  M_{Y,B,C}
-   M_{Y} M_{X,B,C}  \big) + (A \leftrightarrow B,C) \,,
}
governed by deconcatenations of the multiparticle labels. Note that $QM_{1,2,3} =0$ and that $M_{A,B,C}$ is totally symmetric in $A$, $B$ and $C$.

In closed-string amplitudes of multiplicity higher than four, additional zero-mode contributions can arise from the $\Pi^m$
fields in the external vertices. In the simplest case at five points \oneloopMichael, this leads to a single vector index contraction among left- and
right-movers. The aforementioned $d_\alpha d_\beta N^{mn} \rightarrow (\lambda \gamma^{[m})_\alpha (\lambda
\gamma^{n]})_\beta$ prescription identifies contributions of the form ${\cal A}^m_{A} M_{B,C,D}$ to the
left/right-contracting part of the closed-string amplitude.

However, as pointed out in \oneloopMichael\ and generalized in \eombbs, a separate b-ghost contribution
proportional to $\Pi^m d_\a d_\b$ leads to an additional kinematic factor
\eqnn\BRSTw
\eqnn\BRSTQw
$$\eqalignno{
\cW^m_{A,B,C,D}& \equiv {1\over 12}(\l\g_n \cW_A)(\l\g_p \cW_B)(\cW_C\g^{mnp}\cW_D) + (A,B|A,B,C,D) &\BRSTw\cr
Q\cW^m_{A,B,C,D}&= \sum_{XY=A}\bigl(
M_{X}\cW^m_{Y,B,C,D}
- M_{Y}\cW^m_{X,B,C,D}
\bigr) \cr
& \quad{}\quad{}\quad{}- (\l\g^m\cW_A) M_{B,C,D} + (A\leftrightarrow B,C,D) \ . &\BRSTQw 
}$$
The notation $(A_1,{\ldots } , A_p \,|\, A_1,{\ldots} ,A_n)$ will also be used in later sections and
instructs to sum over all possible ways to choose $p$
elements $A_1,A_2,\ldots ,A_p$ out of the set $\{A_1,{\ldots} ,A_n\}$, for a total of ${n\choose p}$ terms.
This yields the following
vector building block,
\eqnn\BRSTm
\eqnn\BRSTQm
$$\eqalignno{
M^m_{A,B,C,D} &\equiv \big[ {\cal A}^m_A M_{B,C,D} + (A \leftrightarrow B,C,D) \big] + \cW^m_{A,B,C,D} &\BRSTm\cr
QM^m_{A,B,C,D} &= \sum_{XY=A} (M_{X} M^m_{Y,B,C,D}
- M_{Y} M^m_{X,B,C,D})\cr
& \quad{}\quad{}\quad{}
+k^m_{A} M_A M_{B,C,D}
+(A \leftrightarrow B,C,D) \,, &\BRSTQm \cr
}$$
which is totally symmetric in $A,B,C,D$. Apart from the last line, the BRST covariant transformation \BRSTQm\ stems
from deconcatenation terms in $Q {\cal K}_B$. This goes back to cancellations between \BRSTQw\ and the first
term of $Q {\cal A}^m_B = (\lambda \gamma^m {\cal W}_B) + \ldots$, see \QBGs.

\subsec Scalar and vector one--loop cohomology

\subseclab\twofive
The simplest kinematic expressions compatible with the one-loop amplitude prescription \multiloop\ are $M_A
M_{B,C,D}$ and $M_A M^m_{B,C,D,E}$ where the Berends--Giele current $M_A$ stems from OPE contractions of the unintegrated vertex $V_i$ with its integrated counterparts. Their covariant BRST transformations motivate to combine them to BRST-closed expressions. Following the experience with scalars
\oneloopbb, BRST invariants are classified in \eombbs\ by a ``leading term'' where a reference leg $i$ is represented through a single-particle unintegrated
vertex $M_i= V_i$
\eqnn\Csc
\eqnn\Cvec
$$\eqalignno{
C_{i|A,B,C} &\equiv M_i M_{A,B,C} + \sum_{E \neq \emptyset} M_{iE } \ldots &\Csc \cr
C^m_{i|A,B,C,D} &\equiv M_i M^m_{A,B,C,D} + \sum_{E \neq \emptyset} M_{iE } \ldots \ .
&\Cvec
}$$
Apart from the explicit leading term, the singled-out label $i$ always enters in a multiparticle Berends--Giele
current. This is formally represented by a sum over (non-empty) words $E$ of external particles which join the reference leg
$i$ in $M_{iE}$. The $\ldots$ in $C_{i|A,B,C}$ ($C^m_{i|A,B,C,D}$) represent linear combinations of
$M_{A,B,C}$ ($M^m_{A,B,C,D}$ and $M_{A,B,C} k_D^m$) such that
\eqn\Qclos{
Q C_{i|A,B,C}  = Q C^m_{i|A,B,C,D}  = 0\,.
}
In later sections, we will encounter plenty of further examples where a leading term $M_i \ldots $ is combined with a
BRST completion made of multiparticle currents $M_{iE}$ (with $E \neq \emptyset$) and ghost-number-two objects. Note
that $C_{i|A,B,C}$ and $C^m_{i|A,B,C,D}$ are totally symmetric in $A,B,C$ and $A,B,C,D$ which follows a general
convention used throughout this work: Whenever multiparticle slots $A,B$ in a subscript are separated by a comma
rather than by a vertical bar as in $\ldots,A |B, \ldots$, then the parental object is understood to be symmetric in $A \leftrightarrow B$.

In \eombbs, the following two observations were exploited to set up a recursive construction of BRST invariants in \Csc\ and \Cvec:
\bigskip
\item{(i)} Nilpotency $Q^2 =0$ implies that also $QM_{A,B,C}$ and $QM^m_{A,B,C,D}$ as given by \QMABC\ and \BRSTQm\ are BRST closed. 
By promoting each $M_i M_{A,B,C}$ and $M_i M^m_{A,B,C,D}$ therein to the corresponding invariant $C_{i|A,B,C}$ and $C^m_{i|A,B,C,D}$,
one arrives at an alternative form of the BRST transformations\foot{At this point, uniqueness of the BRST completions in \Csc\ and
\Cvec\ is assumed. We don't have a rigorous argument to prove this in full generality but rely on ``experimental'' evidence at finite multiplicities.},
\eqnn\QMgen
\eqnn\QMnrew
$$\eqalignno{
Q M_{A , B , C } &=
C_{a_1| a_2\ldots a_{|A|} ,B , C }
- C_{a_{|A|}| a_1\ldots a_{|A|-1} , B , C } \;+\; (A \leftrightarrow B,C) &\QMgen
\cr
Q  M^m_{A,B,C,D} &= C^m_{a_1|a_2\ldots a_{|A|},B,C,D} -  C^m_{a_{|A|}|a_1 \ldots a_{|A|-1},B,C,D} \cr
& \ \ \ \ \
+ \delta_{|A|,1} k_{a_1}^m C_{a_1|B,C,D} + (A \leftrightarrow B,C,D)  \ . &\QMnrew
}$$
Note that $\d_{|A|,1}$ is equal to one when the word $A$ represents a single particle (i.e. $|A|=1$) and zero otherwise.
\item{(ii)} We define a linear concatenation operator
\eqn\concat{
M_B \otimes_{a_1} M_{a_1 a_2\ldots a_{|A|}}  \equiv M_{B a_1 a_2\ldots a_{|A|}} \,,
}
acting on $M_A$ which does not interfere with ghost-number-two superfields such as $M_{A,B,C}$ or $M^m_{A,B,C,D}$.
The deconcatenation formula \QBG\ for $QM_A$ and \Qclos\ imply
\eqnn\MAconcat
\eqnn\MBconcat
$$\eqalignno{
Q(M_i\otimes_j C_{j|A,B,C}) &= M_i C_{j|A,B,C} &\MAconcat \cr
Q(M_i\otimes_j C^m_{j|A,B,C,D}) &= M_i C^m_{j|A,B,C,D} \ , &\MBconcat 
}$$
see subsection \threethree\ for a more detailed and general derivation.

\medskip\noindent
On these grounds, one can show that the following recursions generate BRST invariants for arbitrary multiplicity:
\eqnn\veca
$$\eqalignno{
C_{i|A,B, C} &=
M_iM_{A,B,C} + M_i \otimes \bigl[  C_{a_1| a_2\ldots a_{|A|} ,B , C }
- C_{a_{|A|}| a_1\ldots a_{{|A|}-1} ,B , C }
+ (A \leftrightarrow B,C)\bigr] \cr
C^m_{i|A,B,C,D} &= M_i M^m_{A,B,C,D} + M_i \otimes \big[ C^m_{a_1|a_2\ldots a_{|A|},B,C,D}-C^m_{a_{|A|}|a_1 \ldots a_{|A|-1},B,C,D} \cr 
& \ \ \ \ \ \ \ \ \ \ \ \ \ \ \ \ \ \ \ \ +\d_{|A|,1} k_{a_1}^m  C_{a_1|B,C,D}
  + (A \leftrightarrow B,C,D) \, \big] \,. &\veca 
}$$
Once the leading terms $Q(M_i M_{A,B,C})$ and $Q(M_i M^m_{A,B,C,D})$ are evaluated via (i), they are easily
seen to cancel the BRST variations (ii) of the concatenated terms. 

In \veca\ as well as later equations in this paper, the subscript $j$ of the concatenation $\otimes_j$ in \concat\ is
suppressed and understood to match the reference leg $j$ of subsequent kinematic factor such as $C_{j|\ldots}$ or
$C^m_{j|\ldots}$. In principle, $\otimes$ without further specification does not preserve the Kleiss--Kuijf symmetries \BGsym\ of
the Berends--Giele currents\foot{For example, $M_{132}\neq -M_{123}$ implies that $M_1\otimes M_{32} \neq - M_1\otimes
M_{23}$ even though $M_{32} = - M_{23}$.}. However, this slight ambiguity does not matter in the recursive formulas
\veca\ as long as the objects generated by the recursion are directly used in the next steps without any prior
symmetry manipulations.

The simplest instances of scalar and vector invariants following from the recursions in \veca\ are
\eqnn\Cscex
$$\eqalignno{
C_{1|2,3,4} &\equiv M_1 M_{2,3,4} &\Cscex \cr
C_{1|23,4,5} &\equiv M_1 M_{23,4,5} + 
M_1 \otimes \big[ C_{2|3,4,5} -C_{3|2,4,5} \big]\cr 
&=M_1 M_{23,4,5} + M_{12} M_{3,4,5} - M_{13} M_{2,4,5}  \cr
C^m_{1|2,3,4,5} &\equiv M_1 M^m_{2,3,4,5} + M_1 \otimes  \big[ k_2^m C_{2|3,4,5}+ (2\leftrightarrow 3,4,5) \big] \cr
&=M_1 M^m_{2,3,4,5} + \big[ k_2^m M_{12} M_{3,4,5}+ (2\leftrightarrow 3,4,5) \big] \,.
}$$
The corresponding six-point examples are expanded in appendix \appA, see \sixsc. Seven- and eight-point
examples of $C_{i|A,B,C}$ can be found in \oneloopbb.

The families of scalar and vector invariants $C_{i|A,B,C}$, $C^m_{i|A,B,C,D} $ as well as their recursive construction in \veca\ furnish the first two cells from the left in \figoverview.

\newsec Towards a BRST pseudo-cohomology

\seclab\secthree
In this section, we investigate the applicability of the BRST program in sections \twofour\ and \twofive\ to
tensorial building blocks.  The one-loop prescription of the pure spinor formalism \multiloop\ suggests a natural two-tensor
generalization $M^{mn}_{\ldots}$ of the $M_{A,B,C}$ and $M^m_{A,B,C,D}$ above, but its BRST variation turns out to
involve new classes of superfields. This obstruction is closely related to the pure spinor superspace description
of the hexagon anomaly \anomaly. It leads us to define a pseudo-cohomology as an extension of the standard cohomology
in order to systematically study the multiparticle superfields which play a role in the gauge anomaly of open
superstring amplitudes and its cancellation \GSanomaly.

\subsec Tensorial building blocks $M^{mn}$

\subseclab\threeone
Higher-point loop amplitudes in the closed-string allow for an arbitrary number of $\Pi^m$ zero mode
contractions between left- and right-movers. This motivates the study of higher-rank tensors generalizing \BRSTm\ such as
\eqnn\BGTmn
$$\eqalignno{
M^{mn}_{A,B,C,D,E} &\equiv  2 \big[ {\cal A}_A^{(m} {\cal A}_B^{n)} M_{C,D,E} + (A,B|A,B,C,D,E) \big] \cr
& \ \ \ \ \
+2 \big[  {\cal A}_A^{(m} {\cal W}^{n)}_{B,C,D,E} + (A \leftrightarrow B,C,D,E) \big]\cr
&= \cA^m_A \cW^n_{B,C,D,E} + \cA^n_A M^m_{B,C,D,E} + (A \leftrightarrow B,C,D,E), &\BGTmn\cr
}$$
firstly relevant for the six-point amplitude.
Its first term $\sim {\cal A}_A^{(m} {\cal A}_B^{n)}$ stems from the $\Pi^m \Pi^n d_\alpha d_\beta N_{pq}$ zero-mode coefficient
and its second term $\sim {\cal A}_A^{(m} {\cal W}^{n)}_{B,C,D,E}$ originates from the $b$-ghost sector linear in $\Pi^m$. The BRST variations \QBGs, \BRSTQw\ and \BRSTQm\ for its constituents imply that
\eqnn\BRSTt
$$\eqalignno{
Q M^{mn}_{A,B,C,D,E} &=\delta^{mn}\cY_{A,B,C,D,E} + \Big[  \sum_{XY=A} (M_{X} M^{mn}_{Y,B,C,D,E} 
- M_{Y} M^{mn}_{X,B,C,D,E}) \cr
&\ \ \ \ +2 k_{A}^{(m} M_{A} M^{n)}_{B,C,D,E} +(A \leftrightarrow B,C,D,E) \Big]  ,&\BRSTt
}$$
where the first term is a shorthand for
\eqn\WanonBG{
\cY_{A,B,C,D,E} \equiv \half (\l\g^m \cW_A)(\l\g^n \cW_B)(\l\g^p \cW_C)(\cW_D\g_{mnp}\cW_E)\,.
}
The superfield $\cY_{A,B,C,D,E}$ has ghost-number three and is totally symmetric in $A,B,C,D,E$ due to the pure spinor
constraint. It stems from the term $(\lambda \gamma^{(m} {\cal W}_A)\cW^{n)}_{B,C,D,E}$ in $Q (\cA^{(m}_A \cW^{n)}_{B,C,D,E} )$
where a group-theoretic analysis\foot{The spinors indices in $(\l \g_p)_{[\a_1}(\l\g_q)_{\a_2} (\l
\g^{(m})_{\a_3}\g^{n)pq}_{\a_4\a_5]}$ fall into the tensor product of $[0,0,0,0,3] \ni \lambda^3$ and
$[0,0,0,0,1]^{\wedge 5}= [0,0,0,3,0] \oplus [1,1,0,1,0] \ni W^5$. The LiE program \LiEprogram\ identifies one scalar
$[0,0,0,0,0]$ but no symmetric and traceless $[2,0,0,0,0]$ component in $[0,0,0,0,3] \otimes [0,0,0,0,1]^{\wedge 5}$.
Hence, only the trace with respect to vector indices $m,n$ contributes. We are using standard Dynkin label notation
$[a_1,a_2,a_3,a_4,a_5]$ for $SO(10)$ irreducibles and denote an antisymmetrized $k^{\rm th}$ tensor power by
$[a_1,a_2,a_3,a_4,a_5]^{\wedge k}$.} has been used to replace \eqn\BRSTxx{ (\l \g_p)_{[\a_1} (\l \g_q)_{\a_2} (\l
\g^{(m})_{\a_3} \g^{n)pq}_{\a_4 \a_5]} = {1\over 10} \delta^{mn} (\l \g_p)_{[\a_1} (\l \g_q)_{\a_2} (\l \g_r)_{\a_3}
\g^{pqr}_{\a_4 \a_5]} \ . } Apart from the extra term $\cY_{A,B,C,D,E}$, the BRST variation \BRSTt\ of
$M^{mn}_{A,B,C,D,E}$ is a direct rank-two generalization of  $QM^{m}_{A,B,C,D}$ given in \BRSTQm.

\subsec Pseudo BRST cohomology

\subseclab\threetwo
The pure spinor analysis of the hexagon gauge anomaly \GSanomaly\
showed that the gauge variation of the open superstring six-point amplitude (w.r.t. leg one) is
proportional to \anomaly,
\eqn\kinforan{
{\cal Y}_{2,3,4,5,6} = \half(\l\g^m W_2)(\l\g^n W_3)(\l\g^p W_4)(W_5 \g_{mnp}W_6) \,.
}
Since gauge invariance is related to BRST invariance of the kinematic factors in the amplitudes (see appendix B),
terms of the form \WanonBG\ are expected to describe the BRST anomaly of the amplitudes. They represent obstructions in finding elements in the cohomology.

Therefore it will be convenient to define a {\it pseudo} BRST cohomology in which the variation of pseudo BRST-closed
elements vanish up to anomalous superfields such as \WanonBG. These objects give rise to gauge transformations with
bosonic components proportional to the $\epsilon_{10}$ tensor, see appendix~B.5, so they are suitable to describe the parity odd gauge anomaly of the open superstring. Indeed, it will be shown in \wipG\ that the scalar pseudo BRST cohomology for
six particles discussed in the next section correctly describes the anomaly terms of the six-point one-loop amplitude
which were not included in the discussion of \oneloopbb.

\proclaim Definition 1. A superfield ${\cal Y}$ of ghost-number three or four is called anomalous if it
contains a factor of $\cY_{A,B,C,D,E}$ as in \WanonBG\ with some multiparticle labels $A,B\ldots,E$.
\par

\proclaim Definition 2. Superfields of ghost-number two and three are called pseudo-invariant if their BRST variation
is entirely anomalous. The space of pseudo-invariants is referred to as the {\it pseudo-cohomology}.
\par

\subsec Tools for constructing BRST pseudo-invariants

\subseclab\threethree
The subsequent sections are concerned with a systematic construction of BRST pseudo-invariants. As a driving force for
this endeavor, we generalize the recursions of section \twofive\ to situations with anomalous BRST variations.

\proclaim Lemma 1. Let ${\cal J}$ denote any superfield unaffected by the operation $\otimes$ defined in \concat. Then, concatenations of $M_A {\cal J}$ satisfy
\eqn\lem{
Q(M_i\otimes M_A {\cal J}) = M_i M_A {\cal J} + M_i\otimes Q (M_A {\cal J}) \ .
}
\par
\noindent{\it Proof.} By the deconcatenation formula \QBG\ for $Q M_{A}$, one can identify
$$ 
Q(M_A {\cal J}) = \sum_{XY=A} M_{X} M_{Y} {\cal J} - M_A Q {\cal J} 
$$
in the third line of
$$\eqalignno{
Q(M_i\otimes M_A {\cal J}) &=  Q( M_{iA} {\cal J} )\cr
&= \Big\{ M_i M_A+ \sum_{XY=A} M_{iX} M_{Y} \Big\} {\cal J} - M_{iA} Q {\cal J}
\cr
&= M_i M_A {\cal J}+ M_i \otimes \Big\{ \sum_{XY=A} M_{X} M_{Y} {\cal J} - M_{A} Q {\cal J} \Big\}
\cr
&= M_i M_A {\cal J} + M_i\otimes Q (M_A {\cal J}) \ .&\qed
}$$


\proclaim Corollary 1. Let ${\cal C}$ denote a BRST-invariant superspace expression
\eqn\prereqA{
{\cal C} = \sum_{k} M_{A_k}{\cal J}_k \,, \ \ \ Q {\cal C} = 0
}
where $\otimes$ acts trivially on the ${\cal J}_k$, then
\eqn\corA{
Q(M_i\otimes {\cal C}) = M_i {\cal C} \ .
}
\par
\noindent{\it Proof.} Upon applying Lemma 1 to $M_{A_k}{\cal J}_k$, the second term $M_i \otimes Q(M_{A_k}{\cal J}_k)$
in \lem\ builds up $M_i \otimes Q {\cal C}$ by linearity of $\otimes$ which vanishes by the assumption \prereqA.\hfill\qed


Note that \MAconcat\ and \MBconcat\ are special cases of \corA\ with ${\cal C}\rightarrow C_{j|A,B,C}$ and ${\cal C}\rightarrow C^m_{j|A,B,C,D}$, respectively.

\proclaim Corollary 2. Let ${\cal P}$ denote a BRST-pseudoinvariant superspace expression
\eqn\prereqB{
{\cal P} = \sum_{k} M_{A_k}{\cal J}_k \,, \ \ \ Q {\cal P} = \sum_{l} M_{B_l}{\cal Y}_l 
}
where the ${\cal Y}_l$ are anomalous and $\otimes$ does not act on ${\cal J}_k$ or ${\cal Y}_l$, then the right-hand side of
\eqn\corB{
Q(M_i\otimes {\cal P}) = M_i {\cal P} + M_i\otimes  Q {\cal P}
}
is anomalous up to the first term $M_i {\cal P}$.
\par
\noindent{\it Proof.} Again, apply Lemma 1 to $M_{A_k}{\cal J}_k$, and the second term $M_i \otimes Q(M_{A_k}{\cal
J}_k)$ in \lem\ builds up the expression $M_i \otimes Q {\cal P}$. The latter is anomalous by \prereqB\ since $M_i
\otimes$ action does not alter the anomaly nature of the $M_{B_l}{\cal Y}_l $ in $Q {\cal P}$.\hfill\qed

\subsec Rank-two example of pseudo-cohomology

\subseclab\threefour
As a first example of BRST pseudo-invariants, we derive a rank-two analogue of the recursions in \veca\ for scalar and
vectorial BRST invariants. According to the anomalous transformation \BRSTt\ of the tensorial building block
$M^{mn}_{\ldots}$ in \BGTmn, the resulting tensors $C^{m n}_{i|A,B,C,D,E}$ can at best be pseudo-invariant in the
sense of Definition 2. 

BRST pseudo-completions of rank-two tensors originate from an ansatz
\eqn\Cmndef{
C^{mn}_{i|A,B,C,D,E} \equiv M_i M^{mn}_{A,B,C,D,E} + \sum_{F\neq \emptyset} M_{iF} \cdots
}
similar to \Csc\ and \Cvec. Apart from the leading term $M_i M^{mn}_{A,B,C,D,E}$, particle $i$ always appears in a
multiparticle word $iF$, and the ellipsis represents tensor superfields of the form $M^{mn}_{A,B,C,D,E}, k^{(m}_A
M^{n)}_{B,C,D,E}$ or $k^{(m}_A k^{n)}_B M_{C,D,E}$.

Similar to the expressions \QMgen\ and \QMnrew\ for $QM_{A,B,C}$ and $QM^m_{A,B,C,D}$, one can express $QM^{mn}_{A,B,C,D,E}$ given in \BRSTt\ in terms of pseudoinvariants: Each term containing a factor of $M_i$ in \BRSTt\ signals the leading term of a (pseudo-)invariant, hence:
\eqnn\tsa
$$\eqalignno{
Q  M^{mn}_{A,B,C,D,E} &= \delta^{mn} {\cal Y}_{A,B,C,D,E}
+ \big[ C^{mn}_{a_1|a_2\ldots a_{|A|},B,C,D,E} -  C^{mn}_{a_{|A|}|a_1 \ldots a_{|A|-1},B,C,D,E} \cr
& + \d_{|A|,1} (k_{a_1}^{m} C^{n}_{a_1|B,C,D,E} +k_{a_1}^{n} C^{m}_{a_1|B,C,D,E})+ (A \leftrightarrow B,C,D,E) \big]\,.
&\tsa
}$$
This motivates the following recursion for pseudo-invariants $C^{mn}_{i|A,B,C,D,E}$,
\eqnn\tsb
$$\eqalignno{
C^{mn}_{i|A,B,C,D,E}& = M_i M^{mn}_{A,B,C,D,E} +M_i \otimes \big[  C^{mn}_{a_1|a_2\ldots a_{|A|},B,C,D,E}
-  C^{mn}_{a_{|A|}|a_1 \ldots a_{|A|-1},B,C,D,E} \cr
 &{}+  \d_{|A|,1}   (k_{a_1}^{m} C^{n}_{a_1|B,C,D,E}+k_{a_1}^{n}  C^{m}_{a_1|B,C,D,E})
 + (A \leftrightarrow B,C,D,E)  \big]\,  .
&\tsb \cr
}$$
Their BRST variation is purely anomalous by \tsa\ and \corB\ at ${\cal P} \rightarrow C^{mn}_{\ldots}$. The simplest
example occurs at six points and uses the expression \Cscex\ for $C^m_{2|3,4,5,6}$,
\eqnn\tsd
$$\eqalignno{
C^{mn}_{1|2,3,4,5,6} &= M_1 M^{mn}_{2,3,4,5,6}+ \big[ M_1 \otimes (k_2^{m}  C^{n}_{2|3,4,5,6} +k_2^{n} C^{m}_{2|3,4,5,6} )  + (2 \leftrightarrow 3,4,5,6)  \big] \cr
&= M_1 M^{mn}_{2,3,4,5,6} +  \big[ k_2^{m} M_{12} M^{n}_{3,4,5,6}+k_2^{n} M_{12} M^{m}_{3,4,5,6} + (3\leftrightarrow 4,5,6) \big] & \tsd \cr
& +  \big[ (k_2^{m} k_3^{n} +k_2^{n} k_3^{m}) (M_{123} + M_{132}) M_{4,5,6} + (2,3|2,3,4,5,6) \big] \,.
}$$
The seven-point analogue $C^{mn}_{1|23,4,5,6,7}$ is displayed in appendix \appA, see \sevent.

One can explicitly check their BRST pseudo-invariant nature at low orders,
\eqnn\tse
\eqnn\tsf
$$\eqalignno{
QC^{mn}_{1|2,3,4,5,6} &= {}-\delta^{mn} M_1 {\cal Y}_{2,3,4,5,6}&\tse \cr
QC^{mn}_{1|23,4,5,6,7} &= {}-\delta^{mn} ( M_1 {\cal Y}_{23,4,5,6,7}
+ M_{12} {\cal Y}_{3,4,5,6,7}- M_{13} {\cal Y}_{2,4,5,6,7})  \,. &\tsf
}$$
The general structure of $Q C^{mn}_{i|A,B,C,D,E} = {}-\delta^{mn}(\ldots)$ will be discussed in section \seventwo.
Note that traceless components are BRST-closed,
\eqn\tracelessCmn{
Q \Big( C^{mn}_{i|A,B,C,D,E} - {1\over 10}\delta^{mn} C^{pp}_{i|A,B,C,D,E} \Big) = 0.
}
The family of two-tensor pseudo-invariants constructed via \tsb\ furnishes the third cell in the leading diagonal of
the overview grid in \figoverview. As we will see in the next section, the tools of this section enable to address an infinite tower of
higher-rank generalizations to complete this diagonal.

\newsec Tensor pseudo-cohomology

\seclab\secfour
This section introduces a recursive method to construct higher-rank generalizations of the scalar, vector and
two-tensor BRST (pseudo-)invariants discussed in the previous sections.
As shown below, $C^{m_1 \ldots m_r}_{i|A_1, \ldots, A_{r+3}}$ for $r\ge 2$ is a pseudo-invariant according to 
Definition~2.

\subsec Higher-rank building blocks and anomaly blocks

\subseclab\fourone
Following the logic behind the two-tensor in \BGTmn, we define a building block of arbitrary rank $r$ by extracting
zero modes of $\Pi^{m_1}\ldots \Pi^{m_r} d_\alpha d_{\beta} N_{pq}$ from $r+3$ integrated vertex operators in their
multiparticle Berends--Giele version. Similarly, the $b$-ghost sector proportional to the zero-mode of $\Pi^m$ gives rise to
a second sort of superfield with a factor of ${\cal W}^m_{A,B,C,D}$:
\eqnn\HRa
$$\eqalignno{
M^{m_1\ldots m_r}_{B_1,B_2,\ldots,B_{r+3}} &\equiv r! \big[ M_{B_1,B_2,B_3} \cA_{B_4}^{(m_1} \cA_{B_5}^{m_2} \ldots \cA_{B_{r+3}}^{m_r)} + (B_1,B_2,B_3| B_1,B_2,\ldots, B_{r+3}) \big] \cr
&{} + r! \bigl[ \cW^{(m_1}_{B_1,B_2,B_3,B_4}  \cA_{B_5}^{m_2} \ldots \cA_{B_{r+3}}^{m_r)} + (B_1,\ldots,B_4|
B_1,B_2,\ldots, B_{r+3}) \big] &\HRa
}$$
In order to get a recursive handle on the combinatorics in \HRa, it is convenient to define higher-rank versions of
${\cal W}^m_{A,B,C,D}$ in \BRSTw,
\eqn\Wmsdef{
\cW^{m_1 \ldots m_{r-1}|m_r}_{B_1,B_2, \ldots,B_{r+3}} \equiv  \cA_{B_1}^{m_1} \cW^{m_2 \ldots m_{r-1}|m_r}_{B_2, \ldots, B_{r+3}} + (B_1 \leftrightarrow B_2, \ldots, B_{r+3}).
}
Then, the rank-$r$ building block $M^{m_1\ldots m_r}_{B_1,B_2,\ldots,B_{r+3}}$ in \HRa\ can be written recursively as
\eqn\Tmrecurs{
\eqalign{
M^{m_1\ldots m_r}_{B_1,\ldots,B_{r+3}} &= \cA^{m_1}_{B_1} M^{m_2 \ldots m_r}_{B_2, \ldots, B_{r+3}} +
\cA^{m_r}_{B_1} \cW^{m_{r-1} \ldots m_2|m_1}_{B_2, \ldots, B_{r+3}} + (B_1\leftrightarrow B_2,B_3, \ldots,B_{r+3})\cr
& = \bigl[ \cA^{m_1}_1 M^{m_2 \ldots m_r}_{B_2, \ldots, B_{r+3}} + (B_1\leftrightarrow B_2,B_3, \ldots,B_{r+3})\bigr] + \cW^{m_r m_{r-1} \ldots m_2|m_1}_{B_1,B_2, \ldots,B_{r+3}} \,,
}}
for example \BGTmn\ at rank two and
\eqnn\recrk
$$\eqalignno{
M^{mnp}_{B_1, \ldots,B_6} &= \cA^{m}_{B_1} M^{np}_{B_2, \ldots,B_6} + \cA^{p}_{B_1} \cW^{n|m}_{B_2, \ldots,B_6} +
(B_1\leftrightarrow B_2, \ldots, B_6) &\recrk \cr
M^{mnpq}_{B_1, \ldots,B_7} &= \cA^{m}_{B_1} M^{npq}_{B_2, \ldots,B_7} + \cA^{q}_{B_1} \cW^{pn|m}_{B_2, \ldots,B_7} +
(B_1\leftrightarrow B_2, \ldots, B_7)
}$$
at $r=3,4$. Note that $\cW^{m_1 \ldots m_{r-1}|m_r}_{B_1,B_2, \ldots,B_{r+3}}$ defined in \Wmsdef\ is symmetric in all
its slots $B_i$ but only in its first $r-1$ vector indices $m_i$. That explains the notation $\ldots m_{r-1}|m_r$ in
\Wmsdef.

Also the scalar anomaly building block ${\cal Y}_{A,B,C,D,E}$ defined in \WanonBG\ has a natural higher-rank generalization. It can be defined explicitly in analogy to \HRa,
\eqnn\HRc
$$\eqalignno{
\cY^{m_1\ldots m_r}_{B_1,B_2,\ldots,B_{r+5}} &\equiv r! \cY_{B_1,\ldots,B_5} \cA_{B_6}^{(m_1} \cA_{B_7}^{m_2} \ldots \cA_{B_{r+5}}^{m_r)} + (B_1,\ldots,B_5| B_1,\ldots, B_{r+5}) \ , &\HRc 
}$$
or recursively like \Tmrecurs,
\eqn\HRf{
\cY^{m_1\ldots m_r}_{B_1,B_2,\ldots,B_{r+5}} \equiv \cA_{B_1}^{m_1} \cY^{m_2 \ldots m_r}_{B_2,B_3,\ldots,B_{r+5}} + (B_1\leftrightarrow B_2,B_3,\ldots,B_{r+5})\,.
}
Even though the recursion \HRf\ for
anomaly blocks resembles \Wmsdef\ for $\cW^{m_1 \ldots m_{r-1}|m_r}_{B_1,B_2, \ldots,B_{r+3}}$, the vector indices of
the latter are not entirely carried by $\cA_{B}^m$ superfields. That is why only $\cY^{m_1\ldots
m_r}_{B_1,B_2,\ldots,B_{r+5}} $ is totally symmetric in both $B_i$ and $m_i$.

\subsec Anomalous BRST variations at higher rank

\subseclab\fourthree
Both expressions \HRa\ and \Tmrecurs\ for higher-rank building blocks serve as a starting point to determine their BRST variation
\eqnn\HRb
$$\eqalignno{
Q&M^{m_1\ldots m_r}_{B_1,B_2,\ldots,B_{r+3}} =   {r \choose 2}  \delta^{(m_1 m_2} \cY^{m_3 \ldots m_r)}_{B_1,B_2,\ldots,B_{r+3}} + \Big[  r M_{B_1} k_{B_1}^{(m_1} M^{m_2\ldots m_r)}_{B_2,B_3,\ldots,B_{r+3}} &\HRb \cr
&+ \sum_{XY=B_1} (M_X M^{m_1\ldots m_r}_{Y,B_2,B_3,\ldots,B_{r+3}} - M_Y M^{m_1\ldots m_r}_{X,B_2,B_3,\ldots,B_{r+3}}) + (B_1\leftrightarrow B_2,\ldots, B_{r+3})  \Big] . 
}$$
The $\delta^{mn}$ tensors in the anomalous part are due to the group-theory identity \BRSTxx. The
rank-two example has been given in \BRSTt, and ranks three and four give rise to
\eqnn\HRthree
\eqnn\HRfour
$$\eqalignno{
Q&M^{mnp}_{B_1,B_2,\ldots,B_{6}} =    3\, \delta^{(mn} \cY^{p)}_{B_1,B_2,\ldots,B_{6}}
+ \Big[  3 M_{B_1} k_{B_1}^{(m} M^{np)}_{B_2,B_3,\ldots,B_{6}} &\HRthree \cr
&+ \sum_{XY=B_1} (M_X M^{mnp}_{Y,B_2,B_3,\ldots,B_{6}} - M_Y M^{mnp}_{X,B_2,B_3,\ldots,B_{6}}) + (B_1\leftrightarrow B_2,\ldots, B_{6})  \Big] \cr
Q&M^{mnpq}_{B_1,B_2,\ldots,B_{7}} =    6\, \delta^{(mn} \cY^{pq)}_{B_1,B_2,\ldots,B_{7}}
+ \Big[  4 M_{B_1} k_{B_1}^{(m} M^{npq)}_{B_2,B_3,\ldots,B_{7}} &\HRfour \cr
&+ \sum_{XY=B_1} (M_X M^{mnpq}_{Y,B_2,B_3,\ldots,B_{7}} - M_Y M^{mnpq}_{X,B_2,B_3,\ldots,B_{7}}) + (B_1\leftrightarrow B_2,\ldots, B_{7})  \Big] .
}$$
The recursive approach makes use of the BRST variation
\eqnn\HRww
$$\displaylines{
Q\cW^{m_1\ldots m_{r-1}|m_r}_{B_1,B_2,\ldots,B_{r+3}} =   (r-1) \delta^{m_r(m_1} \cY^{m_2 \ldots
m_{r-1})}_{B_1,B_2,\ldots,B_{r+3}}  \hfil\HRww\hfilneg \cr
 + \Big[  (r-1) M_{B_1} k_{B_1}^{(m_1} \cW^{m_2\ldots m_{r-1})|m_r}_{B_2,B_3,\ldots,B_{r+3}} - (\lambda \gamma^{m_r} \cW_{B_1}) M^{m_1\ldots m_{r-1}}_{B_2,B_3,\ldots,B_{r+3}} \cr
 \ \ \ + \!\!\sum_{XY=B_1}\!\! (M_X \cW^{m_1\ldots m_{r-1}|m_r}_{Y,B_2,\ldots,B_{r+3}} - M_Y \cW^{m_1\ldots m_{r-1}|m_r}_{X,B_2,\ldots,B_{r+3}}) + (B_1\leftrightarrow B_2,\ldots, B_{r+3})  \Big]  \,,
}$$
which generalizes the rank-one variation \BRSTQw\ and specializes as follows at rank $r \leq 3$,
\eqnn\HRwwtwo
\eqnn\HRwwthree
$$\eqalignno{
Q&\cW_{B_1,B_2,\ldots,B_5}^{m|n} = \delta^{mn}\cY_{B_1,B_2,\ldots,B_5} +\bigl[ k_{B_1}^m M_{B_1} \cW^n_{B_2,B_3,B_4,B_5} - (\l\g^n \cW_{B_1})M^m_{B_2,B_3,B_4,B_5} \cr
&\quad{}\ \ \ \ \ \ + \sum_{XY=B_1} (M_X \cW^{m|n}_{Y,B_2,\ldots,B_{5}} - M_Y \cW^{m|n}_{X,B_2,\ldots,B_{5}}) + (B_1\leftrightarrow B_2,B_3,B_4,B_5)\bigr] &\HRwwtwo \cr
Q&\cW_{B_1,B_2,\ldots,B_6}^{mn|p} = 2\delta^{p(m}\cY^{n)}_{B_1,B_2,\ldots,B_6} +\bigl[ 2 k_{B_1}^{(m} M_{B_1} \cW^{n) |p}_{B_2,\ldots,B_6} - (\l\g^p \cW_{B_1})M^{mn}_{B_2,B_3,B_4,B_5,B_6} \cr
&\quad{} \ \ \ \ \ \ + \sum_{XY=B_1} (M_X \cW^{mn|p}_{Y,B_2,\ldots,B_{6}} - M_Y \cW^{mn|p}_{X,B_2,\ldots,B_{6}}) +
(B_1\leftrightarrow B_2,\ldots,B_6)\bigr]\,. &\HRwwthree
}$$

\subsec Recursion for higher rank pseudoinvariants

\subseclab\fourfour
The construction of general BRST pseudo-invariants
\eqnn\HRg
$$\eqalignno{
&C_{i|A_1,A_2,\ldots,A_{r+3}}^{m_1\ldots m_r} \equiv M_i M^{m_1\ldots m_r}_{A_1,A_2,\ldots,A_{r+3}} + \sum_{B\neq \emptyset} M_{iB}\cdots
&\HRg
}$$
generalizes the scalars \Csc, vectors \Cvec\ and two-tensors \Cmndef\ to arbitrary rank. As before, the leading term
$M_i M^{m_1\ldots m_r}_{A_1,A_2,\ldots,A_{r+3}}$ is the only instance where the reference leg~$i$ enters
through a single-particle vertex operator $M_i$. The ellipsis along with multiparticle $M_{iB}$ takes the form
$k_{A_1}^{(m_1} \ldots k_{A_j}^{m_j} M^{m_{j+1}\ldots m_r)}_{A_{j+1},\ldots, A_{r+3}}$ with $j=0,1,\ldots,r$. The role
of $M_i$ as defining the pseudoinvariant \HRg\ leads to the following alternative form of \HRb:
\eqnn\HRh
$$\displaylines{
Q M^{m_1\ldots m_r}_{A_1,A_2,\ldots,A_{r+3}} =
{r \choose 2} \delta^{(m_1 m_2} {\cal Y}^{m_3 \ldots m_r)}_{A_1,A_2,\ldots,A_{r+3}}
+ \big\{ r \delta_{|A_1|,1}  k_{a_1}^{(m_1} C^{m_2\ldots m_r)}_{a_1|A_2,\ldots,A_{r+3}} \hfil\HRh\hfilneg\cr
{}+ C^{m_1\ldots m_r}_{a_1| a_2\ldots a_{|A_1|},A_2,\ldots,A_{r+3}}- C^{m_1\ldots m_r}_{a_{|A_1|}| a_1\ldots a_{|A_1|-1},A_2,\ldots,A_{r+3}}
+(A_1 \leftrightarrow A_2,\ldots,A_{r+3}) \big\} \ .
}$$
This in turn gives rise to a recursion for the pseudo-invariants $C_{i|A_1,A_2,\ldots,A_{r+3}}^{m_1\ldots m_r}$ in
terms of lower-multiplicity representatives of rank $r$ and $r-1$,
\eqnn\HRi
$$\displaylines{
C^{m_1\ldots m_r}_{i|A_1,A_2,\ldots,A_{r+3}} =M_i M^{m_1\ldots m_r}_{A_1,A_2,\ldots,A_{r+3}} + M_i \otimes
 \big\{ r \delta_{|A_1|,1}  k_{a_1}^{(m_1} C^{m_2\ldots m_r)}_{a_1|A_2,\ldots,A_{r+3}} \hfil\HRi\hfilneg\cr
{}+ C^{m_1\ldots m_r}_{a_1| a_2\ldots a_{|A_1|},A_2,\ldots,A_{r+3}}- C^{m_1\ldots m_r}_{a_{|A_1|}| a_1\ldots a_{|A_1|-1},A_2,\ldots,A_{r+3}} +(A_1 \leftrightarrow A_2,\ldots,A_{r+3}) \big\} \ ,
}$$
which reduces to \veca\ and \tsb\ for $r \leq 2$. BRST pseudoinvariance follows from \corB\ at
${\cal P} \rightarrow C^{m_1 m_2\ldots m_r}_{\ldots}$. The anomalous BRST variations entirely reside
in trace components $\sim \delta^{m_i m_j}$ and will be systematically discussed in section~\seventwo.
Similar as before, the traceless components are BRST invariant, e.g.
\eqnn\traceinv
$$\eqalignno{
Q \Big( C^{mnp}_{i|A,B,C,D,E,F} - {1\over4 } \delta^{(mn} C^{p)qq}_{i|A,B,C,D,E,F}  \Big) &=0 \ .
&\traceinv
}$$
The simplest pseudoinvariant of rank greater than two is $C^{mnp}_{1|2,3,4,5,6,7}$, its expansion is displayed in appendix \appA, see \seventt.

The recursion \HRi\ for pseudo-invariants of arbitrary rank completes the leading diagonal of the overview grid in
\figoverview. In the next sections we explore the building blocks and recursions governing the subleading diagonals.

\newsec Towards a refined pseudo-cohomology

\seclab\secfive
The discussion of BRST invariance of the closed-string five-point amplitude in \oneloopMichael\ naturally
led to consider the following combination of superfields\foot{In various places of this section, we will encounter the local
representatives $V_A$, $T_{A,B,C}$, $W^m_{A,B,C,D}$, $T^m_{A,B,C,D}$ and $Y_{A,B,C,D,E}$ of the more frequently-used
Berends--Giele superfields $M_A$, $M_{A,B,C}$, $W^m_{A,B,C,D}$, $M^m_{A,B,C,D}$ and ${\cal Y}_{A,B,C,D,E}$ as
defined by \QBGV, \BRSTMi, \BRSTw, \BRSTm\ and \WanonBG. They follow by trading
any ${\cal K}_B \in \{  \cA^{B}_\alpha, \cA^m_{B} , \cW_{B} ^\alpha, \cF_{B} ^{mn} \}$ in these definitions for the standard
BRST blocks $K_B \in \{ A^{B}_\alpha, A^m_{B} , W_{B} ^\alpha, F_{B} ^{mn} \}$, see section \twothree. Some of their BRST variations are displayed in appendix \appC.}
\eqn\introJ{
k_1^m V_1 T^m_{2,3,4,5}+\big[ V_{12} T_{3,4,5} + (2 \leftrightarrow 3,4,5) \big] + Y_{1,2,3,4,5}
}
which was shown to be BRST exact in the appendix B of \oneloopMichael. Given the appearance of the anomalous
superfield $Y_{1,2,3,4,5}$, it will not be surprising to discover that this particular combination \introJ\ signals a much deeper family of pseudo cohomology elements which will play an important role in
the discussion of anomalous terms in the one-loop open superstring amplitudes \wipG.

\subsec Refined currents

\subseclab\fiveone
It turns out that to extend and generalize the discussion of \introJ\ it is convenient to define the following superfield
\eqn\Jhatdef{
\widehat J_{A|B,C,D,E} \equiv {1 \over 2} (A^m_A T^m_{B,C,D,E}+A^m_A W^m_{B,C,D,E})\,,
}
symmetric in $B,C,D,E$. In view of the special role of the first slot $A$, we refer to such objects as {\it refined} currents. Accordingly, any slot $A|\ldots$ on the left of the vertical bar of the subscript will be referred to as refined. It is not hard to check that the simplest case $\widehat J_{1|2,3,4,5}$ gives rise to \introJ\ under $Q$ variation and that higher-multiplicity
currents satisfy
\eqnn\pseuda
\eqnn\pseudaa
\eqnn\pseudaaa
$$\eqalignno{
Q\widehat J_{1|2,3,4,5} &= k_1^m V_1 T^m_{2,3,4,5}+\big[ V_{12} T_{3,4,5} + (2 \leftrightarrow 3,4,5) \big] + Y_{1,2,3,4,5} &\pseuda \cr
Q\widehat J_{12|3,4,5,6} &= k_{12}^m V_{12} T^m_{3,4,5,6}+\big[ \widehat V_{123} T_{4,5,6} + (3 \leftrightarrow 4,5,6) \big] + Y_{12,3,4,5,6}  \cr
&\quad{} + s_{12}(V_1 \widehat J_{2|3,4,5,6}-V_2 \widehat J_{1|3,4,5,6})&\pseudaa \cr
Q\widehat J_{1|23,4,5,6} &= k_1^m V_1 T^m_{23,4,5,6}-\widehat V_{231} T_{4,5,6}+\big[ V_{14} T_{23,5,6} + (4 \leftrightarrow 5,6) \big] \cr
&\quad{} + Y_{1,23,4,5,6}+ s_{23}(V_2 \widehat J_{1|3,4,5,6}-V_3 \widehat J_{1|2,4,5,6})  \ . &\pseudaaa
}$$
Both $V_{12}$ in \pseuda\ and the hatted superfields $\widehat V_A$ in \pseudaa\ and \pseudaaa\ build up through the recursions \Atwo\ and \Athree\ for $A_{12}^{\alpha}$ and $\widehat A_{123}^{\alpha}$. More generally, the recursion \Arankp\ relates the multiparticle spinor superpotential $A_{\alpha}^B$ to BRST blocks $K_C$ at lower multiplicity $|C|<|B|$ which are generated by $Q\widehat J_{A|B,C,D,E}$. However, the direct output $\widehat{A}_{\alpha}^B$ of the recursion requires redefinitions by BRST trivial components $H_{12\ldots p}\equiv H_{[12\ldots p-1,p]}$ in order to yield the BRST block $A_{\alpha}^B$ subject to Lie symmetries, see \redefsthree\ and \eombbs. The appearance of $\widehat V_B = \lambda^{\alpha} \widehat{A}_{\alpha}^B$ in \pseudaa\ and \pseudaaa\ suggests to redefine
$\widehat J$ by the tensors $H_{ijk} \equiv H_{[ij,k]}$ in \Rthree\ such that their $Q$ variation can be expressed in terms of the BRST block $V_B = \lambda^{\alpha} A_{\alpha}^B$, e.g.
\eqnn\redefJ
$$\eqalignno{
J_{1|2,3,4,5} &\equiv \widehat J_{1|2,3,4,5}\cr
J_{12|3,4,5,6} &\equiv \widehat J_{12|3,4,5,6}
	- \bigl[ H_{[12,3]}\Tijk,4,5,6 + (3\leftrightarrow 4,5,6)\bigr]\,,&\redefJ\cr
J_{1|23,4,5,6} &\equiv \widehat J_{1|23,4,5,6} + H_{[23,1]}T_{4,5,6}\,.
}$$
Generalizations $H_{[A,B]}$ of the redefining superfields are explained in appendix \appD. They give
rise to
\eqn\Jnohatdef{
J_{A|B,C,D,E} \equiv \widehat J_{A|B,C,D,E} - \bigl[ H_{[A,B]}T_{C,D,E} + (A\leftrightarrow B,C,D,E)\bigr] \ ,
}
with the understanding that $H_{[A,B]}=0$ for $|A|=|B|=1$ \eombbs.
After the redefinition \Jnohatdef, the BRST transformation of $J_{A|B,C,D,E}$
contains BRST blocks $V_X$ rather than $\widehat V_X$,
\eqnn\pseudd
$$\eqalignno{
Q J_{A|B,C,D,E}& = k_A^m V_A T^m_{B,C,D,E} + V_{[A,B]} T_{C,D,E} + V_{[A,C]} T_{B,D,E}
\cr
&+ V_{[A,D]} T_{B,C,E} + V_{[A,E]} T_{B,C,D}+Y_{A,B,C,D,E} 
+ {\cal O}(k_i\cdot k_j) \ .&\pseudd
}$$
Appendix \appC\ displays the four inequivalent seven-point examples of \pseudd, including the contact terms 
$ {\cal O}(k_i\cdot k_j)$, see \pseude.
The latter represent the generalization of $s_{12}(V_1 \widehat J_{2|3,4,5,6}-V_2 \widehat J_{1|3,4,5,6})$ in \pseudaa\ and $s_{23}(V_2 \widehat J_{1|3,4,5,6}-V_3 \widehat J_{1|2,4,5,6}) $ in \pseudaaa\ which simplifies once the $J_{A|B,C,D,E}$ in \Jnohatdef\ are converted to Berends--Giele currents ${\cal J}_{A|B,C,D,E}$.

The bracket notation $V_{[A,B]}$ has been explained in the appendix A of \eombbs\ and can be diagrammatically understood from figure \figseven. A few explicit examples are as
follows\foot{As explained in \eombbs, the multiparticle label $B=b_1b_2 \ldots b_{|B|}$ in a BRST block $K_B$ satisfies Lie
symmetries. They can be elegantly incorporated by writing $B=[ \ldots[[[b_1,b_2],b_3],b_4], \ldots,b_{|B|}]$ and using Jacobi identities for iterated brackets. In particular $B=23$
translates into $B=[2,3]$. Furthermore, the translation from bracketed to non-bracketed labels is given by $K_{[
\ldots[[[1,2],3],4], \ldots,|B|]}=K_{123 \ldots|B|}$, for example $K_{[[1,2],3]} = K_{123}$.},
\eqnn\notationV
$$\displaylines{
V_{[1,2]}=V_{12},\quad V_{[12,3]} = V_{123},\quad V_{[12\ldots p-1,p]}=V_{12\ldots p} \hfil\notationV\hfilneg\cr
V_{[1,23]} = V_{123}-V_{132} = -V_{231},\quad  V_{[12,34]} = V_{1234}-V_{1243}=-V_{3412}+V_{3421} \ .
}$$

\subsec Berends--Giele version of refined currents

\subseclab\fivetwo
The Berends--Giele version ${\cal J}_{A|B,C,D,E}$ of refined currents $J_{A|B,C,D,E}$ in \Jnohatdef\ can be
obtained by applying the Berends--Giele map discussed in section \twothree\ to each of its five slots.
The resulting definition\foot{As will become clear in later sections, $\cJ_{A|B,C,D,E}$ should really be denoted
$M_{A|B,C,D,E}$ since many formulas would acquire a more natural interpretation.
However, we use the notation of \JBGdef\ for hysterical reasons.}
\eqn\JBGdef{
{\cal J}_{A|B,C,D,E} \equiv {1 \over 2} (\cA^m_A M^m_{B,C,D,E}+\cA^m_A \cW^m_{B,C,D,E}) - \bigl[ {\cal H}_{[A,B]}M_{C,D,E} + (A\leftrightarrow B,C,D,E)\bigr]
}
incorporates the Berends--Giele version ${\cal H}_{[A,B]}$ of the above superfields $H_{[A,B]}$, see appendix \appD\ and in particular \ArankAM\ for examples. The contact terms in $QJ_{A|B,C,D,E}$ translate into deconcatenations in $Q{\cal J}_{A|B,C,D,E}$ in the same way as contact terms in $QV_B$
are mapped into the deconcatenation formula \QBG\ for $Q M_B$. Moreover, $k_A^m V_A T^m_{B,C,D,E}$ and $Y_{A,B,C,D,E}$ on the right-hand side of \pseudd\ can be
straightforwardly replaced by $k_A^m M_A M^m_{B,C,D,E}$ and ${\cal Y}_{A,B,C,D,E}$, respectively.
Only the four permutations of $V_{[A,B]} T_{C,D,E}$ require closer inspection since
their expansion in terms of $M_{X}M_{C,D,E}$ will introduce explicit
Mandelstam variables. 

\subsubsec The $S[A,B]$ map

At the superfield level, the recursive definition of BRST blocks in \eombbs\ has the structure of a commutator;
$V_{[A,B]} \rightarrow [V_A,V_B]$.
At the level of diagrams, $V_{[A,B]}$ can be interpreted as connecting the off-shell legs in the subdiagrams represented by $V_A$ and $V_B$ through a
cubic vertex, see \eombbs\ and \figseven. Expanding any $V_C$ in terms of Berends--Giele currents $M_C$ gives
rise to a similar diagrammatic interpretation shown in \figseven, i.e. if two Berends--Giele currents $M_A$ and $M_B$ are
attached to a cubic vertex, the resulting diagram is a linear combination of currents $M_C$ at overall multiplicity
$|C|=|A|+|B|$. We denote this linear combination by $M_{S[A,B]}$, where the letter $S$ reminds of a factor of $s_{ij}$ which enters on
dimensional grounds. In other words, the $S[A,B]$ map captures the difference of applying the Berends--Giele
map as described in section \twothree\ to the
multiparticle label $C$ as a whole as compared to applying it simultaneously and individually to $A$ and $B$, where $|C| = |A|+|B|$,
\eqn\SmapVM{
V_{[A,B]} = [V_A,V_B] \ \ \Longrightarrow \ \ M_{S[A,B]} = [M_A,M_B]\,.
}
For example, converting both sides of $V_{12} = [V_1,V_2]$ to Berends--Giele currents leads to $s_{12}M_{12} =
[M_1,M_2]$ and therefore $M_{S[1,2]} = s_{12}M_{12}$. Similarly, converting both sides of $V_{123} = [V_{12},V_3]$ to
Berends--Giele currents gives
\eqn\rankthreeEx{
s_{12}(s_{23}M_{123} - s_{13}M_{213}) = [s_{12} M_{12},M_3] \ \ \Longrightarrow \ \ M_{S[12,3]} = s_{23}M_{123} - s_{13}M_{213} \ .
}
To find $S[1,23]$ one repeats the analysis with
$[V_1,V_{23}] = - [V_{23},V_1]$ and uses the antisymmetry $M_{S[A,B]}=-M_{S[B,A]}$ due to \SmapVM. Following this procedure one obtains,
\eqnn\QEoneEx
$$\eqalignno{
M_{S[1,2]} &= s_{12} M_{12}&\QEoneEx\cr
M_{S[1,23]} &= s_{12} M_{123} - s_{13}M_{132}\cr
M_{S[1,234]} &= s_{12} M_{1234} - s_{13}(M_{1324} + M_{1342}) + s_{14}M_{1432}\cr
M_{S[12,34]} &= - s_{13} M_{2134} + s_{14}M_{2143} + s_{23}M_{1234} - s_{24}M_{1243}\,.
}$$

\ifig\figseven{Diagrammatic interpretation of $M_{S[A,B]}$.}
{\epsfxsize=0.70\hsize\epsfbox{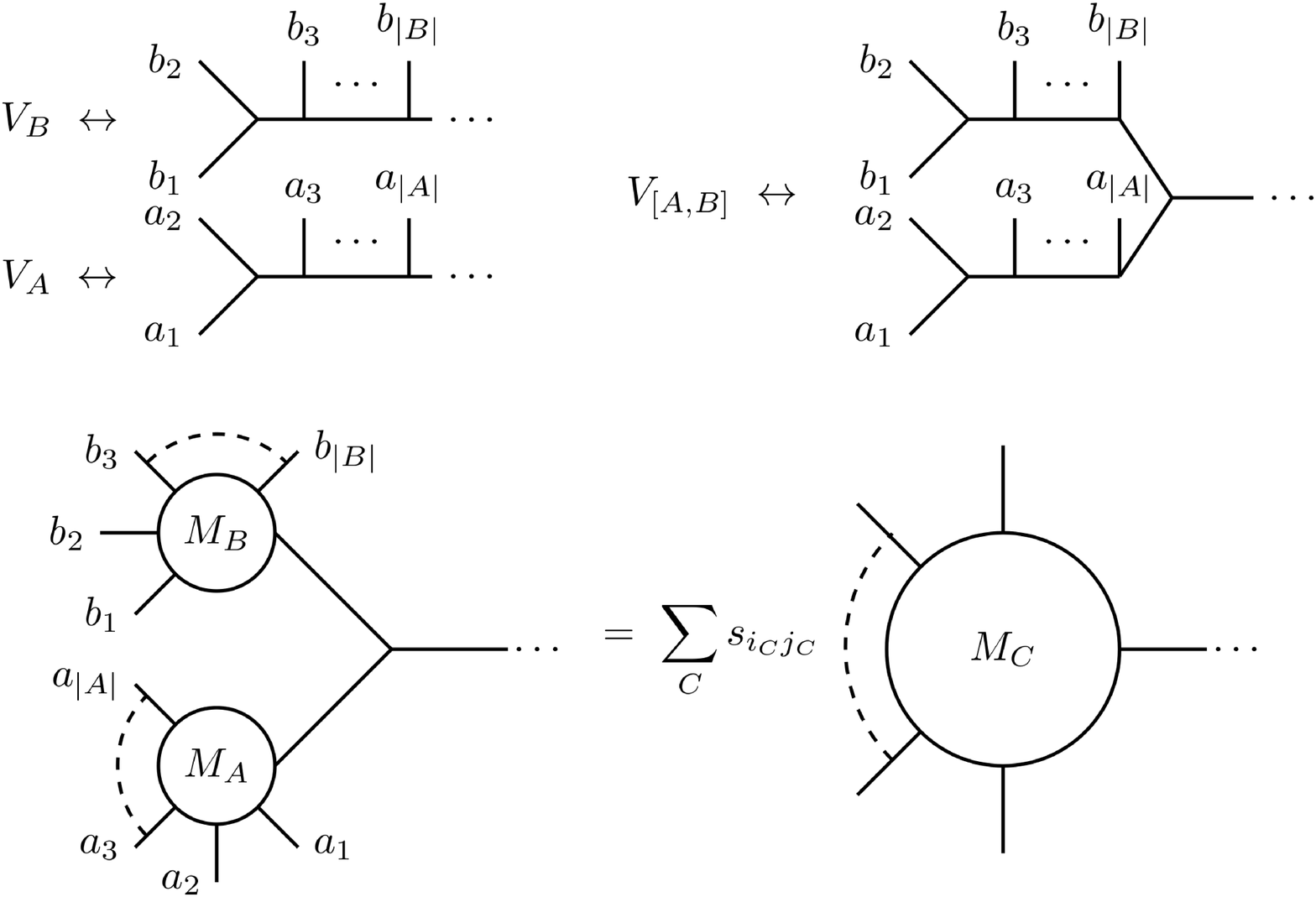}}

It turns out that the general formula for $M_{S[A,B]}$ reads,
\eqn\QEone{
M_{S[A,B]} \equiv \sum_{i=1}^{|A|} \sum_{j=1}^{|B|} (-1)^{i-j+|A|-1}
s_{a_i b_j} M_{(a_1 a_2\ldots a_{i-1} \shuffle a_{|A|} a_{|A|-1}\ldots a_{i+1})a_ib_j
(b_{j-1}\ldots b_2 b_1 \shuffle b_{j+1} \ldots b_{|B|})}  \,.
}
The $S[A,B]$ map has been investigated in appendix B of \eombbs\ in a different context -- it facilitates the expansion of $C_{i|A,B,C}$ in terms of SYM tree subamplitudes.

\subsubsec The BRST variation of refined currents

Let us compare the $M_{S[A,B]}$ in \QEoneEx\ with the BRST transformations of various ${\cal J}_{A|B,C,D,E}$, starting with the trivial five point case,
\eqnn\pseudii
$$\eqalignno{
Q{\cal J}_{1|2,3,4,5} &= k_{1}^m M_{1} M^m_{2,3,4,5}+\big[ s_{12}M_{12} M_{3,4,5} + (2 \leftrightarrow 3,4,5) \big] + {\cal Y}_{1,2,3,4,5} \ . &\pseudii 
}$$
At six points there are two inequivalent partitions of legs in ${\cal J}$ satisfying
\eqnn\pseudi
\eqnn\pseudj
$$\eqalignno{
Q{\cal J}_{12|3,4,5,6} &= k_{12}^m M_{12} M^m_{3,4,5,6}+\big[ (s_{23}M_{123}-s_{13}M_{213}) M_{4,5,6} + (3 \leftrightarrow 4,5,6) \big] \cr
& \ \ \ \ + {\cal Y}_{12,3,4,5,6} + M_1 {\cal J}_{2|3,4,5,6} - M_2 {\cal J}_{1|3,4,5,6} &\pseudi \cr
Q{\cal J}_{1|23,4,5,6} &= k_{1}^m M_{1} M^m_{23,4,5,6}+(s_{12}M_{123}-s_{13}M_{132})M_{4,5,6}+\big[ s_{14}M_{14} M_{23,5,6} + (4 \leftrightarrow 5,6) \big] \cr 
& \ \ \ \ + {\cal Y}_{1,23,4,5,6} +  M_2 {\cal J}_{1|3,4,5,6} -  M_3 {\cal J}_{1|2,4,5,6} \ .  &\pseudj 
}$$
The four inequivalent seven-point examples are displayed in appendix \appC, see \pseudk.

Using the $S[A,B]$ map in \QEone, we can write down a general formula for the BRST variation of refined currents,
\eqnn\pseudoo
$$\eqalignno{
Q&{\cal J}_{A|B,C,D,E} =  {\cal Y}_{A,B,C,D,E} + k_A^m M_A M^m_{B,C,D,E} + \sum_{XY=A} (M_X {\cal J}_{Y|B,C,D,E} - M_Y {\cal J}_{X|B,C,D,E})
 \cr
&+\Bigl[ M_{S[A,B]} M_{C,D,E}+ \sum_{XY=B} (M_{X} {\cal J}_{A|Y,C,D,E}  - M_{Y} {\cal J}_{A|X,C,D,E}) + (B
\leftrightarrow C,D,E) \Bigr] \,. &\pseudoo
}$$
It is amusing that the anomaly building block ${\cal Y}_{A,B,C,D,E}$ is completely symmetric in $A, B,C,D,E$ whereas
${\cal J}_{A|B,C,D,E}$ has a reduced symmetry in $B,C,D,E$ due to the refined slot $A$. Note that the
non-anomalous part of the right-hand side contains the same kind of terms as they appear in $Q M_{A,B,C}$ and $Q
M^m_{A,B,C,D}$, see \QMABC\ and \BRSTQm. This allows to assemble superfields ${\cal J}_{A|B,C,D,E} , \ k^m_F M_A M^m_{B,C,D,E}$ and $s_{ij} M_A M_{B,C,D}$ into BRST-pseudoinvariants, the
details are worked out in following section.

\subsec Scalar pseudo cohomology

\subseclab\fivethree
Scalar BRST pseudoinvariants involving refined currents ${\cal J}_{A|B,C,D,E}$ are defined along the lines
of the pseudoinvariants $C^{m_1\ldots m_r}_{i|A_1,\ldots,A_{r+3}}$
in \HRg\foot{As will become clear in later sections, $P_{i|A|B,C,D,E}$ should really be denoted
$C_{i|A|B,C,D,E}$ since many formulas would acquire a more natural interpretation.
However, we use the notation of \Pdef\ for hysterical reasons.},
\eqn\Pdef{
P_{i|A|B,C,D,E} \equiv M_i {\cal J}_{A|B,C,D,E} + \sum_{F\neq \emptyset} M_{iF} \ldots \ .
}
The leading term $M_i {\cal J}_{A|B,C,D,E}$ furnishes the only instance of a single-particle vertex $M_i$ and therefore
defines the reference leg $i$ as well as the multiparticle labels of the pseudoinvariant $P_{i|A|B,C,D,E}$. The suppressed terms in the $\ldots$ along with a
multiparticle $M_{iF}$ follow from demanding $QP_{i|A|B,C,D,E}$ to be purely anomalous. We determine them by writing
\pseudoo\ in terms of vector invariants from section \twofive\ and pseudoinvariants \Pdef,
\eqnn\pseudop
$$\eqalignno{
Q{\cal J}_{A|B,C,D,E} &= {\cal Y}_{A,B,C,D,E} + \d_{|A|,1} k_{a_1}^m C^m_{a_1|B,C,D,E} & \pseudop\cr
&\quad{}+ P_{a_1 |a_{2} \ldots a_{|A|}|B,C,D,E} - P_{a_{|A|} |a_{1} \ldots a_{|A|-1}|B,C,D,E}   \cr
&\quad{}+ \big[ P_{b_1 |A|b_{2} \ldots b_{|B|},C,D,E} - P_{b_{|B|} |A|b_{1} \ldots b_{|B|-1},C,D,E}  + (B \leftrightarrow
C,D,E)\big]\,.
}$$
As usual, any $M_i$ in \pseudoo\ has been identified as a leading term of some $C^m_{i|A,B,C,D}$ or $P_{i|A|B,C,D,E}$,
see \Cvec\ and \Pdef. Furthermore, equation \pseudop\ can be verified once the explicit form of the
pseudo-invariants $P_{i|A|B,C,D,E}$ to be determined below is plugged in and the result compared to \pseudoo.
By \corB, the non-anomalous terms in \pseudop\ drop out from the BRST variation of the following recursion:
\eqnn\pseudoq
$$\eqalignno{
P_{i|A|B,C,D,E} &= M_i {\cal J}_{A|B,C,D,E} + M_i \otimes \big\{ \d_{|A|,1} k_{a_1}^m C^m_{a_1|B,C,D,E} \cr
&\quad{} +  P_{a_1 |a_{2} \ldots a_{|A|}|B,C,D,E} -  P_{a_{|A|} |a_{1} \ldots a_{|A|-1}|B,C,D,E}  & \pseudoq  \cr
&\quad{} +   \bigl[P_{b_1 |A|b_{2} \ldots b_{|B|},C,D,E} - P_{b_{|B|} |A|b_{1} \ldots b_{|B|-1},C,D,E}  + (B \leftrightarrow C,D,E) \bigr]  \} \ .
}$$
When applied to the simplest six-point example, the recursion yields
\eqnn\pseudos
$$\eqalignno{
P_{1|2|3,4,5,6} &= M_1 {\cal J}_{2|3,4,5,6} + k_2^m M_1 \otimes C^m_{2|3,4,5,6} & \pseudos\cr
&=  M_1 {\cal J}_{2|3,4,5,6} + M_{12} k_2^m M_{3,4,5,6}^m + \bigl[ s_{23} M_{123} M_{4,5,6} + (3 \leftrightarrow 4,5,6)\bigr]  \,,
}$$
and the two inequivalent seven-point analogues are displayed in appendix \appA, see \sevenpt\ and \sevenptt. For
these simple cases, pseudoinvariance is still easy to check explicitly,
\eqnn\IndeedPseudo
$$\eqalignno{
QP_{1|2|3,4,5,6} &=-M_1 {\cal Y}_{2,3,4,5,6}&\IndeedPseudo\cr
QP_{1|23|4,5,6,7} &= -M_1 {\cal Y}_{23,4,5,6,7} - M_{12} {\cal Y}_{3,4,5,6,7} + M_{13} {\cal Y}_{2,4,5,6,7}
\cr
Q P_{1|2|34,5,6,7} &= -M_1 {\cal Y}_{2,34,5,6,7} - M_{13} {\cal Y}_{2,4,5,6,7}+ M_{14} {\cal Y}_{2,3,5,6,7}\,.
}$$
A general discussion of $Q P_{i|A|B,C,D,E}$ is given in the later section \sevenfour. Note that the $P_{i|A|B,C,D,E}$
furnish the first cell of the subleading diagonal in the overview grid in \figoverview.

\subsec Scalar pseudoinvariants versus tensor traces

\subseclab\fivefour
The definition \JBGdef\ of the refined current ${\cal J}_{A|B,C,D,E}$ exhibits a strong similarity to the trace of the
two tensor $M^{mn}_{A,B,C,D,E}$ in \BGTmn. Only the redefinitions by $H_{[A,B]}=-H_{[B,A]}$ terms might pose an obstruction,
but their antisymmetry implies that the difference between $\widehat J_{A|B,C,D,E}$ and $J_{A|B,C,D,E}$ in \Jnohatdef\ drops out upon
symmetrization in $A,B,C,D,E$,
\eqn\symmABCDE{
\widehat J_{A|B,C,D,E} -  J_{A|B,C,D,E} + (A\leftrightarrow B,C,D,E)= 0\,.
}
Similarly, the ${\cal H}_{[A,B]}$ corrections in \JBGdef\ cancel out when symmetrizing their corresponding Berends--Giele
currents and one gets
\eqn\Tmntrace{
\delta_{mn}M^{mn}_{A,B,C,D,E} = 2{\cal J}_{A|B,C,D,E} + (A\leftrightarrow B,C,D,E)\ . }
After multiplication with $M_i$, \Tmntrace\ can be viewed as relating the leading terms of
$\delta_{mn}C^{mn}_{i|A,B,C,D,E}$ and permutations of \Pdef, leading to
\eqn\Cmntrace{
\delta_{mn}C^{mn}_{i|A,B,C,D,E} = 2P_{i|A|B,C,D,E} + (A\leftrightarrow B,C,D,E).
}
In other words, scalar pseudoinvariants $P_{i|A|B,C,D,E} $ describe the tensor trace of
$C^{mn}_{i|A,B,C,D,E}$ in terms of more fundamental objects.
Similarly, it will be shown in the next section that traces of higher-rank pseudo-invariants $\delta_{m_1
m_2}C^{m_1\ldots m_r}_{i|A_1,\ldots,A_{r+3}}$ decompose into tensorial generalizations of
$P_{i|A|B,C,D,E} $.
Starting from rank two, the latter give rise to traces by themselves (corresponding to double
traces of $C^{m_1\ldots m_r}_{i|A_1,\ldots,A_{r+3}}$), and one can anticipate an infinite number of all-rank families
of pseudoinvariants. These are the different diagonals in the overview grid in \figoverview\ where contractions with
$\delta_{mn}$ move any tensorial object downwards to the next diagonal. Since an individual $P_{i|A|B,C,D,E}$ contains
more information than the trace $\delta_{mn}C^{mn}_{i|A,B,C,D,E}$, we refer to the former as belonging to the refined
pseudo-cohomology.

\newsec Generalizing the refined pseudo-cohomology

\seclab\secsix
In this section, we generalize the refined currents \JBGdef\ in two directions: firstly by defining tensorial
counterparts and secondly by increasing the number of refined slots such as the distinguished word $A$ in ${\cal
J}_{A|B,C,D,E}$. Each of these currents gives rise to a pseudoinvariant which can be recursively constructed along the
lines of sections \fourfour\ and \fivethree.

\subsec Higher-rank refined currents and their anomaly

\subseclab\sixone
We define the higher-rank version of the scalar refined current \JBGdef\ by
\eqnn\Jaconee
$$\eqalignno{
\cJ^{m_1 \ldots m_r}_{A|B_1, \ldots, B_{r+4}} &\equiv \half \cA_{A}^p \bigl[
M^{pm_1 \ldots m_r }_{B_1, \ldots,B_{r+4}} +
\cW^{m_1 \ldots m_r | p}_{B_1, \ldots,B_{r+4}}\bigr]\cr
&\qquad{}- \big[ {\cal H}_{[A,B_1]}M^{m_1 \ldots m_r}_{B_2, \ldots,B_{r+4}}
+ (B_1\leftrightarrow B_2,B_3, \ldots, B_{r+4})\big]
&\Jaconee
}$$
in terms of higher-rank building blocks $M^{pm_1 \ldots m_r }_{B_1, \ldots,B_{r+4}}$ and $\cW^{m_1 \ldots m_r |
p}_{B_1, \ldots,B_{r+4}}$ defined in \Tmrecurs\ and \Wmsdef. The redefinition by superfields ${\cal H}_{[A,B_i]}$ as
in appendix \appD\ is necessary to trade the $\widehat V_A$ in its BRST variation for BRST blocks $V_A$. As before, it
vanishes whenever both slots are of single-particle type, i.e. $|A|=|B_i|=1$.

At rank $r \geq 2$, the BRST variation of $\cJ^{m_1 \ldots m_r}_{A|B_1, \ldots, B_{r+4}} $ turns out to involve
anomalous traces in the same way as $QM^{m_1 \ldots m_r }_{B_1, \ldots,B_{r+3}}$ given by \HRb. They are
accompanied by anomalous counterparts of the refined current \Jaconee,
\eqnn\Jaconef
$$\eqalignno{
\cY^{m_1 \ldots m_r}_{A|B_1, \ldots, B_{r+6}} &\equiv \half \cA_{A}^p  
\cY^{pm_1 \ldots m_r }_{B_1, \ldots,B_{r+6}}  - \big[{\cal H}_{[A,B_1]}\cY^{m_1 \ldots m_r}_{B_2, \ldots, B_{r+6}} + (B_1 \leftrightarrow B_2, \ldots, B_{r+6})\big]\,,
&\Jaconef
}$$
whose corrections $\sim {\cal H}_{[A,B_i]}$ are analogous to \Jaconee\ and ensure a BRST variation in terms of $V_A$ rather than $\widehat V_A$, see section \fiveone.

Equipped with the definitions above, we can write down the general BRST variation of higher-rank refined currents,
\eqnn\HRJC
$$\eqalignno{
Q& {\cal J}^{m_1\ldots m_r}_{A|B_1,\ldots,B_{r+4}} = {r\choose 2} \delta^{(m_1 m_2} {\cal Y}^{m_3\ldots m_r)}_{A|B_1,\ldots,B_{r+4}}+  {\cal Y}^{m_1\ldots m_r}_{A,B_1,\ldots, B_{r+4}} + k_A^p M_A M^{pm_1\ldots m_r}_{B_1,\ldots ,B_{r+4}} \cr
&+\Bigl[ M_{S[A,B_1]} M^{m_1\ldots m_r}_{B_2,\ldots,B_{r+4}} + rk_{B_1}^{(m_1}  M_{B_1} {\cal J}^{m_2\ldots m_r)}_{A|B_2,\ldots,B_{r+4}} \cr
&\ \ + \sum_{XY=B_1} (M_{X} {\cal J}^{m_1\ldots m_r}_{A|Y,B_2,\ldots,B_{r+4}}  - M_{Y} {\cal J}^{m_1\ldots m_r}_{A|X,B_2,\ldots,B_{r+4}}) + (B_1 \leftrightarrow B_2,\ldots ,B_{r+4}) \Bigr] \cr
&+ \sum_{XY=A} (M_X {\cal J}^{m_1\ldots m_r}_{Y|B_1,\ldots,B_{r+4}} - M_Y {\cal J}^{m_1\ldots m_r}_{X|B_1,\ldots,B_{r+4}})\ ,
&\HRJC
}$$
see \QEone\ for the map $S[A,B_i]$. For example, in the case of vectors and two-tensors the general formula \HRJC\ yields
\eqnn\HRJCa
\eqnn\HRJCb
$$\eqalignno{
Q& {\cal J}^{m}_{A|B_1,B_2,B_3,B_4,B_{5}} =  {\cal Y}^{m}_{A,B_1,B_2,B_3,B_4, B_{5}} + k_A^p M_A M^{pm}_{B_1,B_2,B_3,B_4,B_{5}} +\Bigl[ M_{S[A,B_1]} M^{m}_{B_2,B_3,B_4,B_{5}} \cr
&+ k_{B_1}^m M_{B_1} {\cal J}_{A|B_2,\ldots,B_{5}}  + \sum_{XY=B_1} (M_{X} {\cal J}^{m}_{A|Y,B_2,\ldots,B_{5}}  - M_{Y} {\cal J}^{m}_{A|X,B_2,\ldots,B_{5}}) + (B_1 \leftrightarrow B_2,\ldots ,B_{5}) \Bigr] \cr
&+  \sum_{XY=A} (M_X {\cal J}^{m}_{Y|B_1,\ldots,B_{5}} - M_Y {\cal J}^{m}_{X|B_1,\ldots,B_{5}})
&\HRJCa \cr
Q& {\cal J}^{mn}_{A|B_1,B_2,B_3,B_4,B_5,B_{6}} =  \delta^{mn} {\cal Y}_{A|B_1,\ldots,B_{6}}+   {\cal Y}^{mn}_{A,B_1,\ldots, B_{6}} + k_A^p M_A M^{pmn}_{B_1,\ldots ,B_{6}} +\Bigl[ M_{S[A,B_1]} M^{mn}_{B_2,\ldots,B_{6}}\cr
&+ 2k_{B_1}^{(m}  M_{B_1} {\cal J}^{n)}_{A|B_2,\ldots,B_{6}} + \sum_{XY=B_1} (M_{X} {\cal J}^{mn}_{A|Y,B_2,\ldots,B_{6}}  - M_{Y} {\cal J}^{mn}_{A|X,B_2,\ldots,B_{6}}) + (B_1 \leftrightarrow B_2,\ldots ,B_{6}) \Bigr]\cr
& + \sum_{XY=A} (M_X {\cal J}^{mn}_{Y|B_1,\ldots,B_{6}} - M_Y {\cal J}^{mn}_{X|B_1,\ldots,B_{6}})  \,.
&\HRJCb
}$$

\subsec Recursion for refined higher-rank pseudoinvariants

\subseclab\sixtwo
Each of the tensorial refined currents in \Jaconee\ can be regarded as the leading term of a tensorial refined pseudoinvariant,
\eqnn\psjac
$$\eqalignno{
P^{m_1\ldots m_r}_{i|A|B_1,\ldots,B_{r+4}} &\equiv M_i \cJ^{m_1 \ldots m_r}_{A|B_1, \ldots, B_{r+4}} + \sum_{C \neq \emptyset} M_{iC}\ldots \ .
&\psjac
}$$
The BRST pseudo-completion through multiparticle $M_{iC}$ along with momenta and ghost number two objects $\cJ^{m_1 \ldots
m_p}_{A|B_1, \ldots, B_{p+4}}$, $M^{m_1\ldots m_p}_{B_1,\ldots,B_{p+3}}$ follows the same logic as explained
below \HRg\ and \Pdef. The recursive construction of the $P^{m_1\ldots m_r}_{i|A|B_1,\ldots,B_{r+4}}$ relies on an alternative form of the BRST variation \HRJC,
\eqnn\HRv
$$\eqalignno{
Q {\cal J}^{m_1\ldots m_r}_{A|B_1,\ldots,B_{r+4}} &=  {r \choose 2} \delta^{(m_1 m_2} {\cal Y}^{m_3 \ldots m_r)}_{A|B_1,\ldots,B_{r+4}} +  {\cal Y}^{m_1\ldots m_r}_{A,B_1,\ldots,B_{r+4}}  + \delta_{|A|,1} k_{a_1}^p C^{pm_1\ldots m_r}_{a_1|B_1,\ldots,B_{r+4}}\cr
&\! \! \! \! \! \! \! \! \! \! \! \! \! \! \! \! \! \! \! \! \! \! \! \! \! \! \! \! \! \! \! \! \! \! \! \! \! \! \! + P^{m_1\ldots m_r}_{a_1|a_2\ldots a_{|A|}|B_1,\ldots,B_{r+4}} - P^{m_1\ldots m_r}_{a_{|A|}|a_1\ldots a_{|A|-1}|B_1,\ldots,B_{r+4}} + \big[ r \delta_{|B_1|,1} k_{b_1}^{(m_1} P^{m_2\ldots m_r)}_{b_1|A| B_{2},\ldots, B_{r+4}} 
\cr
&\! \! \! \! \! \! \! \! \! \! \! \! \! \! \! \! \! \! \! \! \! \! \! \! \! \! \! \! \! \! \! \! \! \! \! \! \! \! \!+  P^{m_1\ldots m_r}_{b_1|A|b_2\ldots b_{|B_1|},B_2,\ldots,B_{r+4}} - P^{m_1\ldots m_r}_{b_{|B_1|}|A|b_1\ldots b_{|B_1|-1},B_2,\ldots,B_{r+4}} + (B_1 \leftrightarrow B_2,\ldots, B_{r+4}) \big] \ .
 &\HRv
}$$
As usual, this follows from isolating the single-particle $M_i$ in \HRJC\ and promoting them to a (pseudo-)invariant
$C^{m_1\ldots m_r}_{i|A_1,\ldots,A_{r+3}}$ or $P^{m_1\ldots m_r}_{i|A|B_1,\ldots,B_{r+4}}$. By \corB\ and \HRv, the
following recursion eliminates any non-anomalous contribution from $QP^{m_1\ldots m_r}_{i|A|B_1,\ldots,B_{r+4}} $,
\eqnn\HRw
$$\eqalignno{
P^{m_1\ldots m_r}_{i|A|B_1,\ldots,B_{r+4}} &=M_i {\cal J}^{m_1\ldots m_r}_{A|B_1,\ldots,B_{r+4}}  + M_i \otimes \Big\{ \delta_{|A|,1} k_{a_1}^p C^{m_1\ldots m_rp}_{a_1|B_1,\ldots,B_{r+4}} + P^{m_1\ldots m_r}_{a_1|a_2\ldots a_{|A|}|B_1,\ldots,B_{r+4}} \cr
&\! \! \! \! \! \! \! \! \! \! \! \! \! \! \! \! \! \! \! \! \! \! \! \! \! \! \! \! \! \! \! \! \! \! \! \! \! \! \!  - P^{m_1\ldots m_r}_{a_{|A|}|a_1\ldots a_{|A|-1}|B_1,\ldots,B_{r+4}} + \big[ r \delta_{|B_1|,1} k_{b_1}^{(m_1} P^{m_2\ldots m_r)}_{b_1|A| B_{2},\ldots, B_{r+4}} +  P^{m_1\ldots m_r}_{b_1|A|b_2\ldots b_{|B_1|},B_2,\ldots,B_{r+4}} 
\cr
&\! \! \! \! \! \! \! \! \! \! \! \! \! \! \! \! \! \! \! \! \! \! \! \! \! \! \! \! \! \! \! \! \! \! \! \! \! \! \!
- P^{m_1\ldots m_r}_{b_{|B_1|}|A|b_1\ldots b_{|B_1|-1},B_2,\ldots,B_{r+4}} + (B_1 \leftrightarrow B_2,\ldots, B_{r+4})
\big] \Big\}\,. &\HRw
}$$
This completes the subleading diagonal in the overview grid in \figoverview. The anomalous BRST variations of \HRw\ are discussed in section \sevenfour.

At rank one, the general recursion \HRv\ reduces to
\eqnn\HRwone
$$\eqalignno{
P^m_{i|A|B,C,D,E,F} &= M_i {\cal J}^m_{A|B,C,D,E,F}
 + M_i \otimes \Big\{  \d_{|A|,1} k_{a_1}^p C_{a_1|B,C,D,E,F}^{pm} +P^m_{a_1|a_2\ldots a_{|A|}|B,C,D,E,F}  \cr
 &- P^m_{a_{|A|}|a_1\ldots a_{|A|-1}|B,C,D,E,F}  + \big[ \delta_{|B|,1} k^m_{b_1} P_{b_1|A|C,D,E,F} + P^m_{b_1|A|b_2\ldots b_{|B|},C,D,E,F} \cr
 &-  P^m_{b_{|B|}|A|b_1\ldots b_{|B|-1},C,D,E,F}+ (B \leftrightarrow C,D,E,F)  \bigr] \Big\} \,,
  &\HRwone }$$
and the simplest vectorial pseudoinvariant $P^m_{1|2|3,4,5,6,7}$ is displayed in appendix~\appA.

\subsec Higher-refinement building blocks

\subseclab\sixthree
The definition \Jaconee\ of refined building blocks can be endowed with a recursive structure which allows to
successively increase the number $d$ of refined slots. In order to do that, first define the generalization of
the tensor $\cW^{m_1 \ldots m_r| p}_{B_1, \ldots,B_{r+4}}$ in \Wmsdef\ for any number of specialized legs,
\eqnn\refWa
$$\eqalignno{
\cW^{m_1 \ldots m_{r-1}| m_r}_{A_1,\ldots,A_d|B_1, \ldots,B_{d+r+3}} &\equiv \half \cA_{A_1}^p \cW^{pm_1 \ldots
m_{r-1}| m_r}_{A_2,\ldots,A_d|B_1, \ldots,B_{d+r+3}} &\refWa\cr
&- \big[{\cal H}_{[A_1,B_1]}\cW^{m_1 \ldots m_{r-1}|m_r}_{A_2, \ldots, A_d|B_2, \ldots,B_{d+r+3}} + (B_1\leftrightarrow B_2,
\ldots, B_{d+r+3})\big]\,.
}$$
Then the recursion for refined currents $\cJ^{m_1 \ldots m_r}_{A_1, \ldots,A_d|B_1, \ldots, B_{d+r+3}}$ of arbitrary
refinement can be immediately written down\foot{We keep both notations for $M^{m_1\ldots m_r}_{B_1,\ldots,B_{r+3}}={\cal J}^{m_1\ldots
m_r}_{B_1,\ldots,B_{r+3}}$ and $C^{m_1\ldots m_r}_{i|B_1,\ldots,B_{r+3}}=P^{m_1\ldots m_r}_{i|B_1,\ldots,B_{r+3}}$ to
make the source of anomalous BRST transformations more transparent in the scattering amplitudes presented in
\refs{\wipG, \wipH}.},
\eqnn\JacobiDefs
$$\eqalignno{
\cJ^{m_1 \ldots m_r}_{B_1, \ldots, B_{r+3}} &\equiv M^{m_1 \ldots m_r }_{B_1, \ldots,B_{r+3}} &\JacobiDefs\cr
\cJ^{m_1 \ldots m_r}_{A_1, \ldots,A_d|B_1, \ldots, B_{d+r+3}} &\equiv \half \cA_{A_1}^p \bigl[
\cJ^{pm_1 \ldots m_r }_{A_2,\ldots, A_d|B_1, \ldots,B_{d+r+3}} +
\cW^{m_1 \ldots m_r| p}_{A_2,\ldots, A_d|B_1, \ldots,B_{d+r+3}}\bigr]\cr
&- \big[{\cal H}_{[A_1,B_1]}\cJ^{m_1 \ldots m_{r}}_{A_2, \ldots, A_d|B_2, \ldots,B_{d+r+3}} + (B_1\leftrightarrow B_2, \ldots, B_{d+r+3})\big]\,.
}$$
Even though it is not manifest from their definitions \refWa\ and \JacobiDefs, the objects $\cW^{m_1 \ldots m_{r-1}|
m_r}_{A_1,\ldots,A_d|B_1, \ldots,B_{d+r+3}}$ and $\cJ^{m_1 \ldots m_r}_{A_1, \ldots,A_d|B_1, \ldots, B_{d+r+3}}$ are
totally symmetric under exchange of refined slots $A_i \leftrightarrow A_j$. Moreover, symmetry in $B_i \leftrightarrow B_j$ is
obviously inherited from $M^{pm_1 \ldots m_r }_{B_1, \ldots,B_{r+4}}$ and $\cW^{m_1 \ldots m_r | p}_{B_1,
\ldots,B_{r+4}}$ in the first step \Jaconee\ of the recursion.

The BRST variation of \JacobiDefs\ involves anomaly building blocks of higher refinement which are defined in analogy
to \refWa,
\eqnn\refWc
$$\eqalignno{
\cY^{m_1 \ldots  m_r}_{A_1,\ldots,A_d|B_1, \ldots,B_{d+r+5}} &\equiv \half \cA_{A_1}^p \cY^{pm_1 \ldots
m_r}_{A_2,\ldots,A_d|B_1, \ldots,B_{d+r+5}} &\refWc\cr
&- \big[{\cal H}_{[A_1,B_1]}\cY^{m_1 \ldots m_{r}}_{A_2, \ldots, A_d|B_2, \ldots,B_{d+r+5}} + (B_1\leftrightarrow B_2, \ldots, B_{d+r+5})\big]\,.
}$$
These definitions give rise to the following formula for the most general case,
\eqnn\HRJCgen
$$\eqalignno{
&\qquad Q {\cal J}^{m_1\ldots m_r}_{A_1,\ldots,A_d | B_1,\ldots,B_{d+r+3}} = {r\choose 2} \delta^{(m_1 m_2} {\cal Y}^{m_3\ldots m_r)}_{A_1,\ldots,A_d  |B_1,\ldots,B_{d+
r+3}} \cr
&+ \big[ {\cal Y}^{m_1\ldots m_r}_{A_2,\ldots,A_d |A_1,B_1,\ldots, B_{d+r+3}} + k_{A_1}^p M_{A_1} {\cal J}^{pm_1\ldots m_r}_{A_2,\ldots,A_d |B_1,\ldots ,B_{d+r+3}} + (A_1 \leftrightarrow A_2,\ldots, A_d) \big] \cr
&+\bigl[  rk_{B_1}^{(m_1}  M_{B_1} {\cal J}^{m_2\ldots m_r)}_{A_1,\ldots,A_d|B_2,\ldots,B_{d+r+3}} + (B_1 \leftrightarrow B_2,\ldots ,B_{d+r+3}) \bigr]  \cr
&+\Big[ M_{S[A_1,B_1]} {\cal J}^{m_1\ldots m_r}_{A_2,\ldots, A_d | B_2,\ldots,B_{d+r+3}} + { A_1 \leftrightarrow A_2,A_3,\ldots, A_d \choose B_1 \leftrightarrow B_2,\ldots ,B_{d+r+3}} \Big] &\HRJCgen \cr
&+ \Big[ \sum_{XY=A_1} (M_{X} {\cal J}^{m_1\ldots m_r}_{Y,A_2,\ldots,A_d  |B_1,\ldots,B_{d+r+3}}  - M_{Y} {\cal J}^{m_1\ldots m_r}_{X,A_2,\ldots,A_d  |B_1,\ldots,B_{d+r+3}}) + (A_1 \leftrightarrow A_2,\ldots ,A_{d}) \Bigr]  \cr
&+ \Big[ \sum_{XY=B_1} (M_{X} {\cal J}^{m_1\ldots m_r}_{A_1,\ldots,A_d  |Y,B_2,\ldots,B_{d+r+3}}  - M_{Y} {\cal J}^{m_1\ldots m_r}_{A_1,\ldots,A_d |X,B_2,\ldots,B_{d+r+3}})  \cr
& \ \ \ \ + (B_1 \leftrightarrow B_2,\ldots ,B_{d+r+3}) \Bigr]  \ .
}$$
Any term of $Q{\cal J}^{m_1\ldots m_r}_{A|B_1,\ldots,B_{r+4}}$ as given by \HRJC\ has a counterpart in \HRJCgen\ at
higher degree $d$ of refinement. Moreover, the three classes of terms ${\cal Y}^{m_1\ldots m_r}_{A,B_1,\ldots,
B_{r+4}}$, $k_A^p M_A M^{pm_1\ldots m_r}_{B_1,\ldots ,B_{r+4}}$ and $M_{S[A,B_1]} M^{m_1\ldots
m_r}_{B_2,\ldots,B_{r+4}}$ in \HRJC\ which release $A$ from the refined slot of ${\cal J}^{m_1\ldots
m_r}_{A|B_1,\ldots,B_{r+4}}$ have multiple images in $Q {\cal J}^{m_1\ldots m_r}_{A_1,\ldots,A_d |
B_1,\ldots,B_{d+r+3}}$ according to $A_1 \leftrightarrow A_2,\ldots ,A_{d}$.

For scalars of refinement $d=2$, for instance,
\eqnn\HOl
$$\eqalignno{
Q&{\cal J}_{A,B | C,D,E,F,G} = {\cal Y}_{A|B,C,D,E,F,G} +{\cal Y}_{B|A,C,D,E,F,G} +M_A k_A^m {\cal J}^m_{B | C,D,E,F,G}  \cr
&+ M_B k_B^m {\cal J}^m_{A | C,D,E,F,G} +  \Big[ M_{S[A,C]} {\cal J}_{B|D,E,F,G} +M_{S[B,C]} {\cal J}_{A|D,E,F,G}  \cr
& \ \ \ + \sum_{XY=C} (M_{X} {\cal J}_{A,B|Y,D,E,F,G} -M_{Y} {\cal J}_{A,B|X,D,E,F,G} )   + (C\leftrightarrow D,E,F,G)   \Big]  \cr
&+ \sum_{XY=A} (M_{X} {\cal J}_{Y,B|C,D,E,F,G} -M_{Y} {\cal J}_{X,B|C,D,E,F,G} ) \cr 
&+ \sum_{XY=B} (M_{X} {\cal J}_{A,Y|C,D,E,F,G} -M_{Y} {\cal J}_{A,X|C,D,E,F,G} ) \ .  &\HOl
}$$

\subsec The general recursion for pseudoinvariants

\subseclab\sixfour
The refined current \JacobiDefs\ of arbitrary rank $r$ and refinement $d$ can be promoted to a pseudoinvariant via
\eqnn\pspsjac
$$\eqalignno{
P^{m_1\ldots m_r}_{i|A_1,\ldots,A_d|B_1,\ldots,B_{d+r+3}} &\equiv M_i \cJ^{m_1 \ldots m_r}_{A_1,\ldots,A_d|B_1,\ldots,B_{d+r+3}} + \sum_{C \neq \emptyset} M_{iC}\ldots \,,
&\pspsjac
}$$
which generalizes \psjac\ to $d >1$ and by convention reduces to $P^{m_1\ldots m_r}_{i|B_1,\ldots,B_{r+3}} \equiv
C^{m_1\ldots m_r}_{i|B_1,\ldots,B_{r+3}}$ at $d=0$. The suppressed companions of the multiparticle $M_{iC}$ are
further instances of momenta and $\cJ^{m_1 \ldots m_p}_{A_1,\ldots,A_q|B_1,\ldots,B_{p+q+3}}$ which have to be chosen
such that $QP^{m_1\ldots m_r}_{i|A_1,\ldots,A_d|B_1,\ldots,B_{d+r+3}}$ is purely anomalous. These contributions are
determined by the following rewriting of \HRJCgen:
\eqnn\JCgen
$$\eqalignno{
Q& {\cal J}^{m_1\ldots m_r}_{A_1,\ldots,A_d | B_1,\ldots,B_{d+r+3}} = {r\choose 2} \delta^{(m_1 m_2} {\cal Y}^{m_3\ldots m_r)}_{A_1,\ldots,A_d  |B_1,\ldots,B_{d+
r+3}}  + \big[ {\cal Y}^{m_1\ldots m_r}_{A_2,\ldots,A_d |A_1,B_1,\ldots, B_{d+r+3}} \cr
&\ \ + \delta_{|A_1|,1} k_{a_1}^p P^{pm_1\ldots m_r}_{a_1|A_2,\ldots,A_d |B_1,\ldots ,B_{d+r+3}}  + P^{m_1\ldots m_r}_{a_1|a_2\ldots a_{|A_1|},A_2,\ldots,A_d |B_1,\ldots ,B_{d+r+3}} \cr
& \ \ -P^{m_1\ldots m_r}_{a_{|A_1|}|a_1\ldots a_{|A_1|-1},A_2,\ldots,A_d |B_1,\ldots ,B_{d+r+3}} + (A_1 \leftrightarrow A_2,\ldots, A_d) \big] \cr
&+\bigl[  r \delta_{|B_1|,1} k_{b_1}^{(m_1}  P^{m_2\ldots m_r)}_{b_1|A_1,\ldots,A_d|B_2,\ldots,B_{d+r+3}} + P^{m_1\ldots m_r}_{b_1|A_1,\ldots, A_d| b_2\ldots b_{|B_1|}, B_2,\ldots, B_{d+r+3}}\cr
& \ \ - P^{m_1\ldots m_r}_{b_{|B_1|}|A_1,\ldots, A_d| b_1\ldots b_{|B_1|-1}, B_2,\ldots, B_{d+r+3}}
+ (B_1 \leftrightarrow B_2,\ldots ,B_{d+r+3}) \bigr]  \ . &\JCgen
}$$
As usual, \corB\ allows to derive a recursion from \JCgen:
\eqnn\pinvgen
$$\eqalignno{
&P^{m_1\ldots m_r}_{i|A_1,\ldots,A_d | B_1,\ldots,B_{d+r+3}} =  M_i {\cal J}^{m_1\ldots m_r}_{A_1,\ldots,A_d | B_1,\ldots,B_{d+r+3}} \cr
& + M_i \otimes \Big\{ \big[ \delta_{|A_1|,1} k_{a_1}^p P^{pm_1\ldots m_r}_{a_1|A_2,\ldots,A_d |B_1,\ldots ,B_{d+r+3}}  + P^{m_1\ldots m_r}_{a_1|a_2\ldots a_{|A_1|},A_2,\ldots,A_d |B_1,\ldots ,B_{d+r+3}} \cr
& \ \ \ \ -P^{m_1\ldots m_r}_{a_{|A_1|}|a_1\ldots a_{|A_1|-1},A_2,\ldots,A_d |B_1,\ldots ,B_{d+r+3}} + (A_1 \leftrightarrow A_2,\ldots, A_d) \big] \cr
&\ \ +\bigl[  r \delta_{|B_1|,1} k_{b_1}^{(m_1}  P^{m_2\ldots m_r)}_{b_1|A_1,\ldots,A_d|B_2,\ldots,B_{d+r+3}} + P^{m_1\ldots m_r}_{b_1|A_1,\ldots, A_d| b_2\ldots b_{|B_1|}, B_2,\ldots, B_{d+r+3}}\cr
& \ \ \ \ - P^{m_1\ldots m_r}_{b_{|B_1|}|A_1,\ldots, A_d| b_1\ldots b_{|B_1|-1}, B_2,\ldots, B_{d+r+3}}
+ (B_1 \leftrightarrow B_2,\ldots ,B_{d+r+3}) \bigr] \Big\}  \ .&\pinvgen 
}$$
This is the most general pseudoinvariant presented in this work, it completes the overview grid in \figoverview. Its anomalous BRST variation will be discussed in section \sevenfour.

In the $d=2$ example of ${\cal J}_{A,B|C,D,E,F,G}$, the variation in \HOl\ can be rewritten as
\eqnn\HOq
$$\eqalignno{
Q{\cal J}_{A,B | C,D,E,F,G} &={\cal Y}_{A|B,C,\ldots,G} + {\cal Y}_{B|A,C,\ldots,G} +\delta_{|A|,1} k_a^m P^{m}_{a|B|C,\ldots,G}+\delta_{|B|,1} k_b^m P^{m}_{b|A|C,\ldots,G} \cr
&+ \big[ P_{a_1|a_2 \ldots a_{|A|},B|C,\ldots,G}-P_{a_{|A|}|a_1 \ldots a_{|A|-1},B|C,\ldots,G} + (A\leftrightarrow B) \big] &\HOq \cr
&+\big[ P_{c_1|A,B|c_2\ldots c_{|C|},D,\ldots,G}-P_{c_{|C|}|A,B|c_1\ldots c_{|C|-1},D,\ldots,G}+ (C\leftrightarrow D,\ldots,G) \big] 
}$$
and converted to the recursion
\eqnn\HOs
$$\eqalignno{
P_{i|A,B | C,D,E,F,G} &= M_i{\cal J}_{A,B | C,D,E,F,G}  + M_i \otimes \Big\{ \delta_{|A|,1} k_a^m P^{m}_{a|B|C,\ldots,G}+\delta_{|B|,1} k_b^m P^{m}_{b|A|C,\ldots,G} \cr
&+ \big[ P_{a_1|a_2 \ldots a_{|A|},B|C,\ldots,G}-P_{a_{|A|}|a_1 \ldots a_{|A|-1},B|C,\ldots,G} + (A\leftrightarrow B) \big] &\HOs \cr
&+\big[ P_{c_1|A,B|c_2\ldots c_{|C|},D,\ldots,G}-P_{c_{|C|}|A,B|c_1\ldots c_{|C|-1},D,\ldots,G}+ (C\leftrightarrow D,\ldots,G) \big]  \Big\} \ .
}$$
The simplest example $P_{1|2,3 | 4,5,6,7,8}$ is displayed in appendix \appA, see \HOu.

\subsec Trace relations among pseudoinvariants

\subseclab\sixfive
In section \fivefour, we have discussed the relation between tensor traces $\delta_{mn} M^{mn}_{A,B,C,D,E}$,
$\delta_{mn} C^{mn}_{i|A,B,C,D,E}$ and the refined objects, ${\cal
J}_{A|B,C,D,E},P_{i|A|B,C,D,E}$. The trace relations \Tmntrace\ and \Cmntrace\ are now
generalized to higher rank $r$ and refinement $d$. 

\proclaim Lemma 2. The following is true,
\eqnn\symmWs
\eqnn\symmBBs
$$\eqalignno{
\delta_{np} {\cal W}^{npm_1\ldots m_{r-1}|m_r}_{A_1,\ldots,A_{d} | B_1,\ldots,B_{d+r+5}} &=
2\cW^{m_1 \ldots m_{r-1}|m_r}_{A_1, \ldots,A_{d},B_1|B_2, \ldots, B_{d+r+5}} + (B_1 \leftrightarrow B_2,\ldots,
B_{d+r+5})\,,\qquad  & \symmWs\cr
\delta_{np} {\cal J}^{npm_1\ldots m_r}_{A_1,\ldots,A_{d} | B_1,\ldots,B_{d+r+5}} &=
2\cJ^{m_1 \ldots m_r}_{A_1, \ldots,A_{d},B_1|B_2, \ldots, B_{d+r+5}} + (B_1 \leftrightarrow B_2,\ldots, B_{d+r+5})  \ . & \symmBBs
}$$\par

\noindent{\it Proof.}
To prove this inductively, first assume that \symmWs\ is true for $d-1$,
\eqn\assumeW{
\delta_{np} {\cal W}^{npm_1\ldots m_{r-1}|m_r}_{A_1,\ldots,A_{d-1} | B_1,\ldots,B_{d+r+4}} =
2\cW^{m_1 \ldots m_{r-1}|m_r}_{A_1, \ldots,A_{d-1},B_1|B_2, \ldots, B_{d+r+4}} + (B_1 \leftrightarrow B_2,\ldots,
B_{d+r+4})\,.
}
From the definition \refWa\ it follows that,
$$\eqalignno{
\delta_{np} {\cal W}^{npm_1\ldots m_{r-1}|m_r}_{A_1,\ldots,A_{d} | B_1,\ldots,B_{d+r+5}} &=
\half  \cA^t_{A_d}\cW^{tppm_1 \ldots m_{r-1}|m_r}_{A_1, \ldots, A_{d-1} | B_1,\ldots,B_{d+r+5}}\cr
&{}-\big[ {\cal H}_{[A_d,B_1]} \cW^{ppm_1 \ldots m_{r-1}|m_r}_{A_1, \ldots, A_{d-1} | B_2,\ldots,B_{d+r+5}} +
(B_1\leftrightarrow B_2, \ldots, B_{d+r+5})\big] \ ,\cr
}$$
and therefore \assumeW\ leads to,
$$\eqalignno{
\delta_{np} {\cal W}^{npm_1\ldots m_{r-1}|m_r}_{A_1,\ldots,A_{d} | B_1,\ldots,B_{d+r+5}} &=
\half \cA^t_{A_d}\, 2\cW^{tm_1 \ldots m_{r-1}|m_r}_{A_1, \ldots, A_{d-1},B_1 | B_2,\ldots,B_{d+r+5}}\cr
&{} - 2\big[ {\cal H}_{[A_d,B_1]} \big(\cW^{m_1 \ldots m_{r-1}|m_r}_{A_1, \ldots, A_{d-1},B_2 | B_3,\ldots,B_{d+r+5}} +
(B_2\leftrightarrow B_3, \ldots, B_{d+r+5})\big) \cr
&\qquad{}+ (B_1\leftrightarrow B_2, \ldots, B_{d+r+5})\big]\cr
&= 2\cW^{m_1 \ldots m_{r-1}|m_r}_{A_1, \ldots,A_{d},B_1|B_2, \ldots, B_{d+r+5}} + (B_1 \leftrightarrow B_2,\ldots,
B_{d+r+5})\, ,
}$$
where in the last line one uses that,
\eqnn\combina
$$\eqalignno{
& {\cal H}_{[A_d,B_1]} \big[\cW^{m_1 \ldots m_{r-1}|m_r}_{A_1, \ldots, A_{d-1},B_2 | B_3,\ldots,B_{d+r+5}} +
(B_2\leftrightarrow B_3, \ldots, B_{d+r+5})\big] + (B_1\leftrightarrow B_2, \ldots, B_{d+r+5})\cr
&= \big[{\cal H}_{[A_d,B_2]}\cW^{m_1 \ldots m_{r-1}|m_r}_{A_1, \ldots, A_{d-1},B_1 | B_3,\ldots,B_{d+r+5}} +
(B_2\leftrightarrow B_3, \ldots, B_{d+r+5})\big]
+ (B_1\leftrightarrow B_2, \ldots, B_{d+r+5})\,.\cr
&&\combina
}$$
Furthermore, it is easy to show that when $d=0$,
\eqn\Wtrace{
\delta_{np}\cW^{npm_1 \ldots m_{r-1}|m_r}_{B_1,B_2 \ldots,B_{r+5}} = 2 \cW^{m_1 \ldots m_{r-1}|m_r}_{B_1|B_2, \ldots,B_{r+5}} +
(B_1\leftrightarrow B_2, \ldots, B_{r+5}) \,,
}
finishing the proof of \symmWs.

To show \symmBBs\ one proceeds similarly by first assuming that it holds for $d-1$,
\eqn\assump{
\delta_{np} {\cal J}^{npm_1\ldots m_r}_{A_1,\ldots,A_{d-1} | B_1,\ldots,B_{d+r+4}} =
2\cJ^{m_1 \ldots m_r}_{A_1, \ldots,A_{d-1},B_1|B_2, \ldots, B_{d+r+4}} + (B_1 \leftrightarrow B_2,\ldots, B_{d+r+4}) \,.
}
A direct application of the definition \JacobiDefs\ leads to
\eqnn\induct
$$\displaylines{
\cJ^{ppm_1 \ldots m_r}_{A_1, \ldots,A_{d}|B_{1},B_{2}, \ldots,B_{d+r+5}} =
\half \cA_{A_d}^q \Bigl[\cJ^{ppm_1 \ldots m_rq}_{A_1, \ldots,A_{d-1}|B_{1},B_{2}, \ldots,B_{d+r+5}}
+ \cW^{ppm_1 \ldots m_r|q}_{A_1, \ldots,A_{d-1}|B_{1},B_{2}, \ldots,B_{d+r+5}}\Bigr]\cr
{}-\big[ {\cal H}_{[A_d,B_1]} \cJ^{ppm_1 \ldots m_{r}}_{A_1, \ldots,A_{d-1}|B_2, \ldots,B_{d+r+5}} +
(B_1\leftrightarrow B_2, \ldots,B_{d+r+5})\big] \ .\hfil\induct\hfilneg
}$$
Now one rewrites \induct\ using \symmWs\ together with the assumption \assump,
\eqnn\inductMiddle
$$\eqalignno{
\cJ^{ppm_1 \ldots m_r}_{A_1, \ldots,A_{d}|B_{1},B_{2}, \ldots,B_{d+r+5}} &=
\half \cA_{A_d}^q \Bigl[ 2\cJ^{m_1 \ldots m_rq}_{A_1, \ldots,A_{d-1},B_{1}|B_{2}, \ldots,B_{d+r+5}}
+ 2 \cW^{m_1 \ldots m_r|q}_{A_1, \ldots,A_{d-1},B_{1}|B_{2}, \ldots,B_{d+r+5}}\Bigr]\cr
&{}- 2\Bigl[ {\cal H}_{[A_d,B_1]} \cJ^{m_1 \ldots m_{r}}_{A_1, \ldots,A_{d-1},B_2|B_3, \ldots,B_{d+r+5}} +
(B_2\leftrightarrow B_3, \ldots,B_{d+r+5})\Bigr]\cr
&{}+ (B_1\leftrightarrow B_2, \ldots,B_{d+r+5}) \ , &\inductMiddle\cr
}$$
to finally obtain
\eqn\inductFinal{
\cJ^{ppm_1 \ldots m_r}_{A_1, \ldots,A_{d}|B_{1},B_{2}, \ldots,B_{d+r+5}} = 2 \cJ^{m_1 \ldots m_r}_{A_1, \ldots,A_d,B_{1}|B_{2},B_{3}, \ldots,B_{d+r+5}} + (B_{1}\leftrightarrow
B_{2},B_{3},\ldots,B_{d+r+5})\,,
}
where a relation analogous to \combina\ has been used to arrive at \inductFinal.
When $d=0$ \symmBBs\ can be easily verified using the definitions \JacobiDefs\ and \Tmrecurs\ since the commutator
drops out due to the sum over permutations,
\eqnn\firstrue
$$\eqalignno{
\cJ^{pp m_1 \ldots m_r}_{B_1,B_2, \ldots,B_{r+5}} &= M^{p m_1 \ldots m_rp}_{B_1,B_2, \ldots,B_{r+5}}&\firstrue\cr
&=\cA^p_{B_1} M^{m_1 \ldots m_r p}_{B_2,B_3, \ldots,B_{r+5}}  + \cA^p_{B_1} \cW^{m_1 \ldots m_r| p}_{B_2,B_3,
\ldots,B_{r+5}} + (B_1\leftrightarrow B_2,B_3, \ldots,B_{r+5})\cr
&= 2 \cJ^{m_1 \ldots m_r}_{B_1|B_2,B_3, \ldots,B_{r+5}} + (B_1\leftrightarrow B_2,B_3, \ldots,B_{r+5}) \ .
}$$
The above manipulations make use of the symmetry properties $M^{pp m_1 \ldots m_r}=M^{p m_1 \ldots m_r p}$ and $\cW^{m_r m_{r-1} \ldots m_1|p} =\cW^{m_1 \ldots
m_r|p}$.\hfill\qed

After multiplication by $M_i$, \firstrue\ and \symmBBs\ relate the leading terms of  pseudoinvariants, so we can directly promote them to their BRST pseudo-completion:
\eqnn\symmCs
$$\eqalignno{
\delta_{np}C^{npm_1\ldots m_r}_{i|B_1,\ldots,B_{r+5}}
 &= 2 P^{m_1 \ldots m_r}_{i|B_1|B_2, \ldots, B_{r+5}} + (B_1 \leftrightarrow B_2,\ldots,B_{r+5}) &\symmCs\cr
\delta_{np} P^{npm_1\ldots m_r}_{i|A_1,\ldots,A_{d} | B_1,\ldots,B_{d+r+5}} &=
2P^{m_1 \ldots m_r}_{i|A_1, \ldots,A_{d},B_1|B_2, \ldots, B_{d+r+5}} + (B_1 \leftrightarrow B_2,\ldots, B_{d+r+5})  \ .
}$$ 
This demonstrates that the family of pseudoinvariants defined in \pspsjac\ and recursively constructed in \pinvgen\ is
closed under the trace operation. This is particularly relevant for their contractions with loop momenta in one-loop
amplitudes, see \wipH. 
 
\newsec Anomalous BRST invariants

\seclab\secseven
This section is devoted to BRST variations of anomaly blocks such as ${\cal Y}_{A,B,C,D,E}$
given by \WanonBG\ as well as its generalization to higher rank and refinement, see \HRf\ and \refWc. We are led to
BRST-invariant ghost-number-four objects built from $M_C
\cY^{m_1\ldots m_r}_{A_1,\ldots, A_d|B_1,\ldots,B_{d+r+5}}$ and momenta. They turn out to
share the grid structure of pseudoinvariants in \figoverview, see \anomgrid\ for an overview and the subsequent
sections for the notation therein.

\ifig\anomgrid{Overview of anomaly invariants. The arrows indicate whenever superfields of different type enter the recursion for the invariants on their right.}
{\epsfxsize=0.80\hsize\epsfbox{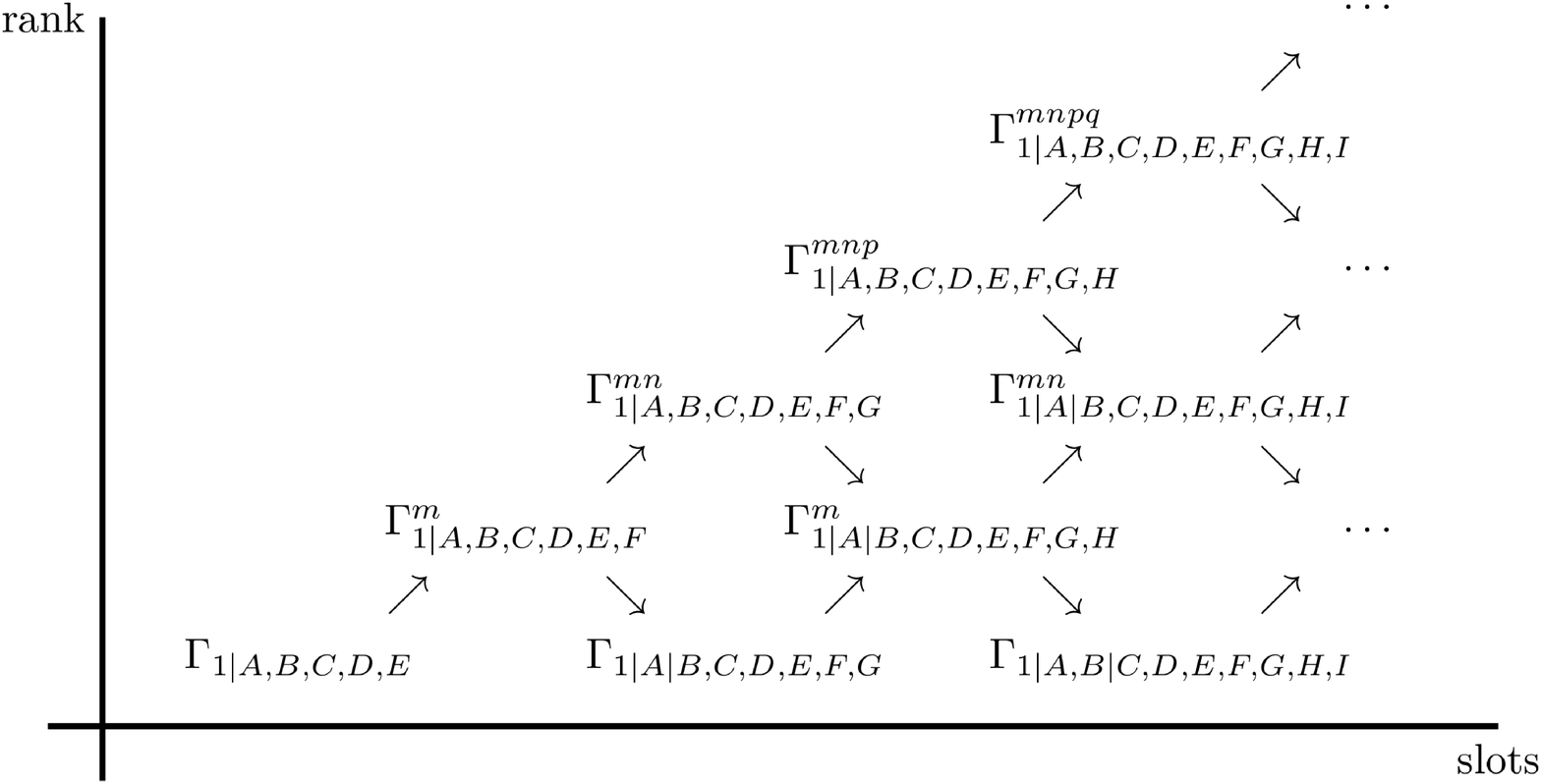}}


These anomaly invariants capture the systematics of anomalous BRST variations of pseudoinvariants. Moreover, we point
out close analogies between the $Q$ action on $\cJ^{m_1\ldots m_r}_{A_1,\ldots, A_d|B_1,\ldots,B_{d+r+3}}$ and
$\cY^{m_1\ldots m_r}_{A_1,\ldots, A_d|B_1,\ldots,B_{d+r+5}}$. This firstly allows to recycle a lot of results from previous
sections and secondly motivates a more abstract viewpoint on the recursion for pseudoinvariants which will prove essential for the subsequent sections.
 
\subsec BRST variation of unrefined anomaly blocks

\subseclab\sevenone
Let us firstly analyze the BRST variations of unrefined anomaly building blocks ${\cal Y}^{m_1\ldots m_r}_{A_1,\ldots,A_{r+5}}$. In the scalar case \WanonBG,
\eqnn\QWanonBG
$$\eqalignno{
Q\cY_{A,B,C,D,E} &=
\sum_{XY=A} (M_{X} \cY_{Y,B,C,D,E}
-  M_{Y} \cY_{X,B,C,D,E})+ (A \leftrightarrow B,C,D,E) &\QWanonBG\cr
}$$
has the same structure as $QM_{A,B,C}$ in \QMABC, and in particular $\cY_{1,2,3,4,5}$ is BRST closed. The pure spinor
constraint guarantees that the first term in $Q {\cal W}_B$ given by \QBGs\ does not contribute. Starting from the
vector building block as in \HRf, we additionally need the group-theoretic identity\foot{This is a consequence of
having no vector representation $[1,0,0,0,0]$ in the decomposition $[0,0,0,0,4]\otimes[0,0,0,0,1]^{\wedge 6}$.}
\multiloop,
\eqnn\group
$$\eqalignno{
(\l \g^m)_{[\a_1} (\l \g_p)_{\a_2}  (\l \g_q)_{\a_3}  (\l \g_r)_{\a_4}  \g^{pqr}_{\a_5 \a_6]} &= 0
 &\group\cr
}$$
to prove that
\eqnn\tsj
$$\eqalignno{
Q {\cal Y}^m_{A,B,C,D,E,F} =& \sum_{XY=A} (M_{X}  {\cal Y}^m_{Y,B,C,D,E,F} - M_{Y}  {\cal Y}^m_{X,B,C,D,E,F}) \cr
& \ \ + k_A^m M_A {\cal Y}_{B,C,D,E,F}  + (A \leftrightarrow B,C,D,E,F)\,,
&\tsj 
}$$
i.e. the first term of $Q {\cal A}_B^m=(\lambda \gamma^m \cW_B) + \ldots$ drops out from $Q {\cal Y}^m_{A,B,C,D,E,F}$.
Note the direct correspondence of \tsj\ with $QM^m_{A,B,C,D}$ given by \BRSTQm. Accordingly, higher-tensor
generalizations of $Q {\cal Y}^m_{A,B,C,D,E,F}$ can be almost literally borrowed from $Q M^{m_1\ldots
m_r}_{B_1,\ldots,B_{r+3}}$ given by \HRb\ except for one simplification: There is no anomaly analogue of the ${\cal
W}^{m_1\ldots m_{r-1}|m_r}_{B_1,\ldots,B_{r+3}}$ tensor in \HRf\ which prevents trace contributions in the following
expression,
\eqnn\HanomRb
$$\eqalignno{
Q\cY^{m_1\ldots m_r}_{B_1,B_2,\ldots,B_{r+5}} &=     \sum_{XY=B_1} (M_X \cY^{m_1\ldots m_r}_{Y,B_2,B_3,\ldots,B_{r+5}} - M_Y \cY^{m_1\ldots m_r}_{X,B_2,B_3,\ldots,B_{r+5}})  \cr
&\ \ \ \ \ +r M_{B_1} k_{B_1}^{(m_1} \cY^{m_2\ldots m_r)}_{B_2,B_3,\ldots,B_{r+5}}  + (B_1\leftrightarrow B_2,\ldots, B_{r+5})   \ .
&\HanomRb
}$$

\subsec Unrefined anomaly invariants

\subseclab\seventwo
We repeat the steps of section \fourfour\ to recursively construct tensorial BRST invariants
$\Gamma^{m_1\ldots}_{i|\ldots}$ at ghost-number four from anomaly blocks. They are defined by a leading term $\sim M_i$,
\eqnn\HanomRc
$$\eqalignno{
\Gamma^{m_1\ldots m_r}_{i|A_1,A_2,\ldots,A_{r+5}} &\equiv M_i \cY^{m_1\ldots m_r}_{A_1,A_2,\ldots,A_{r+5}} + \sum_{B \neq \emptyset} M_{iB}\ldots \,.
&\HanomRc
}$$
The suppressed terms $\ldots$ along with multiparticle $M_{iB}$ are also anomalous and can be found
from a recursion relation. To see this, firstly rewrite \HanomRb\ as follows
\eqnn\HRhf
$$\displaylines{
Q \cY^{m_1\ldots m_r}_{A_1,A_2,\ldots,A_{r+5}} =  r \delta_{|A_1|,1}  k_{a_1}^{(m_1} \Gamma^{m_2\ldots
m_r)}_{a_1|A_2,\ldots,A_{r+5}}\hfil\HRhf\hfilneg\cr
+ \Gamma^{m_1\ldots m_r}_{a_1| a_2\ldots a_{|A_1|},A_2,\ldots,A_{r+5}}
 -  \Gamma^{m_1\ldots m_r}_{a_{|A_1|}| a_1\ldots a_{|A_1|-1},A_2,\ldots,A_{r+5}} +(A_1 \leftrightarrow A_2,\ldots,A_{r+5}) \,, 
}$$
which resembles \HRh\ for $Q M^{m_1\ldots m_r}_{A_1,\ldots,A_{r+3}}$. Together with \corA, this implies BRST
invariance of the recursively-generated objects
\eqnn\HRif
$$\eqalignno{
\Gamma^{m_1\ldots m_r}_{i|A_1,A_2,\ldots,A_{r+5}} &=M_i \cY^{m_1\ldots m_r}_{A_1,A_2,\ldots,A_{r+5}} + M_i \otimes
 \bigl[ r \delta_{|A_1|,1}  k_{a_1}^{(m_1} \Gamma^{m_2\ldots m_r)}_{a_1|A_2,\ldots,A_{r+5}}  &\HRif \cr
&\! \!  \! \!  \! \! \! \! \! \! \! \! \! \! \! \! \! \! \! \! \! \! \! \! \! \! \! \! \! \! \! \! \! \! \! \! \! \! \! \! \! \! \! \! \!
+ \Gamma^{m_1\ldots m_r}_{a_1| a_2\ldots a_{|A_1|},A_2,\ldots,A_{r+5}}- \Gamma^{m_1\ldots m_r}_{a_{|A_1|}| a_1\ldots a_{|A_1|-1},A_2,\ldots,A_{r+5}} 
+(A_1 \leftrightarrow A_2,\ldots,A_{r+5}) \bigr] \ ,
}$$
see \HRi\ for the non-anomalous counterpart. For example,
\eqnn\HRig
$$\eqalignno{
\Gamma_{1|2,3,4,5,6} &=M_1 \cY_{2,3,4,5,6} &\HRig \cr
\Gamma_{1|23,4,5,6,7} &=M_1 \cY_{23,4,5,6,7} 
+ M_1 \otimes \big[ \Gamma_{2|3,4,5,6,7} - \Gamma_{3|2,4,5,6,7}\big] \cr
&=M_1 \cY_{23,4,5,6,7} + M_{12} \cY_{3,4,5,6,7}-  M_{13} \cY_{2,4,5,6,7}   \cr
\Gamma^m_{1|2,3,4,5,6,7} &=M_1 \cY^m_{2,3,4,5,6,7} + M_1\otimes \big[ k_2^m  \Gamma_{2|3,4,5,6,7} + (2\leftrightarrow3,4,5,6,7) \big] \cr
&=M_1 \cY^m_{2,3,4,5,6,7} + \big[ k_2^m M_{12} \cY_{3,4,5,6,7} + (2\leftrightarrow3,4,5,6,7) \big] 
}$$
furnish the anomaly counterparts of $C_{1|2,3,4},C_{1|23,4,5}$ and $C^m_{1|2,3,4,5}$ given in \Cscex.

The anomaly invariants in \HRif\ allow to concisely describe the anomalous BRST variations of pseudoinvariants at ghost-number three:
\eqnn\HRij
$$\eqalignno{
Q C^{m_1\ldots m_r}_{i | A_1,\ldots,A_{r+3}} &= - {r\choose 2 } \delta^{(m_1 m_2}  \Gamma^{m_3\ldots m_r)}_{i | A_1,\ldots,A_{r+3}} \ .&\HRij 
}$$
The anomaly counterpart of this statement is simply
\eqnn\HRik
$$\eqalignno{
Q\Gamma^{m_1\ldots m_r}_{i | A_1,\ldots,A_{r+5}} &= 0 \,, &\HRik
}$$
which follows from $Q^2=0$.
To justify \HRij, it
is sufficient to study the anomalous BRST variation of the leading term $M_i M^{m_1\ldots m_r}_{A_1,\ldots,A_{r+3}}$
in $C^{m_1\ldots m_r}_{i|A_1,\ldots,A_{r+3}}$ and to promote the resulting $M_i \cY^{m_1\ldots
m_p}_{A_1,\ldots,A_{p+5}}$ to their BRST invariant completion in \HanomRc.

\subsec BRST variation of general anomaly blocks

\subseclab\seventhree
The close parallels between BRST manipulations of building blocks $M^{m_1\ldots
m_r}_{A_1,\ldots,A_{r+3}}$ and their anomaly counterparts $\cY^{m_1\ldots m_r}_{A_1,\ldots,A_{r+5}}$ propagate to
their refined versions. This can be seen by comparing the recursions \JacobiDefs\ and \refWc\ for $\cJ^{m_1 \ldots
m_r}_{A_1, \ldots,A_d|B_1, \ldots, B_{d+r+3}}$ and $\cY^{m_1 \ldots m_r}_{A_1,\ldots,A_d|B_1, \ldots,B_{d+r+5}}$,
respectively. The absence of the $\cW^{m_1\ldots m_{r-1}|m_r}_{A_1,\ldots,A_d|B_1,\ldots,B_{d+r+3}}$ tensor in the anomalous case
implies that there is no higher anomaly image of the $\cY$ contribution to $Q \cJ$. In particular, the expression
\HRJC\ for $Q\cJ^{m_1 \ldots m_r}_{A|B_1, \ldots, B_{r+4}}$ implies that 
\eqnn\qanomA
$$\displaylines{
Q {\cal Y}^{m_1\ldots m_r}_{A|B_1,\ldots,B_{r+6}} = k_A^p M_A \cY^{pm_1\ldots m_r}_{B_1,\ldots ,B_{r+6}}  +\Bigl[ M_{S[A,B_1]} \cY^{m_1\ldots m_r}_{B_2,\ldots,B_{r+6}} + rk_{B_1}^{(m_1}  M_{B_1} {\cal Y}^{m_2\ldots m_r)}_{A|B_2,\ldots,B_{r+6}} \cr
 + \sum_{XY=B_1} (M_{X} {\cal Y}^{m_1\ldots m_r}_{A|Y,B_2,\ldots,B_{r+6}}  - M_{Y} {\cal Y}^{m_1\ldots m_r}_{A|X,B_2,\ldots,B_{r+6}}) + (B_1 \leftrightarrow B_2,\ldots ,B_{r+6}) \Bigr]  \cr
+ \sum_{XY=A} (M_X {\cal Y}^{m_1\ldots m_r}_{Y|B_1,\ldots,B_{r+6}} - M_Y {\cal Y}^{m_1\ldots
m_r}_{X|B_1,\ldots,B_{r+6}}) \ . \hfil\qanomA\hfilneg
}$$
More generally, \HRJCgen\ for $Q\cJ^{m_1 \ldots m_r}_{A_1,
\ldots,A_d|B_1, \ldots, B_{d+r+3}}$ leads to
\eqnn\qanomB
$$\displaylines{
Q {\cal Y}^{m_1\ldots m_r}_{A_1,\ldots,A_d | B_1,\ldots,B_{d+r+5}} =  \big[  k_{A_1}^p M_{A_1} {\cal Y}^{pm_1\ldots m_r}_{A_2,\ldots,A_d |B_1,\ldots ,B_{d+r+5}} + (A_1 \leftrightarrow A_2,\ldots, A_d) \big] \cr
+\bigl[  rk_{B_1}^{(m_1}  M_{B_1} {\cal Y}^{m_2\ldots m_r)}_{A_1,\ldots,A_d|B_2,\ldots,B_{d+r+5}} + (B_1 \leftrightarrow B_2,\ldots ,B_{d+r+5}) \bigr]  \hfil\qanomB\hfilneg\cr
+\Big[ M_{S[A_1,B_1]} {\cal Y}^{m_1\ldots m_r}_{A_2,\ldots, A_d | B_2,\ldots,B_{d+r+5}} + { A_1 \leftrightarrow A_2,A_3,\ldots, A_d \choose B_1 \leftrightarrow B_2,\ldots ,B_{d+r+5}} \Big] \cr
+ \Big[\!\! \sum_{XY=A_1}\!\! (M_{X} {\cal Y}^{m_1\ldots m_r}_{Y,A_2,\ldots,A_d  |B_1,\ldots,B_{d+r+5}}  - M_{Y} {\cal Y}^{m_1\ldots m_r}_{X,A_2,\ldots,A_d  |B_1,\ldots,B_{d+r+5}}) + (A_1 \leftrightarrow A_2,\ldots ,A_{d}) \Bigr]  \cr
+ \Big[\!\! \sum_{XY=B_1}\!\! (M_{X} {\cal Y}^{m_1\ldots m_r}_{A_1,\ldots,A_d  |Y,B_2,\ldots,B_{d+r+5}}  - M_{Y} {\cal Y}^{m_1\ldots m_r}_{A_1,\ldots,A_d |X,B_2,\ldots,B_{d+r+5}})
 + (B_1 \leftrightarrow B_2,\ldots ,B_{d+r+5}) \Bigr]  \,. 
}$$
Recall that the $S[A,B]$ map entering $M_{S[A,B_i]}$ is explained in section \fivetwo\ and defined in \QEone. The
${\cal H}_{[A_i,B_j]}$ corrections in the recursion \refWc\ for ${\cal Y}^{m_1\ldots m_r}_{A_1,\ldots,A_d |
B_1,\ldots,B_{d+r+5}}$ ensure that any $M_{S[A,B_i]}$ in \qanomA\ and \qanomB\ is built from BRST blocks $V_C$ rather than
$\widehat V_C$.

\subsec Refined anomaly invariants

\subseclab\sevenfour
Similar to \HanomRc, we introduce refined anomaly invariants with a more general leading term,
\eqnn\HanomRRc
$$\eqalignno{
\Gamma^{m_1\ldots m_r}_{i|A_1,\ldots,A_{d}|B_1,\ldots,B_{d+r+5}} &\equiv M_i \cY^{m_1\ldots m_r}_{A_1,\ldots,A_{d}|B_1,\ldots,B_{d+r+5}} + \sum_{C \neq \emptyset} M_{iC}\ldots \,.
&\HanomRRc
}$$
The BRST completion $\ldots$ along with multiparticle $M_{iC}$ is built from $\cY^{m_1\ldots
m_p}_{A_1,\ldots,A_{q}|B_1,\ldots,B_{p+q+5}} $ and momenta to ensure that $Q\Gamma^{m_1\ldots
m_r}_{i|A_1,\ldots,A_{d}|B_1,\ldots,B_{d+r+5}}=0$. This is the anomaly counterpart
of the pseudoinvariant $P^{m_1\ldots
m_r}_{i|A_1,\ldots,A_{d}|B_1,\ldots,B_{d+r+3}}$ given by \pspsjac. The definition \HanomRRc\ leads to the following
rewriting of \qanomB,
\eqnn\aJCgen
$$\displaylines{
Q {\cal Y}^{m_1\ldots m_r}_{A_1,\ldots,A_d | B_1,\ldots,B_{d+r+5}} = \big[  \delta_{|A_1|,1} k_{a_1}^p \Gamma^{pm_1\ldots m_r}_{a_1|A_2,\ldots,A_d |B_1,\ldots ,B_{d+r+5}} \cr
+ \Gamma^{m_1\ldots m_r}_{a_1|a_2\ldots a_{\mod{A_1}},A_2,\ldots,A_d |B_1,\ldots ,B_{d+r+5}} - \Gamma^{m_1\ldots m_r}_{a_{\mod{A_1}}|a_1\ldots a_{\mod{A_1}-1},A_2,\ldots,A_d |B_1,\ldots ,B_{d+r+5}}
+ (A_1 \leftrightarrow A_2,\ldots, A_d) \big] \cr
+\bigl[  r \delta_{\mod{B_1},1} k_{b_1}^{(m_1} \Gamma^{m_2\ldots m_r)}_{b_1|A_1,\ldots,A_d|B_2,\ldots,B_{d+r+5}} 
+ \Gamma^{m_1\ldots m_r}_{b_1|A_1,\ldots, A_d| b_2\ldots b_{\mod{B_1}}, B_2,\ldots, B_{d+r+5}}\cr
 - \Gamma^{m_1\ldots m_r}_{b_{\mod{B_1}}|A_1,\ldots, A_d| b_1\ldots b_{\mod{B_1}-1}, B_2,\ldots, B_{d+r+5}}
+ (B_1 \leftrightarrow B_2,\ldots ,B_{d+r+5}) \bigr]  \ . \hfil\aJCgen\hfilneg
}$$
This in turn suggests a recursion for the most general anomaly invariant in \HanomRRc,
\eqn\apinvgen{\belowdisplayskip=-3pt\relax
\Gamma^{m_1\ldots m_r}_{i|A_1,\ldots,A_d | B_1,\ldots,B_{d+r+5}} =  M_i {\cal Y}^{m_1\ldots m_r}_{A_1,\ldots,A_d | B_1,\ldots,B_{d+r+5}}
}
$$\eqalignno{
 + M_i \otimes \Big\{ &\big[ \delta_{\mod{A_1},1} k_{a_1}^p \Gamma^{pm_1\ldots m_r}_{a_1|A_2,\ldots,A_d |B_1,\ldots ,B_{d+r+5}}  
+ \Gamma^{m_1\ldots m_r}_{a_1|a_2\ldots a_{\mod{A_1}},A_2,\ldots,A_d |B_1,\ldots ,B_{d+r+5}} \cr
&{} -\Gamma^{m_1\ldots m_r}_{a_{\mod{A_1}}|a_1\ldots a_{\mod{A_1}-1},A_2,\ldots,A_d |B_1,\ldots ,B_{d+r+5}} + (A_1 \leftrightarrow A_2,\ldots, A_d) \big] \cr
&{} +\bigl[  r \delta_{|B_1|,1} k_{b_1}^{(m_1} \Gamma^{m_2\ldots m_r)}_{b_1|A_1,\ldots,A_d|B_2,\ldots,B_{d+r+5}} 
+ \Gamma^{m_1\ldots m_r}_{b_1|A_1,\ldots, A_d| b_2\ldots b_{|B_1|}, B_2,\ldots, B_{d+r+5}}\cr
&{} - \Gamma^{m_1\ldots m_r}_{b_{\mod{B_1}}|A_1,\ldots, A_d| b_1\ldots b_{\mod{B_1}-1}, B_2,\ldots, B_{d+r+5}}
+ (B_1 \leftrightarrow B_2,\ldots ,B_{d+r+5}) \bigr] \Big\}  \,.
}$$
For example
\eqnn\Ypseudos
$$\eqalignno{
\Gamma_{1|2|3,4,5,6,7,8} &= M_1 {\cal Y}_{2|3,4,5,6,7,8} + k_2^m M_1 \otimes \Gamma^m_{2|3,4,5,6,7,8} & \Ypseudos\cr
&=  M_1 {\cal Y}_{2|3,4,5,6,7,8} + M_{12} k_2^m \cY_{3,4,5,6,7,8}^m + \bigl[ s_{23} M_{123} \cY_{4,5,6,7,8} + (3 \leftrightarrow 4,\ldots,8)\bigr]  \ .
}$$
Note that \aJCgen\ and \apinvgen\ resemble the derivation of pseudoinvariants $P^{m_1\ldots m_r}_{i|A_1,\ldots,A_d | B_1,\ldots,B_{d+r+3}} $ via \JCgen\ and \pinvgen, and the example \Ypseudos\ is the anomaly counterpart of $P_{1|2|3,4,5,6}$ given in \pseudos.

The ghost-number-four invariants \apinvgen\ describe the anomalous BRST transformation
\eqnn\HRijk
$$\eqalignno{
Q P^{m_1\ldots m_r}_{i|A_1,\ldots,A_d | B_1,\ldots,B_{d+r+3}}  &= - {r\choose 2 } \delta^{(m_1 m_2}  \Gamma^{m_3\ldots m_r)}_{i | A_1,\ldots,A_{d}|B_1,\ldots,B_{d+r+3}} &\HRijk \cr
& - \big[ \Gamma^{m_1\ldots m_r}_{i | A_2,\ldots,A_{d}|A_1,B_1,\ldots,B_{d+r+3}} + (A_1\leftrightarrow A_2,\ldots, A_d) \big] 
}$$
with anomaly counterpart
\eqnn\aHRijk
$$\eqalignno{
Q \Gamma^{m_1\ldots m_r}_{i|A_1,\ldots,A_d | B_1,\ldots,B_{d+r+5}}  &=0  \ .&\aHRijk
}$$
The former can be see from the BRST variation of the leading term $M_i {\cal J}^{m_1\ldots m_r}_{A_1,\ldots,A_d |
B_1,\ldots,B_{d+r+3}}$ in $P^{m_1\ldots m_r}_{i|A_1,\ldots,A_d | B_1,\ldots,B_{d+r+3}} $ where any $M_i {\cal
Y}^{m_1\ldots m_p}_{A_1,\ldots,A_q | B_1,\ldots,B_{p+q+5}}$ is identified as a leading term in \HanomRRc\ and promoted
to its completion $\Gamma^{m_1\ldots m_p}_{i|A_1,\ldots,A_q | B_1,\ldots,B_{p+q+5}}$.

\subsec Anomaly trace relations

\subseclab\sevenfive
The analysis of trace relations in section \sixfive\ straightforwardly carries over to anomalous building blocks. As before, the
${\cal H}_{[A,B]}$ corrections in the definition \refWc\ of $\cY^{m_1\ldots m_r}_{A_1,\ldots,A_{d}|B_1,\ldots,B_{d+r+5}}$
drop out in the combinations on the right-hand side of
\eqn\symmaBBs{
\delta_{np} {\cal Y}^{npm_1\ldots m_r}_{A_1,\ldots,A_{d} | B_1,\ldots,B_{d+r+7}} =
2\cY^{m_1 \ldots m_r}_{A_1, \ldots,A_{d},B_1|B_2, \ldots, B_{d+r+7}} + (B_1 \leftrightarrow B_2,\ldots, B_{d+r+7})  \,.
}
The inductive proof for the non-anomalous counterpart \symmBBs\ fits to the present setting after trivial adjustments -- adding
two extra slots and suppressing the ${\cal W}^{m_1\ldots}_{\ldots}$ contribution.
Similar to \symmCs, one can uplift $M_i$ times \symmaBBs\ to the BRST completions,
\eqn\symmaCCs{
\delta_{np} \Gamma^{npm_1\ldots m_r}_{i|A_1,\ldots,A_{d} | B_1,\ldots,B_{d+r+7}} =
2 \Gamma^{m_1 \ldots m_r}_{i|A_1, \ldots,A_{d},B_1|B_2, \ldots, B_{d+r+7}} + (B_1 \leftrightarrow B_2,\ldots, B_{d+r+7})\,,
}
consistent with the BRST variations in \HRijk.

\newsec Generalizing the recursion scheme

\seclab\seceight
In sections \secfour\ to \secsix, we have built up a grid of BRST pseudo-invariant objects of ghost number three
whose structure is summarized in \figoverview. As we have seen in section \secseven\ and in particular \anomgrid, the
grid of pseudoinvariants has a straightforward extension to the anomaly sector at ghost-number four with two further
slots. Given the almost identical recursion relations \pinvgen\ and \apinvgen\ for the BRST (pseudo-)invariants
$P^{m_1\ldots m_r}_{i|A_1,\ldots,A_{d} | B_1,\ldots,B_{d+r+3}}$ and $\Gamma^{m_1\ldots m_r}_{i|A_1,\ldots,A_{d} |
B_1,\ldots,B_{d+r+5}}$, it is natural to embed these two cases into a unified framework.

We will do so in section \eightone\ by promoting \pinvgen\ and \apinvgen\ to a ``master recursion''. The latter points
towards further special cases besides $P^{m_1\ldots }_{i| \ldots }$ and $\Gamma^{m_1 }_{i|\ldots}$. In
sections~\eighttwo\ and \eightthree, we are led to two families of ghost-number-two objects, and their anomalous
counterparts at ghost-number three are discussed in subsection~\eightfour\ and \eightfive. Each of these four cases
exhibits a grid structure almost identical to \figoverview\ and \anomgrid. The superficial disparity in the number of
unrefined slots $B_i$ is taken care of as an integer parameter of the master recursion.

As a major benefit of the ghost-number-two families described in sections \eighttwo\ and \eightthree, their BRST
variation generates a rich network of relations among pseudoinvariants, beyond the trace identities in section~\sixfive,
\eqn\schem{
Q(\hbox{\it ghost-number-two  object}) = \hbox{\it ghost-number-three relation}\,.
}
These BRST-exact relations turn out to connect momentum contractions $k_{B_i}^p P^{p m_1 \ldots}_{i|\ldots}$ with pseudoinvariants of lower rank.
For example, the five-point combinations
\eqnn\fivezero
$$\eqalignno{
&k_1^m C^m_{1|2,3,4,5} \, \ , \ \ \  \ \ \ k_2^m C^m_{1|2,3,4,5} + s_{23} C_{1|23,4,5}+ s_{24} C_{1|24,3,5}+ s_{25} C_{1|25,3,4} & \fivezero
}$$
will be identified as BRST-exact if momentum conservation $k_{12345}^m=0$ holds. As will be detailed in sections
\secnine\ and \secten, the master recursion in section \eightone\ systematically constructs the required
ghost-number-two superfields which generate meaningful relations via \schem.

\subsec The master recursion

\subseclab\eightone 
The purpose of this section is to unify the almost identical recursions \pinvgen\ and \apinvgen\
for the BRST (pseudo-)invariants $P^{m_1\ldots m_r}_{i|A_1,\ldots,A_{d} | B_1,\ldots,B_{d+r+3}}$ and anomaly
invariants $\Gamma^{m_1\ldots m_r}_{i|A_1,\ldots,A_{d} | B_1,\ldots,B_{d+r+5}}$. The superficial difference set by the number
three and five of unrefined slots in the simplest constituents $M_{A,B,C}$ and ${\cal Y}_{A,B,C,D,E}$
is described by an integer parameter. This amounts to replacing the superfields by an abstract symbol ${\cal
U}_{A_1,A_2,\ldots,A_N}$ with a variable number~$N$ of slots.

In the same way as $M_{A,B,C}$ and ${\cal Y}_{A,B,C,D,E}$ have been generalized to arbitrary rank~$r$ and refinement
$d$, we introduce formal symbols at all values of~$d$ and $r$,
\eqn\symb{
{\cal J}^{m_1\ldots m_r}_{A_1,\ldots,A_{d} | B_1,\ldots,B_{d+r+3}}, \ {\cal Y}^{m_1\ldots m_r}_{A_1,\ldots,A_{d} | B_1,\ldots,B_{d+r+5}} \ \rightarrow \ {\cal U}^{m_1\ldots m_r}_{A_1,\ldots,A_{d} | B_1,\ldots,B_{d+r+N}} \ .
}
They are defined to be symmetric in $m_i, A_i,B_i$ but not under exchange of $A_i \leftrightarrow B_j$, so they may be
identified with ${\cal J}^{m_1\ldots m_r}_{A_1,\ldots,A_{d} | B_1,\ldots,B_{d+r+3}}$ and
${\cal Y}^{m_1\ldots m_r}_{A_1,\ldots,A_{d} | B_1,\ldots,B_{d+r+5}}$ if $N=3$ and $N=5$, respectively.
In terms of the standard Berends--Giele currents $M_A$ and the symbol in \symb, we recursively define abstract
tensors
\eqnn\master
$$\displaylines{
R^{(N), \, m_1\ldots m_r}_{i|A_1,\ldots,A_d | B_1,\ldots,B_{d+r+N}} \equiv  M_i {\cal U}^{m_1\ldots m_r}_{A_1,\ldots,A_d | B_1,\ldots,B_{d+r+N}} \cr
 + M_i \otimes \Big\{ \big[ \delta_{\mod{A_1},1} k_{a_1}^p R^{(N), \,pm_1\ldots m_r}_{a_1|A_2,\ldots,A_d |B_1,\ldots ,B_{d+r+N}}  
+ R^{(N), \,m_1\ldots m_r}_{a_1|a_2\ldots a_{\mod{A_1}},A_2,\ldots,A_d |B_1,\ldots ,B_{d+r+N}} \cr
 -R^{(N), \,m_1\ldots m_r}_{a_{\mod{A_1}}|a_1\ldots a_{\mod{A_1}-1},A_2,\ldots,A_d |B_1,\ldots ,B_{d+r+N}} + (A_1 \leftrightarrow A_2,\ldots, A_d) \big] \cr
 +\bigl[  r \delta_{|B_1|,1} k_{b_1}^{(m_1} R^{(N), \,m_2\ldots m_r)}_{b_1|A_1,\ldots,A_d|B_2,\ldots,B_{d+r+N}} + R^{(N), \,m_1\ldots m_r}_{b_1|A_1,\ldots, A_d| b_2\ldots b_{|B_1|}, B_2,\ldots, B_{d+r+N}}\cr
 - R^{(N), \,m_1\ldots m_r}_{b_{\mod{B_1}}|A_1,\ldots, A_d| b_1\ldots b_{\mod{B_1}-1}, B_2,\ldots, B_{d+r+N}}
+ (B_1 \leftrightarrow B_2,\ldots ,B_{d+r+N}) \bigr] \Big\}  \,. \hfil\master\hfilneg
}$$
The following two specializations reproduce the known recursions \pinvgen\ and \apinvgen:
\eqnn\masterA
\eqnn\masterB
$$\eqalignno{
P^{m_1\ldots m_r}_{i|A_1,\ldots,A_{d} | B_1,\ldots,B_{d+r+3}} &=
R^{(N=3), \, m_1\ldots m_r}_{i|A_1,\ldots,A_d | B_1,\ldots,B_{d+r+3}}[ {\cal U}^{\ldots}_{\ldots} \rightarrow {\cal J}^{\ldots}_{\ldots} ]  &\masterA \cr
\Gamma^{m_1\ldots m_r}_{i|A_1,\ldots,A_{d} | B_1,\ldots,B_{d+r+5}} &= 
R^{(N=5), \, m_1\ldots m_r}_{i|A_1,\ldots,A_d | B_1,\ldots,B_{d+r+5}}[ {\cal U}^{\ldots}_{\ldots} \rightarrow {\cal Y}^{\ldots}_{\ldots} ] \ .
&\masterB
}$$
The simplest examples for the generalized $R^{(N),\ldots}_{i|\ldots}$ are
\eqnn\masterC
\eqnn\masterD
\eqnn\masterE
\eqnn\masterF
$$\eqalignno{
R^{(N)}_{1|2,3,\ldots,N+1} &= M_1 {\cal U}_{2,3,\ldots,N+1} &\masterC \cr
R^{(N)}_{1|23,4,\ldots,N+2} &= M_1 {\cal U}_{23,4,\ldots,N+2} +M_{12} {\cal U}_{3,4,\ldots,N+2} -M_{13} {\cal U}_{2,4,\ldots,N+2} &\masterD \cr
R^{(N), \, m}_{1|2,3,\ldots,N+2} &= M_1 {\cal U}^m_{2,3,\ldots,N+2} + \big[ k_2^m M_{12} {\cal U}_{3,4,\ldots,N+2} + (2\leftrightarrow 3,4,\ldots,N+2) \big] &\masterE \cr
R^{(N)}_{1|2|3,\ldots,N+3} &= M_1 {\cal U}_{2|3,4,\ldots,N+3} + M_{12} k_2^m {\cal U}^m_{3,4,\ldots,N+3} \cr
& \ \ \ \ \ + \big[ s_{23} M_{123} {\cal U}_{4,\ldots,N+3} + (3\leftrightarrow 4,5,\ldots,N+3) \big] \ .&\masterF
}$$
The right-hand sides obviously specialize to familiar expressions such as
\item{$\bullet$} \Cscex\ and \pseudos\ for
$C_{1|2,3,4},C_{1|23,4,5},$ $C^{m}_{1|2,3,4,5}$ and $P_{1|2|3,4,5,6}$ under \masterA\ 
\item{$\bullet$} \HRig\ and \Ypseudos\
for $\Gamma_{1|2,3,4,5,6},\Gamma_{1|23,4,5,6,7},\Gamma^{m}_{1|2,3,4,5,6,7}$ and $\Gamma_{1|2|3,4,5,6,7,8}$ under
\masterB.

\noindent
In the following sections, we consider the abstract tensors $R^{(N),\ldots}_{i|\ldots}$ in \master\ at values $N=2,4,6$.
In order to accommodate this with the number of slots of ${\cal U} \in \{ {\cal J},{\cal Y} \}$, we have to eliminate
the Berends--Giele currents $M_A$ and adjoin the word $A$ to the slots of the accompanying symbol. In
the non-anomalous case ${\cal U} = {\cal J}$, this gives rise to ghost-number-two objects, and the anomalous choice
${\cal U} = {\cal Y}$ yields ghost number three. Moreover, the word $A$ from the eliminated $M_A$ can become
either a refined or a non-refined slot of the symbol, leading to ${\cal U}_{A,\ldots|\ldots}$ or ${\cal
U}_{\ldots|A,\ldots}$. These two independent choices yield a total of four new families of superfields whose notation
and schematic form is summarized by
\eqnn\masterG
\eqnn\masterH
\eqnn\masterI
\eqnn\masterJ
$$\eqalignno{
D^{ \ldots }_{i|\ldots } &\equiv
R^{(N=2), \,  \ldots }_{i|\ldots}[ M_A {\cal U}^{\ldots}_{\{B_j\} | \{C_j\}} \rightarrow {\cal J}^{\ldots}_{\{B_j\} |A, \{C_j\}} ]  &\masterG \cr
L^{\ldots }_{i| \ldots }  &\equiv
R^{(N=4), \,  \ldots }_{i| \ldots}[ M_A {\cal U}^{\ldots}_{\{B_j\} | \{C_j\}} \rightarrow {\cal J}^{\ldots}_{A, \{B_j\} | \{C_j\}} ] &\masterH \cr
\Delta^{ \ldots }_{i| \ldots } &\equiv
R^{(N=4), \,  \ldots }_{i| \ldots}[ M_A {\cal U}^{\ldots}_{\{B_j\} | \{C_j\}} \rightarrow {\cal Y}^{\ldots}_{\{B_j\} |A, \{C_j\}} ]  &\masterI \cr
\Lambda^{ \ldots }_{i| \ldots }  &\equiv
R^{(N=6), \, \ldots }_{i| \ldots}[ M_A {\cal U}^{\ldots}_{\{B_j\} | \{C_j\}} \rightarrow {\cal Y}^{\ldots}_{A, \{B_j\} | \{C_j\}} ] \ . &\masterJ
}$$
The precise definitions and simplest examples are given in the following subsections.

\subsec The $D$-superfields at ghost-number two

\subseclab\eighttwo 
As a first avenue towards BRST generators at ghost number two, we consider the tensors $R^{(N),\ldots}_{i|\ldots}$ in \master\ at $N=2$ and convert the word $A$ associated with $M_A$ to a non-refined
slot of the associated symbol ${\cal U} \rightarrow {\cal J}$. As sketched in \masterG, this gives rise to the
definition
\eqnn\masterK
$$\eqalignno{
D^{m_1\ldots m_r}_{i| B_1,\ldots,B_d | C_1,\ldots,C_{d+r+2} } &\equiv
R^{(N=2), \, m_1\ldots m_r}_{i| B_1,\ldots,B_d | C_1,\ldots,C_{d+r+2} }[ M_A {\cal U}^{\ldots}_{\{F_j\} | \{G_j\}} \rightarrow {\cal J}^{\ldots}_{\{F_j\} |A, \{G_j\}} ]  \ .&\masterK
}$$
The replacement rule in \masterK\ converts the formal objects \masterC, \masterD, \masterE\ and \masterF\ to
\eqnn\quasie
$$\eqalignno{
D_{1|2,3} &= M_{1,2,3}\,, &\quasie
\cr
D_{1|23,4} &= M_{12,3,4}+M_{1,23,4}+M_{31,2,4}\,,
\cr
D^m_{1|2,3,4} &= M^m_{1,2,3,4}+ k_2^m M_{12,3,4}+k_3^m M_{13,2,4} + k_4^m M_{14,2,3}\,,
\cr
D_{1|2|3,4,5}&={\cal J}_{2|1,3,4,5} + k_2^m M^m_{12,3,4,5} + \big[ s_{23} M_{123,4,5} + (3\leftrightarrow 4,5) \big] \,.
}$$
In the next section \secnine, these ghost-number-two objects and their generalizations are shown to serve as powerful BRST generators in the sense of \schem.

\subsec The $L$-superfields at ghost-number two

\subseclab\eightthree
The $N=4$ version of the $R^{(N),\ldots}_{i|\ldots}$ in \master\ allows to convert the word $A$ associated
with $M_A$ to a refined slot of the associated symbol ${\cal U} \rightarrow {\cal J}$. The precise form of the
definition sketched in \masterH\ is
\eqn\masterL{
L^{m_1\ldots m_r}_{i| B_1,\ldots,B_d | C_1,\ldots,C_{d+r+4} } \equiv
R^{(N=4), \, m_1\ldots m_r}_{i| B_1,\ldots,B_d | C_1,\ldots,C_{d+r+4} }[ M_A {\cal U}^{\ldots}_{\{F_j\} | \{G_j\}}
\rightarrow {\cal J}^{\ldots}_{A,\{F_j\} | \{G_j\}} ]  \,.
}
Starting from the examples in \masterC\ to \masterF, the prescription \masterL\ yields
\eqnn\Lsa
$$\eqalignno{
L_{1|2,3,4,5} &= \cJ_{1|2,3,4,5}&\Lsa\cr
L_{1|23,4,5,6} &= \cJ_{1|23,4,5,6} + \cJ_{12|3,4,5,6} - \cJ_{13|2,4,5,6}\cr
L^m_{1|2,3,4,5,6} &=  \cJ^m_{1|2,3,4,5,6} + \bigl[k_2^m \cJ_{12|3,4,5,6} + (2\leftrightarrow 3,4,5,6)\bigr] \cr
L_{1|2|3,4,5,6,7} &= {\cal J}_{1,2|3,4,5,6,7} +k_2^m {\cal J}^m_{12|3,4,5,6,7} + \big[ s_{23} {\cal J}_{123|4,5,6,7} + (3 \leftrightarrow 4,5,6,7) \big] \ .
}$$
These superfields of ghost-number two serve as another family of BRST generators, see section \secten.

\subsec The $\Delta$-superfields at ghost-number three

\subseclab\eightfour
Another specialization of the $R^{(N),\ldots}_{i|\ldots}$ in \master\ to $N=4$ generates anomalous superfields. In
this case, the word $A$ associated with $M_A$ is adjoined to the non-refined slots of the associated symbol
${\cal U} \rightarrow {\cal Y}$. As sketched in \masterI, this gives rise to the definition
\eqnn\masterM
$$\eqalignno{
\Delta^{m_1\ldots m_r}_{i| B_1,\ldots,B_d | C_1,\ldots,C_{d+r+4} } &\equiv
R^{(N=4), \, m_1\ldots m_r}_{i| B_1,\ldots,B_d | C_1,\ldots,C_{d+r+4} }[ M_A {\cal U}^{\ldots}_{\{F_j\} | \{G_j\}} \rightarrow {\cal Y}^{\ldots}_{\{F_j\} |A, \{G_j\}} ]  \ ,&\masterM
}$$
which can be viewed as the anomaly counterparts of $D^{m_1\ldots m_r}_{i| B_1,\ldots,B_d | C_1,\ldots,C_{d+r+2} } $ in \masterK.

Applying the replacement rule \masterM\ to the examples in \masterC\ to \masterF, one arrives at
\eqnn\Zsa
\eqnn\Zsb
\eqnn\Zsc
\eqnn\Zsd
$$\eqalignno{
\Delta_{1|2,3,4,5} &= \cY_{1,2,3,4,5}&\Zsa\cr
\Delta_{1|23,4,5,6} &= \cY_{1,23,4,5,6} + \cY_{12,3,4,5,6} - \cY_{13,2,4,5,6} &\Zsb\cr
\Delta^m_{1|2,3,4,5,6} &= {\cal Y}^m_{1,2,3,4,5,6} +  \bigl[k_2^m \cY_{12,3,4,5,6} + (2\leftrightarrow 3,4,5,6)\bigr] &\Zsc \cr
\Delta_{1|2|3,4,5,6,7} &= {\cal Y}_{2|1,3,4,5,6,7} + k_2^m {\cal Y}^m_{12,3,4,5,6,7} + \big[ s_{23} {\cal Y}_{123,4,5,6,7} + (3 \leftrightarrow 4,5,6,7) \big] \ ,  &\Zsd
}$$
which can be easily recognized as the anomaly analogues of \quasie.

\subsec The $\Lambda$-superfields at ghost-number three

\subseclab\eightfive Finally, there is a $N=6$ version of the $R^{(N),\ldots}_{i|\ldots}$ in \master\ where
the word $A$ associated with $M_A$ becomes a refined slot of the symbol ${\cal U} \rightarrow {\cal
Y}$. The resulting anomalous superfields were sketched in \masterJ\ and are more cleanly defined as
\eqnn\masterN
$$\eqalignno{
\Lambda^{m_1\ldots m_r}_{i| B_1,\ldots,B_d | C_1,\ldots,C_{d+r+6} } &\equiv
R^{(N=6), \, m_1\ldots m_r}_{i| B_1,\ldots,B_d | C_1,\ldots,C_{d+r+6} }[ M_A {\cal U}^{\ldots}_{\{F_j\} | \{G_j\}} \rightarrow {\cal Y}^{\ldots}_{A,\{F_j\} | \{G_j\}} ]  \ .&\masterN
}$$
This is the anomaly counterpart of the objects $L^{m_1\ldots m_r}_{i| B_1,\ldots,B_d | C_1,\ldots,C_{d+r+4} }$ in \masterL.

Under the prescription in \masterN, the examples in \masterC\ to \masterF\ are mapped to
\eqnn\JRp
\eqnn\JRq
\eqnn\JRr
\eqnn\JRs
$$\eqalignno{
\Lambda_{1|2,3,4,5,6,7}  &=\cY_{1|2,3,4,5,6,7}   &\JRp
\cr
\Lambda_{1|23,4,5,6,7,8}  &= \cY_{1|23,4,5,6,7,8}   + \cY_{12|3,4,5,6,7,8}  - \cY_{13|2,4,5,6,7,8}  &\JRq
\cr
\Lambda^m_{1|2,3,4,5,6,7,8}  &= \cY^m_{1|2,3,4,5,6,7,8} + \big[ k_2^m \cY_{12|3,4,5,6,7,8} + (2 \leftrightarrow 3,\ldots,8) \big]   &\JRr 
\cr
\Lambda_{1|2|3,4,5,6,7,8,9}  &= {\cal Y}_{1,2|3,4,5,6,7,8,9} +k_2^m {\cal Y}^m_{12|3,4,5,6,7,8,9} \cr
& \ \ \ + \big[ s_{23} {\cal Y}_{123|4,5,6,7,8,9} + (3 \leftrightarrow 4,\ldots,9) \big]  \ . &\JRs
}$$
They can be quickly seen to furnish the anomaly counterparts of \Lsa.

\newsec Pseudoinvariant relations for $k_{B}$ momentum contractions

\seclab\secnine
The main concern in this paper is to systematically study the properties of and relations among the pseudoinvariants
$P^{m_1\ldots}_{i|\ldots}$ which carry the polarization dependence of one-loop amplitudes. In the previous section, we have constructed two families of ghost-number-two superfields
whose BRST variations will be demonstrated to generate relations among the $P^{m_1\ldots}_{i|\ldots}$. 

Recall that pseudoinvariants $P^{m_1\ldots}_{i|\ldots}$ single out a reference leg $i$ which always enter through a Berends--Giele current of type $M_{i\ldots}$ and
which is represented by an unintegrated vertex $V_i$ in the one-loop amplitude prescription \multiloop. The
ghost-number-two generators of relations among pseudoinvariants in \schem\ must be carefully chosen in order to
avoid admixtures of pseudoinvariants $P^{m_1\ldots}_{k\neq i |\ldots}$ with a different reference leg $k \neq i$. 

It turns out that both the $D$ superfields from section \eighttwo\ and the $L$ superfields from section \eightthree\
satisfy this criterion, see \masterK\ and \masterL\ for their precise definitions. BRST variations of type $QD$ are
systematically analyzed in the present section, and section \secten\ is devoted to $QL$. We will see how the
ghost-number-three expression for $QD^{m_1\ldots m_r}_{i|A_1,\ldots,A_d | B_1,\ldots,B_{d+r+2}}$ relates momentum
contractions $k_{B_j}^{n}P^{nm_2\ldots m_r}_{i|A_1,\ldots,A_d | B_1,\ldots,B_{d+r+3}}$ to pseudoinvariants at lower
rank. These relations are crucial to translate the SYM one-loop amplitudes presented in
\wipH\ into worldline parametrization and to make contact with their string theory ancestors.

\subsec BRST-exactness versus momentum phase space

\subseclab\nineone
As a starting point, we investigate the $Q$ action on unrefined $D$ superfields $D^{m_1\ldots m_r}_{i|
A_1,\ldots,A_{r+2}}$ such as the simplest examples given in \quasie. For scalars and vectors, we find
\eqnn\JRzero
\eqnn\JRa
$$\eqalignno{
Q D_{i|A,B}  &=0  &\JRzero \cr
Q   D^m_{i|A,B,C}  &=k_{iABC}^m C_{i|A,B,C}  &\JRa 
}$$
with overall momentum $k_{iABC}^m \equiv k^m_i+k^m_A+k^m_B+k^m_C$. This can be verified case by case using the BRST variations
\QMgen\ and \QMnrew\ for each $M_{A,B,C}$ and $M^m_{A,B,C,D}$ occurring in $D_{i|A,B}$ and $D^m_{i|A,B,C} $,
respectively. Analogous methods are used in all the subsequent cases when $Q$ variations are computed.

Contracting \JRa\ with any momentum, one can solve for $C_{i|A,B,C}$,
\eqnn\quasill
$$\eqalignno{
C_{i|A,B,C} &=
Q\left[ { k_m^i D^m_{i|A,B,C}    \over (k_i \cdot k_{iABC})} \right] \,, &\quasill
}$$
i.e. $C_{i|A,B,C}$ is BRST exact unless $k_i \cdot k_{iABC} = 0$. Note, however, that momentum conservation $k^m_{iABC}
= 0$ in an $n$-point amplitude (with $n=1+|A|+|B|+|C|$) implies $k_i \cdot k_{iABC} = 0$ and renders the right-hand
side of \quasill\ ill-defined. Hence, momentum phase space constraints for $n$ massless particles save $C_{i|A,B,C}$ from being BRST
exact and preserve its cohomological nature in $n$ point amplitudes. 

This is analogous to the superspace representation $\sum_{j=1}^{n-2} M_{12 \ldots j} M_{j+1\ldots n-1}V_n$ of color
ordered SYM tree amplitudes \nptTree. This expression can be rewritten as $Q(M_{12 \ldots n-1}V_n)$ as long as the
overall propagator $M_{12 \ldots n-1} \sim s_{12 \ldots n-1}^{-1}$ does not diverge. Again, $n$ particle momentum
conservation implying $s_{12 \ldots n-1}=0 $ is essential to avoid BRST exactness of the tree amplitude.

In both cases, the cohomology nature of BRST-closed kinematic factors crucially depends on vanishing conformal weight
$h\sim s_{12\ldots n}$. Recall that in a topological conformal field theory where $Qb_0 = L_0$, the cohomology at non-zero
conformal weight is empty since every BRST-closed operator would also be BRST-exact \Figueroa,
\eqn\topolog{
Q\phi=0 \ , \ \ \ L_0\phi=h\phi\ , \ \ \ h\neq 0 \ \  \Rightarrow \ \ \phi=Q\Big( {b_0\phi \over h} \Big)\,.
}
Starting from rank two, the $Q$ transformations of $D^{m_1 m_2 \ldots m_r}_{i|A_1,\ldots,A_{r+2}}$ additionally give
rise to anomalous superfields $\Delta^{m_1 \ldots m_p}_{i|A_1,\ldots,A_{p+4}}$ defined in \masterM, e.g.
\eqnn\JRb
$$\eqalignno{
Q D^{mn}_{i|A,B,C,D} &=\delta^{mn} \Delta_{i|A,B,C,D}+ 2k_{iABCD}^{(m} C_{i|A,B,C,D}^{n)}   \,, &\JRb
}$$
and more generally,
\eqn\JRe{
Q D^{m_1 m_2 \ldots m_r}_{i|A_1,\ldots,A_{r+2}}   = { r \choose 2} \delta^{(m_1 m_2}
\Delta^{m_3 m_4 \ldots m_r)}_{i|A_1,\ldots,A_{r+2}}+r  k_{iA_1\ldots A_{r+2}}^{(m_1} C_{i|A_1,\ldots,A_{r+2}}^{m_2 m_3
\ldots m_r)} \,.
}
As a consequence, the identification of BRST-exact quantities crucially depends on the momentum phase space. In case
of momentum conservation $k_{iA_1\ldots A_{r+2}}^{m}=0$, \JRe\ implies\foot{At rank $r=2$ and $r=3$, BRST exactness of
$\Delta_{i|A,B,C,D}$ and $\Delta^m_{i|A,B,C,D,E}$ immediately follows from single traces of \JRe\ at $k_{iA_1\ldots
A_{r+2}}^{m}=0$. Higher rank $r \geq 4$ requires combinations of multiple $\delta_{m_im_j}$ contractions in order to
identify the BRST generator of $\Delta^{m_1 \ldots m_p}_{i|A_1,\ldots,A_{p+4}}$ at any rank.} $Q$ exactness of the
anomalous superfield $\Delta^{m_1 \ldots m_p}_{i|A_1,\ldots,A_{p+4}}$, hence the latter does not contribute to
physical amplitudes at multiplicity $1 + \sum_{j=1}^{p-4} |A_j|$. However, it is important to stress that the hexagon
gauge anomaly superfield $\Delta_{2|3,4,5,6}=\cY_{2,3,4,5,6}$ in the
one-loop six-point amplitude \anomaly\ is {\it not} BRST exact for the momentum phase space of six particles since $k^m_{23456}$
is not zero.

On the other hand, generic momentum configurations with $k_{iA_1\ldots A_{r+2}}^{m}\neq 0$ render the traceless
components of pseudoinvariants $C^{m_1 m_2 \ldots m_r}_{i|A_1,\ldots,A_{r+3}}$ BRST exact. This can be seen from the
traceless projection of \JRe, see \quasiqq\ in appendix \appE\ for the explicit form of the BRST generator for
$C^m_{i|A,B,C,D}$.

The correspondence between the pseudoinvariants $C^{m_1 m_2 \ldots m_r}_{i|A_1,\ldots,A_{r+3}}$ and the
anomaly invariants $\Gamma^{m_1 m_2 \ldots m_r}_{i|A_1,\ldots,A_{r+5}}$ described in section \secseven\ and formalized
in section \eightone\ allows to immediately write down the anomaly correspondent of \JRe:
\eqnn\AJRe
$$\eqalignno{
Q \Delta^{m_1 m_2 \ldots m_r}_{i|A_1,\ldots,A_{r+4}}& =  r  k_{iA_1\ldots A_{r+4}}^{(m_1} \Gamma_{i|A_1,\ldots,A_{r+4}}^{m_2 m_3 \ldots m_r)} \ . &\AJRe
}$$
We exploit that $\Delta^{m_1 m_2 \ldots m_r}_{i|A_1,\ldots,A_{r+4}}$ is the anomaly counterpart of $D^{m_1 m_2 \ldots
m_r}_{i|A_1,\ldots,A_{r+2}} $ which, loosely speaking, does not have a higher anomaly image. Of course, \AJRe\
confirms that momentum conservation $k_{iA_1\ldots A_{r+4}}^{m}=0$ implies BRST closure of $\Delta^{m_1 m_2 \ldots
m_r}_{i|A_1,\ldots,A_{r+4}}$, in lines with the discussion of BRST exactness along with \JRe.

\subsec Momentum contractions of unrefined pseudoinvariants

\subseclab\ninetwo
Refined versions of the $D$ superfields turn out to generate a much richer set of ghost number
three relations than their unrefined counterparts studied in section \nineone. We start by exploring the case of
minimal refinement $d=1$ and will find that $QD^{m_1\ldots m_r}_{i|A|B_1,\ldots,B_{r+3}} $ relates momentum contractions
$\sim k_{A_j}$ of $C^{m_1\ldots m_r}_{i|A_1,\ldots,A_{r+3}}$ to pseudoinvariants at lower rank.

As a first example, consider the inequivalent cases at five- and six-points,
\eqnn\quasjf
$$\eqalignno{
Q D_{1|2|3,4,5}  &=\Delta_{1|2,3,4,5} + k_2^m C_{1|2,3,4,5}^m +  \big[ s_{23} C_{1|23,4,5} + (3 \leftrightarrow 4,5) \big]&\quasjf
\cr
Q D_{1|23|4,5,6}   &=\Delta_{1|23,4,5,6}  + P_{1|3|2,4,5,6} - P_{1|2|3,4,5,6}+ k_{23}^m C_{1|23,4,5,6}^m  \cr
& +\big[ s_{34} C_{1|234,5,6} - s_{24} C_{1|324,5,6} + (4\leftrightarrow 5,6) \big] 
\cr
Q D_{1|4|23,5,6} &=\Delta_{1|23,4,5,6} + k_4^m C_{1|23,4,5,6}^m +s_{24} C_{1|324,5,6} - s_{34} C_{1|234,5,6}\cr
& + s_{45} C_{1|23,45,6} + s_{46} C_{1|23,46,5} \,, 
}$$
where \quasjf\ underpins the second example in \fivezero\ provided that momentum conservation $k_{12345}^m =0$ renders
$\Delta_{1|2,3,4,5}$ BRST exact. 

The combinations of $s_{ij} C_{1|A,B,C}$ can be neatly described using the $S[A,B]$ map in \QEone, see in particular
\QEoneEx\ for examples. The seven-point instances of $QD_{1|A|B,C,D}$ displayed in \JRt\ to \JRw\ support this pattern and
help to identify the appearance of $P_{1|A|B,C,D,E}$ as deconcatenations. These observations lead to the following
generalization,
\eqnn\JRac
$$\eqalignno{
Q D_{i|A|B,C,D} &= \Delta_{i|A,B,C,D} + k_A^m C^m_{i|A,B,C,D}+\sum_{XY=A} ( P_{i|Y |X,B,C,D} -P_{i|X| Y,B,C,D} ) \cr
&\ \ \ \  \ \ \ \  + C_{i| S[A,B],C,D}+ C_{i| S[A,C],B,D} + C_{i| S[A,D],B,C} \ .
 &\JRac
}$$
Note that $QD_{i|A|B,C,D} $ always generates relations for contractions of $C^m_{i|A,B,C,D}$ with the entire
momentum $k_A^m = \sum_{j=1}^{|A|} k_{a_j}^m$ in the slot $A=a_1a_2\ldots a_{|A|}$. This method does not provide any information on contractions with partial slot momenta, e.g.
$k_{a_1}^m C^m_{i|A,B,C,D}$ with $|A|\geq 2$. 

It is natural to repeat the BRST manipulations for vectorial $D$ superfields such as
\eqnn\quasjl
$$\eqalignno{
Q D^m_{1|2|3,4,5,6}  &= \Delta^m_{1|2,3,4,5,6}+ k_p^2 C^{mp}_{1|2,3,4,5,6} + (k_{123456}^m-k_2^m) P_{1|2|3,4,5,6} &\quasjl\cr
&+  s_{23} C^m_{1|23,4,5,6} + s_{24} C^m_{1|24,3,5,6} + s_{25} C^m_{1|25,3,4,6} + s_{26} C^m_{1|26,3,4,5}
\cr
Q  D^{m}_{1|23|4,5,6,7} &=\Delta^m_{1|23,4,5,6,7} + k_{23}^p C^{mp}_{1|23,4,5,6,7}  + (k_{1234567}^m - k_{23}^m ) P_{1|23|4,5,6,7}  \cr
&-  P^m_{1|2|3,4,5,6,7} + P^m_{1|3|2,4,5,6,7} + \big[ s_{34} C^m_{1| 234,5,6,7} - s_{24} C^m_{1| 324,5,6,7} + (4
\leftrightarrow 5,6,7) \big] 
\cr
Q  D^m_{1|4|23,5,6,7}&=\Delta^m_{1|23,4,5,6,7}  + k_{4}^p C^{mp}_{1|23,4,5,6,7} + (k_{1234567}^m-k_4^m) P_{1|4|23,5,6,7}   \cr
&+ s_{24} C^m_{1| 324,5,6,7} - s_{34} C^m_{1| 234,5,6,7} + \big[ s_{45} C^m_{1|23,45,6,7} + (5 \leftrightarrow 6,7)
\big] 
}$$
which can be summarized by a general formula similar to the scalar case in \JRac,
\eqnn\JRad
$$\displaylines{
Q D^m_{i|A|B,C,D,E} = \Delta^{m}_{i|A,B,C,D,E}  + k_A^p C^{pm}_{i|A,B,C,D,E} + (k_{iABCDE}^m-k_A^m) P_{i|A|B,C,D,E} \cr
+\big[ C^m_{i| S[A,B],C,D,E}+ (B \leftrightarrow C,D,E) \big]  +\sum_{XY=A}  ( P^m_{i|Y| X,B,C,D,E} -P^m_{i| X | Y,B,C,D,E} )  \ .
\hfil\JRad\hfilneg
 }$$
Tensorial generalizations at rank $r\geq 2$ additionally involve refined versions of the anomalous $\Delta$ superfields in \masterM. The simplest example occurs at seven points,
\eqnn\JRz
$$\eqalignno{
Q D^{mn}_{1|2|3,4,5,6,7} &= \Delta^{mn}_{1|2,3,4,5,6,7}+\delta^{mn} \Delta_{1|2|3,4,5,6,7}+ k_2^p C^{mnp}_{1|2,3,4,5,6,7}  &\JRz \cr
&+ 2 (k^{(m}_{1234567}-k_2^{(m}) P^{n)}_{1|2|3,4,5,6,7} + \big[ s_{23} C^{mn}_{1|23,4,5,6,7} + (3 \leftrightarrow 4,5,6,7) \big]  \,,
}$$
where $\Delta_{1|2|3,4,5,6,7}$ is given by \Zsd. The structure of the scalar and vector cases \JRac\ and \JRad\ inspires the following generalization of \JRz\ to multiparticle slots:
\eqnn\JRae
$$\eqalignno{
 Q   &D^{mn}_{i|A|B,C,D,E,F}  = \Delta^{mn}_{i|A,B,C,D,E,F} + \delta^{mn} \Delta_{i|A|B,C,D,E,F} + k_A^p C^{mnp}_{i|A,B,C,D,E,F}   \cr
&+ 2(k_{iABCDEF}^{(m} -k_A^{(m} )  P^{n)}_{i|A|B,C,D,E,F}+\big[ C^{mn}_{i| S[A,B],C,D,E,F}+ (B \leftrightarrow C,D,E,F) \big]   \cr
&  +\sum_{XY=A} ( P^{mn}_{i|Y | X,B,C,D,E,F} -P^{mn}_{i|X| Y,B,C,D,E,F} ) \ .&\JRae 
}$$
This in turn allows to infer the BRST variation of $D^{m_1 \ldots}_{i|A|\ldots}$ superfields at generic rank:
\eqnn\JRal
$$\eqalignno{
  Q &D^{m_1 m_2 \ldots m_r}_{i|A|B_1,\ldots,B_{r+3}} = \Delta^{m_1 m_2 \ldots m_r}_{i | A,B_1,\ldots,B_{r+3}} + {r\choose 2} \delta^{(m_1 m_2} \Delta^{m_3\ldots m_r)}_{i|A|B_1,\ldots,B_{r+3}}  + k_A^p C^{pm_1\ldots m_r}_{i|A,B_1,\ldots,B_{r+3}} \cr
&+ r (k_{iAB_1\ldots B_{r+3}}^{(m_1} -k_A^{(m_1} ) P^{m_2\ldots m_r)}_{i|A|B_1,\ldots,B_{r+3}}
+\big[ C^{m_1\ldots m_r}_{i| S[A,B_1],B_2,\ldots,B_{r+3}}+ (B_1 \leftrightarrow B_2,\ldots,B_{r+3}) \big]  \cr
&+\sum_{XY=A} ( P^{m_1\ldots m_r}_{i|Y| X,B_1,\ldots,B_{r+3}} -P^{m_1\ldots m_r}_{i| X| Y,B_1,\ldots,B_{r+3}} ) \ . &\JRal
}$$
Recall that momentum conservation $k_{iA_1\ldots A_{r+4}}^m=0$ implies BRST exactness of the unrefined representatives
$\Delta^{m_1\ldots m_r}_{i|A_1,\ldots,A_{r+4}}$ of the anomalous $\Delta$ superfields. Hence, the latter do not
contribute when the relations \JRac, \JRad, \JRae\ and \JRal\ are applied to physical amplitudes. However, the situation
is completely different for their refined counterparts $\Delta^{m_1\ldots m_r}_{i|A|B_1,\ldots,B_{r+5}}$. As will be
demonstrated in the following, refined $\Delta$ superfields are {\it not} BRST closed, regardless of momentum phase space constraints, so $Q$
exactness can be clearly ruled out. Starting from seven points, refined anomaly superfields $\Delta^{m_1\ldots
m_r}_{i|A|B_1,\ldots,B_{r+5}}$ at ghost-number three cannot be discarded in the discussion of one-loop amplitudes, see
\wipH.

The ghost-number-four expression for $Q\Delta^{m_1\ldots m_r}_{i|A|B_1,\ldots,B_{r+5}}$ can be inferred by analogy
with \JRac\ to \JRal. Since \masterM\ identifies $\Delta^{m_1\ldots m_r}_{i|A|B_1,\ldots,B_{r+5}}$ to be the anomaly
counterpart of $D^{m_1 m_2 \ldots m_r}_{i|A|B_1,\ldots,B_{r+3}}$, the BRST transformation of the former follows from
\JRac\ and \JRal\ upon discarding anomalous terms and converting $C^{m_1\ldots}_{i|\ldots}
,P^{m_1\ldots}_{i|\ldots}\rightarrow \Gamma^{m_1\ldots}_{i|\ldots} $:
\eqnn\AJRac
\eqnn\AJRal
$$\eqalignno{
Q \Delta_{i|A|B,\ldots,F} &= k_A^m \Gamma^m_{i|A,B,\ldots,F}+\sum_{XY=A} (  \Gamma_{i|Y |X,B,\ldots,F} -  \Gamma_{i|X| Y,B,\ldots,F} ) \cr
&\ \ \ \  \   + \big[ \Gamma_{i| S[A,B],C,D,E,F}+ (B\leftrightarrow C,D,E,F) \big] 
 &\AJRac
\cr
  Q\Delta^{m_1 m_2 \ldots m_r}_{i|A|B_1,\ldots,B_{r+5}} & =  k_A^p  \Gamma^{pm_1\ldots m_r}_{i|A,B_1,\ldots,B_{r+5}} + r (k_{iAB_1\ldots B_{r+5}}^{(m_1} -k_A^{(m_1} ) \Gamma^{m_2\ldots m_r)}_{i|A|B_1,\ldots,B_{r+5}}\cr
&\ \ \ \  \ + \big[ \Gamma^{m_1\ldots m_r}_{i| S[A,B_1],B_2,\ldots,B_{r+5}}+ (B_1 \leftrightarrow B_2,\ldots,B_{r+5}) \big]  \cr
&\ \ \ \  \ +\sum_{XY=A} ( \Gamma^{m_1\ldots m_r}_{i|Y| X,B_1,\ldots,B_{r+5}} -\Gamma^{m_1\ldots m_r}_{i| X| Y,B_1,\ldots,B_{r+5}} )  \ . &\AJRal
}$$
Together with expressions for $Q C^{m_1\ldots}_{i| \ldots}$ and $Q P^{m_1\ldots}_{i| \ldots}$ in \HRij\ and \HRijk, one can check BRST closure of the right-hand side of \JRac\ and \JRal.

\subsec Momentum contractions of refined pseudoinvariants

\subseclab\ninethree
The procedure from the previous section is now extended to higher refinement. In the simplest scalar cases at seven-
and eight-points, we find
\eqnn\HOae
$$\eqalignno{
Q D_{1|2,3|4,5,6,7}  &= \Delta_{1|2|3,4,5,6,7}+ \Delta_{1|3|2,4,5,6,7}+ k_3^m P^m_{1|2|3,4,5,6,7}+k_2^m P^m_{1|3|2,4,5,6,7}  \cr
&+ \big[ s_{34} P_{1|2|34,5,6,7} + s_{24} P_{1|3|24,5,6,7} + (4 \leftrightarrow 5,6,7) \big] 
&\HOae \cr
QD_{1|23,4|5,6,7,8} &= \Delta_{1|23|4,5,6,7,8}+ \Delta_{1|4|23,5,6,7,8} + k_{23}^m P^m_{1|4|23,5,6,7,8} + k_4^m P^m_{1|23|4,5,6,7,8}   \cr
&+ \big[ s_{35} P_{1|4|235,6,7,8}-s_{25} P_{1|4|325,6,7,8} +s_{45} P_{1|23|45,6,7,8} + (5 \leftrightarrow 6,7,8) \big] \cr
&- P_{1|2,4|3,5,6,7,8}+ P_{1|3,4|2,5,6,7,8} \cr
Q D_{1|4,5|23,6,7,8} &= \Delta_{1|4|23,5,6,7,8}+ \Delta_{1|5|23,4,6,7,8} + k_4^m P^m_{1|5|23,4,6,7,8}+k_5^m P^m_{1|4|23,5,6,7,8} \cr
&+s_{24} P_{1|5|324,6,7,8} -s_{34} P_{1|5|234,6,7,8}+s_{25} P_{1|4|325,6,7,8} -s_{35} P_{1|4|235,6,7,8} \cr
& + \big[ s_{46} P_{1|5|23,46,7,8}+s_{56} P_{1|4|23,56,7,8}  + (6 \leftrightarrow 7,8) \big]  \,,
}$$
signaling the general rule
\eqnn\HOah
$$\displaylines{
Q D_{i|A,B|C,D,E,F}   = \Delta_{i|A|B,C,D,E,F}+ \Delta_{i|B|A,C,D,E,F} +k_A^m P^m_{i|B|A,C,D,E,F} \cr
+k_B^m P^m_{i|A|B,C,D,E,F}   + \big[  P_{i|A| S[B,C] , D,E,F } +  P_{i|B| S[A,C] , D,E,F }  + (C \leftrightarrow
D,E,F) \big] \hfil\HOah\hfilneg \cr
+ \! \! \sum_{XY=A}  \! \! ( P_{1|Y,B | X,C,D,E,F} -P_{1|X,B| Y,C,D,E,F} )  +  \! \! \sum_{XY=B} \! \! ( P_{1|Y,A |
X,C,D,E,F} -P_{1| X,A| Y,C,D,E,F} )  \, .
}$$
Given the appearance of two different momentum contractions $k_A^m P^m_{i|B|A,\ldots}$ and $k_B^m P^m_{i|A|B,\ldots}$,
\HOah\ can be viewed as a weaker result in comparison to the relations in section \ninetwo\ for a single $k_A^p
C^{pm_1\ldots }_{i|A,\ldots}$. 

Recall that tensorial superfields $D^{m_1\ldots}_{i|A|B_1,\ldots} $ give rise to additional terms $\sim k^m , \delta^{mn}$ absent in the scalar case, see \JRad, \JRae\ and \JRal.
The same kind of contributions appear in the vector and tensor generalization of \HOah, e.g.
\eqnn\allJaco
\eqnn\allJacp
$$\eqalignno{
QD^m_{1|2,3|4,\ldots,8}  &=  \Delta^m_{1|2|3,\ldots,8}+\Delta^m_{1|3|2,4,\ldots,8}+ k_3^p P^{mp}_{1|2|3,\ldots,8}+k_2^p P^{mp}_{1|3|2,4,\ldots,8}  \cr
&+ \big[ s_{34} P^m_{1|2|34,5,\ldots,8} + s_{24} P^m_{1|3|24,5,\ldots,8} + (4\leftrightarrow 5,\ldots,8) \big] \cr
&+ (k_{12345678}^m-k_{23}^m) P_{1|2,3|4,\ldots,8} \,,
&\allJaco \cr
QD^{mn}_{1|2,3|4,\ldots,9} &= \Delta^{mn}_{1|2|3,\ldots,9}+  \Delta^{mn}_{1|3|2,4,\ldots,9}+ \delta^{mn} \Delta_{1|2,3|4,\ldots,9} + k_3^p P^{mnp}_{1|2|3,\ldots,9}  \cr
&+k_2^p P^{mnp}_{1|3|2,4,\ldots,9}+ \big[ s_{34} P^{mn}_{1|2|34,5,\ldots,9} + s_{24} P^{mn}_{1|3|24,5,\ldots,9} + (4\leftrightarrow 5,\ldots,9) \big] \cr
& + 2(k_{123456789}^{(m}-k_{23}^{(m}) P^{n)}_{1|2,3|4,\ldots,9}  \,.
&\allJacp
}$$
This allows to anticipate the multiparticle version at general rank,
\eqnn\allJacr
$$\eqalignno{
Q D^{m_1\ldots m_r}_{i|A,B|C_1,\ldots,C_{r+4}} &= \Delta^{m_1\ldots m_r}_{i|A|B,C_1,\ldots,C_{r+4}} + \Delta^{m_1\ldots m_r}_{i|B|A,C_1,\ldots,C_{r+4}}
+  {r\choose 2} \delta^{(m_1 m_2}   \Delta^{m_3\ldots m_r)}_{i|A,B|C_1,\ldots,C_{r+4}} \cr
& + \Big[  P^{m_1\ldots m_r}_{i|A| S[B,C_1] , C_2,\ldots,C_{r+4} } +  P^{m_1\ldots m_r}_{i|B| S[A,C_1] , C_2,\ldots,C_{r+4} }
+ (C_1 \leftrightarrow C_2,\ldots,C_{r+4}) \Big]\cr
&+ r ( k_{i AB C_1\ldots C_{r+4}}^{(m_1} -k_{AB}^{(m_1} )P^{m_2\ldots m_r)}_{i|A,B|C_1,\ldots ,C_{r+4}}\cr
&+ k_A^p P^{pm_1\ldots m_r}_{i|B|A,C_1,\ldots,C_{r+4}} + k_B^p P^{pm_1\ldots m_r}_{i|A|B,C_1,\ldots,C_{r+4}}\cr
&- \! \! \sum_{XY=A} \! ( P^{m_1\ldots m_r}_{i|X,B | Y,C_1\ldots C_{r+4}} -P^{m_1\ldots m_r}_{i|Y,B| X,C_1\ldots C_{r+4}} )\cr
&-  \! \! \sum_{XY=B}  \!( P^{m_1\ldots m_r}_{i|A,X | Y,C_1\ldots C_{r+4}} -P^{m_1\ldots m_r}_{i|A,Y| X,C_1\ldots
C_{r+4}} )  \,, &\allJacr
}$$
where the deconcatenation terms $\sim \sum_{XY=A,B}$
follow by analogy with \HOah. In comparison to the counterpart \JRal\ of lower refinement, terms of the form $\Delta^{m_1\ldots m_r}_{i|A|B,C_1,\ldots}$, $r k_{A}^{(m_1} P^{m_2\ldots m_r)}_{i|A,B|C_1,\ldots }$,
$k_A^p P^{pm_1\ldots m_r}_{i|B|A,C_1,\ldots}$, $P^{m_1\ldots m_r}_{i|A| S[B,C_1] , C_2,\ldots} $ and
$\sum_{XY=A}\ldots $ are doubled in \allJacr. This suggests the
following BRST variation for $D$ superfields at general refinement,
\eqnn\allJacv
$$\eqalignno{
& QD^{m_1\ldots m_r}_{i|A_1,\ldots,A_d|B_1,\ldots,B_{r+d+2}} = \big[ \Delta^{m_1\ldots m_r}_{i|A_2,\ldots,A_d|A_1,B_1,\ldots,B_{r+d+2}} + (A_1\leftrightarrow A_2,\ldots,A_d) \big]  \cr
&+  {r\choose 2} \delta^{(m_1 m_2}   \Delta^{m_3\ldots m_r)}_{i|A_1,\ldots,A_d|B_1,\ldots,B_{r+d+2}}  + r k_{i B_1 B_2\ldots B_{r+d+2}}^{(m_1} P^{m_2\ldots m_r)}_{i|A_1,\ldots,A_d|B_1,\ldots,B_{r+d+2}} &\allJacv \cr
&+ \Big( k_{A_1}^p P^{pm_1\ldots m_r}_{i|A_2,\ldots,A_d|A_1,B_1,\ldots,B_{r+d+2}} + \big[  P^{m_1\ldots m_r}_{i|A_2,\ldots,A_d| S[A_1,B_1] , B_2,\ldots,B_{r+d+2} }  + (B_1 \leftrightarrow B_2,\ldots,B_{r+d+2}) \big]  \cr
&- \sum_{XY=A_1} ( P^{m_1\ldots m_r}_{i|X,A_2,\ldots,A_d | Y,B_1,\ldots,B_{r+d+2}} -P^{m_1\ldots m_r}_{i|Y,A_2,\ldots,A_d| X,B_1,\ldots,B_{r+d+2}} ) +(A_1\leftrightarrow A_2,\ldots,A_d) \Big) \ .
}$$
Again, we can directly infer the BRST variation of the anomalous counterparts $\Delta^{m_1\ldots }_{i| \ldots}$ by
discarding their appearance in the right-hand side of \allJacv\ and replacing the remaining terms via
$C^{m_1\ldots}_{i|\ldots} ,P^{m_1\ldots}_{i|\ldots}\rightarrow \Gamma^{m_1\ldots}_{i|\ldots} $:
\eqnn\AallJacv
$$\displaylines{
 Q\Delta^{m_1\ldots m_r}_{i|A_1,\ldots,A_d|B_1,\ldots,B_{r+d+4}}  =  r k_{1 B_1 B_2\ldots B_{r+d+4}}^{(m_1}
 \Gamma^{m_2\ldots m_r)}_{i|A_1,\ldots,A_d|B_1,\ldots,B_{r+d+4}} \hfil\AallJacv\hfilneg \cr
+ \Big( k_{A_1}^p \Gamma^{pm_1\ldots m_r}_{i|A_2,\ldots,A_d|A_1,B_1,\ldots,B_{r+d+4}} + \big[ \Gamma^{m_1\ldots m_r}_{i|A_2,\ldots,A_d| S[A_1,B_1] , B_2,\ldots,B_{r+d+4} }  + (B_1 \leftrightarrow B_2,\ldots,B_{r+d+4}) \big]  \cr
- \sum_{XY=A_1} ( \Gamma^{m_1\ldots m_r}_{i|X,A_2,\ldots,A_d | Y,B_1,\ldots,B_{r+d+4}} - \Gamma^{m_1\ldots m_r}_{i|Y,A_2,\ldots,A_d| X,B_1,\ldots,B_{r+d+4}} ) +(A_1\leftrightarrow A_2,\ldots,A_d) \Big) \ .
}$$
Using \AallJacv\ and the $Q$ variation \HRijk\ of the pseudoinvariants, one can verify BRST closure of the right-hand
side of \allJacv. This is a strong consistency check since it requires every single term in \allJacv\ to conspire.

\newsec Pseudoinvariant relations for $k_i$ momentum contractions

\seclab\secten
In the previous section, $D$ superfields defined in \masterK\ were shown to generate relations for $k_{B_{j}}$
contractions of pseudoinvariants. We shall now investigate the second family of ghost-number-two objects, the $L$
superfields defined in \masterL. It turns out that their BRST variations relate contractions of
$P^{m_1\ldots}_{i|\ldots}$ with the momentum $k_i$ of the reference leg $i$ to pseudoinvariants of lower rank.

\subsec $k_i$ contractions of unrefined pseudoinvariants

\subseclab\tenone
This section is devoted to the unrefined superfields $L^{m_1\ldots m_r}_{i|A_1,\ldots,A_{r+4}}$, see \masterL. The BRST variations of the simplest scalars are given by
\eqnn\jacg
$$\eqalignno{
 QL_{1|2,3,4,5} &= \Delta_{1|2,3,4,5} + k_1^m C^m_{1|2,3,4,5} &\jacg \cr
QL_{1|23,4,5,6}&= \Delta_{1|23,4,5,6} +k_1^m C^m_{1|23,4,5,6} +  P_{1|2|3,4,5,6} - P_{1|3|2,4,5,6}\cr
 QL_{1|234,5,6,7}&= \Delta_{1|234,5,6,7}
+k_1^m C^m_{1|234,5,6,7} \cr
&\quad{}+ P_{1|23|4,5,6,7} + P_{1|2|34,5,6,7} - P_{1|34|2,5,6,7} - P_{1|4|23,5,6,7} \cr
QL_{1|23,45,6,7}&=\Delta_{1|23,45,6,7}+k_1^m C^m_{1|23,45,6,7} \cr
&\quad{} + P_{1|2|3,45,6,7} - P_{1|3|2,45,6,7} + P_{1|4|23,5,6,7} - P_{1|5|23,4,6,7} \,,
}$$
and thereby provide relations for $k_i^m C^m_{i|A,B,C,D}$. The explicit form of $L_{1|2,3,4,5}$ and $L_{1|23,4,5,6}$
can be found in \Lsa, and the former underpins the first example in \fivezero\ provided that momentum conservation
$k_{12345}^m =0$ renders $\Delta_{1|2,3,4,5}$ BRST exact.

The examples in \jacg\ suggest the multiparticle pattern,
\eqnn\jack
$$\eqalignno{
QL_{i|A,B,C,D} &= \Delta_{i|A,B,C,D} + k_i^m C^m_{i|A,B,C,D}&\jack
\cr
&\quad{} +\Bigl[ \sum_{XY=A} (P_{i|X|Y,B,C,D}
- P_{i|Y|X,B,C,D} ) + (A\leftrightarrow B,C,D)\Bigr] \,,
}$$
where the anomalous $\Delta_{i|A,B,C,D}$ are defined by \masterM\ and also appear in the relations \JRac\ for different contractions $k^m_A C^m_{i|A,B,C,D}$.

The simplest $Q$ variations of ghost number two vectors $L^m_{i|A,B,C,D,E}$ read
\eqnn\QLmEx
\eqnn\QLmExx
$$\eqalignno{
QL^m_{1|2,3,4,5,6} &= \Delta^m_{1|2,3,4,5,6} + k^n_1 C^{mn}_{1|2,3,4,5,6} +  \big[k^m_2 P_{1|2|3,4,5,6} + (2\leftrightarrow
3,4,5,6)\big]&\QLmEx\cr
QL^m_{1|23,4,5,6,7} &=  \Delta^m_{1|23,4,5,6,7} +k^n_1 C^{mn}_{1|23,4,5,6,7} + \big[ k_4^m P_{1|4|23,5,6,7} +
(4\leftrightarrow 5,6,7) \big]\cr
&\quad{} + k_{23}^m P_{1|23|4,5,6,7} + P^m_{1|2|3,4,5,6,7} - P^m_{1|3|2,4,5,6,7} \,,&\QLmExx
}$$
see \Lsa\ for the expansion of $L^m_{1|2,3,4,5,6}$. The novel class of terms $\sim k^m$ in \QLmEx\ and \QLmExx\ are reproduced by the general formula,
\eqnn\jaczk
$$\eqalignno{
QL^m_{i|A,B,C,D,E} &= \Delta^m_{i|A,B,C,D,E} +k_i^n C^{mn}_{i|A,B,C,D,E} +  \Big[ k_A^m P_{i|A|B,C,D,E}&\jaczk \cr
&\quad{}  +\!\!\sum_{XY=A} (P^m_{i|X|Y,B,C,D,E}-P^m_{i|Y|X,B,C,D,E} )  + (A\leftrightarrow B,C,D,E) \Big].
}$$
As the last explicit example in this section, consider the two-tensor relation,
\eqnn\HRag
$$\eqalignno{
QL^{mn}_{1|2,3,4,5,6,7}&= \Delta^{mn}_{1|2,3,4,5,6,7}  +  \delta^{mn} \Lambda_{1|2,3,4,5,6,7} + k_1^p C^{mnp}_{1|2,3,4,5,6,7}   \cr
&\quad{} + 2 \big[ k_2^{(m}  P^{n)}_{1|2|3,4,5,6,7} +  (2 \leftrightarrow 3,4,5,6, 7) \big] \ ,
&\HRag 
}$$
subject to an anomalous trace with $\Lambda_{1|2,3,4,5,6,7}$ given by \JRp. Its generalization $\Lambda^{m_1\ldots m_r}_{i|A_1,A_2,\ldots,A_{r+6}}$ to
multiparticle slots and higher rank is defined in \masterN\ and finds appearance in the general rank-two relation,
\eqnn\JRn
$$\displaylines{
Q L^{mn}_{i|A,B,C,D,E,F} =  \Delta^{mn}_{i|A,B,C,D,E,F} + \delta^{mn} \Lambda_{i|A,B,C,D,E,F} +k_i^p
C^{mnp}_{i|A,B,C,D,E,F} \hfil\JRn\hfilneg  \cr
+ \Big[\! \sum_{XY=A} (P^{mn}_{i|X|Y,B,C,D,E,F} - P^{mn}_{i|Y|X,B,C,D,E,F} )+ 2 k_A^{(m} P^{n)}_{i|A|B,C,D,E,F} + (A \leftrightarrow B,\ldots,F) \Big] \ .
}$$
The expressions \jack, \jaczk\ and \JRn\ for $QL^{m_1\ldots m_r}_{i|A_1,A_2,\ldots,A_{r+4}}$ at rank $r=0,1,2$ lead to
a natural generalization for higher ranks,
\eqnn\JRs
$$\eqalignno{
Q L^{m_1\ldots m_r}_{i|A_1,\ldots,A_{r+4}} &= \Delta^{m_1\ldots m_r}_{i|A_1,\ldots,A_{r+4}} + {r\choose 2}
\delta^{(m_1m_2} \Lambda^{m_3\ldots m_r)}_{i|A_1,\ldots,A_{r+4}} + k_i^p C^{p m_1\ldots m_r}_{i|A_1,\ldots,A_{r+4}} \cr
&\ \ \ +\Big[ \sum_{XY=A_1} (P^{m_1\ldots m_r}_{i|X|Y,A_2,\ldots ,A_{r+4}} - P^{m_1\ldots m_r}_{i|Y|X,A_2,\ldots ,A_{r+4}})  \cr
&\ \ \ \ \ \ \ +  r k_{A_1}^{(m_1} P^{m_2\ldots m_r)}_{i|A_1|A_2,\ldots,A_{r+4}}  + (A_1 \leftrightarrow A_2,\ldots,A_{r+4}) \Big] \ .
&\JRs 
}$$
Similar to \JRal\ for $k_{A_1}^p C^{p m_1\ldots m_r}_{i|A_1,\ldots,A_{r+4}}$ as derived from $QD^{m_1\ldots}_{i|\ldots}$, two classes of anomalous terms appear in \JRs. The unrefined $\Delta^{m_1\ldots m_r}_{i|A_1,\ldots,A_{r+4}}$ common to both relations are BRST exact under momentum conservation, see \JRe, and can be discarded in the context of amplitudes. However, the second anomalous term $\Lambda^{m_1\ldots m_r}_{i|A_1,\ldots,A_{r+6}}$ in the trace of \JRs\ is not even BRST closed, regardless of the momentum phase space. Its non-vanishing $Q$ variation follows as the anomaly analogue of \jack\ and \JRs, i.e. by dropping the anomalous contributions from the latter and replacing the rest according to $C^{m_1\ldots}_{i|\ldots} ,P^{m_1\ldots}_{i|\ldots}\rightarrow \Gamma^{m_1\ldots}_{i|\ldots} $:
\eqnn\clanomX
\eqnn\clanomZ
$$\eqalignno{
Q\Lambda_{i|A,B,C,D,E,F} &= k_i^m \Gamma^m_{i|A,B,C,D,E,F} &\clanomX \cr
+\Big[ \sum_{XY=A} (&\Gamma_{i|X| Y,B,C,D,E,F} - \Gamma_{i|Y| X,B,C,D,E,F}) + (A \leftrightarrow B,C,D,E,F) \Big]  
\cr
Q\Lambda^{m_1\ldots m_r}_{i|A_1,A_2,\ldots,A_{r+6}} &= k_i^p  \Gamma^{p m_1\ldots m_r}_{i|A_1,\ldots,A_{r+6}} + \Big[ r k_{A_1}^{(m_1} \Gamma^{m_2\ldots m_r)}_{i|A_1|A_2,\ldots,A_{r+6}} &\clanomZ\cr
+\sum_{XY=A_1} (&\Gamma^{m_1\ldots m_r}_{i|X|Y,A_2,\ldots ,A_{r+6}} - \Gamma^{m_1\ldots m_r}_{i|Y|X,A_2,\ldots ,A_{r+6}}) + (A_1 \leftrightarrow A_2,\ldots,A_{r+6}) \Big] 
}$$
 Together with the BRST transformations \HRij\ and \HRijk\ of pseudoinvariants, \clanomX\ and \clanomZ\ allow to check BRST closure of \JRs\ and furnish a strong consistency check on the results in this section.

\subsec $k_i$ contractions of refined pseudoinvariants

\subseclab\tentwo
We next proceed to refined versions $L^{m_1\ldots}_{i|A_1,\ldots,A_d|B_1,\ldots}$ of the ghost-number-two objects under discussion. In the simplest scalar cases,
\eqnn\HOanull
$$\eqalignno{
QL_{1|2|3,4,5,6,7} 
&= \Delta_{1|2|3,4,5,6,7} + \Lambda_{1|2,3,4,5,6,7}+ k_{12}^m P^m_{1|2|3,4,5,6,7}  \cr
&+ \big[ s_{23} P_{1|23|4,5,6,7} + (3 \leftrightarrow 4,5,6,7) \big] &\HOanull   
\cr
QL_{1|23|4,5,6,7,8}
&= \Delta_{1|23|4,5,6,7,8} + \Lambda_{1|23,4,5,6,7,8} + k_{123}^m P^m_{1|23|4,5,6,7,8} \cr
&+\big[ s_{34} P_{1|234|5,6,7,8}-s_{24} P_{1|324|5,6,7,8} + (4 \leftrightarrow 5,6,7,8) \big]  \cr
QL_{1|4|23,5,6,7,8}   &=  \Delta_{1|4|23,5,6,7,8} + \Lambda_{1|23,4,5,6,7,8} + k_{14}^m P^m_{1|4|23,5,6,7,8}   \cr
&+ s_{24} P_{1|324|5,6,7,8} -s_{34} P_{1|234|5,6,7,8}+ \big[ s_{45} P_{1|45|23,6,7,8} + (5 \leftrightarrow 6,7,8) \big]
 \cr
 &+ P_{1|2,4|3,5,6,7,8} - P_{1|3,4|2,5,6,7,8} \,,
}$$
where the expansion of $L_{1|2|3,4,5,6,7}$ is given in \Lsa. This generalizes to
\eqnn\HOac
$$\displaylines{
QL_{i|A|B,C,D,E,F}
= \Delta_{i|A|B,C,D,E,F} + \Lambda_{i|A,B,C,D,E,F}+ k_{iA}^m P^m_{i|A|B,C,D,E,F} \hfil\HOac\hfilneg \cr
   + \Big[ P_{i| S[A,B] |C,D,E,F} + \sum_{XY=B} (P_{i|A,X| Y ,C,D,E,F}-P_{i|A,Y| X ,C,D,E,F}) +  (B \leftrightarrow  C,D,E,F) \Big]    \,.
}$$
Interestingly, \HOac\ contains a contraction of $P^m_{i|\ldots}$ with the combined momentum $k^m_{iA}=k^m_i+k^m_A$
including the refined slot $A$ and does not isolate its $k_i$ contraction. This is analogous to the shortcoming of the
relation \HOah\ to address the two-term combinations $k_A^m P^m_{i|B|A,C,D,E,F}+k_B^m P^m_{i|A|B,C,D,E,F}$ instead of
the individual terms.

Vectorial and tensorial BRST variations exhibit novel terms $\sim k^m,\delta^{mn}$ which are absent for the scalars \HOac, e.g.
\eqnn\allJacx
\eqnn\allJacy
$$\eqalignno{
QL^m_{1|2|3,\ldots,8} &= \Delta^m_{1|2|3,\ldots,8} + \Lambda^m_{1|2,3,\ldots,8} + k_{12}^p P^{pm}_{1|2|3,\ldots,8} \cr
&+\big[ s_{23} P^m_{1|23|4,\ldots,8} + k_3^m P_{1|2,3|4,\ldots,8} + (3\leftrightarrow 4,\ldots 8)\big] \,,
&\allJacx \cr
QL^{mn}_{1|2|3,\ldots,9} &=  \Delta^{mn}_{1|2|3,\ldots,9} +  \Lambda^{mn}_{1|2,3,\ldots,9}+\delta^{mn}  \Lambda_{1|2|3,\ldots,9} + k_{12}^p P^{pmn}_{1|2|3,\ldots,9} \cr
&+\big[ s_{23} P^{mn}_{1|23|4,\ldots,9} + 2 k_3^{(m} P^{n)}_{1|2,3|4,\ldots,9} + (3\leftrightarrow 4,\ldots 9)\big]  \ . 
&\allJacy
}$$
Experience with the scalar counterpart \HOac\ suggests that only the non-refined multiparticle
slots $B_j$ give rise to deconcatenation terms. This leads to the following all-rank generalization:
\eqnn\allJacz
$$\eqalignno{
Q&L^{m_1\ldots m_r}_{i|A|B_1,\ldots,B_{r+5}} = \Delta^{m_1\ldots m_r}_{i|A|B_1,\ldots,B_{r+5}} +  \Lambda^{m_1\ldots m_r}_{i|A,B_1,\ldots,B_{r+5}}+ {r\choose 2} \delta^{(m_1m_2}  \Lambda^{m_3\ldots m_r)}_{i|A|B_1,\ldots,B_{r+5}} &\allJacz  \cr
&\ \ \ \ \ \ \ \ \ \ \ \ \ \ \ \ \ \  + k_{iA}^p P^{pm_1\ldots m_r}_{i|A|B_1,\ldots,B_{r+5}} +\Big[ P^{m_1\ldots m_r}_{i| S[A,B_1] | B_2,\ldots, B_{r+5}} + r k_{B_1}^{(m_1} P^{m_2\ldots m_r)}_{i|A,B_1|B_2,\ldots, B_{r+5}}  \cr
&\ \ \ \ \ \ \ \ \ \ \ \ \ \ \ \ \ \   + \sum_{XY=B_1} (P^{m_1\ldots m_r}_{i|A,X|Y,B_2,\ldots,B_{r+5}} - P^{m_1\ldots m_r}_{i|A,Y|X,B_2,\ldots,B_{r+5}})
 + (B_1\leftrightarrow B_2,\ldots B_{r+5})\Big] \ .
}$$
For the extension of \allJacz\ to $L$ superfields of higher refinement $d$, one can expect
that three classes of terms $\Lambda^{m_1\ldots m_r}_{i|A,B_1,\ldots,B_{r+5}}$, $k_{A}^p P^{pm_1\ldots m_r}_{i|A|B_1,\ldots,B_{r+5}}$ and
$P^{m_1\ldots m_r}_{i| S[A,B_1] | B_2,\ldots, B_{r+5}}$ have to be symmetrized in $A_1\leftrightarrow A_2,\ldots,A_d$.
We therefore propose the following expression for the most general case:
\eqnn\allJacDD
$$\eqalignno{
Q&L^{m_1\ldots m_r}_{i|A_1,\ldots,A_d|B_1,\ldots,B_{r+d+4}}  = \Delta^{m_1\ldots m_r}_{i|A_1,\ldots,A_d|B_1,\ldots,B_{r+d+4}} + {r\choose 2} \delta^{(m_1m_2}  \Lambda^{m_3\ldots m_r)}_{i|A_1,\ldots,A_d|B_1,\ldots,B_{r+d+4}} &\allJacDD
   \cr
&+ \big[ \Lambda^{m_1\ldots m_r}_{i|A_2,\ldots,A_d|A_1,B_1,\ldots,B_{r+d+4}} + (A_1 \leftrightarrow A_2,\ldots,A_d)\big]
+ k_{iA_1 A_2\ldots A_d}^p P^{pm_1\ldots m_r}_{i|A_1,\ldots,A_d|B_1,\ldots,B_{r+d+4}} \cr
& +\Big[  \big\{ P^{m_1\ldots m_r}_{i|A_2,\ldots ,A_d,S[A_1,B_1] | B_2,\ldots, B_{r+d+4}} + (A_1\leftrightarrow A_2,\ldots, A_d) \big\} + r k_{B_1}^{(m_1} P^{m_2\ldots m_r)}_{i|A_1,\ldots,A_d,B_1|B_2,\ldots,B_{r+d+4}} \cr
&+ \sum_{XY=B_1} (P^{m_1\ldots m_r}_{i|A_1,\ldots,A_d,X|Y,B_2,\ldots,B_{r+d+4}} - P^{m_1\ldots m_r}_{i|A_1,\ldots,A_d,Y|X,B_2,\ldots,B_{r+d+4}} )
 + (B_1\leftrightarrow B_2,\ldots B_{r+d+4})\Big] \ .  
}$$
Similar to the correspondence between \JRs\ and \clanomZ, one can infer the BRST variation of $\Lambda^{m_1\ldots
}_{i|\ldots}$ by trading the constituents of \allJacDD\ for their anomaly counterparts,
\eqn\allJacEE{\belowdisplayskip=-3pt\relax
Q\Lambda^{m_1\ldots m_r}_{i|A_1,\ldots,A_d|B_1,\ldots,B_{r+d+6}} = k_{iA_1 A_2\ldots A_d}^p \Gamma^{pm_1\ldots m_r}_{i|A_1,\ldots,A_d|B_1,\ldots,B_{r+d+6}}
}
$$\eqalignno{
&{} +\Big[  \big\{ \Gamma^{m_1\ldots m_r}_{i|A_2,\ldots ,A_d,S[A_1,B_1] | B_2,\ldots, B_{r+d+6}} + (A_1\leftrightarrow A_2,\ldots, A_d) \big\} + r k_{B_1}^{(m_1} \Gamma^{m_2\ldots m_r)}_{i|A_1,\ldots,A_d,B_1|B_2,\ldots,B_{r+d+6}} \cr
&{}+ \sum_{XY=B_1} ( \Gamma^{m_1\ldots m_r}_{i|A_1,\ldots,A_d,X|Y,B_2,\ldots,B_{r+d+6}} - \Gamma^{m_1\ldots m_r}_{i|A_1,\ldots,A_d,Y|X,B_2,\ldots,B_{r+d+6}} )
 + (B_1\leftrightarrow B_2,\ldots B_{r+d+6})\Big] \,.
}$$
Using \allJacEE\ and the BRST transformations \HRij\ and \HRijk\ of pseudoinvariants, one can verify that the
right-hand side of \allJacDD\ is BRST closed. Given the delicate interplay of every single term in \allJacDD, this is a
highly nontrivial consistency check for the results in this section.

\subsec Trace relations and anomaly bookkeeping 

\subseclab\tenthree
The purpose of the above BRST-exact relations is to express momentum contractions of pseudoinvariants in terms of simpler pseudoinvariants at lower rank. However, two classes of obstructions arose, set
by anomalous superfields $\Delta$ and $\Lambda$. Only the unrefined special case $\Delta^{m_1\ldots
m_r}_{i|A_1,\ldots,A_{r+4}}$ was shown to be BRST trivial under momentum conservation, see \JRe, whereas refined $\Delta$
superfields and any $\Lambda$ superfield cannot be discarded in scattering amplitudes. Hence, it is desirable to
identify relations among these anomalous admixtures.

In particular, one might wonder if the trace relations \symmCs\ and \symmaCCs\ found among pseudoinvariants
$P^{m_1\ldots}_{i|\ldots}$ and anomaly invariants $\Gamma^{m_1\ldots}_{i|\ldots}$ carry over to their counterparts
$D$, $L$ and $\Delta$, $\Lambda$ at different ghost-number. Even though they all originate from the same master recursion
\master, there are subtleties under the slot rearrangements $M_A {\cal U}^{\ldots}_{\{B_j\} | \{C_j\}} \rightarrow
{\cal U}^{\ldots}_{\{B_j\} |A, \{C_j\}}$ and $M_A {\cal U}^{\ldots}_{\{B_j\} | \{C_j\}} \rightarrow {\cal
U}^{\ldots}_{A,\{B_j\} | \{C_j\}}$ entering the definitions \masterG\ to \masterJ.

Since the formal symbols ${\cal U}^{m_1\ldots m_r}_{A_1,\ldots,A_d|B_1,\ldots,B_{d+r+N}}$ are eventually identified
with either ${\cal J}$ or ${\cal Y}$, it is safe to impose their trace relations \symmBBs\ and \symmaBBs\ on the
${\cal U}$,
\eqn\symmaBBU{
\delta_{np} {\cal U}^{npm_1\ldots m_{r-2}}_{B_1,\ldots,B_{d} | C_1,\ldots,C_{d+r+N}} =
2{\cal U}^{m_1 \ldots m_{r-2}}_{B_1, \ldots,B_{d},C_1|C_2, \ldots, C_{d+r+N}} + (C_1 \leftrightarrow C_2,\ldots,
C_{d+r+N})  \,.
}
It turns out that this relation is {\it not} preserved under the slot
rearrangement $M_A {\cal U}^{\ldots}_{\{B_j\} | \{C_j\}} \rightarrow {\cal U}^{\ldots}_{\{B_j\} |A, \{C_j\}}$ relevant for $D$ and $\Delta$ (followed by multiplication with $M_A$) since
\eqn\BBU{
\delta_{np} {\cal U}^{npm_1\ldots m_{r-1}}_{B_1,\ldots,B_{d} | A,C_1,\ldots,C_{d+r+N}} \neq
2{\cal U}^{m_1 \ldots m_{r-1}}_{B_1, \ldots,B_{d},C_1|A,C_2, \ldots, C_{d+r+N}} + (C_1 \leftrightarrow C_2,\ldots, C_{d+r+N})  \,.
}
The missing term to restore \symmaBBU\ is easily seen to be $2{\cal U}^{m_1 \ldots m_{r-1}}_{B_1, \ldots,B_{d},A|C_1,C_2, \ldots, C_{d+r+N}} $ which in turn originates from
the alternative slot rearrangement $M_A {\cal U}^{\ldots}_{\{B_j\} | \{C_j\}} \rightarrow {\cal U}^{\ldots}_{A,\{B_j\} | \{C_j\}}$. Since this is the defining map for
$L$ and $\Lambda$, we are led to the following trace relation:
\eqnn\trD
$$\eqalignno{
\delta_{np}&D^{npm_1\ldots m_r}_{i|A_1,\ldots,A_{d} | B_1,\ldots,B_{d+r+4}} = 
2L^{m_1\ldots m_r}_{i|A_1,\ldots,A_{d} | B_1,\ldots,B_{d+r+4}}
&\trD \cr
& \ \ \ \ +
2\big[ D^{m_1 \ldots m_r}_{i|A_1, \ldots,A_{d},B_1|B_2, \ldots, B_{d+r+4}} + (B_1 \leftrightarrow B_2,\ldots, B_{d+r+4}) \big] \ .
}$$ 
The mixing between $L$ and $D$ superfields propagates to their anomalous counterparts $(D,L)\rightarrow (\Delta,\Lambda)$:
\eqnn\trDelta
$$\eqalignno{
\delta_{np}&\Delta^{npm_1\ldots m_r}_{i|A_1,\ldots,A_{d} | B_1,\ldots,B_{d+r+6}} = 
2\Lambda^{m_1\ldots m_r}_{i|A_1,\ldots,A_{d} | B_1,\ldots,B_{d+r+6}}
&\trDelta \cr
& \ \ \ \ +
2\big[ \Delta^{m_1 \ldots m_r}_{i|A_1, \ldots,A_{d},B_1|B_2, \ldots, B_{d+r+6}} + (B_1 \leftrightarrow B_2,\ldots, B_{d+r+6}) \big] \ .
}$$ 
Note that \trDelta\ and the BRST variation of \trD\ are consistent with the expressions \allJacv\ and \allJacDD\ for $QD^{m_1\ldots}_{i|\ldots}$ and $QL^{m_1\ldots}_{i|\ldots}$.

The slot rearrangement $M_A {\cal U}^{\ldots}_{\{B_j\} | \{C_j\}} \rightarrow {\cal U}^{\ldots}_{A,\{B_j\} | \{C_j\}}$ entering
the definition of $L$ and $\Lambda$ preserves the trace relations \symmaBBU\ and bypasses the subtlety in \BBU.
Hence, traces of $L$ and $\Lambda$ fall into the same pattern found for $P^{m_1\ldots}_{i|\ldots}$ and
$\Gamma^{m_1\ldots}_{i|\ldots}$ in \symmCs\ and \symmaCCs,
\eqnn\trLL
\eqnn\trLambda
$$\eqalignno{
\delta_{np}L^{npm_1\ldots m_r}_{i|A_1,\ldots,A_{d} | B_1,\ldots,B_{d+r+6}} &= 
2\big[ L^{m_1 \ldots m_r}_{i|A_1, \ldots,A_{d},B_1|B_2, \ldots, B_{d+r+6}} + (B_1 \leftrightarrow B_2,\ldots,
B_{d+r+6}) \big] \cr&&\trLL\cr
\delta_{np}\Lambda^{npm_1\ldots m_r}_{i|A_1,\ldots,A_{d} | B_1,\ldots,B_{d+r+8}} &= 
2\big[ \Lambda^{m_1 \ldots m_r}_{i|A_1, \ldots,A_{d},B_1|B_2, \ldots, B_{d+r+8}} + (B_1 \leftrightarrow B_2,\ldots,
B_{d+r+8}) \big] \ .\cr &&\trLambda
}$$ 
Again, one can verify consistency of \trLambda\ and the BRST variation
of \trLL\ by means of the expression \allJacDD\ for $QL^{m_1\ldots}_{i|\ldots}$.

As a main benefit of this discussion, the trace relations \trDelta\ and \trLambda\ are useful in
manipulating anomalous ghost-number-three contributions to scattering amplitudes. In particular, we can take advantage of the
decoupling of unrefined objects $\Delta^{m_1\ldots m_r}_{i|A_1,\ldots,A_{r+4}}$ as well as its traces and discard the
following right-hand sides:
\eqnn\traceDD
$$\eqalignno{
{1\over 2}\delta_{np}\Delta^{npm_1\ldots m_r}_{i| B_1,\ldots,B_{r+6}} &= \Lambda^{m_1\ldots m_r}_{i| B_1,\ldots,B_{r+6}} +
\big[ \Delta^{m_1 \ldots m_r}_{i|B_1|B_2, \ldots, B_{r+6}} + (B_1 \leftrightarrow B_2,\ldots, B_{r+6}) \big] \cr
{1\over 4}\delta_{np}\delta_{qr}\Delta^{npqr m_1\ldots m_r}_{i| B_1,\ldots,B_{r+8}} &= 
\big[ \Lambda^{m_1 \ldots m_r}_{i|B_1|B_2, \ldots, B_{r+8}} + (B_1 \leftrightarrow B_2,\ldots, B_{r+8}) \big]&\traceDD \cr
&+ \big[ \Delta^{m_1 \ldots m_r}_{i|B_1,B_2|B_3, \ldots, B_{r+8}} + (B_1,B_2| B_1, B_2,\ldots, B_{r+8}) \big] \ .
}$$ 
Generalizations to multitraces of $\Delta^{m_1\ldots m_r}_{i|A_1,\ldots,A_{r+4}}$ are straightforward.

\subsec The web of relations between ghost-number two and four

\subseclab\tenfour
In the last sections we have constructed a variety of superfields and derived a rich set of relations among them. The
recursions which led to pseudoinvariants $P$ in section \secsix\ and to anomaly invariants $\Gamma$ in section
\secseven\ were unified to a master recursion \master\ in section \seceight. As detailed in sections \eighttwo\ to \eightfive,
the master recursion points towards four further replicae $D$, $L$, $\Delta$ and $\Lambda$ of the family of $P$ and $\Gamma$
which can be visualized in grids similar to \figoverview\ and \anomgrid.

The BRST-exact relations presented in this section and section \secnine\ mediate between the six families of
superfields as visualized in \roadmap. Since BRST action increases the ghost number by one, we arrange
the superfields according to their ghost number. As a second coordinate for the roadmap of superfields, we take the
number $N$ of multiparticle slots in the scalar and unrefined representatives, see section \eightone.

\ifig\roadmap{Overview of superfield families. Diagonal lines point towards the images of BRST variations, and horizontal lines remind of trace relations.}
{\epsfxsize=0.80\hsize\epsfbox{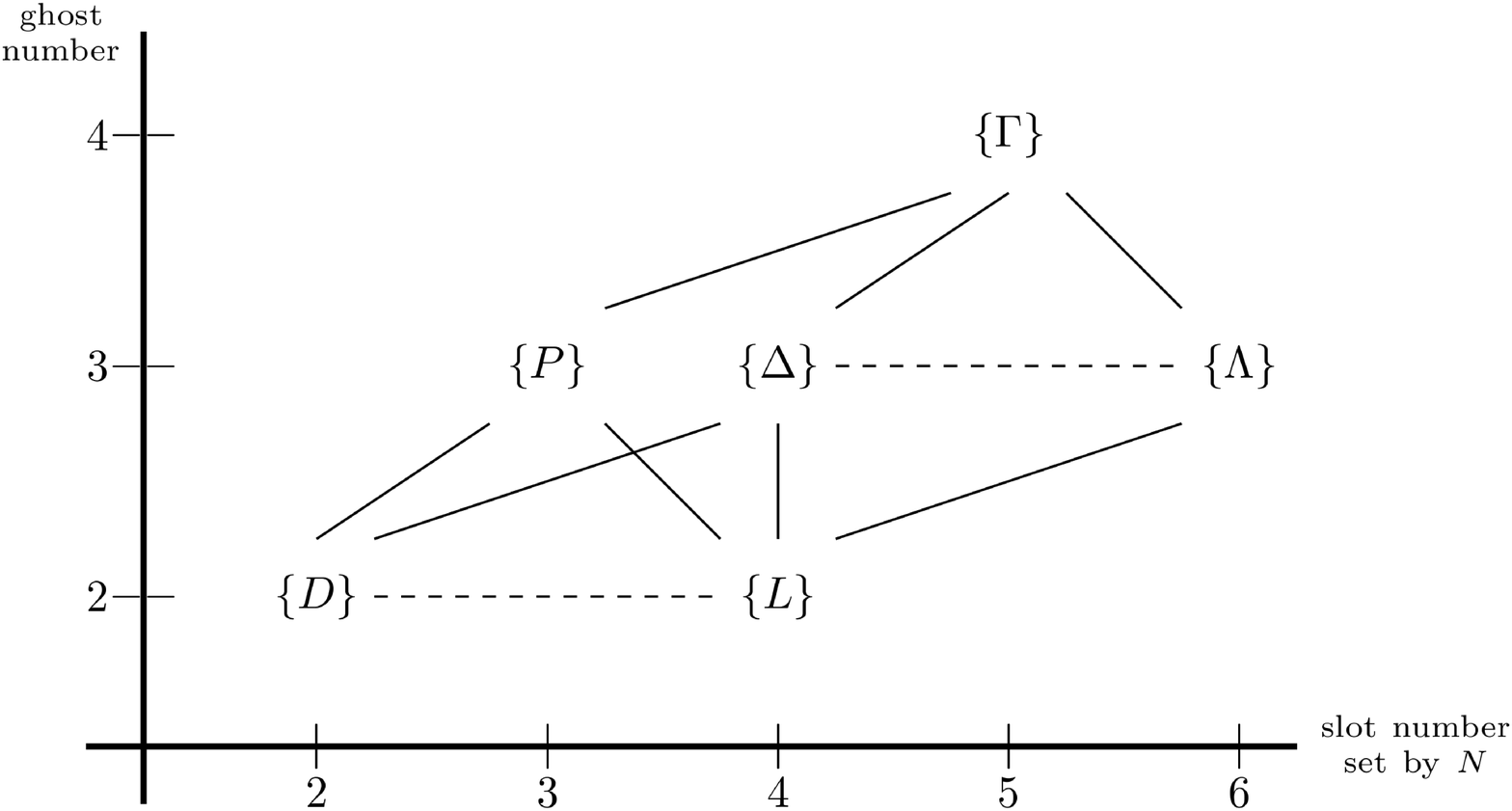}}


In \roadmap, the BRST variations of all the six families are represented by solid lines, the arrows pointing towards
the image of higher ghost number. The underlying expressions are given in
\item{$\bullet$} \allJacv, \allJacDD\ and \HRijk\ for BRST
action on the non-anomalous fields $D,L$ and $P$
\item{$\bullet$} \AallJacv\ and \allJacEE\ for their anomalous counterparts
$\Delta$ and $\Lambda$

\noindent
whereas $Q\Gamma=0$. Horizontal dashed arrows additionally remind of the mixing of
$D\leftrightarrow L$ as well as $\Delta\leftrightarrow \Lambda$ under the trace relations \trD\ and \trDelta.


\newsec Canonicalizing pseudoinvariants

\seclab\seceleven
In the previous sections, we have systematically derived two families of relations among pseudoinvariants
$P^{m_1\ldots }_{i| \ldots}$ at fixed reference leg $i$. This section is devoted to a different class of relations
which mediates between different choices of the reference particle $i$. It will be demonstrated that any $P^{m_1\ldots
}_{k| \ldots}$ with $k \neq i$ can be expressed in terms of $P^{m_1\ldots }_{i| \ldots}$, up to a constrained set of
anomaly terms. This rearrangement will be referred to as {\it canonicalization}.

The anomalous admixtures in the canonicalization process reflect the property of the hexagon gauge anomaly in both
field and string theory to break the permutation symmetry of one-loop amplitudes at multiplicity $n \geq 6$. Apart from this
anomaly subtlety, however, we learn that the set of pseudoinvariants $P^{m_1\ldots m_r}_{i| \ldots}$ at fixed
reference leg $i$ spans the same space of kinematic factors as would be obtained by any other choice of reference leg $k\neq i$. This is a necessary condition for the independence of string amplitudes on the choice of the unintegrated vertex $i$.

The methods parallel the procedure in section \secnine\ and \secten, i.e. we identify suitable BRST generators at
ghost-number two to derive relations among ghost-number-three objects. The $Q$ generators to trade different choices
of the reference leg $i$ in $P^{m_1\ldots }_{i| \ldots}$ turn out to be minor modifications of the ghost-number-two
superfields of types $D$ and $L$, see section \seceight. 

We start by discussing the canonicalization of scalar invariants in detail. This serves as a motivation of
certain operations which capture the structure of the canonicalization process and easily carry over to more general
pseudoinvariants.

\subsec Canonicalizing scalar invariants

\subseclab\elevenone
As pointed out in \oneloopbb\ through a few examples at low multiplicity, any scalar invariant $C_{k|A,B,C}$ as given
in \veca\ can be cast into a basis of $C_{i|D,E,F}$ with $i \neq k$. We shall now present the general solution for
arbitrary $A,B,C$ and thereby develop the maps and notation to extend the procedure to higher tensors and to refined
pseudoinvariants.

The trial and error method in \oneloopbb\ to canonicalize $C_{2|A,B,C}$ towards $C_{1|D,E,F}$ leads to the following
expressions at multiplicity $n\leq 6$ (suppressing the laborious case $C_{2|34,56,1}$):
\eqnn\patternNa
\eqnn\patternNb
\eqnn\patternNc
\eqnn\patternNd
\eqnn\patternNf
\eqnn\patternNe
$$\eqalignno{
C_{2|1,3,4}&= -Q(M_{12,3,4}) + C_{1|2,3,4} &\patternNa
\cr
C_{2|13,4,5}&= -Q(M_{132,4,5}) + C_{1|32,4,5} &\patternNb
\cr
C_{2|1,34,5}&= -Q(M_{12,34,5} + M_{123,4,5}- M_{124,3,5}) + C_{1|2,34,5}  + C_{1|23,4,5}  - C_{1|24,3,5}  &\patternNc
\cr
C_{2|134,5,6}&= -Q(M_{1342,5,6} ) + C_{1|342,5,6}    &\patternNd
\cr
C_{2|13,45,6}&= -Q(M_{132,45,6} + M_{1324,5,6} - M_{1325,4,6} )+ C_{1|32,45,6} + C_{1|324,5,6} - C_{1|325,4,6} &\patternNf \cr
C_{2|1,345,6}&= -Q(M_{12,345,6} + M_{1234,5,6}+M_{1254,3,6} - M_{1235,4,6} - M_{1253,4,6} + M_{123,45,6} + M_{125,43,6})\cr
&\mkern-40mu {} + C_{1|2,345,6} +  C_{1|234,5,6}+C_{1|254,3,6} - C_{1|235,4,6}
- C_{1|253,4,6} + C_{1|23,45,6} + C_{1|25,43,6} &\patternNe
}$$
These examples can be easily verified using the form \QMgen\ of $QM_{A,B,C}$. The following three observations on
\patternNa\ to \patternNe\ guide the way towards a general solution for both the BRST ancestor and the set of
$C_{1|D,E,F}$ which appear in the canonicalization of $C_{2|\ldots}$:
\bigskip
\item{(i)} Each term of the form $-QM_{1D,E,F}$ in the BRST generator is accompanied by a corresponding invariant $C_{1|D,E,F}$. By assuming this pattern to hold in general, knowledge of the BRST generator already determines the canonicalization in terms of $C_{1|D,E,F}$.
\item{(ii)} Suppose particle $1$ appears in the left hand side as $C_{2|1A,B,C}$ (where $A$ can be empty), then each term of the BRST generator is of the form $M_{1AD,E,F}$. In other words, the entire slot $1A$ of the desired reference leg is concatenated with a pattern of $M_{\ldots D,E,F}$.
\item{(iii)} This pattern of $M_{\ldots D,E,F}$ must contain the information on the remaining labels $2,B,C$. The superfield $D_{2|B,C}$ as defined in \masterK\ is the natural object to do so, and indeed, its concatenation \concat\ with $M_{1A}$ reproduces the above BRST generators, e.g.
\eqnn\patternOa
\eqnn\patternOb
\eqnn\patternOc
$$\eqalignno{
M_1 \otimes D_{2|34,5}  &= M_{12,34,5} + M_{123,4,5}- M_{124,3,5}  &\patternOa
\cr
M_{13} \otimes D_{2|45,6}  &= M_{132,45,6} + M_{1324,5,6} - M_{1325,4,6}  &\patternOb\cr
M_1 \otimes D_{2|345,6}  &= M_{12,345,6} + M_{1234,5,6}+M_{1254,3,6} - M_{1235,4,6}  \cr
& \ \ \ \ \ \ - M_{1253,4,6} + M_{123,45,6} + M_{125,43,6}  &\patternOc
}$$
for \patternNc, \patternNf\ and \patternNe, respectively. The $D$ superfields in the first two cases are given in \quasie\ whereas $D_{2|345,6}$ can be inferred from  $C_{1|234,5,6}$ in \sixsc.

\medskip
\noindent So one can promote the sample canonicalizations of $C_{2|1A,B,C}$ in \patternNa\ to \patternNe\ to the following
general formula\foot{To support the plausibility of the canonicalization prescription in \patternQ, note that any term
in the concatenation product $M_{iA} \otimes D_{k|B,C} $ takes the form $M_{iA\ldots,D,E}$. BRST action then
generates one term $C_{i|A\ldots,D,E}$ and up to five others, see \QMgen. The map $\wp_{i}$ makes sure that the
$C_{i|A\ldots,D,E}$ contribution in $-QM_{iA\ldots,D,E}$ is compensated. Other invariants of the form $C_{l\neq
i|F,G,H}$ largely cancel thanks to the fine-tuned arrangement of slots governed by the recursive origin \master\ of
$D_{k|B,C}$. Only one term $C_{k|iA,B,C}$ with reference leg $\neq i$ will emerge from the $Q$ action on the leading
term $D_{k|B,C} \rightarrow M_{k,B,C}$, as required by the left-hand side of \patternQ.},
\eqnn\patternQ
$$\eqalignno{
C_{k|iA,B,C} &=(\wp_{i}-Q) (M_{iA} \otimes D_{k|B,C}  ) \ .&\patternQ
}$$
The ``pseudoinvariantization'' map $\wp_i$ in \patternQ\ is defined to transform $M_{iA,B,C} \rightarrow C_{i|A,B,C}$ according to observation (i) and, more generally,
\eqnn\patternQa
\eqnn\patternQb
$$\eqalignno{
\wp_i( {\cal J}^{m_1\ldots m_r}_{B_1,\ldots,B_d | iA,C_2,\ldots,C_{d+r+3}} ) &\equiv P^{m_1\ldots m_r}_{i|B_1,\ldots,B_d | A,C_2,\ldots,C_{d+r+3}}
&\patternQa
\cr
\wp_i( {\cal J}^{m_1\ldots m_r}_{iA,B_2,\ldots,B_d | C_1,\ldots,C_{d+r+3}} ) &\equiv P^{m_1\ldots m_r}_{i|A,B_2,\ldots,B_d | C_1,\ldots,C_{d+r+3}} \ .
&\patternQb
}$$
Loosely speaking, $\wp_i$ in \patternQa\ and \patternQb\ removes particle $i$ from the leading position of a word in ${\cal J}^{m_1\ldots m_r}_{B_1,\ldots,B_d | C_1,\ldots,C_{d+r+3}}$ and converts it into the reference leg of a pseudoinvariant. Its remaining labels are those of the above current with particle $i$ removed.

Further applications of \patternQ\ can be found in appendix \appX, see \canexA\ to \canexF.

\subsec Canonicalizing unrefined pseudoinvariants

\subseclab\eleventwo
In order to canonicalize tensorial pseudoinvariants $C_{k|iA,B_2,\ldots,B_{r+3}}^{m_1\ldots m_r}$ to the form
$C_{i|C_1,\ldots,C_{r+3}}^{m_1\ldots m_r}$, it is tempting to simply replace $D_{k|B,C} \rightarrow D^{m_1\ldots
m_r}_{k|B_2,\ldots,B_{r+3}}$ in the scalar prescription \patternQ. However, $D$ superfields at rank $r\geq 2$
additionally generate anomalous terms such as $\Delta^{m_1\ldots m_r}_{i|A_1,\ldots,A_4}$ in \JRb\ and \JRe. In order
to have a well-defined notion of $M_{iA} \otimes \Delta^{m_1\ldots m_r}_{k|B_1,\ldots,B_{r+4}}$ and $M_{iA} \otimes
D^{m_1 \ldots m_r}_{k|B_1,\ldots,B_{r+2}}$ (needed in the subsequent), we extend the
concatenation operation to anomaly building blocks and refined currents,
\eqnn\concaYtt
\eqnn\concaYttt
\eqnn\concatt
\eqnn\concattt
$$\eqalignno{
M_{iA} \otimes {\cal Y}^{m_1\ldots m_r}_{kB,B_2,\ldots,B_{d}|C_1,\ldots,C_{d+r+5}} &\equiv {\cal Y}^{m_1\ldots m_r}_{iAkB,B_2,\ldots,B_{d}|C_1,\ldots,C_{d+r+5}} &\concaYtt
\cr
M_{iA} \otimes {\cal Y}^{m_1\ldots m_r}_{B_1,\ldots,B_{d}|kC,C_2,\ldots,C_{d+r+5}} &\equiv {\cal Y}^{m_1\ldots
m_r}_{B_1,\ldots,B_{d}|iAkC,C_2,\ldots,C_{d+r+5}}  &\concaYttt\cr
 M_{iA} \otimes {\cal J}^{m_1\ldots m_r}_{kB,B_2,\ldots,B_{d}|C_1,\ldots,C_{d+r+3}} &\equiv {\cal J}^{m_1\ldots m_r}_{iAkB,B_2,\ldots,B_{d}|C_1,\ldots,C_{d+r+3}} &\concatt
 \cr
 M_{iA} \otimes {\cal J}^{m_1\ldots m_r}_{B_1,\ldots,B_{d}|kC,C_2,\ldots,C_{d+r+3}} &\equiv {\cal J}^{m_1\ldots m_r}_{B_1,\ldots,B_{d}|iAkC,C_2,\ldots,C_{d+r+3}} \ . &\concattt
}$$
As before, the instruction to concatenate the word $iA$ with $kB$ and $kC$ is clear from the reference leg $k$ of the parental $D^{m_1\ldots}_{k|\ldots}$.
The anomaly concatenations $M_{iA} \otimes \Delta^{m_1\ldots m_r}_{k|B_1,\ldots,B_{r+4}}$ serve to compensate for the anomalous part of $Q(M_{iA} \otimes D^{m_1\ldots}_{k|\ldots} )$:
\eqnn\patternU
$$\eqalignno{
C^{m_1 m_2 \ldots m_r}_{k|iA,B_2,\ldots,B_{r+3}} &=(\wp_i-Q)
(M_{iA} \otimes D^{m_1 m_2 \ldots m_r}_{k|B_2,B_3,\ldots ,B_{r+3}}  )  &\patternU\cr
& \ \ \ + {r\choose 2} \delta^{(m_1m_2} (M_{iA} \otimes \Delta^{m_3 m_4 \ldots m_r)}_{k|B_2,B_3,\ldots,B_{r+3}}) \ .
}$$
The cancellation of anomalous superfields on the right-hand side can be understood as follows: The anomalous term in
$Q M^{m_1\ldots m_r}_{B_1,\ldots,B_{r+3}} = {r\choose 2} \delta^{(m_1 m_2} \cY^{m_3\ldots m_r)}_{B_1,\ldots, B_{r+3}}
+ \ldots$ preserves the structure of the slots $B_k$. In a truncation to anomaly building blocks, BRST action and the
concatenation through $M_{iA} \otimes$ commute. Since $\delta^{(m_1m_2} \Delta^{m_3 m_4 \ldots
m_r)}_{k|B_2,B_3,\ldots,B_{r+3}}$ in the second line of \patternU\ can be traced back to $QD^{m_1 m_2 \ldots
m_r}_{k|B_2,B_3,\ldots ,B_{r+3}} $, see \JRe, any anomalous contribution effectively originates from a ``commutator''
of the operations $M_{iA} \otimes$ and $Q$ acting on $D^{m_1 m_2 \ldots m_r}_{k|B_2,B_3,\ldots ,B_{r+3}}$.

Two simple examples of the general procedure in \patternU\ are given by,
\eqnn\patternR
\eqnn\patternT
$$\eqalignno{
C^m_{2|1,3,4,5}&= (\wp_{1}-Q)(M_{1} \otimes D^m_{2|3,4,5}  ) \cr
&=-Q\Big(M^m_{12,3,4,5}+\big[k_3^m M_{123,4,5}+(3\leftrightarrow 4,5) \big] \Big) \cr
& \quad{} + C^m_{1|2,3,4,5}+[k_3^m C_{1|23,4,5}+(3\leftrightarrow 4,5)]  
&\patternR \cr
C^{mn}_{2|1,3,4,5,6}&= (\wp_{1}-Q)(M_{1} \otimes D^{mn}_{2|3,4,5,6}  ) + \delta^{mn} (M_1 \otimes \Delta_{2|3,4,5,6})
\cr
&=-Q \Big( M^{mn}_{12,3,4,5,6} + \big[ 2 k_3^{(m} M^{n)}_{123,4,5,6} + (3\leftrightarrow 4,5,6) \big] \cr
& \qquad{} + \big[ 2 k^{(m}_3 k^{n)}_4 (M_{1234,5,6}+M_{1243,5,6}) + (3,4|3,4,5,6) \big] \Big) \cr
 &\quad{} + \delta^{mn} {\cal Y}_{12,3,4,5,6}  + C^{mn}_{1|2,3,4,5,6}  + \big[ 2 k_3^{(m} C^{n)}_{1|23,4,5,6} + (3\leftrightarrow 4,5,6) \big] \cr
 & \quad{} + \big[ 2 k^{(m}_3 k^{n)}_4 (C_{1|234,5,6} +C_{1|243,5,6} )+ (3,4|3,4,5,6) \big]  \ .&\patternT
}$$
Further examples can be found in appendix \appX, see \canexG.

\subsec Canonicalizing non-refined slots in refined pseudoinvariants

\subseclab\eleventhree
Only minor modifications are required to generalize the above canonicalization rules to refined pseudoinvariants
$P^{m_1 m_2 \ldots m_r}_{k|A_1,\ldots,A_d|B_1,\ldots ,B_{d+r+3}} $, as long as the preferred reference leg~$i$ resides
in a unrefined slot $B_j$. By analogy with the first line of \patternU, it is natural to expect a BRST generator of
the form $M_{iA} \otimes D^{m_1 m_2 \ldots m_r}_{k|B_1,\ldots,B_d|C_2,\ldots,C_{d+r+3}} $.

Similar to the unrefined tensors, the BRST variation of refined $D$ superfields incorporates anomalous $\Delta$ superfields, see \allJacv.
In order to address them, recall that the anomalous part of $Q\cJ^{m_1 m_2 \ldots m_r}_{A_1,\ldots,A_d|B_1,\ldots ,B_{d+r+3}}$ given in \JCgen\ has
an unmodified slot structure. Hence, we can neglect the concatenation with $M_{iA}$ and compensate the anomalous part
of $Q(M_{iA} \otimes D^{m_1 m_2 \ldots m_r}_{k|B_1,\ldots,B_d|C_2,\ldots}) $ using the corresponding
terms of $M_{iA} \otimes QD^{m_1 m_2 \ldots m_r}_{k|B_1,\ldots,B_d|C_2,\ldots}$. This reasoning motivates the last two lines of
\eqnn\patternUU
$$\eqalignno{
P^{m_1 m_2 \ldots m_r}_{k|B_1,\ldots,B_d | iA,C_2,\ldots,C_{d+r+3}} &=(\wp_i-Q)
(M_{iA} \otimes D^{m_1 m_2 \ldots m_r}_{k|B_1,\ldots,B_d|C_2,\ldots ,C_{d+r+3}} )  \cr
& \quad{} + M_{iA} \otimes \Big\{ {r\choose 2} \delta^{(m_1m_2}  \Delta^{m_3 m_4 \ldots m_r)}_{k|B_1,\ldots,B_d|C_2,C_3,\ldots,C_{d+r+3}} &\patternUU  \cr
& \qquad{} + \big[ \Delta^{m_1 m_2 \ldots m_r}_{k|B_2,\ldots,B_d|B_1,C_2,C_3,\ldots,C_{d+r+3}}  + (B_1\leftrightarrow B_2,\ldots,B_d) \big] \Big\} 
}$$
and guarantees that they cancel the anomalous contributions from the first line.

As the simplest application of \patternUU,
\eqnn\patternUA
$$\eqalignno{
P_{2|3|1,4,5,6} &= (\wp_1 - Q) (M_1 \otimes D_{2|3|4,5,6}   ) + M_1 \otimes \Delta_{2|3,4,5,6} &\patternUA  \cr
&={} - Q \Big( \cJ_{3|12,4,5,6} + k_3^m M_{123,4,5,6}^m + \big[ s_{34} M_{1234,5,6} + (4\leftrightarrow 5,6) \big] \Big) \cr
&\quad{} + \cY_{12,3,4,5,6} + P_{1|3|2,4,5,6} + k_3^m C_{1|23,4,5,6}^m + \big[ s_{34} C_{1|234,5,6} + (4\leftrightarrow 5,6) \big] \ ,
}$$
and more involved cases are displayed in appendix \appX, see \canexJ\ to \canexM.

\subsec Canonicalizing refined slots in refined pseudoinvariants

\subseclab\elevenfour
A different canonicalization procedure is needed when the preferred reference label $i$ resides
in a refined slot $A_j$ of a pseudoinvariant $P^{m_1 m_2 \ldots m_r}_{k|A_1,\ldots,A_d|B_1,\ldots
,B_{d+r+3}}$ at $d\neq 0$. In order to gain intuition for suitable BRST generators, consider the following examples:
\eqnn\patternJA
$$\eqalignno{
P_{2|1|3,4,5,6} = &{}- Q {\cal J}_{12|3,4,5,6} + {\cal Y}_{12,3,4,5,6} + P_{1|2|3,4,5,6} &\patternJA \cr
P_{2|13|4,5,6,7} = &{}- Q {\cal J}_{132|4,5,6,7} + {\cal Y}_{132,4,5,6,7} + P_{1|32|4,5,6,7} 
\cr
P_{2|1|34,5,6,7} =&{}- Q( {\cal J}_{12|34,5,6,7} +{\cal J}_{123|4,5,6,7}-{\cal J}_{124|3,5,6,7}) \cr
&{}+  {\cal Y}_{12,34,5,6,7} +{\cal Y}_{123,4,5,6,7}-{\cal Y}_{124,3,5,6,7} \cr
&{} +  P_{1|2|34,5,6,7} +P_{1|23|4,5,6,7}-P_{1|24|3,5,6,7}  \,.
}$$ 
The appearance of anomalous contributions on the right-hand side is not surprising in view of the examples \patternT\
and \patternUA. In the present cases, however, the BRST generators are entirely built from refined building blocks
$\cJ$ at $d\neq 0$. This is a defining property of the $L$ superfields defined in \masterL. Indeed, \patternJA\
is consistently described by
\eqn\patternJH{
P_{k |iA|B,C,D,E} = (\wp_i-Q) (M_{iA} \otimes L_{k | B,C,D,E }) + M_{iA} \otimes \Delta_{k|B,C,D,E} \ .
}
Similar to \patternU\ and \patternUU, the anomalous part of the BRST generator (i.e. the first term
of $QL_{k | B,C,D,E } = \Delta_{k|B,C,D,E} + \ldots$ in \jack) is manually compensated by the last
term in \patternJH, see the arguments in the previous sections \eleventwo\ and \eleventhree.

In the tensorial generalization of \patternJH, the anomalous traces in the expression \JRs\
for $QL^{\ldots }_{k | \ldots } = \delta^{\ldots} \Lambda^{\ldots}_{k|\ldots}+\ldots$ have to be taken into account. This leads to the second line of
\eqnn\patternJHH
$$\eqalignno{
P^{m_1\ldots m_r}_{k |iA|B_1,\ldots,B_{r+4}} &= (\wp_i-Q) (M_{iA} \otimes L^{m_1\ldots m_r}_{k | B_1,\ldots,B_{r+4} }) \cr
& \quad{} + M_{iA} \otimes \Big\{ {r\choose 2} \delta^{(m_1 m_2} \Lambda^{m_3\ldots m_r)}_{k|B_1,\ldots,B_{r+4}}+\Delta^{m_1\ldots m_r}_{k|B_1,\ldots,B_{r+4}} \Big\} \,.
&\patternJHH
}$$
The structure of the following vector and tensor examples resembles the canonicalization of $C^m_{2|1,3,4,5}$ and $C^{mn}_{2|1,3,4,5,6}$ performed in \patternR\ and \patternT, respectively:
\eqnn\patternRJ
\eqnn\patternTJ
$$\eqalignno{
P^m_{2|1|3,4,5,6,7}&= (\wp_{1}-Q)(M_{1} \otimes L^m_{2|3,4,5,6,7}) + M_1 \otimes \Delta^m_{2|3,4,5,6,7} \cr
&=
{}-Q\Big(\cJ^m_{12|3,4,5,6,7}+\big[k_3^m \cJ_{123|4,5,6,7}+(3\leftrightarrow 4,5,6,7) \big] \Big) \cr
& \ \ \ \ + \cY^m_{12,3,4,5,6,7}+\big[k_3^m \cY_{123,4,5,6,7}+(3\leftrightarrow 4,5,6,7) \big] \cr
& \ \ \ \ + P^m_{1|2|3,4,5,6,7}+\big[k_3^m P_{1|23|4,5,6,7}+(3\leftrightarrow 4,5,6,7) \big]
&\patternRJ \cr
P^{mn}_{2|1|3,4,5,6,7,8}&= (\wp_{1}-Q)(M_{1} \otimes L^{mn}_{2|3,4,5,6,7,8}) + M_1 \otimes (  \delta^{mn} \Lambda_{2|3,4,\ldots,8} + \Delta^{mn}_{2|3,4,\ldots,8} )
\cr
&={}-Q \Big(
\cJ^{mn}_{12|3,4,5,6,7,8} + \big[ 2 k_3^{(m} \cJ^{n)}_{123|4,5,6,7,8} + (3\leftrightarrow 4,5,6,7,8) \big] \cr
& \ \ \ \ + \big[ 2 k^{(m}_3 k^{n)}_4 (\cJ_{1234|5,6,7,8}+\cJ_{1243|5,6,7,8}) + (3,4|3,4,5,6,7,8) \big]
 \Big) \cr
 &\ \ \ \ + \delta^{mn} {\cal Y}_{12|3,4,5,6,7,8} + \cY^{mn}_{12,3,4,5,6,7,8} + \big[ 2 k_3^{(m} \cY^{n)}_{123,4,5,6,7,8} + (3\leftrightarrow 4,5,6,7,8) \big] \cr
& \ \ \ \ + \big[ 2 k^{(m}_3 k^{n)}_4 (\cY_{1234,5,6,7,8}+\cY_{1243,5,6,7,8}) + (3,4|3,4,5,6,7,8) \big] \cr
& \ \ \ \ + P^{mn}_{1|2|3,4,5,6,7,8} + \big[ 2 k_3^{(m} P^{n)}_{1|23|4,5,6,7,8} + (3\leftrightarrow 4,5,6,7,8) \big] \cr
& \ \ \ \ + \big[ 2 k^{(m}_3 k^{n)}_4 (P_{1|234|5,6,7,8}+P_{1|243|5,6,7,8}) + (3,4|3,4,5,6,7,8) \big] \ .&\patternTJ
}$$
Finally, the canonicalization prescription \patternJHH\ can be generalized to higher refinement. We follow the usual
logic and manually remove the anomalous contributions from the natural BRST generator, see \allJacDD\ for
$QL^{m_1\ldots m_r}_{k| B_2,\ldots,B_d | C_1,\ldots,C_{d+r+3} }$:
\eqnn\patternJHHH
$$\eqalignno{
&P^{m_1\ldots m_r}_{k |iA,B_2,\ldots,B_d|C_1,\ldots,C_{d+r+3}} = (\wp_i-Q) (M_{iA} \otimes L^{m_1\ldots m_r}_{k| B_2,\ldots,B_d | C_1,\ldots,C_{d+r+3} })\cr
& \ \ + M_{iA} \otimes \Big\{ {r\choose 2} \delta^{(m_1 m_2} \Lambda^{m_3\ldots m_r)}_{k|B_2,\ldots,B_d|C_1,\ldots,C_{d+r+3}}+
\Delta^{m_1\ldots m_r}_{k|B_2,\ldots,B_d|C_1,\ldots,C_{d+r+3}} \cr
& \ \ \ \ \ \ \ \ \ \ \ \ \ \ \ \ \ \ \ + \big[ \Lambda^{m_1\ldots m_r}_{k| B_3,\ldots, B_d| B_2,C_1,\ldots,C_{d+r+3}}  + (B_2\leftrightarrow B_3,\ldots,B_d) \big] \Big\} \ .
&\patternJHHH
}$$
The simplest application occurs at eight-points,
\eqnn\patternJHeight
$$\eqalignno{
P_{2|1,3|4,5,6,7,8} &= (\wp_1-Q)(M_1\otimes L_{2|3|4,5,6,7,8}) + M_1\otimes (\Delta_{2|3|4,5,6,7,8} +
\Lambda_{2|3,4,5,6,7,8})  \cr
&={}- Q \Big( \cJ_{12,3|4,5,6,7,8} + k_3^m \cJ^m_{123|4,5,6,7,8} + \big[ s_{34} \cJ_{1234|5,6,7,8} + (4\leftrightarrow 5,6,7,8) \big] \Big) \cr
& \ \ \ + \cY_{12|3,4,5,6,7,8} + \cY_{3|12,4,5,6,7,8} + k_3^m \cY^m_{123,4,5,6,7,8}\cr
& \ \ \ + \big[ s_{34} \cY_{1234,5,6,7,8} + (4\leftrightarrow 5,6,7,8) \big]  &\patternJHeight \cr
& \ \ \ + P_{1|2,3|4,5,6,7,8} + k_3^m P^m_{1|23|4,5,6,7,8} + \big[ s_{34} P_{1|234|5,6,7,8} + (4\leftrightarrow
5,6,7,8) \big] \,.
}$$
With the most general canonicalization prescriptions \patternUU\ and \patternJHHH, any pseudoinvariant
$P^{m_1\ldots}_{k|\ldots}$ with reference leg $k \neq i$ can be rewritten in terms of $P^{m_1\ldots}_{i|\ldots}$. The
anomalous extra terms built from concatenations of superfields $\Delta$ and $\Lambda$ signal the breakdown of
permutation symmetry in anomalous one-loop amplitudes, see \refs{\wipG, \wipH}.

\newsec Conclusion and outlook

\seclab\sectwelve

\noindent As explained in the Introduction, the result of a multiloop superstring scattering amplitude in the pure
spinor formalism can be written in terms of pure spinor superspace expressions in the cohomology of the BRST charge.
This realization was the guiding principle which led us to consider the general structures for one-loop amplitudes
presented in this work. The claim is that the kinematic factors considered in the previous sections form a convenient
and  complete set of building blocks which make manifest the BRST cohomology properties of
one-loop amplitudes in pure spinor superspace.

Saturation of zero-modes in the pure
spinor prescription implies that the external vertices in the four-point 
amplitude contribute through a term proportional to $d_\a d_\b N^{mn}$ 
\multiloop. The measures defined in \multiloop\ summarize the net effect 
of
zero-mode integrations by the following rule,
\eqn\rule{
d_\a d_\b N^{mn} \rightarrow (\l\g^{[m})_\a (\l\g^{n]})_\b\,.
}
The resulting kinematic factor of the open superstring four-point amplitude \multiloop\ is written as $V_1 T_{2,3,4}$
in the notation of equation \TABC\ and can be checked to be in the BRST cohomology. Its multiparticle generalization
found in \oneloopbb\ incorporates the contributions from OPE singularities through the basic structure $V_A T_{B,C,D}$
which is most elegantly described using the BRST blocks from \eombbs. The reduction of the associated worldsheet
integrals \oneloopbb\ organizes these superfields into BRST-invariants such as the scalar 
\eqn\firststep{
C_{1|23,4,5}= M_{1}M_{23,4,5} + M_{12}M_{3,4,5} - M_{13}M_{2,4,5}
}
in the five-point amplitude. The trial-and-error construction
of scalar BRST-invariants up to multiplicity eight \oneloopbb\ was improved to a systematic and recursive procedure in
\eombbs, see section \sectwo. Since their origin is ultimately related to the one-loop zero-mode pattern \rule\ from
the one-loop amplitude prescription, these BRST invariants encode the manifestly gauge-invariant pieces of the
$N$-point one-loop superstring amplitudes of \oneloopbb.

However, each topology of open superstring amplitudes at genus one is anomalous for $N\geq 6$ external legs, and the
cancellation of the anomaly relies on an interplay between the cylinder and the M\"obius strip \GSanomaly. Therefore the
manifestly gauge-invariant form of the amplitudes in \oneloopbb\ could not be the complete answer. Finding the missing
anomalous terms from their BRST properties was one of the main goals of this paper and led to the concept of
pseudo-cohomology introduced in section \secthree.

As discussed in the main body, a more general class of superfields $\cJ$ extending the prescription \rule\ gives rise
to a recursive procedure to construct anomalous superfields $P^{mnp \ldots}_{i| \ldots}$, called BRST
pseudoinvariants. As a defining property, their BRST variation takes the form $V_A (\l\g^m W_B)(\l\g^n W_C)(\l\g^p 
W_D)(W_E\g_{mnp}W_F)$ which generalizes the hexagon anomaly $\epsilon_{10} F^5$ to superspace and to higher number of
external particles. The bosonic components of several pseudoinvariants can
be downloaded from the website \WWW.

Therefore, BRST cohomology considerations point towards superfields with the correct properties to describe the
anomalous parts of one-loop open superstring amplitudes which were not considered in \oneloopbb. The methods to
generate these pseudoinvariants are natural extensions of the well-tested recursion \eombbs\ for scalar 
BRST-invariants \oneloopbb. In an upcoming work \wipG\ these pseudoinvariants will be assembled into six-point one-loop
amplitudes of the open and closed superstring, the analogous treatment of higher multiplicity is left for the future. 

The field theory limit of superstring amplitudes is composed of scalar and tensorial Feynman integrals. The underlying
degeneration limit of the worldsheet reorganizes the scalar kinematic factors of the superstring such that loop
momenta contract tensorial BRST pseudoinvariants. The recursive construction of this work naturally includes
superfields of arbitrary tensor rank and motivates kinematic companions for loop momenta. Their precise appearance in
one-loop amplitudes of SYM will be detailed in upcoming work \wipH. The matching of worldsheet and momentum space
representations of one-loop amplitudes requires a precise control of the momentum contractions of pseudoinvariants. As
shown in sections \seceight\ to \secten, this problem is addressed by cohomology considerations which will allow to 
identify the difference of the two representations as BRST exact \wipH.

Tensorial pseudoinvariants also play an essential role for closed string amplitudes and capture their contributions
beyond the naive doubling of open string worldsheet correlators. As will be demonstrated in \wipG, the tensorial
kinematic factors in this work provide a compact description of the interactions between left- and right-moving
degrees of freedom. From a field-theory perspective, this points towards a squaring relation between the numerators
of Feynman integrals in SYM and supergravity amplitudes. It would be interesting to realize the BCJ duality between
color and kinematics \BCJ\ through the pseudoinvariants of this work.

After finding the recursive formulas for pseudoinvariants of arbitrary orders, the natural question to ask is how
these pseudoinvariants can be derived from the pure spinor multiloop amplitude prescription\foot{Some terms in the
scalar $P_{i| \ldots}$ can be explained as the leftovers of the partial fraction manipulations described in
\oneloopbb. This applies to both the single-pole contribution from iterated OPEs and to spurious double-pole
singularities which can be removed via integration by parts.} \multiloop. This is a challenge for the future. We
suspect that the solution involves a careful treatment of OPE contractions between the b-ghost and the external
vertices. The combinatorics must be such that spurious OPE singularities combine to local functions on the worldsheet
(which are regular for all values of $z_i-z_j$ and denoted by $f_{ij}$ in \wipH).  This might bypass the subtleties
related to b-ghost singularities pointed out in \WittenBghost. Given that the b-ghost is a source of technical
difficulties in amplitude calculations with the pure spinor superstring, a first principles explanation for the
pseudoinvariants constructed in this paper might shed new light into this difficult corner of the formalism.

\bigskip
\noindent{\bf Acknowledgements:} We thank Marco Chiodaroli and Song He for useful comments on the draft. CRM and OS acknowledge
financial support by the European Research Council Advanced Grant No. 247252 of Michael Green. CRM and OS cordially
thank the Institute of Advanced Study at Princeton for hospitality during advanced stages of this work. OS is grateful
to the Department of Applied Mathematics and Theoretical Physics of the University of Cambridge for hospitality during
various stages of this work. This work was supported in part by National Science Foundation Grant No. PHYS-1066293 and
the hospitality of the Aspen Center for Physics.

\appendix{A}{Examples of BRST pseudoinvariants}

\applab\appA
\noindent
This appendix gathers recursively generated expansions of various (pseudo-)invariants. Their component expansions can
be found at the website \WWW.


At six points, the recursions \veca\ for scalar and vector invariants yield
\eqnn\sixsc
$$\eqalignno{
C_{1|234,5,6} &= M_{1}M_{234,5,6} + M_1  \otimes \big[ C_{2|34,5,6}  - C_{4|23,5,6} \big] \cr
&= M_1 M_{234,5,6} + M_{12}M_{34,5,6} + M_{123}M_{4,5,6} - M_{124}M_{3,5,6}&\sixsc\cr
&\quad{}- M_{14}M_{23,5,6} - M_{142}M_{3,5,6} + M_{143}M_{2,5,6} \cr
C_{1|23,45,6} &= M_1M_{23,45,6} + M_1  \otimes \big[ C_{2|45,3,6} - C_{3|45,2,6}
+  C_{4|23,5,6} - C_{5|23,4,6} \big] \cr
&= M_1 M_{23,45,6} + M_{12}M_{45,3,6} - M_{13}M_{45,2,6} + M_{14}M_{23,5,6} - M_{15}M_{23,4,6}\cr
&\quad{}+M_{124}M_{3,5,6} - M_{134}M_{2,5,6}+ M_{142}M_{3,5,6} - M_{152}M_{3,4,6} 
\cr
&\quad{}-M_{125}M_{3,4,6} + M_{135}M_{2,4,6} - M_{143}M_{2,5,6} + M_{153}M_{2,4,6}  \cr
C^m_{1|23,4,5,6} &= M_1 M^m_{23,4,5,6} + M_1 \otimes \big[ C^m_{2|3,4,5,6} - C^m_{3|2,4,5,6} + \{ k_4^m  C_{4|23,5,6} + (4\leftrightarrow 5,6) \} \big] \cr
&= M_1 M^m_{23,4,5,6} + M_{12} M^m_{3,4,5,6} - M_{13} M^m_{2,4,5,6} + k^m_3 M_{123}M_{4,5,6} - k^m_2 M_{132}M_{4,5,6} 
\cr
&\quad{}+\big[ k^m_4 M_{14}M_{23,5,6} + (M_{124}+M_{142})M_{3,5,6} - (M_{134}+M_{143})M_{2,5,6} + (4\leftrightarrow 5,6)\bigr] \ . 
}$$
The five-point invariants $C_{1|23,4,5}$ and $C^m_{1|2,3,4,5}$ entering \sixsc\ are given by \Cscex.


The recursions \tsb\ and \HRi\ for tensorial pseudoinvariants gives rise to
\eqnn\sevent
\eqnn\seventt
$$\eqalignno{
C^{mn}_{1|23,4,5,6,7} &= M_1 M^{mn}_{23,4,5,6,7} + M_1 \otimes \big[  C^{mn}_{2|3,4,5,6,7} - C^{mn}_{3|2,4,5,6,7} + 2 \{  k_4^{(m}  C^{n)}_{4|23,5,6,7} + (4\leftrightarrow 5,6,7) \} \big]  \cr
&= M_1 M^{mn}_{23,4,5,6,7} + M_{12} M^{mn}_{3,4,5,6,7} - M_{13} M^{mn}_{2,4,5,6,7} + 2(k_3^{(m} M_{123} - k_2^{(m} M_{132}) M^{n)}_{4,5,6,7}   \cr
&+ 2 \big[ k_4^{(m} k_5^{n)} \big\{  (M_{145}+M_{154}) M_{23,6,7} + (M_{1245}+  {\rm symm}(2,4,5)) M_{3,6,7} \cr
& \ \ \ \ -(M_{1345}+ {\rm symm}(3,4,5)) M_{2,6,7}   \big\}  + (4,5|4,5,6,7) \big] \cr
&+ \big[ 2k_4^{(m} \big\{ M_{14} M^{n)}_{23,5,6,7}  + (M_{124}+M_{142}) M_{3,5,6,7}^{n)}\cr
& \ \ \ \ - (M_{134}+M_{143}) M_{2,5,6,7}^{n)} - k_2^{n)} (M_{1432}+M_{1342}+M_{1324}) M_{5,6,7}   \cr 
& \ \ \ \ + k_3^{n)} (M_{1423}+M_{1243}+M_{1234}) M_{5,6,7}\big\} + (4 \leftrightarrow 5,6,7) \big] 
&\sevent \cr
C^{mnp}_{1|2,3,4,5,6,7} &= M_1 M^{mnp}_{2,3,4,5,6,7} +  M_1 \otimes \big[ 3 k_2^{(m} C^{np)}_{2|3,4,5,6,7}  + (2 \leftrightarrow 3,4,5,6,7)  \big] \cr
&= M_1M_{2,3,4,5,6,7}^{mnp} + \big[ 3k_2^{(m} M_{12} M_{3,4,5,6,7}^{np)}  + ( 2 \leftrightarrow 3,4,5,7) \big] \cr
&+  \big[ 6k_2^{(m} k_3^{n} (M_{123}+M_{132}) M_{4,5,6,7}^{p)} + ( 2,3| 2,3,4,5,6,7) \big]  \cr
&+  \big[ 6k_2^{(m} k_3^n k_4^{p)}   (M_{1234}+{\rm symm}(2,3,4)) M_{5,6,7} + ( 2,3,4|2,3,4,5,6,7) \big]  \ ,
&\seventt
}$$
where $C^{mn}_{1|2,3,4,5,6}$ is given by \tsd.


One can extract the following seven-point pseudoinvariants from the recursion in \pseudoq\ where the expansion of $P_{1|2|3,4,5,6}$ is given by \pseudos:
\eqnn\sevenpt
\eqnn\sevenptt
$$\eqalignno{
P_{1|23|4,5,6,7} &= M_1 {\cal J}_{23|4,5,6,7} +  M_1 \otimes\bigl[ P_{2|3|4,5,6,7} -  P_{3|2|4,5,6,7}\bigr]\cr
&= M_1 {\cal J}_{23|4,5,6,7}  +  M_{12} {\cal J}_{3|4,5,6,7} - M_{13} {\cal J}_{2|4,5,6,7} + k_3^m M_{123} M^m_{4,5,6,7}\cr
&\quad{} - k_2^m M_{132} M_{4,5,6,7}^m + \big[ (s_{34} M_{1234}-s_{24}M_{1324}) M_{5,6,7} + (4\leftrightarrow 5,6,7) \big] &\sevenpt \cr
P_{1|2|34,5,6,7} &= M_1 {\cal J}_{2|34,5,6,7} + M_1 \otimes\bigl[ P_{3|2|4,5,6,7} - P_{4|2|3,5,6,7} + k_2^m
C^m_{2|34,5,6,7}\bigr] \cr
&= M_1 {\cal J}_{2|34,5,6,7} + M_{13} {\cal J}_{2|4,5,6,7} - M_{14} {\cal J}_{2|3,5,6,7} \cr
&\quad{}- s_{23} (M_{1243}+M_{1423}) M_{5,6,7} + s_{24} (M_{1234}+M_{1324}) M_{5,6,7} \cr
&\quad{} + k_2^m ( M_{12} M^m_{34,5,6,7}  + (M_{123}+M_{132}) M^m_{4,5,6,7} - (M_{124}+M_{142}) M^m_{3,5,6,7}) \cr
&\quad{} + \big[ s_{25} (M_{125} M_{34,6,7}+(M_{1325}+M_{1235}+M_{1253}) M_{4,6,7} \cr
&\quad{} \ \ \ -(M_{1425}+M_{1245}+M_{1254})M_{3,6,7}) + (5\leftrightarrow 6,7) \big]  \ .&\sevenptt
}$$
The recursion \HRwone\ for vector pseudoinvariants yields the following seven-point example:
\eqnn\vecps
$$\eqalignno{
P^m_{1|2|3,4,5,6,7} &= M_1 {\cal J}^m_{2|3,4,5,6,7} + M_1 \otimes \Big\{ k_{2}^p  C_{2|3,4,5,6,7}^{pm} + \big[  k^m_{3} P_{3|2|4,5,6,7} + (3 \leftrightarrow 4,5,6,7)  \big]  \Big\} \cr
&= M_1 {\cal J}^m_{2|3,4,5,6,7}  + \big[ k_3^m \big\{ M_{13} {\cal J}_{2|4,5,6,7}  + (M_{123}+M_{132}) k_2^p M^p_{4,5,6,7} \big\}+ (3\leftrightarrow 4,5,6,7) \big]  \cr
&\ \ \ + k_2^p M_{12} M^{pm}_{3,4,5,6,7} + \big[ s_{23} \big\{ M_{123} M^m_{4,5,6,7} + k_4^m(M_{1234}+M_{1243}+M_{1423}) M_{5,6,7} \cr
& \ \ \ \ \ \ + k_5^m(M_{1235}+M_{1253}+M_{1523}) M_{4,6,7} + k_6^m(M_{1236}+M_{1263}+M_{1623}) M_{4,5,7}  \cr
& \ \ \ \ \ \  + k_7^m(M_{1237}+M_{1273}+M_{1723}) M_{4,5,6} 
\big\} + (3\leftrightarrow 4,5,6,7) \big]   \ .
&\vecps
}$$

%

The simplest pseudoinvariant of refinement $d>1$ is generated by the recursion \HOs:
\eqnn\HOu
$$\eqalignno{
&P_{1|2,3 | 4,5,6,7,8} = M_1 \cJ_{2,3 | 4,5,6,7,8}  + M_1 \otimes \big[ k_2^m P^m_{2|3|4,5,6,7,8}+  k_3^m P^m_{3|2|4,5,6,7,8} \big]
 \cr
 & \ \ = M_1 \cJ_{2,3 | 4,5,\ldots,8}+ M_{12} k_m^2 \cJ^m_{3|4,5,\ldots,8} + M_{13} k_m^3 \cJ^m_{2|4,5,\ldots,8} + (M_{123}+M_{132}) k_m^2 k_n^3 M^{mn}_{4,5,6,7,8}  \cr
& \ \ \ + \big[ s_{24} M_{124}\cJ_{3|5,6,7,8} +  s_{34} M_{134} \cJ_{2|5,6,7,8}   + s_{34} (M_{1234}+M_{1324}+M_{1342}) k^m_2  M^m_{5,6,7,8} \cr
& \ \ \ \ \ \ + s_{24} (M_{1324}+M_{1234}+M_{1243}) k^m_3  M^m_{5,6,7,8}  + (4 \leftrightarrow 5,6,7,8)  \big]
\cr
& \ \ \ + \big[ s_{24} s_{35} (M_{12435}+M_{13245}+M_{13524}+M_{12345}+M_{12354}+M_{13254}) M_{6,7,8} \cr
& \ \ \ \ \ \  + s_{25} s_{34} (M_{12534}+M_{13254}+M_{13425}+M_{12354}+M_{12345}+M_{13245}) M_{6,7,8} \cr
& \ \ \ \ \ \  + (4,5|4,5,6,7,8) \big]
& \HOu
}$$

\appendix{B}{Gauge transformations versus BRST transformations}

\applab\appB
\noindent
The purpose of this appendix is to clarify the relation between gauge transformations and BRST variations. As
mentioned below \RankOneEOM, the response of the superfields in ten-dimensional SYM to a gauge transformation $\d_i$
in particle $i$ is given by
\eqn\gaugeA{
\d_i A^i_\a = D_\a \omega_i , \ \ \ \d_i A^i_m = k^i_m \omega_i , \ \ \ \d_i W_i^\alpha= \d_i F_i^{mn} = 0 \,,
}
with some scalar superfield $\omega_i$. In the following, we infer the gauge transformation of multiparticle
superfields from \gaugeA\ using their recursive definition presented in \eombbs\ and reviewed in section \twotwo. This in turn
determines the action of $\d_i$ on the complete set of building blocks for one-loop amplitudes as well as their
(pseudo-)invariant combinations. In particular, we will arrive at a dictionary to translate anomalous BRST variations
at ghost-number four to the corresponding anomalous gauge variations at ghost-number three. This is a convenient approach to component expansions of the hexagon gauge anomaly in one-loop amplitudes of
multiplicity $n \geq 6$.

\subsec Gauge variations of multiparticle superfields

The recursive definitions of the rank-two superfields $K_{12} \in\{ A^{12}_\a, A^{12}_m ,W_{12}^\a,F^{12}_{mn} \}$ in \Atwo\ allows
to infer their gauge variation from \gaugeA,
\eqn\gaugeB{
\eqalign{
\d_1 A^{12}_\a & =  D_\alpha \omega_{1|2} + (k_1 \cdot k_2) \omega_1 A_\a^2 \cr
\d_1 A^{12}_m & = k_{12}^m \omega_{1|2} + (k_1 \cdot k_2) \omega_1 A_2^m}\qquad\eqalign{
\d_1 W_{12}^\a &= (k_1 \cdot k_2) \omega_1 W_2^\alpha  \cr
\d_1 F^{12}_{mn} &= (k_1 \cdot k_2) \omega_1 F_2^{mn}.
}}
We have introduced a shorthand for the multiparticle gauge scalar,
\eqn\gaugeC{
\omega_{1|2}\equiv - {1\over 2} \omega_1 (k_1\cdot A_2),
}
to unify the two-particle expressions in $\d_1 A^{12}_\a$ and $\d_1 A^{12}_m$. The two-particle gauge transformations
\gaugeB\ reproduce the single particle pattern \gaugeA\ with $\omega_1 \rightarrow \omega_{1|2}$ and enrich it by
contact terms $\sim (k_1 \cdot k_2)$. This closely mimics the appearance of contact terms in the two-particle
equations of motion \EOMAtwo. 

It is straightforward to work out the multiparticle gauge transformation at $|B|>2$ using the recursion for $K_B$ as described in
section \twotwo. The structure of contact terms in gauge transformations up to multiplicity three is captured by the
following variations for the unintegrated vertex $V_B \equiv \lambda^\alpha A_\alpha^{B}$,
\eqnn\gaugeC
\eqnn\gaugeD
\eqnn\gaugeE
$$\eqalignno{
\delta_1 V_1 &= Q \omega_1 \,, \ \ \ \delta_1 V_{12} = Q \omega_{1|2} + (k_1 \cdot k_2) \omega_1 V_2
&\gaugeC \cr
\delta_1 V_{123} &= Q \omega_{1|23} + (k_1 \cdot k_2)(\omega_1 V_{23} + \omega_{1|3} V_2 ) + (k_{12} \cdot k_3) \omega_{1|2} V_3
&\gaugeD \cr
\delta_1 V_{231} &= Q \omega_{23|1} + (k_2 \cdot k_3)(\omega_{1|3} V_{2} - \omega_{1|2} V_3 ) - (k_1\cdot k_{23}) \omega_{1} V_{23} \ .
&\gaugeE
}$$
This parallels the contact terms in \EOMAthree\ and defines additional multiparticle gauge scalars
\eqnn\gaugeF
$$\eqalignno{
\omega_{1|23} &\equiv - {1\over 2} \omega_{1|2} (k_{12}\cdot A_3) - {1\over 6}\big[\omega_1(k^1\cdot A^{23}) +
\omega_{1|3}(k^3\cdot A^2) - \omega_{1|2}(k^2\cdot A^3)\big] \,, &\gaugeF\cr
\omega_{23|1} &\equiv  {1\over 2} \omega_{1} (k_{1}\cdot A_{23}) - {1\over 6}\big[\omega_1(k^1\cdot A^{23}) +
\omega_{1|3}(k^3\cdot A^2) - \omega_{1|2}(k^2\cdot A^3)\big]\,,
}$$
where the terms proportional to ${1\over6}$ come from the corrections $H_{ijk}$ of \redefsthree. As a consistency check of
\gaugeF\ one can show that $\d_1(V_{123}+V_{231}+V_{312}) = 0$ after using $\d_1V_{312} = {}- \d_1V_{132}$ due to the
rank-two Lie symmetry in the first two labels.

\subsec Gauge variations of Berends--Giele currents $M_A$

As detailed in section \twothree, a convenient basis of multiparticle fields $K_B$ is furnished by Berends--Giele
currents ${\cal K}_B$, represented by calligraphic letters. In a cubic graph interpretation of multiparticle fields
$K_B$ shown in \figtwo, Berends--Giele currents ${\cal K}_B$ assemble the diagrams of a color-ordered SYM tree
including $|B|-1$ propagators. The dictionary up to multiplicity four is given in \BGtwothree\ and \BGfour.

For the Berends--Giele current $M_B = \lambda^\alpha {\cal A}^B_\alpha$ associated with the unintegrated vertex $V_B$, \gaugeC\ to \gaugeE\ translate into
\eqnn\gaugeV
\eqnn\gaugeW
$$\eqalignno{
\d_1 M_1 &= Q \Omega_{1} \,, \ \ \ \d_1 M_{12} = Q \Omega_{1|2}  + \Omega_1 M_2
&\gaugeV \cr
\d_1 M_{123} &= Q \Omega_{1|23}+ \Omega_{1|2} M_3 + \Omega_1 M_{23}
&\gaugeW
}$$
with Berends--Giele gauge scalars
\eqn\gaugeX{
\Omega_{1} \equiv \omega_{1} \,, \ \ \ \Omega_{1|2}  \equiv { \omega_{1|2} \over s_{12} }  \,, \ \ \ \Omega_{1|23}  \equiv
 { \omega_{1|23} \over  s_{12} s_{123} } - { \omega_{23|1} \over s_{23} s_{123} } \ .
}
With suitable multiparticle generalizations $\Omega_{1|23\ldots p}$ of \gaugeX, one can directly write down a closed
formula for the gauge transformations of $M_B$,
\eqn\gaugeX{
\delta_1 M_{12\ldots p} = \sum_{j=1}^{p-1} \Omega_{1|23\ldots j} M_{j+1 \ldots p} + Q \Omega_{1|23\ldots p} \ .
}
With this form of $\delta_1 M_{12\ldots p}$, the superspace representation $\sum_{j=1}^{n-2} M_{12\ldots j}
M_{j+1\ldots n-1}M_n$ of the SYM tree amplitude \nptTree\ can be easily checked to be gauge invariant up to BRST
exact terms,
\eqn\gaugetree{
\delta_1 \Big( \sum_{j=1}^{n-2} M_{12\ldots j} M_{j+1\ldots n-1}M_n \Big) = Q\Big( \sum_{j=1}^{n-2} \Omega_{1|2\ldots j} M_{j+1\ldots n-1}M_n \Big) \ .
}

\subsec Gauge variations of ghost number two building blocks

Similarly to $\delta_1 M_{12\ldots p}$ in \gaugeX, the Berends--Giele currents ${\cal A}_B^m,\cW_B^\alpha$ and ${\cal F}^{mn}_B$ give rise to gauge transformations
\eqnn\gaugeS
\eqnn\gaugeT
\eqnn\gaugeU
$$\eqalignno{
\d_{1} {\cal A}_{12\ldots p}^m &= k_{12\ldots p}^m \Omega_{1|23\ldots p} + \sum_{j=1}^{p-1} \Omega_{1|23\ldots j} {\cal A}_{ j+1\ldots p}^m
&\gaugeS \cr
\d_{1} {\cal W}_{12\ldots p}^\alpha &=  \sum_{j=1}^{p-1} \Omega_{1|23\ldots j} {\cal W}_{ j+1\ldots p}^\alpha
&\gaugeT \cr
\d_{1} {\cal F}_{12\ldots p}^{mn} &=  \sum_{j=1}^{p-1} \Omega_{1|23\ldots j} {\cal F}_{ j+1\ldots p}^{mn}
&\gaugeU
}$$
which resemble their BRST variations \QBGs. With \gaugeS\ to \gaugeU, one can straightforwardly compute the gauge
variations of all the building blocks $M^{m_1\ldots m_r}_{B_{1},\ldots,B_{r+3}}$ and $\cJ^{m_1\ldots
m_r}_{A_1,\ldots,A_d|B_{1},\ldots,B_{d+r+3}}$ introduced in \Tmrecurs\ and \JacobiDefs, respectively. The simplest
examples $M_{A,B,C}$ and $M^m_{A,B,C,D}$ are defined in \BRSTMi\ and \BRSTm, and their gauge variation
\eqnn\gaugeQ
\eqnn\gaugeR
$$\eqalignno{
\d_{1} M_{12\ldots p,B,C} &=   \sum_{j=1}^{p-1} \Omega_{1|23\ldots j} M_{ j+1\ldots p,B,C}
&\gaugeQ \cr
\d_{1} M^m_{12\ldots p,B,C,D}&=  k_{12\ldots p}^m \Omega_{1|23\ldots p} M_{B,C,D}+\sum_{j=1}^{p-1} \Omega_{1|23\ldots j} M^m_{ j+1\ldots p,B,C,D}
&\gaugeR
}$$
resembles the contributions from a single slot to the BRST variations \QMABC\ and \BRSTQm. The variation of tensors or refined building blocks
\eqnn\gaugeN
\eqnn\gaugeO
\eqnn\gaugeP
$$\eqalignno{
\d_{1} M^{mn}_{12\ldots p,B,C,D,E} &=  2 k_{12\ldots p}^{(m} \Omega_{1|23\ldots p} M^{n)}_{B,C,D,E}+ \sum_{j=1}^{p-1} \Omega_{1|23\ldots j} M^{mn}_{ j+1\ldots p,B,C,D,E}
&\gaugeN \cr
\d_{1} \cJ_{12\ldots p|B,C,D,E}&=  k_{12\ldots p}^m \Omega_{1|23\ldots p} M^m_{B,C,D,E} + \big[ \Omega_{S[12\ldots p ,B]} M_{C,D,E} + (B\leftrightarrow C,D,E) \big] \cr
& \ \ \ \ +\sum_{j=1}^{p-1} \Omega_{1|23\ldots j} {\cal J}_{ j+1\ldots p|B,C,D,E}
&\gaugeO \cr
\d_{1} \cJ_{B|12\ldots p,C,D,E}&=  - \Omega_{S[12\ldots p,B]} M_{C,D,E} +\sum_{j=1}^{p-1} \Omega_{1|23\ldots j} \cJ_{B| j+1\ldots p,C,D,E}
&\gaugeP
}$$
does not reproduce the anomalous terms $\cY_{A,B,C,D,E}$ present in the BRST variations \BRSTt\ and \pseudoo. The
$S[A,B]$ map is defined in \QEone\ and $\Omega_{S[12\ldots p ,B]}$ is understood to be arranged in the form
$\Omega_{1|\ldots}$. 

As general rule, $\delta_i\cJ^{m_1\ldots m_r}_{A_1,\ldots,A_d|B_{1},\ldots,B_{d+r+3}}$ can be reconstructed from those
terms in $Q\cJ^{m_1\ldots m_r}_{A_1,\ldots,A_d|B_{1},\ldots,B_{d+r+3}}$ given in \HRJCgen\ where particle $i$ appears
in a current $M_{iC}$. The gauge variation follows by replacing $M_{iC}\rightarrow \Omega_{i|C}$ and discarding any
other term in the BRST variation. The same prescription applies to anomaly building blocks $\cY^{m_1\ldots
m_r}_{A_1,\ldots,A_d|B_{1},\ldots,B_{d+r+5}}$ which are recursively defined by \refWc, see \WanonBG\ for the
simplest scalar $\cY_{A,B,C,D,E}$ and \qanomB\ for the general BRST transformation.

\subsec Gauge variations of pseudoinvariants

The above gauge variations of $M_B$ and $\cJ^{m_1\ldots m_r}_{A_1,\ldots,A_d|B_{1},\ldots,B_{d+r+3}}$ are sufficient to study the anomalous gauge
transformations of pseudoinvariants. This in turn allows to probe the hexagon anomaly in field theory and string
theory, see \refs{\wipH,\wipG} for details. All the pseudoinvariants $P^{m_1\ldots}_{i|\ldots}$ constructed by the recursion
\pinvgen\ are combinations of $M_{iA} {\cal J}$ where ${\cal J}$ represents any ghost number two building block
$\cJ^{m_1\ldots m_r}_{A_1,\ldots,A_d|B_{1},\ldots,B_{d+r+3}}$, possibly adjoined by momenta. This makes the gauge variation in the reference leg
$i$ particularly convenient to study: By \gaugeX, we have
\eqn\gaugeM{
\delta_i M_{iA} {\cal J} = \sum_{j=0}^{|A|-1} \Omega_{i|a_1 a_2 \ldots a_j} M_{a_{j+1} \ldots a_{|A|}} {\cal J} - \Omega_{i|A} Q{\cal J} + Q (\Omega_{i|A} {\cal J}) \ .
}
Up to the last BRST-exact term, this is closely related to the BRST transformation
\eqn\gaugeL{
Q M_{iA} {\cal J} = \sum_{j=0}^{|A|-1} M_{ia_1 a_2 \ldots a_j} M_{a_{j+1} \ldots a_{|A|}} {\cal J} - M_{iA} Q{\cal J} 
}
upon interchanging $\Omega_{i|B} \leftrightarrow M_{iB}$ for any $|B|=0,1,\ldots,|A|$, i.e.
\eqn\gaugeK{
\delta_i M_{iA} {\cal J} = (Q M_{iA} {\cal J}) \big|_{ M_{iB}\rightarrow \Omega_{i|B} } + Q(\Omega_{i|A} {\cal J})
 \ .
}
This can be applied term by term to the pseudoinvariants $P^{m_1\ldots}_{i|\ldots}$ obtained from \pinvgen,
\eqnn\gaugeJ
$$\eqalignno{
\delta_i P^{m_1\ldots m_r}_{i|A_1,\ldots,A_d|B_{1},\ldots,B_{d+r+3}} &= (QP^{m_1\ldots m_r}_{i|A_1,\ldots,A_d|B_{1},\ldots,B_{d+r+3}} ) \big|_{ M_{iC}\rightarrow \Omega_{i|C} } \cr
& \ \ \ \ \ \ + Q\Big(P^{m_1\ldots m_r}_{i|A_1,\ldots,A_d|B_{1},\ldots,B_{d+r+3}} \big|_{ M_{iC}\rightarrow \Omega_{i|C} } \Big) \,, &\gaugeJ
}$$
where the BRST transformations are given by \HRijk. For example, the anomalous $Q$ variations of the simplest pseudoinvariants at six and seven points
\eqnn\gaugeIA
\eqnn\gaugeIB
\eqnn\gaugeIC
$$\eqalignno{
Q P_{1|2|3,4,5,6} &= {}- M_1 \cY_{2,3,4,5,6} &\gaugeIA \cr
QP_{1|23|4,5,6,7} &= {}- M_1 \cY_{23,4,5,6,7} - M_{12} \cY_{3,4,5,6,7}+  M_{13} \cY_{2,4,5,6,7} &\gaugeIB \cr
QP^m_{1|2|3,4,5,6,7} &={}- M_1 \cY^m_{2,3,4,5,6,7} -  \big[ k_2^m M_{12} \cY_{3,4,5,6,7} + (2\leftrightarrow3,4,5,6,7) \big] &\gaugeIC
}$$
translate into gauge variations
\eqnn\gaugeHA
\eqnn\gaugeHB
\eqnn\gaugeHC
$$\eqalignno{
\delta_1 P_{1|2|3,4,5,6} &= {}- \Omega_1 \cY_{2,3,4,5,6}  +Q(\ldots) &\gaugeHA \cr
\delta_1 P_{1|23|4,5,6,7} &= {}- \Omega_1 \cY_{23,4,5,6,7} - \Omega_{1|2} \cY_{3,4,5,6,7}+  \Omega_{1|3} \cY_{2,4,5,6,7}  +Q(\ldots) &\gaugeHB \cr
\delta_1 P^m_{1|2|3,4,5,6,7} &={}- \Omega_1 \cY^m_{2,3,4,5,6,7} -  \big[ k_2^m \Omega_{1|2} \cY_{3,4,5,6,7} + (2\leftrightarrow3,\ldots,7) \big]  +Q(\ldots)  \ .&\gaugeHC
}$$
Gauge variations $\delta_k P_{i|\ldots}^{m_1\ldots}$ beyond the reference leg $i$ can still be obtained by trading any $M_{kB}$ in the BRST variation $Q P_{i|\ldots}^{m_1\ldots}$ for $\Omega_{k|B}$, e.g.
\eqnn\gaugeXXB
$$\eqalignno{
\delta_2 P_{1|2|3,4,5,6} &= Q(\ldots)
 \cr
\delta_2 P_{1|23|4,5,6,7} &= {}\Omega_{2|1} \cY_{3,4,5,6,7} +Q(\ldots) &\gaugeXXB \cr
\delta_2 P^m_{1|2|3,4,5,6,7} &={}\Omega_{2|1} k_2^m {\cal Y}_{3,4,5,6,7}  +Q(\ldots) \ .
}$$
These expressions of ghost number three can be evaluated in components using the methods in \PSS.
As shown in \anomaly, the bosonic components of \gaugeHA\ yield a Levi-Civita contraction of five gluon field
strengths, i.e. $\sim \epsilon_{m_1 n_1\ldots m_5 n_5} k_2^{m_1} e_2^{n_1} \ldots k_6^{m_5} e_6^{n_5} $. We will argue in the next section that any anomalous superfield has parity odd bosonic components.

\subsec Parity odd nature of multiparticle anomaly tensors

The component evaluation of $\langle (\l\g^m W_2)(\l\g^n W_3)(\l\g^p W_4)(W_5 \g_{mnp}W_6) \rangle$ using the
prescription $\langle \lambda^3 \theta^5 \rangle =1$ \psf\ is particularly simple for five external bosons: The
lowest bosonic component in the superfields occurs at order $W_i^\alpha \rightarrow -{1\over 4} (\gamma_{mn}
\theta)^{\alpha} f_i^{mn}$ (with $f_i^{mn}=2k_i^{[m} e_i^{n]}$ in terms of the gluon polarization vector $e_i^n$), so
the contribution from five factors of $W_i$ to the order $\theta^5$ is unique. The single-particle instance ${\cal
Y}_{2,3,4,5,6}$ of the anomaly superfields in \WanonBG\ therefore reduces to the correlator
\eqn\podd{
\langle (\l\g^m \gamma^{a_1 b_1} \theta)(\l\g^n \gamma^{a_2 b_2} \theta)(\l\g^p \gamma^{a_3 b_3} \theta)(\theta \gamma^{a_4 b_4}  \g_{mnp}\gamma^{a_5 b_5} \theta) \rangle = {1\over 45} \epsilon^{a_1 b_1 a_2 b_2\ldots a_5 b_5} \ ,
}
which has been evaluated in \anomaly\ and shown to flip sign under spacetime parity. 

It turns out that the same correlator \podd\ governs the bosonic components of a generic ${\cal Y}_{A,B,C,D,E}$ built from multiparticle superfields $W^\alpha_A$. 
This follows from the two-form nature of the lowest bosonic component in the $\t$-expansion of the BRST blocks,
\eqn\lowWF{
W^{\alpha}_A = {}-{1\over 4} (\gamma_{mn} \theta)^{\alpha} f_A^{mn} + {\cal O}(\theta^3) \ , \ \ \ F^{mn}_A =  f_A^{mn} + {\cal O}(\theta^2)  \ ,
}
where the two-particle instance of the bosonic field strength $f^{mn}_A$ is given by,
\eqn\ftwo{
{1\over 2}f_{12}^{mn} \equiv  k_{12}^{[m}e_2^{n]} (e_1\cdot k_2) - k_{12}^{[m}e_1^{n]} (e_2\cdot k_1) - k_1^{[m} k_2^{n]} (e_1\cdot e_2) - e_1^{[m} e_2^{n]} (k_1\cdot k_2)  \ .
}
The appearance of $f^{mn}_A$ in both superfields in \lowWF\ is an inevitable consequence of the multiparticle equation
of motion for $D_{\alpha} W^{\beta}_A$, see \EOMAtwo\ and \EOMAthree\ for $|A|=2,3$ and \QBGs\ for the
Berends--Giele version at general multiplicity. The contact terms $\sim A_{\alpha}^B W^{\beta}_{C}$
in the multiparticle equations of motion do not contribute at zero'th order in $\theta$ since both factors are fermionic with lowest gluon
contributions at order $\theta^1$.

The correlator \podd\ and the leading $\theta$ behavior in \lowWF\ for the bosonic part of $W^{\alpha}_A$ imply the gluon component
\eqn\anomcomp{
\langle {\cal Y}_{A,B,C,D,E} \rangle = {1\over 45} \left( -{1\over 4} \right)^5 \epsilon_{a_1 b_1 a_2 b_2\ldots a_5 b_5} f^{a_1 b_1}_A f^{a_2 b_2}_B\ldots f^{a_5 b_5}_E 
}
for the scalar and unrefined anomaly superfield ${\cal Y}_{A,B,C,D,E}$. Its generalization to higher rank or
refinement simply adjoins superspace factors of ${\cal A}_B^m$, see \HRf\ and \refWc. The latter can only contribute
through their $\theta=0$ component since ${\cal Y}_{A,B,C,D,E}$ has a minimum contribution of five thetas for external
bosons. The same is true for the $\Omega_{i|C}$ superfields due to gauge transformations in particle $i$. Hence, the
gluon components of an anomalous gauge transformation $\langle \Omega_{i|C} \cY^{m_1\ldots m_r}_{A_1,\ldots,A_d|
B_1,\ldots,B_{d+r+5}} \rangle$ are proportional to the $\epsilon_{10}$ tensor generated by the correlator \podd.

\appendix{C}{BRST variations of miscellaneous superfields}
\applab\appC

\noindent In this appendix we display explicit BRST variations of various superfields that were omitted from the main
text.

\subsec BRST variations before the Berends--Giele map

Even though we emphasized the simpler BRST transformations of the Berends--Giele version of the various building blocks in the main body
of this work, it is still convenient to know the explicit $Q$ variations of those building blocks prior to the
application of the Berends--Giele map in \BGtwothree\ and \BGfour.

The precursor of the Berends--Giele recursion \Tmrecurs\ for $M^{m_1 \ldots m_r}_{B_1, \ldots, B_{r+3}}$ is based on the expression \TABC\ for $T_{A,B,C}$ as well as
\eqn\Wmndef{
W^m_{A,B,C,D} \equiv {1\over 12}(\l\g_n W_A)(\l\g_p W_B)(W_C\g^{mnp}W_D) + (A,B|A,B,C,D)\,.
}
For the higher rank generalizations
\eqnn\WWmndef
$$\eqalignno{
W^{m_1 \ldots m_{r-1}|m_r}_{B_1,B_2, \ldots,B_{r+3}} &\equiv A_{B_1}^{m_1} W^{m_2 \ldots m_{r-1}|m_r}_{B_2, \ldots, B_{r+3}} + (B_1 \leftrightarrow B_2, \ldots, B_{r+3})
&\WWmndef
\cr
T^{m_1\ldots m_r}_{B_1,B_2,\ldots,B_{r+3}} &\equiv A^{m_1}_{B_1} T^{m_2 \ldots m_r}_{B_2, \ldots, B_{r+3}} +
A^{m_r}_{B_1} W^{m_{r-1} \ldots m_2|m_1}_{B_2, \ldots, B_{r+3}} + (B_1\leftrightarrow B_2,B_3, \ldots,B_{r+3})\,, 
}$$
one can show that
\eqnn\QTs
$$\eqalignno{
QT^m_{1,2,3,4} & = k^m_1 V_1 T_{2,3,4} + (1 \leftrightarrow 2,3,4) &\QTs\cr
QT^m_{12,3,4,5} & =  \big[ k_{12}^m V_{12}T_{3,4,5} + (12\leftrightarrow 3,4,5)\big] + (k^1\cdot k^2)\big(V_1 T^m_{2,3,4,5} - V_2 T^m_{1,3,4,5}\big)\cr
QT^m_{123,4,5,6} & = \big[ k_{123}^m V_{123}T_{4,5,6} + (123\leftrightarrow 4,5,6)\big] \cr
&\quad{} + (k^1\cdot k^2)\big[ V_1 T^m_{23,4,5,6} + V_{13}T^m_{2,4,5,6} - (1\leftrightarrow 2)\big]\cr
&\quad{} + (k^{12}\cdot k^3)\big[ V_{12}T^m_{3,4,5,6} - (12\leftrightarrow 3)\big]\cr
QT^{mn}_{1,2,3,4,5} & = \big[ 2k^{(m}_{1}V_1 T^{n)}_{2,3,4,5} + (1\leftrightarrow 2,3,4,5)\big] + \d^{mn}Y_{1,2,3,4,5}\cr
QT^{mn}_{12,3,4,5,6} & = \big[ 2k^{(m}_{12}V_{12} T^{n)}_{3,4,5,6} + (12\leftrightarrow 3,4,5,6)\big] + \d^{mn}Y_{12,3,4,5,6}\cr
&\quad{}+ (k^1\cdot k^2) \big[ V_1 T^{mn}_{2,3,4,5,6} - (1\leftrightarrow 2)\big]\cr
QT^{mnp}_{1,2,3,4,5,6} & = 3\d^{(mn}Y^{p)}_{1,2,3,4,5,6} + \big[ 3 V_1 k_1^{(m}T^{np)}_{2,3,4,5,6} + (1\leftrightarrow
2,3,4,5,6)\big]\,. \cr
}$$
Similarly, the BRST variations of refined currents \Jnohatdef\ can be computed to be
 \eqnn\pseude
 $$\eqalignno{
 QJ_{1|23,45,6,7} &=  k_{1}^m V_{1} T^m_{23,45,6,7} + \big[ V_{[1,23]} T_{45,6,7} + (23\leftrightarrow 45,6,7)\big]  + Y_{1,23,45,6,7}  &\pseude \cr
 &\quad{} + (k^2\cdot k^3)\big[V_2 J_{1|3,45,6,7}- (2\leftrightarrow 3)\big] + (k^4\cdot k^5)\big[ V_4 J_{1|23,5,6,7}-
 (4\leftrightarrow 5)\big]  \cr
 QJ_{12|34,5,6,7} &= k_{12}^m V_{12} T^m_{34,5,6,7}+ \big[V_{[12,34]}T_{5,6,7} + (34\leftrightarrow 5,6,7)\big] + Y_{12,34,5,6,7}  \cr
 &\quad{} + (k^1\cdot k^2)\big[ V_1 J_{2|34,5,6,7}- (1\leftrightarrow 2)\big]
 + (k^3\cdot k^4)\big[ V_3 J_{12|4,5,6,7}- (3\leftrightarrow 4)\big] \cr
 QJ_{123|4,5,6,7} &=  k_{123}^m V_{123} T^m_{4,5,6,7} +\big[ V_{[123,4]} T_{5,6,7} + (4 \leftrightarrow 5,6,7) \big] + Y_{123,4,5,6,7} \cr
 & \quad{} + (k^1\cdot k^2) \big[V_1 J_{23|4,5,6,7}+ V_{13} J_{2|4,5,6,7} - (1\leftrightarrow 2)\big] \cr
 & \quad{} + (k^{12}\cdot k^3) \big[V_{12} J_{3|4,5,6,7}- (12\leftrightarrow 3)\big] \cr
 QJ_{1|234,5,6,7} &= k_1^m V_1 T^m_{234,5,6,7} + \big[V_{[1,234]} T_{5,6,7} + (234 \leftrightarrow 5,6,7) \big] + Y_{1,234,5,6,7}\cr
 & \quad{} + (k^2\cdot k^3)\big[V_2 J_{1|34,5,6,7} + V_{24} J_{1|3,5,6,7} - (2\leftrightarrow 3)\big] \cr
 & \quad{} + (k^{23}\cdot k^4) \big[ V_{23} J_{1|4,5,6,7}- (23\leftrightarrow 4)\big] \ .
 }$$

\subsec BRST variations after the Berends--Giele map

After applying the Berends--Giele map to the refined currents from \pseude, their BRST variations become
 \eqnn\pseudk
 $$\eqalignno{
 Q{\cal J}_{1|23,45,6,7} &= k_{1}^m M_{1} M^m_{23,45,6,7} &\pseudk \cr
 &\quad{} +(s_{12}M_{123}-s_{13}M_{132}) M_{45,6,7} + (s_{14}M_{145}-s_{15} M_{154}) M_{23,6,7}\cr
 & \quad{} +s_{16} M_{16} M_{23,45,7} + s_{17} M_{17} M_{23,45,6}
 + {\cal Y}_{1,23,45,6,7} \cr
 & \quad{} + M_2 {\cal J}_{1|3,45,6,7} - M_3 {\cal J}_{1|2,45,6,7} + M_4 {\cal J}_{1|23,5,6,7} - M_5 {\cal J}_{1|23,4,6,7}  \cr
 Q{\cal J}_{12|34,5,6,7} &= k_{12}^m M_{12} M^m_{34,5,6,7}\cr
 &\quad{} +
 (s_{23}M_{1234}-s_{13} M_{2134} - s_{24} M_{1243} + s_{14} M_{2143})M_{5,6,7} \cr
 & \quad{} + \big[ (s_{25}M_{125}-s_{15}M_{215}) M_{34,6,7} + (5 \leftrightarrow 6,7) \big] + {\cal Y}_{12,34,5,6,7} \cr
 & \quad{} + M_1 {\cal J}_{2|34,5,6,7} - M_2 {\cal J}_{1|34,5,6,7}+ M_3 {\cal J}_{12|4,5,6,7} - M_4 {\cal J}_{12|3,5,6,7} \cr
 Q{\cal J}_{123|4,5,6,7} &= k_{123}^m M_{123} M^m_{4,5,6,7}\cr
 &\quad{} +\big[ (s_{34}M_{1234}-s_{24}(M_{1324}+M_{3124}) + s_{14} M_{3214}) M_{5,6,7} + (4 \leftrightarrow 5,6,7) \big] \cr
 & \quad{} + {\cal Y}_{123,4,5,6,7} + M_{12} {\cal J}_{3|4,5,6,7} - M_3 {\cal J}_{12|4,5,6,7} + M_1 {\cal J}_{23|4,5,6,7} - M_{23} {\cal J}_{1|4,5,6,7} \cr
 Q{\cal J}_{1|234,5,6,7} &= k_{1}^m M_{1} M^m_{234,5,6,7}\cr
 &\quad{}-(s_{12} M_{4321} - s_{13} (M_{2431}+M_{4231}) + s_{14} M_{2341}) M_{5,6,7} \cr
 & \quad{} +\big[ s_{15} M_{15} M_{234,6,7} + (5 \leftrightarrow 6,7) \big] + {\cal Y}_{1,234,5,6,7}  \cr
 & \quad{} + M_{23} {\cal J}_{1|4,5,6,7} - M_4 {\cal J}_{1|23,5,6,7} 
 + M_{2} {\cal J}_{1|34,5,6,7} - M_{34} {\cal J}_{1|2,5,6,7} \ .
 }$$
The following relations are useful to derive \pseudk\ from \pseude:
 \eqnn\pseudnn
 \eqnn\pseudnnn
 \eqnn\pseudnnnn
 \eqnn\pseudnnnnn
 $$\eqalignno{
 T_{12} &=s_{12} M_{12} , \ \ \ \ \ \ T_{123} = s_{12}(s_{23} M_{123} - s_{13} M_{213}) &\pseudnn \cr
 T_{1234}-T_{1243} &= s_{12}s_{34}(s_{23} M_{1234}-s_{13}M_{2134} - s_{24} M_{1243} +s_{14} M_{2143}) &\pseudnnn \cr
 T_{1234} &= s_{12}\big[s_{23}s_{24}(M_{1234}+M_{1243}) - s_{13} s_{14}(M_{2134}+M_{2143})+ s_{23} s_{34}M_{1234} \cr
 & \ \ - s_{13} s_{34} M_{2134} + s_{14} s_{23} M_{3214} - s_{13} s_{24} M_{3124}\big] \ . &\pseudnnnn
 }$$

\appendix{D}{The $H$ superfields in the redefinition of refined currents}

\applab\appD 
\noindent 
This appendix provides a general definition for the superfields $H_{[A,B]}$ and ${\cal H}_{[A,B]}$ relevant for the redefinition of refined currents ${\cal J}^{m_1\ldots
m_r}_{A_1,\ldots,A_d|B_1,\ldots,B_{d+r+3}}$.

\subsec The $H_{[A,B]}$ tensors from BRST blocks $V_C$

As mentioned in section \twotwo\ and detailed in \eombbs, the recursive construction of BRST blocks $K_B \in \{
A_{\alpha}^B,A^m_B, W^\alpha_B, F_B^{mn} \}$ requires redefinitions by BRST trivial quantities to maintain the Lie
symmetries. We define $\widehat V_{[A,B]}$ through a generalization of the recursion in \Arankp\ to situations where
both $|A|\neq1$ and $|B|\neq1$:
\eqn\ArankAB{
\widehat V_{[A,B]} \equiv {}- \half\bigl[ V_A (k_A \cdot A_B) + A^{A}_m (\lambda \g^m W_B) - (A \leftrightarrow B)\bigr]
}
There are two obstructions to express $\widehat V_{[A,B]}$ as a linear combination of BRST blocks $V_C$ at multiplicity $|C|=|A|+|B|$:
\item{(i)} Generic contributions to $Q\widehat V_{[A,B]}$ have the form $s_{ij} V_C \widehat V_{[D,E]}$. They can be corrected to
$s_{ij} V_C V_{[D,E]}$ by subtracting combinations of $s_{ij} V_C H_{[D,E]}$ for some scalar superfields $H_{[D,E]}$ to be defined in the next step.
\item{(ii)} After the above subtraction, the modified $\widehat V_{[A,B]}$ must still be shifted by a BRST exact quantity $Q H_{[A,B]}$ before
it can be expressed in a basis of BRST blocks $V_C $. If $B=b_1$, i.e. $|B|=1$, this amounts to enforcing the Lie
symmetries at multiplicity $|A|+1$ by adding $QH_{[A,b_1]}$. The latter was denoted by $H_{[A,b_1]}\equiv
H_{a_1a_2\ldots a_{|A|}b_1}$ in \eombbs.

\noindent Let us illustrate the recursive nature of these points through the simplest examples at multiplicity $|A|+|B|\leq 5$. Cases with $|B|=1$ have been discussed in \eombbs,
 \eqnn\ArankAC
 \eqnn\ArankAD
 \eqnn\ArankAE
 $$\eqalignno{
\widehat V_{[12,3]} &=V_{[12,3]} + QH_{[12,3]}  &\ArankAC \cr
\widehat V_{[123,4]} &=V_{[123,4]} + (k_{12} \cdot k_3) H_{[12,4]} V_3 + (k_1\cdot k_2) (H_{[13,4]} V_2-H_{[23,4]} V_1) + QH_{[123,4]} &\ArankAD \cr
 \widehat V_{[1234,5]} &=V_{[1234,5]} + (k_{123} \cdot k_4) H_{[123,5]}V_4
 + (k_{12} \cdot k_3) (H_{[124,5]} V_3+H_{[12,5]} V_{34}-H_{[34,5]} V_{12}  ) \cr
 &+ (k_1\cdot k_2) (H_{[134,5]} V_{2} + H_{[13,5]} V_{24} + H_{[14,5]} V_{23} - H_{[24,5]} V_{13} - H_{[23,5]} V_{14} - H_{[234,5]} V_{1})
\cr
& + QH_{[1234,5]} \ , &\ArankAE
 }$$
where the corrections by $QH_{[12\ldots p-1,p]}$ are explained in (ii) and the remaining terms are due to step (i), see
\eombbs\ for a closed formula. Requiring $V_{[12\ldots p-1,p]}=V_{12\ldots p}$ to satisfy the Lie symmetries at rank
$p$ turns \ArankAC\ to \ArankAE\ into a recursive procedure to determine $H_{[12\ldots p-1,p]}=H_{12\ldots p}$ and
$V_{12\ldots p}$ \eombbs.

Cases with $|B|\neq 1$ introduce new classes of corrections $H_{[A,B]}$:
 \eqnn\ArankAF
 \eqnn\ArankAG
 $$\eqalignno{
\widehat V_{[12,34]} &=V_{[12,34]} + (k_{1} \cdot k_2) (H_{[34,2]} V_1-H_{[34,1]} V_2) \cr
&+ (k_3\cdot k_4) (H_{[12,3]} V_4 - H_{[12,4]} V_3 ) + QH_{[12,34]} &\ArankAF \cr
 \widehat V_{[123,45]} &=V_{[123,45]} + (k_{12} \cdot k_3) (H_{[12,45]}V_3 - H_{[3,45]}V_{12}) \cr
 &+ (k_1\cdot k_2) ( H_{[13,45]}V_{2} +H_{[1,45]}V_{23}  -  H_{[2,45]}V_{13} -H_{[23,45]}V_{1}  ) \cr
 &+   (k_4\cdot k_5) (H_{[123,4]} V_{5} - H_{[123,5]} V_{4} ) + QH_{[123,45]} \ , &\ArankAG
 }$$
The bracket notation $V_{[A,B]}$ on the right-hand side represents linear combinations of BRST blocks such as
$V_{[12,34]} = V_{1234}-V_{1243}$ or $V_{[123,45]} =V_{12345} -V_{12354}$, see appendix A of \eombbs\ for more
details. They follow by identifying $V_{[A,B]}$ with
the cubic diagram depicted in \figseven\ and expanding the latter in terms of a multiperipheral basis in \figtwo. The
required diagram manipulations are depicted in \figtriplet\ and can be though of as the kinematic dual of the Jacobi
identity $f^{e[ab} f^{c]de}=0$ among color tensors along the lines of \BCJ. The $S[A,B]$ map defined in \QEone\ will efficiently
address the conversion of $V_{[A,B]} \rightarrow V_C$ once the participating superfields are transformed into a basis
of Berends--Giele currents.

\ifig\figtriplet{Triplet of subdiagrams whose color representatives sum to zero by virtue of the Jacobi identity $f^{e[ab} f^{c]de}=0$.
The expansion of the above $V_{[A,B]}$ in a basis of BRST blocks $V_C$ can be understood using the same vanishing statement for triplets
of diagrams. The latter allows to expand the diagrammatic representative for $V_{[A,B]}$ shown in \figseven\ in terms of multiperipheral
trees depicted in \figtwo\ and described by $V_C$.}
{\epsfxsize=0.70\hsize\epsfbox{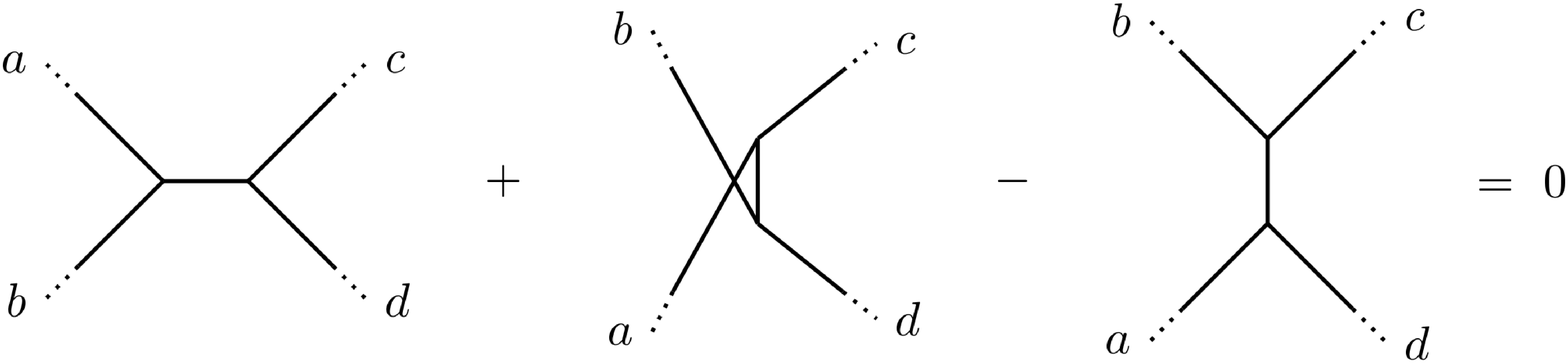}}

With this understanding of the $V_{[A,B]}$ on the right-hand side of \ArankAC\ to \ArankAG, the redefining tensors
$QH_{[A,B]}$ with $|B|\neq 1$ can be obtained recursively. In order to bypass the inconvenience of ``inverting'' the
BRST charge, we next present a setup to determine the $H_{[A,B]}$ directly.

\subsec The $H_{[A,B]}$ tensors from BRST blocks $A^m_C$

Also the BRST block $\widehat A^m_{12\ldots p}$ in its hatted version before $H_{[B,C]}$ modifications is defined
recursively in \eombbs. The expression given for $\widehat A^m_{12\ldots p}\equiv \widehat A^m_{[12\ldots p-1,p]}$ can
be straightforwardly generalized to $\widehat A^m_{[B,C]}$ with $|C|\neq1$:
\eqn\ArankAH{
\widehat A^m_{[B,C]} \equiv \half\bigl[ A_B^p F_C^{pm} + A_C^m (k_C\cdot A_B) - (B \leftrightarrow C)\bigr] + (W_B \gamma^m W_C) \ .
}
As shown in \eombbs\ for $|C|=1$, the redefinitions of $\widehat A^m_{[B,C]}$ and $\widehat V_{[B,C]}$ are mapped into each other
by exchanging the relevant BRST blocks $V_D \leftrightarrow A_D^m$ and trading $Q \leftrightarrow k_{BC}^m = k_B^m+k_C^m$. Up
to multiplicity four, this converts \ArankAC, \ArankAD\ and \ArankAF\ into
 \eqnn\ArankAI
 \eqnn\ArankAJ
 \eqnn\ArankAK
 $$\eqalignno{
\widehat A^m_{[12,3]} &=A^m_{[12,3]} + k^m_{123} H_{[12,3]}  &\ArankAI \cr
\widehat A^m_{[123,4]} &=A^m_{[123,4]} + (k_{12} \cdot k_3) H_{[12,4]} A^m_3 \cr
&+ (k_1\cdot k_2) (H_{[13,4]} A^m_2-H_{[23,4]} A^m_1) + k_{1234}^m H_{[123,4]} &\ArankAJ
\cr
\widehat A^m_{[12,34]} &=A^m_{[12,34]} + (k_{1} \cdot k_2) (H_{[34,2]} A^m_1-H_{[34,1]} A^m_2) \cr
&+ (k_3\cdot k_4) (H_{[12,3]} A^m_4 - H_{[12,4]} A^m_3 ) + k^m_{1234}H_{[12,34]} \ . &\ArankAK
}$$
As emphasized in \eombbs, knowledge of $k^m_{BC} H_{[B,C]}$ is a more convenient starting point to solve for the scalar $H_{[B,C]}$ as compared to $Q H_{[B,C]}$.

\subsec The ${\cal H}_{[A,B]}$ tensors from Berends--Giele currents ${\cal K}_C$

A convenient basis of multiparticle SYM fields $K_C$ to construct BRST (pseudo-)invariants is furnished by the Berends--Giele currents, see
section \twothree. The contact terms in the above formulae turn out to simplify once we transform the
superfields involved according to
 \eqnn\ArankAM
 $$\eqalignno{
{\cal H}_{[12,3]}&= {H_{[12,3]} \over s_{12}} \ , \ \ \  {\cal H}_{[12,34]}= {H_{[12,34]} \over s_{12}s_{34}} \ , \ \ \ {\cal H}_{[123,4]}= {H_{[123,4]} \over s_{12}s_{123}} +  {H_{[321,4]} \over s_{23}s_{123}}  \cr
{\cal H}_{[123,456]}&= {1\over s_{123} s_{456}} \Big( {H_{[123,456]} \over s_{12} s_{45}} + {H_{[321,456]} \over s_{23} s_{45}} + {H_{[123,654]} \over s_{12} s_{56}} + {H_{[321,654]} \over s_{23} s_{56}} \Big)  \ . &\ArankAM
}$$
This amounts to applying the map in \BGtwothree\ and \BGfour\ separately to $B$ and $C$ in $H_{[B,C]}$. The resulting ${\cal H}_{[B,C]}$ are the natural superfields to describe the redefinitions of
 \eqnn\ArankAO
 \eqnn\ArankAP
 $$\eqalignno{
\widehat {\cal V}_{[B,C]} &\equiv - \half\bigl[ M_B (k_B \cdot \cA_C) + \cA^{B}_m (\lambda \g^m \cW_C) - (B \leftrightarrow C)\bigr] &\ArankAO \cr
\widehat \cA^m_{[B,C]} &\equiv \half\bigl[ \cA_B^p {\cal F}_C^{pm} + \cA_C^m (k_C\cdot \cA_B) - (B \leftrightarrow C)\bigr] + (\cW_B \gamma^m \cW_C)   \ . &\ArankAP
}$$
In order to obtain the Berends--Giele images of the BRST blocks $V_D$ and $A_D^m$ with Lie symmetries, we have to modify $\widehat {\cal V}_{[A,B]}$ and $\widehat \cA^m_{[B,C]}$ via
 \eqnn\ArankAQ
 \eqnn\ArankAR
 $$\eqalignno{
\widehat {\cal V}_{[B,C]} &\equiv M_{S[B,C]}  + \sum_{XY=B} ( {\cal H}_{[X,C]} M_Y - {\cal H}_{[Y,C]} M_X) \cr
&+ \sum_{XY=C} ( {\cal H}_{[B,X]} M_Y - {\cal H}_{[B,Y]} M_X) + Q {\cal H}_{[B,C]}  &\ArankAQ \cr
\widehat {\cal A}^m_{[B,C]} &\equiv {\cal A}^m_{S[B,C]}  + \sum_{XY=B} ( {\cal H}_{[X,C]} \cA^m_Y - {\cal H}_{[Y,C]} \cA^m_X) \cr
&+ \sum_{XY=C} ( {\cal H}_{[B,X]} \cA^m_Y - {\cal H}_{[B,Y]} \cA^m_X) + k^m_{BC} {\cal H}_{[B,C]} \ . &\ArankAR
}$$
Note that ${\cal H}_{[B,C]} =0$ whenever $|B|=|C|=1$. Comparison of \ArankAQ\ and \ArankAR\ with the above examples
(say \ArankAG\ or \ArankAK) reveals two benefits of the basis of Berends--Giele currents: Firstly, the pattern of BRST
blocks on the right-hand side without an accompanying factor of ${\cal H}_{[B,C]}$ can be described by the $S[B,C]$
map defined in \QEone. Secondly, the contact terms in \ArankAG\ or \ArankAK\ are converted to simple deconcatenations.
Since $M_{S[B,C]}$ and ${\cal A}^m_{S[B,C]}$ are known in terms of BRST blocks $V_D$ and $A_D^m$ of multiplicity
$|D|=|B|+|C|$ \eombbs, one can view \ArankAR\ as a constructive definition of ${\cal H}_{[B,C]}$.

\appendix{E}{On BRST exact relations among pseudoinvariants}

\applab\appE

\subsec BRST generator of $C^m_{1|A,B,C,D}$

According to the discussion in section \nineone\ and in particular \JRe, the traceless components of $C^{m_1\ldots
m_r}_{i|A_1,\ldots,A_{r+3}}$ are BRST exact. However, it is difficult to extract the
BRST generators, so we will explicitly carry out the analysis for vectors $C^m_{i|A,B,C,D}$.

When contracting \JRb\ with momenta $k^m_r k^n_r$ of any particle $r=i$ or $r \in A,B,C,D$, the on-shell condition $k_r^2$ decouples
the anomalous term $\sim \delta^{mn}$, and we obtain the BRST generator for the corresponding momentum contraction,
\eqnn\quasiq
$$\eqalignno{
Q \left[ { k^m_r k^n_r D^{mn}_{i|A,B,C,D}    \over 2 (k_r \cdot k_{iABCD})}  \right]  &= k_{r}^m C_{i|A,B,C,D}^m \ . &\quasiq
}$$
Plugging this back into the $k_i^n$ contraction of \JRb\ with its trace subtracted:
\eqnn\quasiqq
$$\eqalignno{
&C^m_{i|A,B,C,D} = Q   { 1 \over (k_i \cdot k_{iABCD}) } \Big[ 
k_i^p D^{pm}_{i|A,B,C,D} - {1\over 10} k_i^m \delta_{np}D^{np}_{i|A,B,C,D} \cr
& \ \ -  
{ k_{iABCD}^m k_i^n k_i^p D^{np}_{i|A,B,C,D}  \over 2(k_i \cdot k_{iABCD}) } + {1 \over 10} k_i^m \! \! \! \sum_{r\in i,A,B,C,D} \! \! { k_r^n k_r^p D^{np}_{i|A,B,C,D}  \over (k_r \cdot k_{iABCD})}
 \Big]  \ . &\quasiqq
}$$
Similar to \quasill, the right-hand side is ill-defined if momentum conservation $k^m_{iABCD}=0$ is imposed, so the
vector invariant $C^m_{i|A,B,C,D}$ is {\it not} BRST-exact in the momentum phase space of $1+|A|+|B|+|C|+|D|$ massless
particles. The BRST generator for traceless tensors of rank $r \geq 2$ can be found by the same method.

\subsec Seven-point momentum contractions of $C^m_{1|A,B,C,D}$

The general formula \JRac\ for $QD_{i|A|B,C,D}$ specializes to the following BRST exact relations at seven-points:
\eqnn\JRt
\eqnn\JRu
\eqnn\JRv
\eqnn\JRw
$$\eqalignno{
Q D_{1|234|5,6,7}&  =\Delta_{1|234,5,6,7} + k_{234}^m C_{1|234,5,6,7}^m - P_{1|2|34,5,6,7} - P_{1|23|4,5,6,7} \cr
+ &P_{1|34|2,5,6,7}  + P_{1|4|23,5,6,7}+ \big[- s_{25} C_{1| 5234,6,7} - s_{45}C_{1| 5432,6,7}   \cr
& \ \ + s_{35} (C_{1|5324,6,7} + C_{1|5342,6,7}) + (5 \leftrightarrow 6,7) \big]&\JRt
\cr
Q D_{1|5|234,6,7} & =\Delta_{1|234,5,6,7} + k_{5}^m C_{1|234,5,6,7}^m +s_{56} C_{1|234,56,7}+s_{57} C_{1|234,57,6} \cr
+ &\big[s_{25} C_{1| 5234,6,7}  - s_{35} (C_{1|5324,6,7} + C_{1|5342,6,7})+ s_{45}C_{1| 5432,6,7} \big]&\JRu
\cr
Q D_{1|23|45,6,7} & =\Delta_{1|23,45,6,7} + k_{23}^m C_{1|23,45,6,7}^m - P_{1|2|3,45,6,7} + P_{1|3|2,45,6,7} \cr
+ & \big[ s_{25} C_{1| 3254,6,7} -s_{24} C_{1| 3245,6,7} -s_{35} C_{1| 2354,6,7} +s_{34} C_{1| 2345,6,7}  \big]  \cr
+&\big[ s_{36} C_{1|236,45,7} -s_{26} C_{1|326,45,7} + (6 \leftrightarrow 7) \big]&\JRv
\cr
Q D_{1|6|23,45,7} & =\Delta_{1|23,45,6,7} + k_{6}^m C_{1|23,45,6,7}^m + (s_{26} C_{1|326,45,7}-s_{36} C_{1|236,45,7}) \cr
+ & (s_{46} C_{1|546,23,7}-s_{56} C_{1|456,23,7}) + s_{67} C_{1|23,45,67} \ .
&\JRw
}$$

\appendix{F}{Examples of the canonicalization procedure}

\applab\appX
\noindent This appendix gathers further applications of the canonicalization procedure in section~\seceleven. We suppress the
BRST generators since they can be reconstructed from the right-hand side and do not contribute to amplitudes.
%

The canonicalization prescription \patternQ\ for scalar invariants implies that 
\eqnn\canexA
\eqnn\canexB
\eqnn\canexC
\eqnn\canexD
\eqnn\canexE
\eqnn\canexF
$$\eqalignno{
 C_{2| {1}, {3 4}, {5 6}} &= C_{1|{2}, {3 4}, {5 6}} + C_{1| {2 3}, {5 6}, {4}}  - 
 C_{1|{2 4}, {5 6}, {3}} + C_{1| {2 5}, {3 4}, {6}}  - 
 C_{1|{2 6}, {3 4}, {5}}   \cr
 & + C_{1| {2 35}, {6}, {4}}  - 
 C_{1| {2 36}, {5}, {4}}- C_{1|{245}, {6}, {3}}  + 
 C_{1| {2 46}, {5}, {3}}  +C_{1|{2 5 3}, {4}, {6}} \cr
 & - 
 C_{1| {2 5 4}, {3}, {6}}  - C_{1| {26 3}, {4}, {5}}  + 
 C_{1| {2 6 4}, {3}, {5}} + Q(\ldots)
&\canexA 
\cr
 C_{2|{1 3 4 5}, {6}, {7}}&=
 C_{1|{3 4 5 2}, {6}, {7}} + Q(\ldots)
 &\canexB
\cr
C_{2| {1}, {3 4 5 6}, {7}} &= C_{1| {2}, {3 4 5 6}, {7}} + C_{1| {2 3}, {4 5 6}, {7}} -   C_{1| {26}, {3 4 5}, {7}} +  C_{1|{23 4}, {5 6}, {7}} - 
  C_{1| {23 6}, {4 5}, {7}} -  C_{1| {2 6 3}, {45}, {7}} \cr
  &+ 
  C_{1| {2 6 5}, {3 4}, {7}} +  C_{1|{2 3 4 5}, {6}, {7}} - 
  C_{1| {2 3 4 6}, {5}, {7}} -  C_{1|{2 3 6 4}, {5}, {7}} + 
  C_{1| {2 3 6 5}, {4}, {7}} -  C_{1| {2 6 3 4}, {5}, {7}} \cr
  &+ 
  C_{1| {2 6 3 5}, {4}, {7}} +  C_{1| {2 6 5 3}, {4}, {7}} - 
  C_{1|{2 6 5 4}, {3}, {7}}+ Q(\ldots)
&\canexC
\cr
C_{2| {1 34}, {5 6}, {7}}&=
 C_{1| {3 4 2}, {5 6}, {7}} +  C_{1| {342 5}, {6}, {7}} - 
  C_{1| {3 4 2 6}, {5}, {7}}+ Q(\ldots)
&\canexD
\cr
 C_{2| {1 3}, {45 6}, {7}} &= C_{1| {3 2}, {4 5 6}, {7}} +  C_{1| {3 2 4}, {5 6}, {7}} - 
  C_{1|{3 2 6}, {4 5}, {7}} +  C_{1| {3 2 4 5}, {6}, {7}}\cr
  & - 
 C_{1| {3 2 4 6}, {5}, {7}} -  C_{1| {3 2 6 4}, {5}, {7}} + 
 C_{1| {3 2 6 5}, {4}, {7}}+ Q(\ldots)
&\canexE
\cr
 C_{2| {1 3}, {4 5}, {6 7}} &=
 C_{1| {3 2}, {4 5}, {6 7}} + C_{1| {3 2 4}, {6 7}, {5}} - 
 C_{1| {3 2 5}, {6 7}, {4}} +  C_{1| {3 2 6}, {4 5}, {7}} - 
  C_{1| {3 2 7}, {4 5}, {6}} \cr
  &+  C_{1| {3 2 4 6}, {7}, {5}} - 
  C_{1| {3 2 4 7}, {6}, {5}} -  C_{1| {3 2 5 6}, {7}, {4}} + 
  C_{1| {3 2 5 7}, {6}, {4}} +  C_{1| {3 2 6 4}, {5}, {7}} \cr
  &- 
 C_{1| {3 2 6 5}, {4}, {7}} -  C_{1| {3 2 7 4}, {5}, {6}} + 
  C_{1| {3 2 7 5}, {4}, {6}}+ Q(\ldots) \ .
&\canexF
}$$
Except for the more laborious $C_{2|345,67,1}$, \canexA\ to \canexF\ and the opening examples of section~\elevenone\ cover all canonicalizations of scalars $C_{2|A,B,C}$ up to multiplicity seven.


Next, we apply the canonicalization rule \patternU\ to vectors and tensors:
 \eqnn\canexG
$$\eqalignno{
C^m_{2| {1 3}, {4}, {5},6}&= C^m_{1|32,4,5,6} + \big[ k_4^m C_{1|324,5,6} +(4\leftrightarrow 5,6) \big]+ Q(\ldots)
&\canexG
\cr
 C^m_{2| {1 }, {34}, {5},6} &=  C^m_{1|2,34,5,6}+C^m_{1|23,4,5,6}-C^m_{1|24,3,5,6}+k_4^m C_{1|234,5,6}-k_3^m C_{1|243,5,6} + Q(\ldots) \cr
 &+ \big[ k_5^m (C_{1| {2 5}, {3 4}, {6}}+C_{1| {235}, {4}, {6}}  + C_{1| {253}, {4}, {6}}   - 
C_{1|{2 4 5}, {3}, {6}}- C_{1| {254}, {3}, {6}})   +(5\leftrightarrow 6) \big]
\cr
C^m_{2| {134}, {5}, {6}, {7}}&= C^m_{1|342,5,6,7} + \big[ k_5^m C_{1|3425,6,7} + (5\leftrightarrow 6,7) \big]
+ Q(\ldots)
\cr
C^m_{2| {13}, {4 5}, {6}, {7}}&= C^m_{1|32,45,6,7} +C^m_{1|324,5,6,7} -C^m_{1|325,4,6,7} + k_5^m C_{1|3245,6,7} - k_4^m C_{1|3254,6,7} + Q(\ldots) \cr
&+ \big[ k_6^m (C_{1|326,45,7}+C_{1|3264,5,7}+C_{1|3246,5,7}-C_{1|3265,4,7}-C_{1|3256,4,7}) + (6\leftrightarrow 7) \big]
\cr
 C^{mn}_{2| {1 3}, 4,5,6,7} &= C^{mn}_{1|32,4,5,6,7} + \delta^{mn} {\cal Y}_{132,4,5,6,7} + 2 \big[ k_4^{(m} C^{n)}_{1|324,5,6,7} + (4\leftrightarrow 5,6,7) \big] \cr
 &+ 2 \big[ k_4^{(m} k_5^{n)} (C_{1|3245,6,7} + C_{1|3254,6,7}) + (4,5|4,5,6,7) \big] 
+ Q(\ldots)
\cr
C^{mn}_{2|1,34,5,6,7} &= C^{mn}_{1| {2}, {3 4}, {5}, {6}, {7}} + 
 C^{mn}_{1| {23}, {4}, {5}, {6}, {7}} - 
 C^{mn}_{1| {2 4}, {3}, {5}, {6}, {7}} + \delta^{mn}(  \cY_{{12}, {34}, {5}, {6}, {7}} + 
 \cY_{{1 2 3}, {4}, {5}, {6}, {7}}- \cY_{{124}, {3}, {5}, {6}, {7}}) \cr
 &+ 2 \big[ k_5^{(m} (C^{n)}_{1|{25}, {34}, {6}, {7}} + 
C^{n)}_{1|{235}, {4}, {6}, {7}} + 
 C^{n)}_{1| {25 3}, {4}, {6}, {7}} - 
 C^{n)}_{1| {24 5}, {3}, {6}, {7}} - 
 C^{n)}_{1|{2 5 4}, {3}, {6}, {7}}) + (5\leftrightarrow 6,7) \big] \cr
 & +2 k_4^{(m} C^{n)}_{ 1| {234}, {5}, {6}, {7}}-2 k_3^{(m} C^{n)}_{ 1| {243}, {5}, {6}, {7}}  + 2 \big[ k_5^{(m} \big\{ k_4^{n)} (C_{1|{2345}, {6}, {7}} + C_{1|{2354}, {6}, {7}} + C_{1| {2534}, {6}, {7}}) \cr
 & \ \ - k_3^{n)} (C_{1|{2435}, {6}, {7}} + C_{1|{2453}, {6}, {7}} + C_{1| {2543}, {6}, {7}})  + (5\leftrightarrow 6,7) \big] \cr
 &+ 2 \big[ k_5^{(m} k_{6}^{n)} \big\{ C_{1|256,34,7}+C_{1|265,34,7} + (C_{1|2356,4,7} + {\rm symm}(3,5,6) ) \cr
 & \ \ -(C_{1|2456,3,7} + {\rm symm}(3,5,6) )
 \big\} + (5,6|5,6,7) \big] + Q(\ldots) \ .
}$$
The more laborious vectors $C^m_{2|1,345,6,7}$ and $C^m_{2|1,34,56,7}$ at multiplicity seven are omitted.
%

Finally, the following pseudoinvariants are canonicalized using \patternUU:
 \eqnn\canexJ
\eqnn\canexK
\eqnn\canexL
\eqnn\canexM
$$\eqalignno{
P_{2| {3}| {1 4}, {5}, {6}, {7}}&= P_{1|3|42,5,6,7} + {\cal Y}_{142,3,5,6,7} + k_3^m C^m_{1|423,5,6,7} \cr
&+ \big[ s_{35} C_{1|4235,6,7} + (5\leftrightarrow 6,7) \big]
+ Q(\ldots)
&\canexJ
\cr
P_{2| {3 4}| {1}, {5}, {6}, {7}} &= P_{1|34|2,5,6,7}+P_{1|4|23,5,6,7}-P_{1|3|24,5,6,7}+\cY_{12,34,5,6,7}+\cY_{123,4,5,6,7} \cr
&-\cY_{124,3,5,6,7}+ \big[ s_{45} C_{1|2345,6,7} - s_{35} C_{1|2435,6,7} + (5\leftrightarrow 6,7) \big]
\cr
&+k_4^m C^m_{1|234,5,6,7}-k_3^m C^m_{1|243,5,6,7} + Q(\ldots)
&\canexK
\cr
P_{2|3|1,45,6,7} &= P_{1|3|2,45,6,7} + P_{1|3|24,5,6,7} - P_{1|3|25,4,6,7} +\cY_{12,3,45,6,7} + \cY_{124,3,5,6,7} \cr
& - \cY_{125,3,4,6,7}+ \big[ s_{35} ( C_{1|2345,6,7} + C_{1|2435,6,7} ) - (4\leftrightarrow 5) \big] \cr
& + k_3^m ( C^m_{1| {23}, {4 5}, {6}, {7}} + 
C^m_{1| {2 3 4}, {5}, {6}, {7}} - 
 C^m_{1| {23 5}, {4}, {6}, {7}} + 
 C^m_{1|  {2 4 3}, {5}, {6}, {7}} - 
C^m_{1| {25 3}, {4}, {6}, {7}}) \cr
&+ \big[ s_{36} (C_{1| {236}, {4 5}, {7}}  + 
C_{1|{2 3 4 6}, {5}, {7}}+ 
C_{1|{2 3 6 4}, {5}, {7}}  + 
C_{1| {2 4 3 6}, {5}, {7}} \cr
& \ \ \  - 
C_{1|{2 3 5 6}, {4}, {7}} - 
C_{1| {2 3 6 5}, {4}, {7}} - 
C_{1|{2 5 3 6}, {4}, {7}}) +(6\leftrightarrow 7) \big]
+ Q(\ldots)
&\canexL
\cr
P^m_{2|3|1,4,5,6,7} &= P^m_{1|3|2,4,5,6,7} + {\cal Y}^m_{12,3,4,5,6,7} + \big[k_4^m (P_{1|3|24,5,6,7} + \cY_{124,3,5,6,7} ) + (4\leftrightarrow5,6,7) \big] \cr
&+ \big[ s_{34} C^m_{1|234,5,6,7} + k_4^m k_3^p (C^p_{1|234,5,6,7}+C^p_{1|243,5,6,7}) + (4\leftrightarrow 5,6,7) \big] \cr
&+ k_3^p C^{pm}_{1|23,4,5,6,7} + k_3^m \cY_{123,4,5,6,7}  + \big[ s_{34}\big\{ k_5^m(C_{1|2345,6,7}+C_{1|2354,6,7}+C_{1|2534,6,7}) \cr
& \ \ + k_6^m(C_{1|2346,5,7}+C_{1|2364,5,7}+C_{1|2634,5,7}) \cr
& \ \  + k_7^m(C_{1|2347,5,6}+C_{1|2374,5,6}+C_{1|2734,5,6}) \big\} + (4\leftrightarrow 5,6,7) \big] + Q(\ldots) \ .
&\canexM
}$$

\listrefs

\bye